# Ultrafast Dynamics of Strongly Correlated Fermions — Nonequilibrium Green Functions and Selfenergy Approximations

## N. Schlünzen, S. Hermanns, M. Scharnke, and M. Bonitz

Institut für Theoretische Physik und Astrophysik, Christian-Albrechts-Universität zu Kiel, D-24098 Kiel, Germany

**Abstract.** This article presents an overview on recent progress in the theory of nonequilibrium Green functions (NEGF). We discuss applications of NEGF simulations to describe the femtosecond dynamics of various finite fermionic systems following an excitation out of equilibrium. This includes the expansion dynamics of ultracold atoms in optical lattices following a confinement quench and the excitation of strongly correlated electrons in a solid by the impact of a charged particle. NEGF, presently, are the only *ab-initio* quantum approach that is able to study the dynamics of correlations for long times in two and three dimensions. However, until recently, NEGF simulations have mostly been performed with rather simple selfenergy approximations such as the second-order Born approximation (SOA). While they correctly capture the qualitative trends of the relaxation towards equilibrium, the reliability and accuracy of these NEGF simulations has remained open, for a long time.

Here we report on recent tests of NEGF simulations for finite lattice systems against exact-diagonalization and density-matrix-renormalization-group benchmark data. The results confirm the high accuracy and predictive capability of NEGF simulations—provided selfenergies are used that go beyond the SOA and adequately include strong correlation and dynamical-screening effects. With an extended arsenal of selfenergies that can be used effectively, the NEGF approach has the potential of becoming a powerful simulation tool with broad areas of new applications including strongly correlated solids and ultracold atoms. The present review aims at making such applications possible. To this end we present a selfcontained introduction to the theory of NEGF and give an overview on recent numerical applications to compute the ultrafast relaxation dynamics of correlated fermions. In the second part we give a detailed introduction to selfenergies beyond the SOA. Important examples are the third-order approximation, the $GW$ approximation, the $T$-matrix approximation and the fluctuating-exchange approximation. We give a comprehensive summary of the explicit selfenergy expressions for a variety of systems of practical relevance, starting from the most general expressions and the Feynman diagrams, and including also the important cases of diagonal basis sets, the Hubbard model and the differences occuring for bosons and fermions. With these details, and information on the computational effort and scaling with the basis size and propagation duration, an easy use of these approximations in numerical applications is made possible.

## Contents

## 1. Introduction

Strong correlation effects, arising when the interaction energy of a many-particle system exceeds the single-particle energy, are ubiquitous in nature and laboratory systems. Examples are the interior of dwarf stars or giant planets, the quark–gluon plasma, e.g. Refs. [1, 2] or electrons in strongly correlated materials, e.g. Ref. [3]. In classical systems strong correlations exist e.g. in electrolytes [4], in ultracold plasmas [5, 6], or in complex plasmas where they lead to fluid-like or crystalline behavior of charge particles, for an overview see Ref. [7]. Even though there exist many similarities in the static and dynamic properties between classical and quantum systems [8], the latter have a number of peculiarities and require dedicated theoretical approaches. Therefore, in the present article, we will concentrate only on quantum systems.

### 1.1. Strong correlations in quantum systems

In recent years strong correlations in quantum systems have come into the focus in a variety of fields. The first example are dense plasmas as they exist in the interior of giant planets, dwarf stars or neutron stars. Similar conditions are also generated in the laboratory by compression of matter by means of shock waves, ion beams or high-intensity lasers [9, 10]. This typically leads to situations where the electrons are quantum degenerate whereas the heavy particles exhibit only weak quantum behavior. This peculiar state of highly excited nonideal matter has been termed "warm dense matter" or high-energy density matter, e.g. Ref. [11]. The range of electron densities where correlation effects are important is characterized by values of the Brueckner parameter exceeding unity, i.e. $r_s = \bar{r}/a_B > 1$, where $\bar{r}$ denotes the mean interparticle distance and $a_B$ the Bohr radius. In warm dense matter in thermodynamic equilibrium, temperatures are in the range of $0.1 \lesssim \Theta = \frac{T}{T_F} \lesssim 10$ (with the Fermi temperature $T_F$) which means that electrons are highly excited and ground-state approaches fail. Here, the method of choice are first-principle approaches such as path-integral Monte Carlo simulations [12–14], for a recent overview, see Ref. [15].

The second example are condensed-matter systems where strong electronic correlations are of high importance in many materials, e.g. Refs. [3, 16]. Examples are transition metals and their oxids, rare-earth metals or cuprate superconductors. Here the standard mean-field description fails and correlated approaches such as dynamical mean-field theory [16] or Hubbard-type model Hamiltonians, e.g. [17] are being used.

The third example of strong correlation effects are ultracold fermionic and bosonic atoms. In particular ultracold atoms in optical lattices have allowed one to study correlation effects experimentally with unprecedented accuracy, e.g. Ref. [18]. Moreover, with the advent of atomic microscopes even single-site spatial resolution has been achieved [19–21].

*1.2. Nonequilibrium correlation dynamics following rapid external excitation.*

There is a large a variety of excitation scenarios that drive a many-body system rapidly out of equilibrium. This includes excitation by laser pulses—from the infrared, over the optical and ultraviolet to the x-ray range. Time-resolved optical diagnostics (pump–probe spectroscopy) has evolved as a powerful experimental tool to probe the time evolution of atoms, molecules and materials that has been covered in many textbooks. Another method that provides spatially localized excitations is the impact of charged particles that may lead to surface modifications, heating or excitations of the electronic degrees of freedom, e.g. Refs. [22–24]. For correlated atoms in optical lattices, additional excitation schemes have been developed. This includes rapid changes of confinement potentials (confinement quench) [25, 26], rapid changes of the pair interaction (interaction quench) via Feshbach resonance [27] or periodic modulation of the lattice depth (lattice-modulation spectroscopy), e.g. Refs. [28–30].

All these methods have seen a rapid development in recent years and allow for accurate diagnostic of the time evolution of many-body systems. This, on the other hand, requires extensive theory developments in order to achieve detailed comparisons with and explanation of experimental observations.

## 1.3. Theoretical approaches to computing nonequilibrium dynamics in correlated quantum systems.

The theoretical approaches that have been applied most extensively in the field of correlated lattice systems are exact diagonalization (CI) [31–33], density-matrix renormalization group (DMRG) methods [34–36], diagrammatic Monte Carlo [37–39], real-time quantum Monte Carlo (RTQMC) [40, 41], reduced-density-matrix approaches [42–44], and time-dependent density-functional theory (TDDFT) [45–49]. However, each of these methods has fundamental problems and limitations. CI faces an exponential increase of the CPU time with the system size and applies only for small systems. RTQMC can only treat short evolution times due to the dynamic fermion sign problem. DMRG is accurate at strong coupling but has difficulties at moderate and weak coupling and is, moreover, restricted to 1D systems, e.g. Refs. [50]. Finally, TDDFT has no dimensional restrictions, but it is not able to accurately treat electronic correlations in a systematic way. Besides, the simulations usually involve the adiabatic approximation which neglects memory effects and may make the results unreliable. Presently, there are intense activities underway to improve each of these approaches.

## 1.4. Nonequilibrium Green Functions (NEGF)

There exists an independent approach to the dynamics of correlated systems that originates in quantum-field theory. It is based on nonequilbirium Green functions (NEGF) that were introduced by Keldysh [51] and Baym and Kadanoff [52]. This approach has been extremely successful and extensively applied in many fields of physics, including semiconductor optics [53–56], semiconductor quantum transport [57–60], nuclear physics [61–63], laser plasmas [64, 65], high-energy physics [66–68], and small atoms and molecules [69–71]. For text-book discussions, see Refs. [53, 54, 72–74].

NEGF have only recently been applied to finite correlated lattice systems out of equilibrium [49, 73, 75, 76]. This method is not suffering from most of the limitations of the other approaches and has achieved remarkable results. Benchmarks against CI simulations for small systems, cold atom experiments [26] and DMRG data [50] have shown impressive accuracy of the approach for many observables, for details see Sec. 3. Of course there is a price to pay: NEGF methods are complicated and computationally very expensive. A recent overview on the NEGF results for the dynamics of fermionic lattice systems can be found in Ref. [77], and a recent overview on different NEGF applications is given in Ref. [78].

At this point it is useful to have a look at the conceptual basis of nonequilbrium Green functions. This approach is internally consistent. It obeys conservation laws and the dynamics are time-reversible [79]. NEGF simulations depend on a single input quantity—the selfenergy $\Sigma$ (this is analogous to DFT which depends only on the accuracy of the exchange–correlation potential). Would $\Sigma$ be known exactly, the NEGF method would be exact. In practice, of course, aside from a few model cases, the exact $\Sigma$ is not known and one has to resort to approximations. In the majority of applications to closed correlated many-body systems (neglecting the coupling to phonons or other bosonic degrees of freedom) just two approximations are used: the Hartree–Fock selfenergy and the second-order Born approximation that incorporates correlations to lowest order. These approximations are well studied and their numerical application can be considered routine.

At the same time, the excellent quantitative agreement with benchmark data that was mentioned above could only be achieved by applying more complex selfenergy approximations that adequateley take into account both, the coupling strength and the filling (density) of the system. However, even though a number of improved approximations such as the $T$-matrix selfenergy, that describes strong coupling and bound-state formation, or the $GW$ approximation, that describes dynamical screening, are known for half a century, their application is often still very challenging. Unfortunately, in most publications the presentation of these approximations is rather sketchy, and often does not include all details about the spin degrees of freedom or general basis representations. Moreover, there is a high need for additional approximations, for instance, selfenergies that couple dynamical screening and strong coupling (including the FLEX approximation) or perturbation results that go beyond the second-Born approximation that have occasionally been used in the literature, but usually not under nonequilibrium conditions.

Thus, limited availability of a broad class of selfenergy approximations, including their representations for commonly used situations, can be considered a major bottleneck for further progress in nonequilibrium Green functions and their applications to many fields of many-body physics. It is a goal of the present review, to fill this gap.

*1.5. Outline of this review*

This article is organized as follows. In Section 2 we present a brief but selfcontained introduction into the concepts of nonequilibrium Green functions including the equations of motion for the NEGF—the Keldysh–Kadanoff–Baym equations. This is followed by an introduction to the Hubbard model for strongly correlated systems and the transformation of the NEGF into a Hubbard basis. We then introduce the selfenergy $\Sigma$ and the two main approaches for deriving approximations for $\Sigma$: the first is based on an expansion in terms of the bare pair interaction whereas the second uses the screened interaction, as the basic ingredient (Hedin's equations). We then present an overview of the main selfenergy approximations that follow from those two schemes. This is followed, in Sec. 3, by a summary of representative numerical applications to the dynamics of strongly correlated fermions under various excitation conditions which illustrate the performance of the different approximations for $\Sigma$. In the second part of the review that contains sections 4 and 5 we return to the governing equations for the selfenergy where the former (latter) is devoted to the expansion in terms of unscreened (screened) pair potentials. In each of the two sections the relevant approximations for $\Sigma$ will be presented first in a general form which is then specified to various practically relevant representations including the Hubbard basis. Finally, a summary and outlook is presented in Sec. 6.

## 2. Basics of nonequilibrium Green functions

This section gives an overview about the theoretical foundations of the NEGF method and focuses on the interconnection between and classification of common approximation schemes. As far as we are aware, it provides the first comprehensive overview of the relevant equations in a fully general basis representation[ii]. From this, the common cases of a diagonal basis such as the coordinate basis and the Hubbard basis for fermions and bosons are deduced. Alongside the development of the theory, the numerical scaling of the different approximation techniques will be detailed to enable a suitable choice with respect to the achievable simulation duration and basis size. In Section 2.1, the representation of states of indistinguishable quantum particles such as electrons in the so-called *Fock space* is discussed. The underlying notion of the *second quantization* allows for a suitable description of the dynamics for these particles in terms of canonical

---

[ii]For the particle–particle $T$-matrix approximation, a thorough derivation for a general basis set was presented in [77].

operators which perform the creation and annihilation in a chosen basis comprised of single-particle orbitals. Section 2.2 explores several possible sets of basis functions and their numerical suitability for different classes of systems. As a special case, the description of bosons and fermions in the basis set of the Hubbard model [80] is described.

For general time-dependent problems, it turns out to be advantageous to work on a complex time-contour (Schwinger–Keldysh contour), that is introduced in Section 2.4. The central quantity on the time-contour—the single-particle Green function—which gives access to all single-particle observables, the single-particle spectrum and some two-particle quantities, is defined in Section 2.5. The equations of motion for the Green functions are a set of integro-differential equations, which are mutually coupled, constituting a hierarchy between Green functions of different particle number, the *Martin–Schwinger hierarchy* (MSH). A suitable reformulation of the MSH has been given in [81], where a set of five contour quantities is introduced, which also obey coupled equations of motion, the solutions of which yield the same Green function as the solution of the MSH. The representations of these equations in a general basis set are given in Section 2.8. Since the exact solution of either set of equations is numerically impossible for most realistic systems, approximation techniques have to be employed. The approaches presented in this work are based on the common building block of the so-called *selfenergy* the purpose of which is to capture all relevant many-body effects. How it can be approximately determined using both perturbative and non-perturbative methods is detailed at the end of this section.

### 2.1. Dynamics of indistinguishable quantum particles in second quantization

The physical properties of all quantum particles are determined by their nature as excitations of an underlying field. These fields are quantized, i.e., they can only accommodate an integral number of elementary excitations, which are identified with the quantum particles. If only a single particle is excited, its state can be described by a wavefunction $\left| \Psi \right\rangle$ defined on a single-particle Hilbert space $\mathcal{H}$ over the field of complex numbers $\mathbb{C}$, which is assumed to be of finite dimension[iii]. For excitations of more than one particle, the indistinguishability of quantum particles has to be taken into account properly. Experimentally, it has been found that quantum particles either carry *bosonic* or *fermionic* statistics, i.e., obey either the Fermi–Dirac [82, 83] or the Bose–Einstein [84]

---

[iii]In practice, this does not constitute a restriction, since the Hilbert space is either already of finite dimension or has to be approximated as such anyway to make a numerical treatment possible.

distribution. The group of fermions, which all have half-integer spin, contains the quarks and leptons, such as the electron, whereas phonons, $W$- and $Z$ gauge-particles, gluons and the recently experimentally verified Higgs particle are bosons. Particles that are composed of elementary fermions or bosons can be of either bosonic[iv] (e.g. mesons, pions, kaons, excitons, biexcitons) or fermionic (baryons [68], nucleons, trions etc.) type, depending on the number of fermions involved. In the theoretical description, the spin statistics amounts to the many-body wavefunction being totally symmetric, for bosons, or totally anti-symmetric, for fermions with respect to interchange of two particles. How these statistics are conveniently built into the description of the many-body system, is detailed in the following.

To be able to treat states of varying particle number on an equal footing, it is convenient to define the so-called Fock space $\mathcal{F}_\sigma^{\mathcal{H}}$ induced by the single-particle Hilbert space $\mathcal{H}$ as the (completion of the) direct sum of (anti-)symmetrized n-fold tensor products of $\mathcal{H}$,

$$\mathcal{F}_\sigma^{\mathcal{H}} = \overline{\bigoplus_{n=0}^{\infty} S_\sigma \mathcal{H}^{\otimes n}} = \overline{\mathbb{C} \oplus \mathcal{H} \oplus S_\sigma(\mathcal{H} \otimes \mathcal{H}) \oplus \dots} \quad, \tag{1}$$

with

$$\mathcal{H}^{\otimes n} = \overbrace{\mathcal{H} \otimes \mathcal{H} \otimes \dots \otimes \mathcal{H}}^{\text{n times}} \quad \text{for all } n \in \mathbb{N}_0\,. \tag{2}$$

The operator $S_\sigma$ symmetrizes or anti-symmetrizes tensors for bosonic ($\sigma = +$) or fermionic ($\sigma = -$) particles. To define its action, it is suitable to fix a single-particle orbital basis of $\mathcal{H}$,

$$\mathcal{B}^{\text{sp}} = \left\{ \left| b_i \right\rangle, i \in I \right\}, \tag{3}$$

for an index set $I$ of cardinality $\dim_{\mathcal{H}}$. With this, for every $n \in \mathbb{N}_0$ and basis elements $\left| b_1 \right\rangle, \dots, \left| b_n \right\rangle \in \mathcal{B}^{\text{sp}}$, the action of $S_\sigma$ on the standard tensor product is given by

$$S_+\left(\left| b_1 \right\rangle \otimes \dots \otimes \left| b_n \right\rangle\right) \tag{4}$$
$$= \frac{1}{\sqrt{n!}} \sum_{s \in \text{Sym}_n} \left| b_{s(1)} \right\rangle \otimes \dots \otimes \left| b_{s(n)} \right\rangle =: \left| b_1 \right\rangle \circ \dots \circ \left| b_n \right\rangle$$

---

[iv]Note that the Bose character is only approximate, and deviations may appear on short length scales on the order of the interparticle distance.

and

$$S_-\Big(\big|b_1\big\rangle \otimes \ldots \otimes \big|b_n\big\rangle\Big) \tag{5}$$
$$= \frac{1}{\sqrt{n!}} \sum_{s \in \mathrm{Sym}_n} \mathrm{sign}\big(s\big) \big|b_{s(1)}\big\rangle \otimes \ldots \otimes \big|b_{s(n)}\big\rangle =: \big|b_1\big\rangle \wedge \ldots \wedge \big|b_n\big\rangle\,,$$

for bosons and fermions, respectively. Note that it is sufficient to define the (anti-)symmetrization operator only for basis elements, since it is linear. For example, a general fermionic anti-symmetrized state $\big|\Psi_2^-\big\rangle$ on the 2-fold tensor product $\mathcal{H} \otimes \mathcal{H}$ is of the form

$$\big|\Psi_2^-\big\rangle = \sum_{i<j\in I} c_{ij}\big|b_i\big\rangle \wedge \big|b_j\big\rangle \quad \text{for } \big|b_i\big\rangle, \big|b_j\big\rangle \in \mathcal{B}^{\mathrm{sp}}\,, \tag{6}$$

for $c_{ij} \in \mathbb{C}$. Here, the antisymmetric tensor product $\big|b_i\big\rangle \wedge \big|b_j\big\rangle$ is given in terms of the standard tensor product as

$$\big|b_i\big\rangle \wedge \big|b_j\big\rangle = \frac{1}{2}\Big(\big|b_i\big\rangle \otimes \big|b_j\big\rangle + \big(-1\big)\big|b_j\big\rangle \otimes \big|b_i\big\rangle\Big)\,. \tag{7}$$

Note that $\big|b_i\big\rangle \wedge \big|b_i\big\rangle = 0$, which reflects that, due to the Pauli exclusion principle, no two fermions can occupy the same state. With this, a general state in the Fock space $\mathcal{F}_\sigma^{\mathcal{H}}$, a *Fock state* $\big|\Psi^\sigma\big\rangle$, which is a superposition of states with a different number of particles, can be written as

$$\big|\Psi^\sigma\big\rangle = c_0|0\rangle \oplus \sum_{i\in I} c_i\big|b_i\big\rangle \oplus \sum_{i\leq j\in I} c_{ij}\big|b_i\big\rangle \otimes_\sigma \big|b_j\big\rangle \oplus \ldots\,, \tag{8}$$

for $c_0, c_i, c_{ij} \in \mathbb{C}$, where the short-hand notation

$$\otimes_\sigma = \begin{cases} \circ & \text{for bosons}\,, \\ \wedge & \text{for fermions}\,, \end{cases} \tag{9}$$

has been introduced. The first state, $|0\rangle$, is the vacuum state, which is the state of zero physical particles and of the lowest possible energy, $E^{\mathrm{vac}}$—in the context of this article, $E^{\mathrm{vac}} = 0$ is assumed[v].

With the concept of Fock states that are suitable to describe systems with a varying particle number, it is most natural to define operators that create $\big(\hat{c}^\dagger\big[\big|b_i\big\rangle\big] =: \hat{c}_i^\dagger\big)$ or remove $\big(\hat{c}\big[\big|b_i\big\rangle\big] =: \hat{c}_i\big)$ a particle in a given single-particle orbital $\big|b_i\big\rangle$. To characterize their action, it is sufficient to define the action on all (anti-)symmetrized $n$-particle

---

[v]In quantum chromodynamics and quantum electrodynamics, the lowest energy state may not have zero energy and allow for quantum fluctuations [85].

subspaces of $\mathcal{F}_\sigma^\mathcal{H}$ defined in a fashion similar to Eq. (2),

$$\mathcal{H}^{\otimes_\sigma{}^n} = \overbrace{\mathcal{H} \otimes_\sigma \mathcal{H} \otimes_\sigma \ldots \mathcal{H}}^{\text{n times}},\tag{10}$$

as

$$\hat{c}_i^\dagger \overbrace{\left(\left|b_1\right\rangle \otimes_\sigma \ldots \otimes_\sigma \left|b_n\right\rangle\right)}^{\in\mathcal{H}^{\otimes_\sigma{}^n}} = \overbrace{\left|b_i\right\rangle \otimes_\sigma \left|b_1\right\rangle \otimes_\sigma \ldots \otimes_\sigma \left|b_n\right\rangle}^{\in\mathcal{H}^{\otimes n+1}}$$

and

$$\hat{c}_i \overbrace{\left(\left|b_1\right\rangle \otimes_\sigma \ldots \otimes_\sigma \left|b_n\right\rangle\right)}^{\in\mathcal{H}^{\otimes_\sigma{}^n}}$$

$$= \sum_{k\in I}(-\sigma)^k \left\langle b_i \middle| b_k\right\rangle \left|b_1\right\rangle \otimes_\sigma \ldots \underbrace{\otimes_\sigma \left|\cancel{b_k}\right\rangle \cancel{\otimes_\sigma}}_{\in\mathcal{H}^{\otimes n-1}} \ldots \otimes_\sigma \left|b_n\right\rangle.\tag{11}$$

With these equations, the (anti-)commutator between the creation operators and annihilation operators as well as between one creation and one annihilation operator for fermions (bosons) is easily worked out,

$$\left[\hat{c}_i^\dagger, \hat{c}_j^\dagger\right]_\pm = 0,\quad \left[\hat{c}_i, \hat{c}_j\right]_\pm = 0,\quad \left[\hat{c}_i, \hat{c}_j^\dagger\right]_\pm = \left\langle i \middle| j\right\rangle.\tag{12}$$

Note that we used a general description that allows for a non-orthogonal set, $\{\left|i\right\rangle\}$, of single-particle basis states. In the special case of an orthonormal basis, $\left\langle i \middle| j\right\rangle = \delta_{i,j}$, and one recovers, in the final expression $\delta_{i,j}$ which is familiar from many text books.

The creation and annihilation operators form a basis for all operators acting on the space $\mathcal{F}_\sigma^\mathcal{H}$. For instance, general single-particle and two-particle operators $\hat{O}^{(1)}$, $\hat{O}^{(2)}$ are given as linear superpositions[vi]

$$\hat{O}^{(1)} = \sum_{mn} o_{mn}^{(1)} \hat{c}_m^\dagger \hat{c}_n,\tag{13}$$

$$\hat{O}^{(2)} = \sum_{mnpq} o_{mnpq}^{(2)} \hat{c}_m^\dagger \hat{c}_n^\dagger \hat{c}_p \hat{c}_q,\tag{14}$$

where the matrix elements are

$$o_{mn}^{(1)} = \left\langle b_m \middle| \hat{o}^{(1)} \middle| b_n\right\rangle,\qquad o_{mnpq}^{(2)} = \left\langle b_m b_n \middle| \hat{o}^{(2)} \middle| b_p b_q\right\rangle.\tag{15}$$

As a special case, the Hamiltonian, which carries the specific geometries of the studied systems as well as any external (time-dependent) potentials and forces driving the

---

[vi]From now on, if not stated otherwise, all sums run over the complete basis set.

dynamics transforms to

$$\hat{H}\left(t\right) = \underbrace{\sum_{mn} h_{mn} \hat{c}_m^\dagger \hat{c}_n}_{\hat{H}^0} + \underbrace{\frac{1}{2} \sum_{mnpq} w_{mnpq} \hat{c}_m^\dagger \hat{c}_n^\dagger \hat{c}_q \hat{c}_p}_{\hat{W}} + \underbrace{\sum_{mn} f_{mn}\left(t\right) \hat{c}_m^\dagger \hat{c}_n}_{\hat{F}(t)}, \qquad (16)$$

containing the single-particle part $\hat{H}^0$, the interaction $\hat{W}$ and the time-dependent single-particle excitation part $\hat{F}\left(t\right)$. Since all quantities discussed in this section are formulated in terms of the single-particle basis $\mathcal{B}^{\mathrm{sp}}$ (cf. Eq. (3)), its suitable choice is vital for the numerical implementation to achieve the best possible performance. A strategy for the selection of a set of basis functions is detailed in the next section.

## *2.2. Choice of the one-particle basis*

Selecting a single-particle basis (cf. Eq. (3)) constitutes the first step in the process of the theoretical modeling of a system. With the basis, elements $\left|\Psi\right\rangle$ of the single-particle Hilbert space $\mathcal{H}$ can be expanded as

$$\left|\Psi\right\rangle = \sum_{i \in I} b_i \left|b_i\right\rangle, \qquad (17)$$

where $I$ is an index set of cardinality $\dim_{\mathcal{H}}$. For Hilbert spaces of infinite dimension, $I$ has to be substituted by a finite set $I'$ to make a numerical treatment possible, which renders Eq. (17) only approximately valid. For the formulation of the Hamiltonian, according to Eq. (16), the matrix elements $h_{km}$, $w_{klmn}$, $f_{km}\left(t\right)$ have to be specified. Once they are given in the natural basis of the studied system, they can be transformed into another single-particle basis $\mathcal{C}^{\mathrm{sp}} = \left\{\left|c_j\right\rangle, j \in J\right\}$ by

$$h_{km}^{\mathcal{C}} = \sum_{r=1}^{\dim_{\mathcal{H}}} \sum_{s=1}^{\dim_{\mathcal{H}}} b_{rk}^* h_{rs}^{\mathcal{B}} b_{sm}, \qquad (18)$$

with the expansion of the new basis functions $\left|c_i\right\rangle$ in terms of the old $\left|b_i\right\rangle$ given as

$$\left\langle c_i\right| = \sum_{r=1}^{\dim_{\mathcal{H}}} b_{ri}^* \left\langle b_r\right|, \qquad (19)$$

$$\left|c_i\right\rangle = \sum_{s=1}^{\dim_{\mathcal{H}}} \left|b_s\right\rangle b_{si}, \qquad (20)$$

with the transformation matrix elements

$$b_{si} = \left\langle b_s\middle|c_i\right\rangle, \quad b_{ri}^* = \left\langle b_r\middle|c_i\right\rangle^*. \qquad (21)$$

With these transformations, the basis can be chosen to suit the numerical needs. To this end, two criteria can be formulated which characterize how well numerically tractable a set of basis functions is. First, it should consist of as few basis functions as possible to achieve the accuracy demanded, i.e., it describes single-particle orbitals that are as close as possible to the true orbitals occupied by the particles. To work out the other criterion, one notices that, according to Eq. (16), the interaction—a central quantity in any exact treatment as well as the selfenergy approximations discussed later in this article—is represented by a fourth-order tensor $w_{klmn}$ in a general basis. This structure is numerically prohibitive since it involves at least a scaling of $\mathcal{O}\left(N_{\mathrm{b}}^4\right)$, where $N_b$ is the dimension of the basis set. Fortunately, the interaction tensor can be brought into a diagonal representation, where it is characterized by a second-order tensor, i.e., is of the structure

$$w_{klmn} = \delta_{kn}\delta_{lm}w_{kl} \,. \tag{22}$$

In practice, this diagonalization can be achieved by choosing a quadrature rule for the integrals involved in the computation of the interaction matrix elements (cf. Eq. (15)) and construction of a (finite-element) discrete variable representation upon it [73, 86, 87]. For details, the reader is referred to Ref. [70, 88], where various aspects of different choices of quadratures and their implementation are discussed. Accordingly, the second criterion is that the basis functions are chosen such that the interaction matrix elements are (approximately) diagonal in the sense of Eq. (22). Unfortunately, both criteria are often "orthogonal" to each other, and the user has to choose between them. While physically motivated basis sets achieve a good representation with only a small number of basis functions, they entail a dense fourth-order tensorial structure of the interaction matrix elements. In contrast, discrete-variable-representation basis sets provide the latter in diagonal form, but the basis functions are "general purpose" and the worse representation of physical states requires their number to be comparably large. As a rule of thumb, it can be stated that, for small systems, which require only few basis functions, physical basis sets are preferable while, for large systems, for instance in the description of photoemission experiments on atoms, molecules or solide [87, 89, 90], a grid-based approach is often favorable. In both cases, a close look at the structure of the equations at hand provides a more thorough basis for the decision. As an example, looking ahead to Eqs. (189) and (191) vs. Eqs. (176) and (183), the index structure of the selfenergy—the central quantity in Green function based calculations—for example, in the important second-order Born (2B) approximation, looks like (omitting time arguments and scalar factors)

$$\Sigma_{ij}^{(2),\text{diagonal}} \sim \sum_{np} G_{in} w_{ip} G_{np} G_{pj} w_{nj} \pm \sum_{pr} G_{ij} w_{ip} G_{rp} G_{pr} w_{rj}\,, \tag{23}$$

$$\Sigma_{ij}^{(2)} \sim \sum_{mnpqrs} G_{mn} w_{ipqm} G_{sp} G_{qr} \big( w_{nrjs} \pm w_{rnjs} \big)\,, \tag{24}$$

in a diagonal basis vs. general basis representation. At first glance, the diagonal case of Eq. (23) suggests a scaling of $\mathcal{O}\big(N_{\mathrm{b}}^4\big)$ stemming from the two external indices $i, j$ and the summation over the two internal indices, whereas in the case of full interaction, according to Eq. (24), the summation over six internal indices $(m, n, p, q, r, s)$ prompts a scaling of $\mathcal{O}\big(N_{\mathrm{b}}^8\big)$, which would strongly favor the former over the latter. A quick reordering, though, lets one rewrite Eqs. (23) and (24) as

$$\Sigma_{ij}^{(2),\text{diagonal}} \sim \sum_n G_{in} w_{nj} \sum_p w_{ip} G_{np} G_{pj} \pm G_{ij} \sum_p w_{ip} \sum_r G_{pr} G_{rp} w_{rj}\,, \tag{25}$$

$$\Sigma_{ij}^{(2)} \sim \sum_{mpq} w_{ipqm} \sum_s G_{sp} \sum_r G_{qr} \sum_n G_{mn} \big( w_{nrjs} \pm w_{rnjs} \big)\,. \tag{26}$$

This elucidates that, for diagonal interaction, $\Sigma_{ij}^{(2),\text{diagonal}}$ indeed scales as $\mathcal{O}\big(N_{\mathrm{b}}^4\big) + \mathcal{O}\big(N_{\mathrm{b}}^3\big) = \mathcal{O}\big(N_{\mathrm{b}}^4\big)$, whereas, $\Sigma_{ij}^{(2)}$ scales as $\mathcal{O}\big(N_{\mathrm{b}}^5\big) + \mathcal{O}\big(N_{\mathrm{b}}^5\big) + \mathcal{O}\big(N_{\mathrm{b}}^5\big) + \mathcal{O}\big(N_{\mathrm{b}}^5\big) = \mathcal{O}\big(N_{\mathrm{b}}^5\big)$, in contrast[vii]. Thus, the preferable basis choice strongly depends on the respective basis sizes needed.

### 2.3. The Hubbard model

Since it plays an important role underlying the applications in Sec. 3, the special case of the Hubbard basis and the associated Hubbard model is briefly discussed here. The Hubbard model has been introduced by *John Hubbard*, a British physicist, in 1963 [80] to describe the physics—especially the transition between conducting and insulating behavior—of electrons in narrow energy bands of solid-state systems such as transition-metal oxides. At the heart of the Hubbard model is the observation that, in narrow *d*- and *f*-bands, the electrons are mostly located at the nuclei—where they interact—and only rarely move between different positions on the lattice. Therefore, Hubbard proposed to describe these systems in terms of "sites" between which the electrons "hop" with a given amplitude $J$. At each site, which, in the model, contains one orbital for spin-up and one orbital for spin-down orientation, the electrons experience a repulsion by electrons in

---

[vii]As one notices, the ordering of the terms in Eqs. (25) and (26) is not unique but there exists no ordering which results in a better scaling.

the other orbital of strength $U$. Accordingly, the Hubbard model can be described by the generic Hamiltonian, cf. Eq. (16), with matrix elements (written in the representation in terms of spin-orbitals $|i\alpha\rangle$, with site $i$ and spin $\alpha$),

$$h_{i\alpha j\beta} = -J\delta_{\langle ij\rangle}\delta_{\alpha\beta} - \mu\delta_{ij}\delta_{\alpha\beta}\hat{c}_{i\alpha}^\dagger\hat{c}_{i\alpha}\,, \tag{27}$$

$$w_{i\alpha j\beta k\gamma l\delta} = U\delta_{il}\delta_{\alpha\delta}\delta_{jk}\delta_{\beta\gamma}\delta_{ij}\,, \tag{28}$$

$$f_{i\alpha j\beta}\big(t\big) = \delta_{ij}\delta_{\alpha\beta}f_{i\alpha}\big(t\big)\,, \tag{29}$$

where $\delta_{\langle ij\rangle} = 1$, exactly if the sites $i,j$ are nearest neighbors. The term

$$-\mu\delta_{ij}\delta_{\alpha\beta}\hat{c}_{i\alpha}^\dagger\hat{c}_{i\alpha} \tag{30}$$

describes the chemical energy induced by a chemical potential $\mu$. The time-dependent excitation matrices $f_{i\alpha j\beta}\big(t\big)$ for all processes considered in this work are both, on-site $\big(\delta_{ij}\big)$ and spin-conserving $\big(\delta_{\alpha\beta}\big)$. Inserting Eqs. (27), (28) and (29) into the general form of Eq. (16), one arrives at

$$\begin{aligned}
\hat{H}\big(t\big) = &-J\sum_{m\epsilon n\zeta}\delta_{\langle mn\rangle}\delta_{\epsilon\zeta}\hat{c}_{m\epsilon}^\dagger\hat{c}_{n\zeta} + \frac{U}{2}\sum_{m\epsilon n\zeta p\eta q\theta}\delta_{mq}\delta_{\epsilon\theta}\delta_{np}\delta_{\zeta\eta}\delta_{mn}\hat{c}_{m\epsilon}^\dagger\hat{c}_{n\zeta}^\dagger\hat{c}_{p\eta}\hat{c}_{q\theta} \\
&+ \sum_{m\epsilon n\zeta}\delta_{mn}\delta_{\epsilon\zeta}f_{m\epsilon n\zeta}\big(t\big)\hat{c}_{m\epsilon}^\dagger\hat{c}_{n\zeta} - \mu\sum_{m\epsilon}\hat{c}_{m\epsilon}^\dagger\hat{c}_{m\epsilon} \\
= &-J\sum_{\langle m,n\rangle}\sum_\epsilon\hat{c}_{m\epsilon}^\dagger\hat{c}_{n\epsilon} + \frac{U}{2}\sum_m\sum_{\epsilon\zeta}\hat{c}_{m\epsilon}^\dagger\hat{c}_{m\zeta}^\dagger\hat{c}_{m\zeta}\hat{c}_{m\epsilon} \\
&+ \sum_{m\epsilon}f_{m\epsilon}\big(t\big)\hat{c}_{m\epsilon}^\dagger\hat{c}_{m\epsilon} - \mu\sum_{m\epsilon}\hat{c}_{m\epsilon}^\dagger\hat{c}_{m\epsilon}\,.
\end{aligned} \tag{31}$$

The following results differ for the cases of fermions and bosons, respectively, so we provide both cases separately. With the canonical anti-commutation relations, cf. Eq. (12), for **bosons**, the interaction term can be rewritten as

$$\begin{aligned}
\hat{W}_{\text{bosons}}^{\text{Hubbard}} &= \frac{U}{2}\sum_m\sum_{\epsilon\zeta}\hat{c}_{m\epsilon}^\dagger\hat{c}_{m\zeta}^\dagger\hat{c}_{m\zeta}\hat{c}_{m\epsilon} = \frac{U}{2}\sum_m\sum_{\epsilon\zeta}\hat{c}_{m\epsilon}^\dagger\hat{c}_{m\zeta}^\dagger\hat{c}_{m\epsilon}\hat{c}_{m\zeta} \\
&= \frac{U}{2}\sum_m\sum_{\epsilon\zeta}\hat{c}_{m\epsilon}^\dagger\hat{c}_{m\epsilon}\hat{c}_{m\zeta}^\dagger\hat{c}_{m\zeta} - \frac{U}{2}\sum_m\sum_\epsilon\hat{c}_{m\epsilon}^\dagger\hat{c}_{m\epsilon} \\
&=: \frac{U}{2}\sum_m\sum_{\epsilon\zeta}\hat{n}_{m\epsilon}\hat{n}_{m\zeta} - \frac{U}{2}\sum_m\sum_\epsilon\hat{n}_{m\epsilon} \\
&= \frac{U}{2}\sum_m\sum_{\epsilon\neq\zeta}\hat{n}_{m\epsilon}\hat{n}_{m\zeta} + \frac{U}{2}\sum_m\sum_\epsilon\hat{n}_{m\epsilon}\big(\hat{n}_{m\epsilon} - 1\big)\,.
\end{aligned} \tag{32}$$

The special case of **spin-0 bosons** results in the Bose–Hubbard interaction,

$$\hat{W}_{\text{bosons,0}}^{\text{Hubbard}} = \frac{U}{2}\sum_m\hat{n}_m\big(\hat{n}_m - 1\big)\,, \tag{33}$$

and the corresponding Bose–Hubbard Hamiltonian (without time-dependent excitation),

$$\hat{H}_{\text{spin-0}}^{\text{Bose–Hubbard}} = -J \sum_{\langle m,n \rangle} \hat{c}_m^\dagger \hat{c}_n + \frac{U}{2} \sum_m \hat{n}_m \big(\hat{n}_m - 1\big) - \mu \sum_m \hat{n}_m \,. \tag{34}$$

Next consider **fermions**. Now, due to the Pauli exclusion principle, Eq. (28) can be rewritten as

$$w_{i\alpha j\beta k\gamma l\delta} = U \delta_{il} \delta_{\alpha\delta} \delta_{jk} \delta_{\beta\gamma} \delta_{ij} \bar{\bar{\delta}}_{\alpha\beta} \,, \tag{35}$$

with $\bar{\bar{\delta}}_{\alpha\beta} = 1 - \delta_{\alpha\beta}$. Consequently, the interaction part $\hat{W}$ of the Hamiltonian becomes

$$\begin{aligned}
\hat{W}_{\text{fermions}}^{\text{Hubbard}} &= \frac{U}{2} \sum_m \sum_{\epsilon \neq \zeta} \hat{c}_{m\epsilon}^\dagger \hat{c}_{m\zeta}^\dagger \hat{c}_{m\zeta} \hat{c}_{m\epsilon} = -\frac{U}{2} \sum_m \sum_{\epsilon \neq \zeta} \hat{c}_{m\epsilon}^\dagger \hat{c}_{m\zeta}^\dagger \hat{c}_{m\epsilon} \hat{c}_{m\zeta} \\
&= \frac{U}{2} \sum_m \sum_{\epsilon \neq \zeta} \hat{c}_{m\epsilon}^\dagger \hat{c}_{m\epsilon} \hat{c}_{m\zeta}^\dagger \hat{c}_{m\zeta} = \frac{U}{2} \sum_m \sum_{\epsilon \neq \zeta} \hat{n}_{m\epsilon} \hat{n}_{m\zeta} \,.
\end{aligned} \tag{36}$$

For the special case of **spin-$\frac{1}{2}$ fermions**, this expression simplifies to

$$\hat{W}_{\text{fermions},1/2}^{\text{Hubbard}} = \frac{U}{2} \sum_m \big(\hat{n}_{m\uparrow} \hat{n}_{m\downarrow} + \hat{n}_{m\downarrow} \hat{n}_{m\uparrow}\big) = U \sum_m \hat{n}_{m\uparrow} \hat{n}_{m\downarrow} \tag{37}$$

and the (Fermi–)Hubbard Hamiltonian (again without time-dependent excitation) is given by

$$\hat{H}_{\text{spin-1/2}}^{\text{Fermi–Hubbard}} = -J \sum_{\langle m,n \rangle} \sum_{\epsilon \in \{\uparrow,\downarrow\}} \hat{c}_{m\epsilon}^\dagger \hat{c}_{n\epsilon} + U \sum_m \hat{n}_{m\uparrow} \hat{n}_{m\downarrow} - \mu \sum_m \big(\hat{n}_{m\uparrow} + \hat{n}_{m\downarrow}\big) \,.$$

One notices that the Hubbard interaction $w_{i\alpha j\beta k\gamma l\delta}$ is highly diagonal, which is very advantageous for the numerical treatment—a property which has contributed greatly to the recurring popularity of the Hubbard model in computational physics in the last decade, e.g. Refs. [26, 49, 76, 91–94]. Accordingly, for the example of the second-order selfenergy that was presented above in Eq. (23) and which will be treated in full detail in Sec. 4, the expression in the Hubbard basis reads, cf. Eq. (198),

$$\begin{aligned}
&\Sigma_{i\downarrow(\uparrow)j\downarrow(\uparrow)}^{(2),2,0,\text{Hubbard},f,1/2}\big(z_1, z_2\big) \\
&= \pm \big(\mathrm{i}\hbar\big)^2 G_{i\downarrow(\uparrow)j\downarrow(\uparrow)}\big(z_1, z_2\big) U\big(z_1\big) G_{i\uparrow(\downarrow)j\uparrow(\downarrow)}\big(z_1, z_2\big) G_{j\uparrow(\downarrow)i\uparrow(\downarrow)}\big(z_2, z_1\big) U\big(z_2\big) \,.
\end{aligned} \tag{38}$$

This expression only scales as $\mathcal{O}\big(N_b^2\big)$, since it involves no matrix multiplications, compared to the scaling for a general basis, with $\mathcal{O}\big(N_b^5\big)$, and of $\mathcal{O}\big(N_b^4\big)$, in a diagonal basis. Here, the arguments $z_{1,2}$ denote times that are situated on the Schwinger–Keldysh contour that naturally emerges in nonequilibrium quantum statistics and which we introduce next.

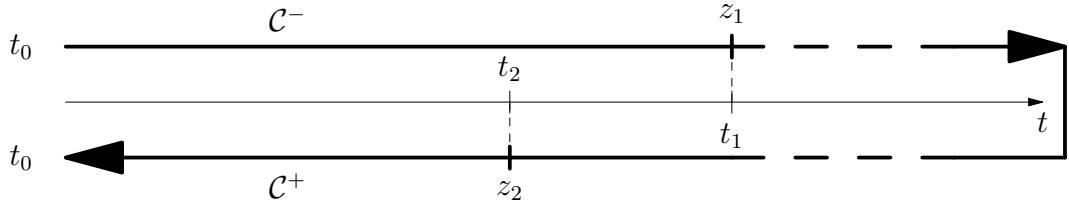

**Figure 1.** Schwinger–Keldysh contour $\mathcal{C}$. The forward-branch $\mathcal{C}^-$ extends from the initial time $t_0$ to the current time $t$, bends and leads back to $t_0$ along the backward $\mathcal{C}^+$-branch. Note that the projections of the contour times $z_1 < z_2$ on the real axis obey the inverse relationship $t_1 > t_2$.

### 2.4. Time-dependence of observables and the Schwinger–Keldysh time-contour

The purpose of the formalism of second quantization, introduced in the last section, is to provide a suitable framework for the description of quantum many-particle systems, in particular, for time-dependent processes. Here, one is mostly interested in the expectation values of operators of the form of Eqs. (13) and (14), at any given time $t$. With the time-dependent many-particle wavefunction $\left|\Psi(t)\right\rangle$, i.e. the solution of the time-dependent Schrödinger equation, the expectation value can be computed as

$$O(t) = \left\langle \Psi(t)\left|\hat{O}(t)\right|\Psi(t)\right\rangle \tag{39}$$
$$= \left\langle \Psi_0\left|\hat{\mathcal{T}}_\mathrm{a}\left\{\exp\left(\frac{1}{\mathrm{i}\hbar}\int_t^{t_0}\mathrm{d}\bar{t}\,\hat{H}(\bar{t})\right)\right\}\hat{O}(t)\hat{\mathcal{T}}_\mathrm{c}\left\{\exp\left(-\frac{1}{\mathrm{i}\hbar}\int_t^{t_0}\mathrm{d}\bar{t}\,\hat{H}(\bar{t})\right)\right\}\right|\Psi_0\right\rangle,$$

where the operators $\hat{\mathcal{T}}_\mathrm{c}\left(\hat{\mathcal{T}}_\mathrm{a}\right)$ are the (anti-)chronological time-ordering superoperators, which rearrange the operators acted on such that the latest (earliest) times are moved to the left-hand side to account for (anti-)causality. A more concise formulation can be achieved by introducing an oriented contour $\mathcal{C}$ which starts from $t_0$, extends to the turning point $t$ and then reaches back to $t_0$,

$$\mathcal{C} = \underbrace{(t_0, t)}_{\mathcal{C}^-} \bigoplus \underbrace{(t, t_0)}_{\mathcal{C}^+}, \tag{40}$$

with a forward branch $\mathcal{C}^-$ and a backward branch $\mathcal{C}^+$, depicted in Fig. 1. Henceforth, a general time on the contour $\mathcal{C}$ will be denoted as $z$ and $z_\pm$ to refer to a time lying on one of the branches. Accordingly, an operator $\hat{O}$ can be extended to the contour, having possibly different values on both branches,

$$\hat{O}(z) = \begin{cases} \hat{O}_-(z) & \text{if } z \in \mathcal{C}^- \\ \hat{O}_+(z) & \text{if } z \in \mathcal{C}^+ \end{cases}. \tag{41}$$

With this definition, one can define a contour time-ordering superoperator $\hat{\mathcal{T}}_\mathcal{C}$ which moves operators at later contour times ahead of operators at earlier contour times. As a consequence, its action agrees with that of $\hat{\mathcal{T}}_c$, for all times $z_- \in \mathcal{C}^-$, and with that of $\hat{\mathcal{T}}_a$, for all times $z_+ \in \mathcal{C}^+$. Furthermore, time integrals are extended in a natural way to the contour by defining

$$\int_{z_1}^{z_2} \mathrm{d}\bar{z}\, \hat{O}(\bar{z}) := \begin{cases} \int_{t_1}^{t_2} \mathrm{d}\bar{t}\, \hat{O}_-(\bar{t}) & \text{if } z_1, z_2 \in \mathcal{C}^- \\ \int_{t_1}^{t} \mathrm{d}\bar{t}\, \hat{O}_-(\bar{t}) + \int_t^{t_2} \mathrm{d}\bar{t}\, \hat{O}_+(\bar{t}) & \text{if } z_1 \in \mathcal{C}^-, z_2 \in \mathcal{C}^+ \\ \int_{t_1}^{t_2} \mathrm{d}\bar{t}\, \hat{O}_+(\bar{t}) & \text{if } z_1, z_2 \in \mathcal{C}^+ \end{cases},$$

assuming $z_1$ is later than $z_2$. Using the contour integral, one can reformulate Eq. (39) for operators which have the same value on both branches, i.e., $\hat{O}_- = \hat{O}_+ =: \hat{O}_\pm$, as

$$O(t) = \left\langle \Psi_0 \middle| \hat{\mathcal{T}}_\mathcal{C} \left\{ \exp\left( \frac{1}{\mathrm{i}\hbar} \int_{\mathcal{C}^+} \mathrm{d}\bar{z}\, \hat{H}(\bar{z}) \right) \hat{O}_\pm(t) \exp\left( \frac{1}{\mathrm{i}\hbar} \int_{\mathcal{C}^-} \mathrm{d}\bar{z}\, \hat{H}(\bar{z}) \right) \right\} \middle| \Psi_0 \right\rangle, \tag{42}$$

which, taking into account the action $\hat{\mathcal{T}}_\mathcal{C}$, can be further simplified to

$$O(t) = \left\langle \Psi_0 \middle| \hat{\mathcal{T}}_\mathcal{C} \left\{ \exp\left( \frac{1}{\mathrm{i}\hbar} \int_{\mathcal{C}} \mathrm{d}\bar{z}\, \hat{H}(\bar{z}) \right) \hat{O}_\pm(t) \right\} \middle| \Psi_0 \right\rangle. \tag{43}$$

On both branches, the contour Hamiltonian $\hat{H}(\bar{z})$ is set equal to its definition in Eq. (16) for the corresponding real-time argument.

An undesirable feature of the introduced contour is that it seemingly depends on the value of $t$. This can be remedied by extending the contour to $t \to \infty$, which leaves all expressions, in particular Eq. (43), invariant, since the additional two integral parts cancel. The corresponding contour is depicted in Fig. 2. Finally, one notices that Eq. (43) is also true for all contour times $z$,

$$O(z) = \left\langle \Psi_0 \middle| \hat{\mathcal{T}}_\mathcal{C} \left\{ \exp\left( \frac{1}{\mathrm{i}\hbar} \int_{\mathcal{C}} \mathrm{d}\bar{z}\, \hat{H}(\bar{z}) \right) \hat{O}(z) \right\} \middle| \Psi_0 \right\rangle. \tag{44}$$

The contour $\mathcal{C}$ was introduced by L. Keldysh in 1964 [51] who showed that, with this modified time axis all expressions of ground state and thermodynamic Green functions, including Feynman's diagram technique, are naturally transferred to arbitrary nonequilibrium situations. The historical context of the development of this method of real-type (Keldysh) Green functions has been reviewed by Keldysh himself, for details see Ref. [95].

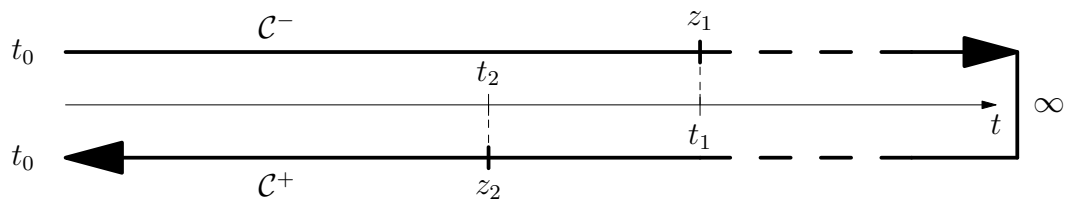

**Figure 2.** Schwinger–Keldysh contour $\mathcal{C}$ extended to $\infty$. The forward-branch $\mathcal{C}^-$ spans from the initial time $t_0$ to $\infty$, bends and leads back to $t_0$ along the backward $\mathcal{C}^+$-branch.

### 2.5. Nonequilibrium Green functions and their equations of motion

To compute time-dependent operator expectation values, there are two immediate choices at hand, following Eq. (39): one can either solve the first or the second line. The first option requires the solution of the equation of motion for the time-dependent wavefunction $\left|\Psi(t)\right\rangle$, which is the Schrödinger equation. This is the road taken by wavefunction-based methods like full configuration interaction [96, 97], multiconfigurational time-dependent Hartree–Fock [88, 98, 99], generalized active-space configuration interaction [90, 100], exact diagonalization [91], coupled-cluster methods [101] and density-matrix renormalization group based approaches [102–107].

The other way is to follow the second line and to work with the (known) initial wavefunction $\left|\Psi_0\right\rangle^{viii}$ and develop an equation of motion for the term

$$\hat{\mathcal{T}}_{\mathcal{C}}\left\{\exp\left(\frac{1}{i\hbar}\int_{\mathcal{C}}\mathrm{d}\bar{z}\,\hat{H}(\bar{z})\right)\hat{O}(z)\right\},\tag{45}$$

according to Eqs. (43) and (44), respectively. Approaches relying on this method are, among others, time-dependent Hartree–Fock [108], reduced-density-matrix theory [42, 109], density-functional theory [49, 110–112], dynamical mean-field theory (DMFT) [113–116] and the method of Green functions [26, 70, 71, 76, 77, 92, 94, 117–125], which is the topic of this article. In principle, both approaches are equivalent and yield the same results. The main difference is the set of available approximation techniques and, foremost, the numerical scaling behavior with respect to the maximal simulation time, particle number, basis size and interaction strength. The wavefunction-based methods, in general, can cope with huge basis sets with a number of basis functions, depending on the system at hand, ranging from thousands to millions and interaction strengths from weak to strong coupling. Additionally, they offer a linear scaling of the numerical effort with the simulation time. The trade-off is the exponential scaling of the numerical effort with the particle number rendering the simulation of systems with more than a few particles impossible [90, 100].

---

$^{viii}$It will be shown in Section 2.11 that, actually, the knowledge of the ideal, i.e., non-interacting, initial state is sufficient.

In contrast, the second group of methods, which relies on the equation of motion for the creation and annihilation operators, are not limited by the particle number. The scaling with the basis size is worse compared to the other group but still polynomial and the scaling with the total simulation time is at least quadratic for methods going beyond Hartree–Fock (which has a linear scaling). Apart from DMFT, which is also good for very strong interactions but can simulate only short time-spans, all methods of the second group, including Green functions, are mostly suited for small interaction strengths. In the following, the theory behind the Green functions method will be summarized. For a more in-detail derivation, see, e.g., Refs. [72, 73]. In the following section, the definition of the Green functions, their equations of motion and the determination of time-dependent observables from them will be discussed.

The direct computation of the time-dependent values of operators according to Eq. (44) involves the evaluation of the time-ordered exponential, which is impractical apart from very small basis sizes due to the dimensionality of the Hamiltonian. One strategy to bypass the direct evaluation of the exponential is to introduce the contour Heisenberg picture, which will be described in the following. Similar as for standard time, one can define the time-evolution operator $\hat{U}\left(z_1, z_2\right)$ on the contour,

$$\hat{U}\left(z_1, z_2\right) = \begin{cases} \hat{\mathcal{T}}_{\mathcal{C}}\left\{\exp\left(\frac{1}{i\hbar}\int_{\mathcal{C}}\mathrm{d}\bar{z}\,\hat{H}\left(\bar{z}\right)\right)\right\} & \text{if } z_1 \text{ later than } z_2\,, \\ \hat{\mathcal{T}}_{\mathcal{C}}^{\mathrm{a}}\left\{\exp\left(\frac{1}{i\hbar}\int_{\mathcal{C}}\mathrm{d}\bar{z}\,\hat{H}\left(\bar{z}\right)\right)\right\} & \text{if } z_1 \text{ earlier than } z_2\,, \end{cases} \tag{46}$$

where, in the second line, the anti-chronological time-ordering operator $\hat{\mathcal{T}}_{\mathcal{C}}^{\mathrm{a}}$ has been introduced, which places operators with later contour times to the right. The contour time-evolution operator has the usual properties, i.e, fulfills

$$i\hbar \frac{\mathrm{d}}{\mathrm{d}z_1}\hat{U}\big(z_1, z_0\big) = \hat{H}\big(z_1\big)\hat{U}\big(z_1, z_0\big)\,, \tag{47}$$

$$i\hbar \frac{\mathrm{d}}{\mathrm{d}z_1}\hat{U}\big(z_0, z_1\big) = -\hat{U}\big(z_0, z_1\big)\hat{H}\big(z_1\big)\,. \tag{48}$$

With this, Eq. (44) can be cast into the form

$$O\big(z_1\big) = \Big\langle \Psi_0 \Big| \hat{U}\big(z_{0^+}, z_{0^-}\big)\hat{U}\big(z_{0^-}, z_1\big)\hat{O}\big(z_1\big)\hat{U}\big(z_1, z_{0^-}\big) \Big| \Psi_0 \Big\rangle\,, \tag{49}$$

where $z_{0^-}$ and $z_{0^+}$ represent the start (end) of the contour. Eq. (49) suggests to introduce the contour Heisenberg picture

$$\hat{O}_{\mathrm{H}}\big(z_1\big) := \hat{U}\big(z_{0^-}, z_1\big)\hat{O}\big(z_1\big)\hat{U}\big(z_1, z_{0^-}\big)\,, \tag{50}$$

with the equation of motion

$$i\hbar \frac{\mathrm{d}}{\mathrm{d}z_1}\hat{O}_{\mathrm{H}}\big(z_1\big) = \Big[\hat{O}_{\mathrm{H}}\big(z_1\big), \hat{H}_{\mathrm{H}}\big(z_1\big)\Big]_- + \partial_{z_1}\hat{O}_{\mathrm{H}}\big(z_1\big)\,. \tag{51}$$

Using the commutator relations, cf. Eq. (12), the contour equations of motion for the canonical creation and annihilation operators for systems described by the Hamiltonian in Eq. (16) are readily found,

$$\begin{aligned}
i\hbar \frac{\mathrm{d}}{\mathrm{d}z_1}\hat{c}_i\big(z_1\big) &= \sum_n \big(h_{in}\big(z_1\big) + f_{in}\big(z_1\big)\big)\hat{c}_n\big(z_1\big) \\
&\quad + \sum_{npq} w_{inpq}\big(z_1\big)\hat{c}_n^\dagger\big(z_1\big)\hat{c}_p\big(z_1\big)\hat{c}_q\big(z_1\big)\,,
\end{aligned} \tag{52}$$

$$\begin{aligned}
-i\hbar \frac{\mathrm{d}}{\mathrm{d}z_1}\hat{c}_i^\dagger\big(z_1\big) &= \sum_m \hat{c}_m^\dagger\big(z_1\big)\big(h_{mi}\big(z_1\big) + f_{mi}\big(z_1\big)\big) \\
&\quad + \sum_{mnp} \hat{c}_m^\dagger\big(z_1\big)\hat{c}_n^\dagger\big(z_1\big)\hat{c}_p\big(z_1\big)w_{mnpi}\big(z_1\big)\,,
\end{aligned} \tag{53}$$

where

$$\hat{c}\big(z_1\big) := \hat{c}_{\mathrm{H}}\big(z_1\big)\,, \quad \hat{c}^\dagger\big(z_1\big) := \hat{c}_{\mathrm{H}}^\dagger\big(z_1\big)\,. \tag{54}$$

These equations can be used to derive equations for operator correlators, such as already encountered in Eq. (44). For $N$ operators, they are of the form

$$\hat{k}\big(z_1 \ldots z_N\big) = \hat{\mathcal{T}}_{\mathcal{C}}\Big\{\hat{O}_1\big(z_1\big)\ldots\hat{O}_N\big(z_N\big)\Big\}\,. \tag{55}$$

Remembering that any operator can be expressed in terms of the canonical operators, a special role is played by the correlators of these operators. From Eqs. (13) and (14), it is evident that especially those with the same number of creation and annihilation operators are of interest, since they give direct access to observables. Thus it is useful to define the correlator of $N$ annihilation and creation operators,

$$
\begin{aligned}
&\hat{G}^{(N)}_{i_1\ldots i_N\, j_1\ldots j_N}\big(z_1\ldots z_N, z_1'\ldots z_N'\big) \\
&\quad := \frac{1}{\big(\mathrm{i}\hbar\big)^N}\hat{\mathcal{T}}_{\mathcal{C}}\Big\{\hat{c}_{i_1}\big(z_1\big)\ldots\hat{c}_{i_N}\big(z_N\big)\hat{c}^{\dagger}_{j_1}\big(z_1'\big)\ldots\hat{c}^{\dagger}_{j_N}\big(z_N'\big)\Big\},
\end{aligned}
\tag{56}
$$

with $2N$ contour time arguments. Using some contour calculus, not repeated here (for details see Refs. [72, 77]), and the contour Heisenberg equations, one can derive their equations of motion, which couple the $N$-particle correlator to the $(N-1)$ and $(N+1)$ particle correlators,

$$
\begin{aligned}
&\sum_l\Big[\mathrm{i}\hbar\frac{\mathrm{d}}{\mathrm{d}z_k}\delta_{i_k l}-h_{i_k l}\big(z_k\big)\Big]\hat{G}^{(N)}_{i_1\ldots l\ldots i_N\, j_1\ldots j_N}\big(z_1\ldots z_N, z_1'\ldots z_N'\big) \\
&\quad = \pm\mathrm{i}\hbar\sum_{lmn}\int_{\mathcal{C}}\mathrm{d}\bar{z}\, w_{i_k lmn}\big(z_k,\bar{z}\big)\hat{G}^{(N+1)}_{i_1\ldots m\ldots i_N n\, j_1\ldots j_N m}\big(z_1\ldots z_N,\bar{z},z_1'\ldots z_N',\bar{z}\big) \\
&\qquad +\sum_p\big(\pm\big)^{k+p}\delta_{i_k j_p}\delta_{\mathcal{C}}\big(z_k,z_p'\big) \\
&\qquad \hat{G}^{(N-1)}_{i_1\ldots\not{i_k}\ldots i_N\, j_1\ldots\not{j_p}\ldots j_N}\big(z_1\ldots\not{z}_k\ldots z_N, z_1'\ldots\not{z}_p'\ldots z_N'\big),
\end{aligned}
\tag{57}
$$

$$
\begin{aligned}
&\sum_l\hat{G}^{(N)}_{i_1\ldots i_N\, j_1\ldots l\ldots j_N}\big(z_1\ldots z_N, z_1'\ldots z_N'\big)\Big[-\mathrm{i}\hbar\frac{\overleftarrow{\mathrm{d}}}{\mathrm{d}z_k'}\delta_{l j_k}-h_{l j_k}\big(z_k'\big)\Big] \\
&\quad = \pm\mathrm{i}\hbar\sum_{lmn}\int_{\mathcal{C}}\mathrm{d}\bar{z}\,\hat{G}^{(N+1)}_{i_1\ldots i_N n\, j_1\ldots l\ldots j_N m}\big(z_1\ldots z_N,\bar{z},z_1'\ldots z_N',\bar{z}\big)w_{lm j_k n}\big(\bar{z},z_k'\big) \\
&\qquad +\sum_p\big(\pm\big)^{k+p}\delta_{i_p j_k}\delta_{\mathcal{C}}\big(z_p,z_k'\big) \\
&\qquad \hat{G}^{(N-1)}_{i_1\ldots\not{i_p}\ldots i_N\, j_1\ldots\not{j_k}\ldots j_N}\big(z_1\ldots\not{z}_p\ldots z_N, z_1'\ldots\not{z}_k'\ldots z_N'\big).
\end{aligned}
\tag{58}
$$

The expectation value of the operator $\hat{G}^{(N)}$ in the initial state $\Psi_0$ yields the $N$-particle Green function $G^{(N)}$,

$$
G^{(N)} = \big\langle\Psi_0\big|\hat{G}^{(N)}\big|\Psi_0\big\rangle.
\tag{59}
$$

Note that in Eqs. (57) and (58), the bare interaction is written as a two-time quantity—a generalization that would become important e.g. in the context of retarded and advanced relativistic potentials. However, in this work, the bare interaction is always considered single-time-dependent, i.e.,

$$
w_{ijkl}\big(z_1,z_2\big) = \delta_{\mathcal{C}}\big(z_1,z_2\big)w_{ijkl}\big(z_1\big).
\tag{60}
$$

Nevertheless, the two-time structure of $w$ is often used for the illustration via Feynman diagrams (see Section 2.8).

The equations of motion for the Green functions are directly generated from the equations for the underlying operators by taking the expectation value, which corresponds to replacing all correlator operators in Eqs. (57) and (58) by the respective Green functions,

$$\hat{G}^{(N)} \longrightarrow G^{(N)}. \tag{61}$$

These mutually coupled equations form a hierarchy, the *Martin–Schwinger hierarchy* [126]. The solution of the full hierarchy gives access to all observables of the studied system and, by virtue of the connections to the $(N-1)$-particle and $(N+1)$-particle spaces, also spectral information is available. Thus, as a subset, the solution of the hierarchy incorporates the solution of the $N$-particle Schrödinger equation. Unfortunately and as expected, the effort for the full solution of the hierarchy also scales exponentially with the particle number. For the one-particle Green function $G^{(1)}$, which will be simply called the Green function $G$ in the following, the equations of motion, the *Keldysh–Kadanoff–Baym equations* (KBE), read

$$\sum_l \left[ i\hbar \frac{d}{dz_1} \delta_{il} - h_{il}(z_1) \right] G_{lj}(z_1, z_2) \tag{62}$$
$$= \delta_{\mathcal{C}}(z_1, z_2) \delta_{ij} \pm i\hbar \sum_{lmn} w_{ilnm}(z_1) G^{(2)}_{mnjl}(z_1, z_1, z_2, z_{1+}),$$

$$\sum_l G_{il}(z_1, z_2) \left[ -i\hbar \frac{\overleftarrow{d}}{dz_2} \delta_{lj} - h_{lj}(z_2) \right] \tag{63}$$
$$= \delta_{\mathcal{C}}(z_1, z_2) \delta_{ij} \pm i\hbar \sum_{lmn} G^{(2)}_{inlm}(z_1, z_{2-}, z_2, z_2) w_{lmnj}(z_2).$$

Note that the short-hand notation $z_\pm := z \pm \epsilon$ ($\epsilon \to +0$) has been introduced here to facilitate the correct ordering of the operators under $\hat{\mathcal{T}}_{\mathcal{C}}$. One notices that even the determination of the one-particle Green functions requires the solution of all other hierarchy equations as well, due to the coupling to the two-particle Green function (which, in turn couples to the three-particle Green function, and so on).

### 2.6. Definition of the selfenergy

To decouple the Martin–Schwinger hierarchy, approximations are necessary. This requires to find a functional relation of $G^{(n)}$ in terms of $G^{(n-1)}$ that is based on physical considerations about the dominant processes. Alternatively, one can apply perturbation theory in terms of the particle interaction. If the knowledge of the single-particle Green function is sufficient for the physical problem at hand, it is suitable to introduce the so-called single-particle *selfenergy* $\Sigma$, which allows one to (formally) decouple the time-evolution of the Green function from those of the $(N > 1)$-particle Green functions and obtain a closed equation for the one-particle Green function. The selfenergy is implicitly defined as

$$\pm \, \mathrm{i}\hbar \sum_{lmn} w_{ilnm}(z_1) G_{mnjl}^{(2)}(z_1, z_1, z_2, z_{1+}) =: \sum_l \int_{z_3} \Sigma_{il}(z_1, z_3) G_{lj}(z_3, z_2) \,, \tag{64}$$

$$\pm \, \mathrm{i}\hbar \sum_{lmn} G_{inlm}^{(2)}(z_1, z_{2-}, z_2, z_2) w_{lmnj}(z_1) =: \sum_l \int_{z_3} G_{il}(z_1, z_3) \Sigma_{lj}(z_3, z_2) \,. \tag{65}$$

With this, Eqs. (62) and (63) transform into

$$\sum_l \left[ \mathrm{i}\hbar \frac{\mathrm{d}}{\mathrm{d}z_1} \delta_{il} - h_{il}(z_1) \right] G_{lj}(z_1, z_2) \tag{66}$$
$$= \delta_{\mathcal{C}}(z_1, z_2) \delta_{ij} + \sum_l \int_{z_3} \Sigma_{il}(z_1, z_3) G_{lj}(z_3, z_2) \,,$$

$$\sum_l G_{il}(z_1, z_2) \left[ -\mathrm{i}\hbar \frac{\overleftarrow{\mathrm{d}}}{\mathrm{d}z_2} \delta_{lj} - h_{lj}(z_2) \right] \tag{67}$$
$$= \delta_{\mathcal{C}}(z_1, z_2) \delta_{ij} + \sum_l \int_{z_3} G_{il}(z_1, z_3) \Sigma_{lj}(z_3, z_2) \,.$$

These equations contain the two main quantities in Green functions theory, both depending on two contour times $z_1, z_2$: the (single-particle) selfenergy $\Sigma(z_1, z_2)$ (which is a functional of $G$) and the (single-particle) Green function $G(z_1, z_2)$ itself. Before turning to the self-consistent determination of $\Sigma[G]$ in Sec. 2.8, and several approximation strategies thereof in Sec. 4, a mapping technique for single-particle contour quantities onto real-time quantities is detailed in Section 2.7.

## 2.7. Keldysh–Kadanoff–Baym equations (KBE)

For the actual computation of expressions containing integrals and products of contour quantities, a mapping to ordinary real-time quantities has to be used. A suitable technique has been provided by *Langreth* and *Wilkins* [127]. Since, in this work, only single-particle correlators like the (single-particle) Green function and selfenergy are of concern, the following technique will only deal with terms of the form [cf. Eq. (55)],

$$k(z_1, z_2) = \left\langle \Psi_0 \middle| \hat{\mathcal{T}}_{\mathcal{C}} \left\{ \hat{O}_1(z_1) \hat{O}_2(z_2) \right\} \middle| \Psi_0 \right\rangle, \tag{68}$$

with the restriction that the operators have to obey

$$\hat{O}_- = \hat{O}_+\,, \tag{69}$$

i.e., they have the same values for contour arguments on the upper and lower branch. The appearance of the contour-ordering operator $\hat{\mathcal{T}}_{\mathcal{C}}$ in Eq. (68) suggest to split $k$ into

$$
\begin{aligned}
k\big(z_1, z_2\big) \\
= \delta_{\mathcal{C}}\big(z_1, z_2\big)k^{\delta}\big(z_1\big) + \Theta_{\mathcal{C}}\big(z_1, z_2\big)k^{>}\big(z_1, z_2\big) + \Theta_{\mathcal{C}}\big(z_2, z_1\big)k^{<}\big(z_1, z_2\big)\,,
\end{aligned} \tag{70}
$$

with

$$k^{>}\big(z_1, z_2\big) = \Big\langle \Psi_0 \Big| \hat{O}_1\big(z_1\big)\hat{O}_2\big(z_2\big) \Big| \Psi_0 \Big\rangle\,, \tag{71}$$

$$k^{<}\big(z_1, z_2\big) = \pm\Big\langle \Psi_0 \Big| \hat{O}_2\big(z_2\big)\hat{O}_1\big(z_1\big) \Big| \Psi_0 \Big\rangle\,, \tag{72}$$

where the $\pm$ stands for bosonic/fermionic operators. Both functions, less and greater, obey

$$
\begin{aligned}
k^{\gtrless}\big(z_{1+}, z_2\big) = k^{\gtrless}\big(z_{1-}, z_2\big)\,, \\
k^{\gtrless}\big(z_1, z_{2+}\big) = k^{\gtrless}\big(z_1, z_{2-}\big)\,.
\end{aligned} \tag{73}
$$

Therefore, only two linearly independent quantities remain and it is thus natural to define the so-called real-time less and greater Keldysh components

$$k^{>}\big(t_1, t_2\big) := k\big(t_{1+}, t_{2-}\big) = k^{>}\big(t_{1+}, t_{2-}\big)\,, \tag{74}$$

$$k^{<}\big(t_1, t_2\big) := k\big(t_{1-}, t_{2+}\big) = k^{<}\big(t_{1-}, t_{2+}\big) \tag{75}$$

and the $\delta$-component

$$
\begin{aligned}
k^{\delta}\big(t_1\big) := k\big(t_{1-}, t_{1-}\big) = k\big(t_{1+}, t_{1+}\big) \\
= k^{\delta}\big(t_{1+}, t_{1+}\big) = k^{\delta}\big(t_{1-}, t_{1-}\big)\,,
\end{aligned} \tag{76}
$$

where $t_{1/2\pm}$ are the projections on the backward/forward branch of the contour, and the relations are depicted in Fig. 3. For convenience, two more (redundant) components, the retarded and advanced component, can be defined as

$$k^{\mathcal{R}}\big(t_1, t_2\big) = \delta\big(t_1, t_2\big)k^{\delta}\big(t_1\big) + \Theta\big(t_1, t_2\big)\big[k^{>}\big(t_1, t_2\big) - k^{<}\big(t_1, t_2\big)\big]\,, \tag{77}$$

$$k^{\mathcal{A}}\big(t_1, t_2\big) = \delta\big(t_1, t_2\big)k^{\delta}\big(t_1\big) + \Theta\big(t_2, t_1\big)\big[k^{<}\big(t_1, t_2\big) - k^{>}\big(t_1, t_2\big)\big]\,. \tag{78}$$

With these components, the real-time expressions for two common concatenations of Keldysh functions, i.e., functions satisfying Eqs. (70) and (73), the convolution and the

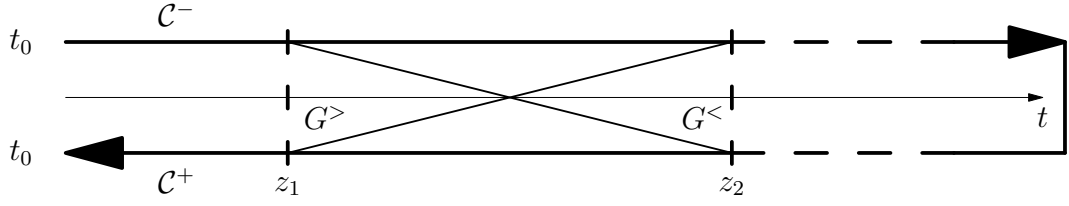

**Figure 3.** Subordinated Green functions on $\mathcal{C}$ with the forward branch $\mathcal{C}^-$ and the backward branch $\mathcal{C}^+$. The positions of the two time arguments of $G\left(z_1, z_2\right)$ for the $\gtrless$-components, which can lie an both parts of the contour, are depicted.

product, can be worked out. For the convolution

$$c\left(z_1, z_2\right) = \int_{\mathcal{C}} \mathrm{d}z_3 \, a\left(z_1, z_3\right) b\left(z_3, z_2\right), \tag{79}$$

one has

$$c^{\gtrless}\left(t_1, t_2\right) = \int_{t_0}^{t_1} \mathrm{d}t_3 \, a^{\mathcal{R}}\left(t_1, t_3\right) b^{\gtrless}\left(t_3, t_2\right) + \int_{t_0}^{t_2} \mathrm{d}t_3 \, a^{\gtrless}\left(t_1, t_3\right) b^{\mathcal{A}}\left(t_3, t_2\right) \tag{80}$$

and

$$c^{\mathcal{R}}\left(t_1, t_2\right) = \int_{t_2}^{t_1} \mathrm{d}t_3 \, a^{\mathcal{R}}\left(t_1, t_3\right) b^{\mathcal{R}}\left(t_3, t_2\right), \tag{81}$$

$$c^{\mathcal{A}}\left(t_1, t_2\right) = \int_{t_1}^{t_2} \mathrm{d}t_3 \, a^{\mathcal{A}}\left(t_1, t_3\right) b^{\mathcal{A}}\left(t_3, t_2\right). \tag{82}$$

For the product of type

$$c\left(z_1, z_2\right) = a\left(z_1, z_2\right) b\left(z_2, z_1\right), \tag{83}$$

with $a^{\delta} = 0 = b^{\delta}$, one arrives at

$$c^{\gtrless}\left(t_1, t_2\right) = a^{\gtrless}\left(t_1, t_2\right) b^{\lessgtr}\left(t_2, t_1\right) \tag{84}$$

and

$$\begin{aligned}
c^{\mathcal{R}/\mathcal{A}}\left(t_1, t_2\right) &= a^{\mathcal{R}/\mathcal{A}}\left(t_1, t_2\right) b^{<}\left(t_2, t_1\right) + a^{<}\left(t_1, t_2\right) b^{\mathcal{A}/\mathcal{R}}\left(t_2, t_1\right) \\
&= a^{\mathcal{R}/\mathcal{A}}\left(t_1, t_2\right) b^{>}\left(t_2, t_1\right) + a^{>}\left(t_1, t_2\right) b^{\mathcal{A}/\mathcal{R}}\left(t_2, t_1\right),
\end{aligned} \tag{85}$$

while for the product of type

$$c\left(z_1, z_2\right) = a\left(z_1, z_2\right) b\left(z_1, z_2\right), \tag{86}$$

with $a^{\delta} = 0 = b^{\delta}$, one has

$$c^{\gtrless}\left(t_1, t_2\right) = a^{\gtrless}\left(t_1, t_2\right) b^{\gtrless}\left(t_1, t_2\right) \tag{87}$$

and

$$c^{\mathcal{R}/\mathcal{A}}(t_1, t_2) = a^{\mathcal{R}/\mathcal{A}}(t_1, t_2) b^<(t_1, t_2) + a^>(t_1, t_2) b^{\mathcal{R}/\mathcal{A}}(t_1, t_2)$$
$$= a^{\mathcal{R}/\mathcal{A}}(t_1, t_2) b^>(t_1, t_2) + a^<(t_1, t_2) b^{\mathcal{R}/\mathcal{A}}(t_1, t_2). \tag{88}$$

With these definitions, the KBE in component representation read

$$\sum_l \left[ i\hbar \frac{d}{dt_1} \delta_{il} - h_{il}(t_1) \right] G_{lj}^{\gtrless}(t_1, t_2) \tag{89}$$

$$= \sum_l \int_{t_0}^{t_1} dt_3 \, \Sigma_{il}^{\mathcal{R}}(t_1, t_3) G_{lj}^{\gtrless}(\bar{t}, t') + \sum_l \int_{t_0}^{t_2} dt_3 \, \Sigma_{il}^{\gtrless}(t_1, t_3) G_{lj}^{\mathcal{A}}(\bar{t}, t')$$

$$= \sum_l \int_{t_0}^{t_1} dt_3 \left( \Sigma_{il}^>(t_1, t_3) - \Sigma_{il}^<(t_1, t_3) \right) G_{lj}^{\gtrless}(\bar{t}, t')$$

$$+ \sum_l \int_{t_0}^{t_2} dt_3 \, \Sigma_{il}^{\gtrless}(t_1, t_3) \left( G_{lj}^<(\bar{t}, t') - G_{lj}^>(\bar{t}, t') \right)$$

and

$$\sum_l G_{il}^{\gtrless}(t_1, t_2) \left[ -i\hbar \frac{\overleftarrow{d}}{dt_2} \delta_{lj} - h_{lj}(t_2) \right] \tag{90}$$

$$= \sum_l \int_{t_0}^{t_1} dt_3 \, G_{il}^{\mathcal{R}}(t_1, t_3) \Sigma_{lj}^{\gtrless}(t_3, t_2) + \sum_l \int_{t_0}^{t_2} dt_3 \, G_{il}^{\gtrless}(t_1, t_3) \Sigma_{lj}^{\mathcal{A}}(t_3, t_2)$$

$$= \sum_l \int_{t_0}^{t_1} dt_3 \left( G_{il}^>(t_1, t_3) - G_{il}^<(t_1, t_3) \right) \Sigma_{lj}^{\gtrless}(t_3, t_2)$$

$$+ \sum_l \int_{t_0}^{t_2} dt_3 \, G_{il}^{\gtrless}(t_1, t_3) \left( \Sigma_{lj}^<(t_3, t_2) - \Sigma_{lj}^>(t_3, t_2) \right).$$

Note the missing $\delta_{\mathcal{C}}$ in the $\gtrless$-components of Eqs. (89) and (90) compared to Eqs. (66) and (67), which, as it is a time-diagonal function, only enters the retarded and advanced components.

## 2.8. Basic equations for deriving selfenergy approximations

In this brief section, a coupled set of equations of motions for five dynamical quantities, two of which are the Green function and the selfenergy, is summarized. It has been first presented by *Lars Hedin* in 1965 [81] in association with the *GW* method, which will be discussed in more detail in Section 5. If solved exactly, the set of Hedin's equations yields the same $G$ as the solution of the Martin–Schwinger hierarchy[ix] and provides

---

[ix]Although, to the knowledge of the authors, no strict proof exists that shows the equivalence of the solutions for $G$ of Hedin's equation versus that from the Martin–Schwinger hierarchy, both approaches agree for all practically relevant approximations.

multiple starting points for approximate solution schemes. To determine the solution for $G\big(z_1, z_2\big)$, its equations of motion, the KBE, cf. Eqs. (66) and (67), have to be solved. This can be either done directly in their differential form, or in the integral form, which reads

$$G_{ij}\big(z_1, z_2\big) = G_{ij}^{(0)}\big(z_1, z_2\big) + \tag{91}$$
$$+ \int_{\mathcal{C}} \mathrm{d}z_3 \mathrm{d}z_4 \sum_{mn} G_{im}^{(0)}\big(z_1, z_3\big) \Sigma_{mn}\big(z_3, z_4\big) G_{nj}\big(z_4, z_2\big),$$

with the reference Green function $G^{(0)}$ that is the solution of the ideal pair of equations

$$\sum_{l} \left[ \mathrm{i}\hbar \frac{\mathrm{d}}{\mathrm{d}z_1} \delta_{il} - h_{il}\big(z_1\big) \right] G_{lj}\big(z_1, z_2\big) = \delta_{\mathcal{C}}\big(z_1, z_2\big) \delta_{ij}, \tag{92}$$

$$\sum_{l} G_{il}\big(z_1, z_2\big) \left[ -\mathrm{i}\hbar \frac{\overleftarrow{\mathrm{d}}}{\mathrm{d}z_2} \delta_{lj} - h_{lj}\big(z_2\big) \right] = \delta_{\mathcal{C}}\big(z_1, z_2\big) \delta_{ij}. \tag{93}$$

Note that $G^{(0)}$ does not refer to zero particles, but to the property that is of zeroth order with respect to the interaction $w$.

At this point, a more compact notation is introduced that focuses on the time structure of the upcoming quantities and uses the corresponding Feynman diagrams to exemplify the underlying connections. Thereto, the basis indices are skipped and the contour-time arguments are replaced by bare numbers ($z_1 \mapsto 1$). The occuring integrations are implicitly determined by times, the corresponding vertices of which are fully connected (i.e. two Green functions and one interaction or an equivalent connectivity state). As it is usually done in the context of Feynman diagrams the bare interaction is used as a two-time quantity, cf. Eq. (60). This notation will be used extensively in Section 4 and 5 to simplify the derivations of the selfenergy approximations. For Eq. (91) this notation reads as follows,

$$G\big(1, 2\big) = G^{(0)}\big(1, 2\big)$$
$$+ G^{(0)}\big(1, 3\big) \Sigma\big(3, 4\big) G\big(4, 2\big)$$

$$\tag{94}$$

Equation (91) [and (94), respectively] is referred to as the *Dyson equation* for the one-particle Green function. Comparing the KBE, cf. Eqs. (66) and (67), to the Dyson equation, cf. Eq. (91), the question may arise whether the solution of one or the other is numerically more favorable. Realizing that the determination of $G^{(0)}$ via Eqs. (92) and (93) is of $\mathcal{O}\left(N_{\mathrm{b}}^2\right)$ and $\mathcal{O}\left(N_{\mathrm{t}}^2\right)$, whereas the solution of the full $G$ via Eq. (91) involves two separable time integrations and matrix multiplications, it is of order $\mathcal{O}\left(N_{\mathrm{b}}^3\right)$ and $\mathcal{O}\left(N_{\mathrm{t}}^3\right)$, which is the same scaling as the solution of the KBE, although the prefactors are higher for the Dyson equation. In Sec. 4, though, it will be shown that in an expansion of $\Sigma$ and, particularly, $G$ with respect to the order of the interaction, only the Dyson equation allows for a strict order-per-order expansion scheme.

Both, the Dyson equation and the KBE, depend on the knowledge of the selfenergy $\Sigma$. It can be decomposed into two parts[x],

$$\Sigma_{ij}\left(z_1, z_2\right) = \Sigma_{ij}^{\mathrm{H}}\left(z_1, z_2\right) + \Sigma_{ij}^{\mathrm{xc}}\left(z_1, z_2\right), \tag{95}$$

with the (static) time-diagonal Hartree part, $\Sigma^{\mathrm{H}}$,

$$\Sigma_{ij}^{\mathrm{H}}\left(z_1, z_2\right) = \pm \mathrm{i}\hbar\delta_{\mathcal{C}}\left(z_1, z_2\right) \sum_{mn} w_{mijn}\left(z_1\right) G_{nm}\left(z_1, z_{1^+}\right), \tag{96}$$

and the exchange–correlation part, $\Sigma^{\mathrm{xc}}$. **To determine $\boldsymbol{\Sigma^{\mathrm{xc}}}$, there exist two commonly used equivalent formally exact approaches. Approach I regards the selfenergy as a functional of the bare interaction**, $\Sigma^{\mathrm{xc}} = \Sigma^{\mathrm{xc}}\left[w\right]$, **whereas approach II treats it as a functional of the screened interaction $\boldsymbol{W}$, i.e. $\boldsymbol{\Sigma^{\mathrm{xc}} = \Sigma^{\mathrm{xc}}\left[W\right]}$, where the screening arises from the dynamic redistribution of the other particles in the system.** Both techniques rely on a so-called *vertex function*, named either $\Lambda$ or $\Gamma$, in the two cases, which involves the derivatives of either $\Sigma$ or $\Sigma^{\mathrm{xc}}$ with respect to $G$ to determine the vertex function and, with it, $\Sigma^{\mathrm{xc}}$. With the coupled equations for $\Sigma^{\mathrm{xc}}$ and the vertex function, both approaches yield a systematic means to generate all selfenergy terms by iteration. We now summarize both approaches.

I.) **With the bare interaction**, $w$, one has

$$\Sigma_{ij}^{\mathrm{xc}}\left(z_1, z_2\right) = \mathrm{i}\hbar \sum_{mpq} w_{ipqm}\left(z_1\right) \int_{\mathcal{C}} \mathrm{d}z_3 \sum_n G_{mn}\left(z_1, z_3\right) \Lambda_{nqpj}\left(z_3, z_2, z_1\right). \tag{97}$$

The *bare vertex* $\Lambda$ is self-consistently given as the solution of

$$\begin{aligned}
\Lambda_{ijkl}\left(z_1, z_2, z_3\right) &= \delta_{\mathcal{C}}\left(z_1, z_{2^+}\right)\delta_{\mathcal{C}}\left(z_3, z_2\right)\delta_{ik}\delta_{jl} \\
&+ \int_{\mathcal{C}} \mathrm{d}z_4 \mathrm{d}z_5 \sum_{mn} \frac{\delta \Sigma_{il}\left(z_1, z_2\right)}{\delta G_{mn}\left(z_4, z_5\right)} \int_{\mathcal{C}} \mathrm{d}z_6 \sum_p G_{mp}\left(z_4, z_6\right) \\
&\quad \int_{\mathcal{C}} \mathrm{d}z_7 \sum_q G_{qn}\left(z_7, z_5\right) \Lambda_{pjkq}\left(z_6, z_7, z_3\right).
\end{aligned} \tag{98}$$

---

[x]The same decomposition is used in density-functional theory

In the compact notation this set of equations becomes,

$$\Sigma(1,2) = \pm i\hbar\delta(1,2)w(1,3)G(3,3^+)$$
$$+ i\hbar w(1,3)G(1,4)\Lambda(4,2,3)$$

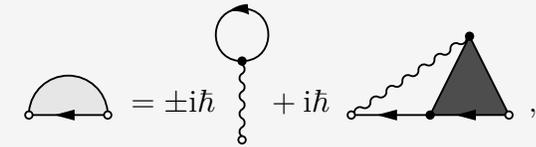

(99)

$$\Lambda(1,2,3) = \delta(1,2^+)\delta(3,2)$$
$$+ \frac{\delta\Sigma(1,2)}{\delta G(4,5)}G(4,6)G(7,5)\Lambda(6,7,3)$$

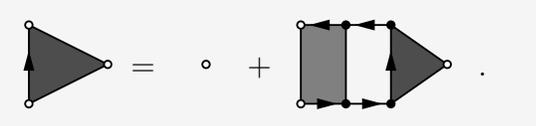

(100)

II. Using the **screened interaction, $W$**, as a basis for the expansion, the exchange–correlation selfenergy reads, cf. Eq. (97),

$$\Sigma_{ij}^{\mathrm{xc}}(z_1, z_2) = i\hbar \int_{\mathcal{C}} \mathrm{d}z_3 \sum_{mpq} W_{ipqm}(z_1, z_3) \times \tag{101}$$
$$\int_{\mathcal{C}} \mathrm{d}z_4 \sum_n G_{mn}(z_1, z_4)\Gamma_{nqpj}(z_4, z_2, z_3),$$

where $W$ obeys

$$W_{ijkl}(z_1, z_2) = W_{ijkl}^{\mathrm{bare}}(z_1, z_2) + W_{ijkl}^{\mathrm{ns}}(z_1, z_2), \tag{102}$$

with the bare interaction

$$W_{ijkl}^{\mathrm{bare}}(z_1, z_2) = \delta_{\mathcal{C}}(z_1, z_2)w_{ijkl}(z_1), \tag{103}$$

and the non-singular (ns) induced part

$$W_{ijkl}^{\mathrm{ns}}(z_1, z_2) = \sum_{mn} w_{imnl}(z_1) \int_{\mathcal{C}} \mathrm{d}z_3 \sum_{pq} P_{nqpm}(z_1, z_3)W_{pjkq}(z_3, z_2). \tag{104}$$

The occurring polarizability $P$ is given by

$$P_{ijkl}(z_1, z_2) = \pm i\hbar \int_{\mathcal{C}} \mathrm{d}z_3 \sum_m G_{im}(z_1, z_3) \times \tag{105}$$
$$\int_{\mathcal{C}} \mathrm{d}z_4 \sum_n G_{nl}(z_4, z_1)\Gamma_{mjkn}(z_3, z_4, z_2).$$

The screened vertex function $\Gamma$—which $\Sigma^{\mathrm{xc}}$ and $P$ depend on—is governed by

$$\Gamma_{ijkl}(z_1, z_2, z_3) = \delta_{\mathcal{C}}(z_1, z_{2^+})\delta_{\mathcal{C}}(z_3, z_2)\delta_{ik}\delta_{jl} + \tag{106}$$

$$+ \int_{\mathcal{C}} \mathrm{d}z_4 \mathrm{d}z_5 \sum_{mn} \frac{\delta\Sigma_{il}^{\mathrm{xc}}(z_1, z_2)}{\delta G_{mn}(z_4, z_5)} \int_{\mathcal{C}} \mathrm{d}z_6 \sum_p G_{mp}(z_4, z_6)$$

$$\int_{\mathcal{C}} \mathrm{d}z_7 \sum_q G_{qn}(z_7, z_5)\Gamma_{pjkq}(z_6, z_7, z_3).$$

To summarize, Hedin's equations are repeated in the compact notation,

$$\Sigma(1,2) = \pm i\hbar\delta(1,2)w(1,3)G(3,3^+)$$
$$+ i\hbar W(1,3)G(1,4)\Gamma(4,2,3)$$

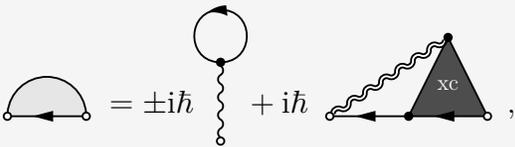

$$(107)$$

$$W(1,2) = w(1,2)$$
$$+ w(1,3)P(3,4)W(4,2)$$

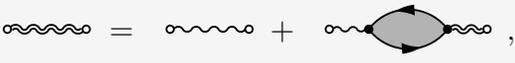

$$(108)$$

$$P(1,2) = \pm i\hbar G(1,3)G(4,1)\Gamma(3,4,2)$$

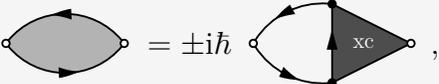

$$(109)$$

$$\Gamma(1,2,3) = \delta(1,2^+)\delta(3,2)$$
$$+ \frac{\delta\Sigma^{\mathrm{xc}}(1,2)}{\delta G(4,5)}G(4,6)G(7,5)\Gamma(6,7,3)$$

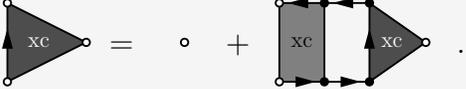

$$(110)$$

## 2.9. Summary of selfenergy approximations

We now list the selfenergies that will be discussed in this paper and briefly summarize their respective strengths and weaknesses. For approach I.) that starts with the bare interaction, Eq. (97), we will consider:

– The **particle–particle $T$-matrix** approximation (TPP)
  The TPP selfenergy sums up the diagrams of the Born series. This process is computationally expensive, which, therefore, restricts the applicability range of the approximation to systems of moderate basis size. The TPP is a moderate- to strong-coupling approximaton, that becomes exact in the limit of low (large) density. It, thus, performes best away from half-filling.

– The **particle–hole $T$-matrix** approximation (TPH)

The TPH selfenergy sums up a series of particle–hole diagrams, which is of comparable numercial complexity as the TPP. It is specifically designed to describe systems around half-filling, i.e., where the particle and hole densities are close to each other. For these cases, it provides accurate results for moderate to strong interaction strengths. For the application to electronic Hubbard systems, the TPH will later be called **electron–hole $T$-matrix** approximation (TEH).

For approach II.) that starts with the screened interaction, Eq. (101), we will consider:

– The **Hartree–Fock** (HF) approximation

The HF selfenergy results from a perturbative expansion up to first order in the interaction. It is equivalent to a description on the mean-field level. Due to its simplicity, it is numerically easy to use and applicable to large systems and long simulation times. However, it only gives accurate results in the weak-coupling regime.

– The **second-order (Born)** approximation (SOA)

The SOA selfenergy consists of all diagrams up to second order in the interaction. It provides the easiest way to include correlation effects in a NEGF calculation. Due to its basic structure, the combination with the GKBA (see Section 2.10) leads to a favorable numerical scaling, which opens its applicability to a wide range of systems. The SOA gives accurate result for weak to moderate coupling strengths.

– The **third-order** approximation (TOA)

The TOA selfenergy combines all possible selfenergy contributions up to third order in the interaction. It is much more involved than the SOA rendering the simulations numerically costly. Thus, the applicability range of the TOA is restricted to problems with a moderate basis size. In return, the TOA remains accurate even in the regime of moderate to strong coupling.

– The **$GW$** approximation (GWA)

The $GW$ selfenergy provides the easiest way to decribe dynamical-screening effects by summing up the polarization-bubble diagram series. The resummation process is computationally demanding which narrows the class of the systems that can be treated, although there are some scaling advantages for problems that require a general (i.e. with non-diagonal interaction matrix) basis set. The GWA can be considered a moderate- to strong-coupling approximation, which is particularly accurate around half filling, where the contributions of particles and holes coincide.

| Abbreviation | Selfenergy | |
|:---:|:---|:---:|
| HF | Hartree–Fock approximation: $\Sigma = \Sigma^{\mathrm{H}} + \Sigma^{\mathrm{F}}$ | Sec. 4.1 |
| SOA | Second-order approximation: $\Sigma = \Sigma^{(2)}$ | Sec. 4.2 |
| TOA | Third-order approximation: $\Sigma = \Sigma^{(3)}$ | Sec. 4.3 |
| GWA | $GW$ approximation : $\Sigma = \Sigma^{GW}$ | Sec. 5.2 |
| TPP | Particle–particle $T$-matrix approximation : $\Sigma = \Sigma^{T^{\mathrm{pp}}}$ | Sec. 5.3 |
| TPH | Particle–hole $T$-matrix approximation : $\Sigma = \Sigma^{T^{\mathrm{ph}}}$ | Sec. 5.3 |
| FLEX | Fluctuating-exchange approximation: $\Sigma = \Sigma^{\mathrm{FLEX}}$ | Sec. 5.5 |

**Table 1.** Main selfenergy approximations, abbreviations and section where the approximation is being introduced and discussed.

Finally, a combination of some of the above results leads to:

– The **fluctuating-exchange** approximation (FLEX)
The FLEX selfenergy merges the diagram series of the TPP, the TPH and the GWA. It, therefore, has the highest computational demands of the presented selfenergy approximations. By combining the advantages of its ingredients, it is applicable for all filling factors and up to strong interaction strengths.

An overview of the selfenergies and the abbreviations that are being used is given in Table 1. The detailed derivation of these expressions will be given later in Section 4 and 5. A thorough comparison of the respective performance of the presented selfenergy approximations is given in Section 3.

## 2.10. The generalized Kadanoff–Baym Ansatz

To compute the time-dependent single-particle Green functions, either the KBE, cf. Eqs. (66) and (67), or the Dyson equation, cf. Eq. (91) have to be solved, which both scale cubically with respect to the time duration. An approximate way to transform the scaling to a quadratic one, has been proposed by Lipavský *et al.* and was named *generalized Kadanoff–Baym ansatz* (GKBA), for details about the derivation see Ref. [128] by Lipavský *et al.* and Refs. [121, 123, 124, 129]. The approximation starts from an exact reformulation of the Dyson equation, the less-component of which reads

$$
\begin{aligned}
G_{ij}^<\big(t_1, t_2\big) = &-\mathrm{i}\hbar \sum_k G_{ik}^{\mathcal{R}}\big(t_1, t_2\big) G_{kj}^<\big(t_2, t_2\big) \\
&+ \int_{t_2}^{t_1} \mathrm{d}t_3 \int_{t_0}^{t_2} \mathrm{d}t_4 \sum_{kl} G_{ik}^{\mathcal{R}}\big(t_1, t_3\big) \Sigma_{kl}^<\big(t_3, t_4\big) G_{lj}^{\mathcal{A}}\big(t_4, t_2\big) \\
&+ \int_{t_2}^{t_1} \mathrm{d}t_3 \int_{t_0}^{t_2} \mathrm{d}t_4 \sum_{kl} G_{ik}^{\mathcal{R}}\big(t_1, t_3\big) \Sigma_{kl}^{\mathcal{R}}\big(t_3, t_4\big) G_{lj}^<\big(t_4, t_2\big) \\
&+ \mathrm{i}\hbar G_{ik}^<\big(t_1, t_1\big) G_{kj}^{\mathcal{A}}\big(t_1, t_2\big) \\
&- \int_{t_0}^{t_1} \mathrm{d}t_3 \int_{t_1}^{t_2} \mathrm{d}t_4 \sum_{kl} G_{ik}^{\mathcal{R}}\big(t_1, t_3\big) \Sigma_{kl}^<\big(t_3, t_4\big) G_{lj}^{\mathcal{A}}\big(t_4, t_2\big) \\
&- \int_{t_0}^{t_1} \mathrm{d}t_3 \int_{t_1}^{t_2} \mathrm{d}t_4 \sum_{kl} G_{ik}^<\big(t_1, t_3\big) \Sigma_{kl}^{\mathcal{A}}\big(t_3, t_4\big) G_{lj}^{\mathcal{A}}\big(t_4, t_2\big),
\end{aligned}
\tag{111}
$$

and analogously for the greater component. The GKBA approximates these terms by only retaining the non-integral contributions, which can be considered a simultaneous perturbative expansion of $G$ with respect to $\Sigma$ and to the spectral structure conveyed by the off-diagonal elements of both quantities. It reads,

$$
\begin{aligned}
G_{ij}^{\gtrless}\big(t_1, t_2\big) &= -\mathrm{i}\hbar \sum_k \left\{ G_{ik}^{\mathcal{R}}\big(t_1, t_2\big) G_{kj}^{\gtrless}\big(t_2, t_2\big) + G_{ik}^{\gtrless}\big(t_1, t_1\big) G_{kj}^{\mathcal{A}}\big(t_1, t_2\big) \right\} \\
&= \sum_k A_{ik}\big(t_1, t_2\big) \left\{ \Theta\big(t_1, t_2\big) G_{kj}^{\gtrless}\big(t_2, t_2\big) + \Theta\big(t_2, t_1\big) G_{kj}^{\gtrless}\big(t_1, t_1\big) \right\},
\end{aligned}
\tag{112}
$$

where, in the second line, the spectral function

$$
A_{ij}\big(t_1, t_2\big) = \mathrm{i}\hbar \left\{ G_{ij}^>\big(t_1, t_2\big) - G_{ij}^<\big(t_1, t_2\big) \right\},
\tag{113}
$$

has been introduced. The approximated $\gtrless$-components are used in the right-hand sides of the KBE which, thereby, need to be propagated only along the time diagonal. To achieve the overall reduction to a quadratic scaling, though, the GKBA has to be accompanied by a second-order selfenergy and another approximation concerning the retarded and advanced components, which, unapproximated, obey equations of similar complexity as the original KBE, i.e., with cubic scaling. In this work, the propagators, and with that the spectral function, will be approximated on the HF level. Another possibility, employed in Ref. [130], is to use approximate correlated propagators. The GKBA has several important benefits: It preserves the causal structure of the KBE and it conserves important constants of motion, whenever the chosen selfenergy approximation does [94]. Further, it cures certain damping-induced artifacts for small systems [76, 118], an example of which will be further explored in Section 3.4. For a recent discussion, see Ref. [24].

### 2.11. Interacting initial state

To compute the time-evolution of the single-particle Green function according to Eqs. (66) and (67), the initial state represented by $G\left(t_0, t_0\right)$ has to be calculated. It is determined by the environment of the system. If the system is isolated, i.e., is described by a pure state, $G\left(t_0, t_0\right)$ is the fully interacting initial state. For a system embedded into a bath with which it exchanges particles or energy, the initial state is strongly influenced by the equilibrium between degrees of freedom of the system and the bath. Under the assumption that the interaction between both is weak and dominantly uncorrelated a suitable ensemble, for instance the canonical or grand-canonical ensemble, determines the occupation of the energy levels in the initial state of the system. For both cases of systems, whether connected to a bath or isolated, there exist several methodologies to generate the interacting initial state, some of which will be detailed in the next sections including the method of adiabatic switch-on of the interaction in Section 2.11.2, which is used throughout this work.

### 2.11.1. Extension of the contour to finite temperatures

One possibility to include the description of the interacting initial state, in equilibrium with a bath or isolated, is to augment the original contour, comprised of a forward and a backward branch $\mathcal{C}^+$, $\mathcal{C}^-$, by a "vertical" branch $\mathcal{C}^{\mathrm{M}}$ of complex time arguments ranging along the imaginary axis from $z_0$ to $z_0 - \mathrm{i}\hbar\beta$. Here $\beta$ is the inverse temperature of the bath [or equal to $\infty$ for an isolated system at zero temperature]. The reasoning behind this can be understood by considering the following observations for quantum systems in contact with an environment. The simplest way to treat the interaction of the system with the environment is statistically, i.e. by assigning bath-induced weights $w_n$ (i.e., probabilities, with $0 \leq w_n \leq 1$ and $\sum_n w_n = 1$) of finding the system in one of its eigenstates $|n\rangle$. With this, the ensemble average of an observable $\hat{O}\left(t_0\right)$ in such a mixed state is defined as

$$O\left(t_0\right) = \sum_n w_n \left\langle n \left| \hat{O}\left(t_0\right) \right| n \right\rangle. \tag{114}$$

Note that Eq. (114) is a natural generalization of a pure state $\left|n\right\rangle = \left|\Psi_k\right\rangle$, to which it reduces if $w_n = \delta_{k,n}$. With Eq. (114), the statistical density-matrix operator $\hat{\rho}$ can be defined[xi],

$$\hat{\rho} = \sum_n w_n \left|n\right\rangle\left\langle n\right|, \tag{115}$$

with which Eq. (114) can be rewritten as

$$O\left(t_0\right) = \mathrm{Tr}\left[\hat{\rho}\hat{O}\left(t_0\right)\right]. \tag{116}$$

The trace Tr is to be understood as acting on the full Fock space $\mathcal{F}_\sigma^{\mathcal{H}}$. For the grand-canonical ensemble (GCE), which describes a system which exchanges energy (characterized by inverse temperature $\beta$) and particles (characterized by the chemical potential $\mu$) with its environment, the density-matrix operator $\hat{\rho}$ reads

$$\hat{\rho} = \frac{\exp\left(-\beta\hat{H}_{\mathrm{M}}\right)}{Z^{\mathrm{GCE}}}, \tag{117}$$

with the corresponding Hamiltonian, $\hat{H}_{\mathrm{M}} = \hat{H} - \mu\hat{N}$, and the partition function $Z^{\mathrm{GCE}} = \mathrm{Tr}\left[\exp\left(-\beta\hat{H}_{\mathrm{M}}\right)\right]$. For the GCE, Eq. (116) becomes

$$O\left(t_0\right) = \frac{\mathrm{Tr}\left[\exp\left(-\beta\hat{H}_{\mathrm{M}}\right)\hat{O}\left(t_0\right)\right]}{Z^{\mathrm{GCE}}}. \tag{118}$$

With this result, Eq. (44) can be specialized to

$$O\left(z\right) = \frac{\mathrm{Tr}\left[\exp\left(-\beta\hat{H}_{\mathrm{M}}\right)\hat{\mathcal{T}}_{\mathcal{C}}\left\{\exp\left(\frac{1}{\mathrm{i}\hbar}\int_{\mathcal{C}}\mathrm{d}z_3\,\hat{H}\left(z_3\right)\hat{O}\left(z\right)\right)\right\}\right]}{\mathrm{Tr}\left[\exp\left(-\beta\hat{H}_{\mathrm{M}}\right)\right]}. \tag{119}$$

Using

$$\hat{\mathcal{T}}_{\mathcal{C}}\left\{\exp\left(\frac{1}{\mathrm{i}\hbar}\int_{\mathcal{C}}\mathrm{d}z_3\,\hat{H}\left(z_3\right)\right)\right\} = \hat{1}, \tag{120}$$

and introducing the vertical part of the contour $\mathcal{C}^{\mathrm{M}}$ running from $z_0$ to $z_0 - \mathrm{i}\hbar\beta$ with the identity

$$\exp\left(-\beta\hat{H}_{\mathrm{M}}\right) = \exp\left(-\frac{\mathrm{i}}{\hbar}\int_{\mathcal{C}^{\mathrm{M}}}\mathrm{d}z_3\,\hat{H}_{\mathrm{M}}\right), \tag{121}$$

---

[xi]The concept of the density operator was introduced by Landau and von Neumann. For a general nonequilibrium approach, see Ref. [131].

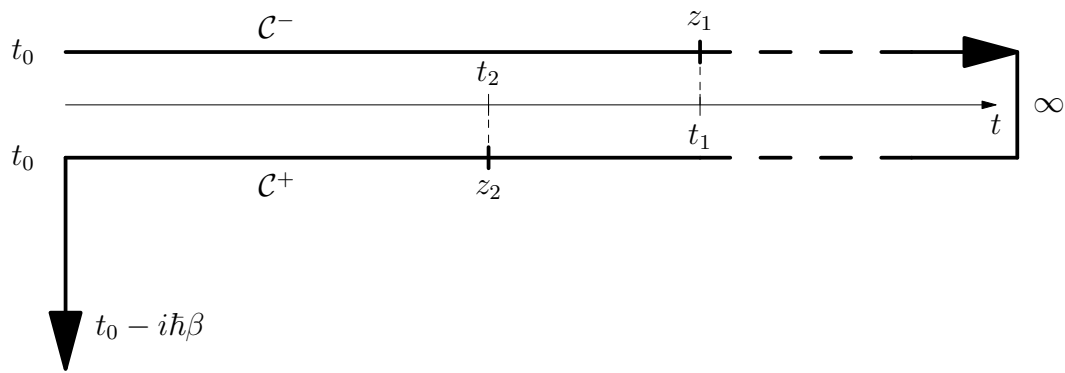

**Figure 4.** Schwinger–Keldysh contour $\mathcal{C}$ extended to the imaginary axis. The forward-branch $\mathcal{C}^-$ spans from the initial time $t_0$ to $\infty$, bends and leads back to $t_0$ along the backward $\mathcal{C}^+$-branch. Then it continues along the imaginary branch $\mathcal{C}^{\mathrm{M}}$ to $t_0 - \mathrm{i}\hbar\beta$, where $\beta$ is the inverse temperature.

we arrive at

$$O(z) = \tag{122}$$
$$\frac{\mathrm{Tr}\left[\exp\left(-\frac{\mathrm{i}}{\hbar}\int_{\mathcal{C}^{\mathrm{M}}}\mathrm{d}z_3\,\hat{H}_{\mathrm{M}}\right)\hat{\mathcal{T}}_{\mathcal{C}}\left\{\exp\left(-\frac{\mathrm{i}}{\hbar}\int_{\mathcal{C}}\mathrm{d}z_3\,\hat{H}(z_3)\right)\hat{O}(z)\right\}\right]}{\mathrm{Tr}\left[\exp\left(-\frac{\mathrm{i}}{\hbar}\int_{\mathcal{C}^{\mathrm{M}}}\mathrm{d}z_3\,\hat{H}_{\mathrm{M}}\right)\hat{\mathcal{T}}_{\mathcal{C}}\left\{\exp\left(-\frac{\mathrm{i}}{\hbar}\int_{\mathcal{C}}\mathrm{d}z_3\,\hat{H}(z_3)\right)\right\}\right]}.$$

From this structure, the contour extension idea can be directly derived. If one defines

$$\hat{H}\big|_{\mathcal{C}^{\mathrm{M}}} \equiv \hat{H}_{\mathrm{M}}, \quad \hat{O}\big|_{\mathcal{C}^{\mathrm{M}}} \equiv \hat{O}(t_0) \tag{123}$$

and redefines the contour $\mathcal{C}$ as

$$\mathcal{C} = \mathcal{C}^- \bigoplus \mathcal{C}^+ \bigoplus \mathcal{C}^{\mathrm{M}}, \tag{124}$$

so that every point on the vertical track is defined as "later" than all points on the forward and backward branches, Eq. (122) can be recast as

$$O(z) = \frac{\mathrm{Tr}\left[\hat{\mathcal{T}}_{\mathcal{C}}\left\{\exp\left(-\frac{\mathrm{i}}{\hbar}\int_{\mathcal{C}}\mathrm{d}z_3\,\hat{H}(z_3)\right)\hat{O}(z)\right\}\right]}{\mathrm{Tr}\left[\hat{\mathcal{T}}_{\mathcal{C}}\left\{\exp\left(-\frac{\mathrm{i}}{\hbar}\int_{\mathcal{C}}\mathrm{d}z_3\,\hat{H}(z_3)\right)\right\}\right]}. \tag{125}$$

The corresponding contour is depicted in Fig. 4. With this, definition (125) correctly reproduces the time-dependent expectation values in accordance with Eq. (44) for $z \in \mathcal{C}^- \bigoplus \mathcal{C}^+$ and the ensemble average for $z \in \mathcal{C}^{\mathrm{M}}$, agreeing with Eq. (118). Note though that this treatment of the system–bath interaction is only valid for times smaller than its relaxation time as the bath only directly influences the initial state and not any time-dependent excitations during the propagation [72].

### 2.11.2. Adiabatic switch-on of interactions

If one is mainly interested in the evolution of isolated systems described by a pure state, a suitable procedure is the generation of the non-interacting state of the system, which is known for most systems and a subsequent sufficiently slow ramp-up of the interaction strength from zero to the desired value. Provided the *Gell-Mann–Low theorem* holds [132], which assures the existence of some limites, and the non-interacting ground state is non-degenerate, it follows that the state of the system after switch-on of the interaction is an eigenstate of the fully interacting Hamiltonian. It remains to be checked—e.g., by comparison with other methods—that it is the ground state. Under the adiabatic-switching protocol, the Hamiltonian of Eq. (16) is replaced by

$$
\hat{H}^{\mathrm{AS}}(t) = \underbrace{\sum_{mn} h_{mn} \hat{c}_m^\dagger \hat{c}_n}_{\hat{H}^0} + \underbrace{\frac{1}{2} f^{\mathrm{AS}}(t) \sum_{mnpq} w_{mnpq} \hat{c}_m^\dagger \hat{c}_n^\dagger \hat{c}_p \hat{c}_q}_{\hat{W}^{\mathrm{AS}}(t)} \tag{126}
$$
$$
+ \underbrace{\sum_{mn} f_{mn}(t) \hat{c}_m^\dagger \hat{c}_n}_{\hat{F}(t)},
$$

where the monotonically increasing switching function $f^{\mathrm{AS}} : \mathbb{R} \longrightarrow [0,1]$ satisfies

$$
\lim_{t \to -\infty} f^{\mathrm{AS}}(t) = 0, \quad f^{\mathrm{AS}}(t) = 1, \quad \text{for } t \geq t_0. \tag{127}
$$

To achieve a high fidelity of the final state the switch-on process has to be performed slow enough and as smooth as possible. Here, the use of the function [133]

$$
f_{\mathrm{AS}}^{\tau, t_{\mathrm{H}}}(t) = \exp\left( -\frac{A_{t_{\mathrm{H}}}^\tau}{t_1/(2t_{\mathrm{H}})} \exp\left( \frac{B_{t_{\mathrm{H}}}^\tau}{t_1/(2t_{\mathrm{H}}) - 1} \right) \right),
$$
$$
B_{t_{\mathrm{H}}}^\tau = \frac{t_{\mathrm{H}}}{\tau \ln 2} - \frac{1}{2}, \qquad A_{t_{\mathrm{H}}}^\tau = \frac{\ln 2}{2} \exp\left( 2 B_{t_{\mathrm{H}}}^\tau \right), \tag{128}
$$

is superior compared to an approach based on a Fermi function, since it provides a very small relative change for values near the beginning and the end of the switch-on process. The free parameters $\tau$ and $t_{\mathrm{H}}$, the halftime, control the steepness and the duration of the switch-on. Using the adiabatic-switching methodology, the time-contour attains the form depicted in Fig. 5.

A third option to include initial correlation is via an additional collision integral or selfenergy term [24, 131, 134, 135]. For completeness, we also mention that similar

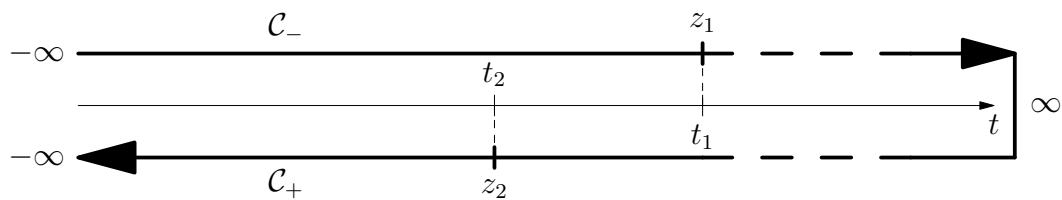

**Figure 5.** Schwinger–Keldysh contour $\mathcal{C}^{\text{AS}}$ with adiabatic switch-on of the interactions. The starting and end point is now $-\infty$.

problems arise by use of the GKBA [131].

## 3. Applications: Numerical results for fermionic lattice systems

This section discusses some applications of the approximation strategies detailed in the previous section. All simulations described in this section have been performed for spin-1/2 fermions in the Hubbard model, cf. Section 2.3, for zero temperature. After providing an overview of the algorithm for the numerical solution of the KBE in Section 3.1, results for the correlated ground state of Hubbard nano-clusters are presented in Section 3.3. The ground-state energy and spectral function are used as benchmarks to compare the performance of the selfenergy approximations listed in 2.9 for different filling factors and interaction strengths. Subsequently, in Section 3.4, the behavior of the approximation strategies in the simulation of the time-dependent response of Hubbard clusters to external excitations is studied. A special focus lies on excitations that strongly drive the system out of equilibrium. There, the occurrence of a particular weakness of selfconsistent approaches, the so-called correlation-induced damping [118], is analyzed for all approximations and it is demonstrated how it can be overcome to a large extent by application of the GKBA, in Section 3.4.

### 3.1. Algorithm for the solution of the Keldysh–Kadanoff–Baym equations (KBE)

This section gives an overview of the algorithm to calculate the solution of the KBE, cf. Eqs. (89) and (90), for spin-1/2 fermions. Both equations need to be equipped with the initial value $G^{\gtrless}(t_0, t_0)$, which are the Green functions of the—in general, correlated—initial state. These Green functions are, in turn, generated from the ones of the ideal ground state, $G^{(0),\gtrless}$, via the *adiabatic-switching method* described in Section 2.11.2. The $\gtrless$-components of the ideal Green function directly follow from the ideal one-particle density matrix $n^{(0)}$ via the relations

$$G_{ij}^{(0),<} = -\frac{1}{i\hbar} n_{ji}^{(0)}, \tag{129}$$

$$G_{ij}^{(0),>} = \frac{1}{i\hbar}\left(\delta_{ij} - n_{ji}^{(0)}\right), \tag{130}$$

where the second follows from the more general relation for an arbitrary time $t$,

$$G_{ij}^{>}\left(t,t\right) = \frac{1}{i\hbar}\delta_{ij} + G_{ij}^{<}\left(t,t\right). \tag{131}$$

Since, here, only zero-temperature applications are considered, the ideal fermionic density matrix can be found by diagonalization of the single-particle part of the Hamiltonian, $\hat{H}^0$, cf. Eq. (16). The density matrix $n^{\hat{H}^0}$ in the eigenbasis of $\hat{H}^0$ is diagonal and, for $N$ particles, reads (sorted by the eigenvalues, starting from the smallest one)

$$n_{ij}^{\hat{H}^0} = \delta_{ij}\begin{cases} 1\,, & \text{if } i \leq N \\ 0\,, & \text{else} \end{cases}. \tag{132}$$

After transformation into the Hubbard basis, this yields the ideal density matrix $n_{ji}^{(0)}$ and, via Eqs. (129) and (130), the components of the ideal Green function. Using $G_{ij}^{(0),\gtrless}$ as initial values, following Eq. (126), the KBE are propagated along both time-directions simultaneously switching on the interaction with the switching function, cf. Eq. (128), i.e., the solutions of

$$\sum_l \left[i\hbar\frac{d}{dt_1}\delta_{il} - h_{il}\left(t_1\right)\right]G_{lj}^{\gtrless}\left(t_1,t_2\right) \tag{133}$$

$$= \sum_l \int_{t_s}^{t_1} dt_3 \left[\Sigma_{il}^{>}\left(t_1,t_3\right) - \Sigma_{il}^{<}\left(t_1,t_3\right)\right]G_{lj}^{\gtrless}\left(\bar{t},t'\right)$$

$$+ \sum_l \int_{t_s}^{t_2} dt_3\, \Sigma_{il}^{\gtrless}\left(t_1,t_3\right)\left[G_{lj}^{<}\left(\bar{t},t'\right) - G_{lj}^{>}\left(\bar{t},t'\right)\right],$$

and

$$\sum_l G_{il}^{\gtrless}\left(t_1,t_2\right)\left[-i\hbar\frac{\overleftarrow{d}}{dt_2}\delta_{lj} - h_{lj}\left(t_2\right)\right] \tag{134}$$

$$= \sum_l \int_{t_s}^{t_1} dt_3 \left[G_{il}^{>}\left(t_1,t_3\right) - G_{il}^{<}\left(t_1,t_3\right)\right]\Sigma_{lj}^{\gtrless}\left(t_3,t_2\right)$$

$$+ \sum_l \int_{t_s}^{t_2} dt_3\, G_{il}^{\gtrless}\left(t_1,t_3\right)\left[\Sigma_{lj}^{<}\left(t_3,t_2\right) - \Sigma_{lj}^{>}\left(t_3,t_2\right)\right],$$

for all values $t_1, t_2 \in \left[t_s, t_0\right]$, are computed, where $t_s$ is the starting time of the adiabatic switching. The interaction matrix in the selfenergy terms for the chosen approximation, on the right-hand sides are replaced by [cf. Section 4 and 5]

$$w_{ijkl} \longrightarrow f_{\mathrm{AS}}^{\tau, t_{\mathrm{H}}}(t) w_{ijkl}, \qquad (135)$$

where the switching parameters $t_{\mathrm{H}}$ and $\tau$ are chosen such that the resulting state is converged with respect to the relevant observables. In practice, values of $\tau = 19.0 \, J^{-1}$ and $t_{\mathrm{H}} = 25.0 \, J^{-1}$ have been found sufficient for all calculations. After the switching is completed, the system is in the correlated ground state[xii] described by $G^{\gtrless}(t_0, t_0)$. The information about the correlations in the system is encoded in the values of the Green functions for all time-points during the switching. That is why all integrals occurring in the solution of the KBE for physically relevant times have to extend along the whole time-plane including the adiabatic-switching part, i.e., Eqs. (133) and (134) have to be used also for $t_1, t_2 > t_0$. It has to be noted that any time-dependent excitation of the system has to occur after the switching is finished and that the values of $G$ with at least one argument in the switching region cannot be used for the determination of observables, as detailed in Section 3.2.

For the numerical solution of the KBE, standard approaches for the solution of ordinary differential equations (ODEs), such as Runge–Kutta methods [136], can be employed upon appropriately discretizing the two-time plane. Further, an integration routine for the right-hand sides of Eqs. (133) and (134) is required. Here, the approach detailed in Ref. [77] has been employed for all calculations. The accuracy of the simulations can be monitored by verifying the conservation laws, in particular of the density and total energy [137] and time-reversibility [79]. As a general note, the use of higher-order methods for the solution of the ODEs and the integrals, i.e., methods where the error scales with a high power of the time step such that time steps of the order of $10^{-2}$ to $10^{-1}$ are possible, is especially advisable to achieve performance and accuracy, since the right-hand sides of the KBE are numerically very expensive.

### 3.2. Important time-dependent observables

This section briefly describes how important physical observables can be obtained from the time-dependent Green functions and additionally from the right-hand sides of the KBE, cf. Eqs. (133) and (134), the so-called collision integrals. As detailed in the Eq. (129), the less-component of the Green function of the interacting ground state is

---

[xii]As already pointed out in Section 2.11.2, it has to be checked externally, e.g., by comparison with other methods, that the final state of the adiabatic switching is indeed the ground state.

directly linked to the density matrix. This relation also holds true for the time-dependent density matrix $n_{ij}(t)$, which describes the single-particle response of the system subject to a time-dependent excitation, reading

$$G_{ij}^<(t,t) = -\frac{1}{i\hbar}n_{ji}(t).$$ (136)

Apart from the time-dependent occupations of the Hubbard sites, which are given by the diagonal elements of the time-dependent density matrix, the latter also permits the calculation of several energy contributions:

– the kinetic energy,

$$E_{\text{kin}}(t) = \text{Re}\left(\sum_{mn} h_{mn} n_{nm}(t)\right),$$ (137)

– the energy induced by a time-dependent excitation $f_{mn}(t)$,

$$E_{\text{ex}}(t) = \text{Re}\left(\sum_{mn} f_{mn}(t) n_{nm}(t)\right).$$ (138)

– the mean interaction energy is also available, but it requires the full two-time Green functions,

$$\begin{aligned}
E_{\text{int}}(t) = & -\frac{i\hbar}{2}\sum_{kl}\int_{t_0}^t d\bar{t}\left[\Sigma_{kl}^>(t,\bar{t}) - \Sigma_{kl}^<(t,\bar{t})\right]G_{lk}^<(\bar{t},t) \\
& + \sum_{kl}\int_{t_0}^t d\bar{t}\,\Sigma_{kl}^<(t,\bar{t})\left[G_{lk}^<(\bar{t},t) - G_{lk}^>(\bar{t},t)\right] \\
= & -\frac{i\hbar}{2}\sum_{kl}\int_{t_0}^t d\bar{t}\left\{\Sigma_{kl}^>(t,\bar{t})G_{lk}^<(\bar{t},t) - \Sigma_{kl}^<(t,\bar{t})G_{lk}^>(\bar{t},t)\right\}.
\end{aligned}$$ (139)

This expression originates from the trace over the right-hand side of Eq. (89)—the collision integral. The possibility to compute the interaction energy (a two-particle quantity) from a single-particle function is a unique feature of the Green functions approach (in contrast, in reduced-density-operator theory this requires the two-particle density operator [131]).

The availability of information off the time diagonal in the one-particle NEGF of the Green function allows, furthermore, to gain insight into the $(N + 1)$- and $(N - 1)$- particle spaces by means of the single-particle spectral function, already encountered in Eq. (113). If an $N$-particle system is prepared in the ground state $\left|\Psi_0^{(N)}\right\rangle$ with energy $E_0^{(N)}$ (or any other $N$-particle energy eigenstate), e.g., via adiabatic switching, and afterwards propagated without additional excitations, the greater/less-components of the two-time Green functions obey

$$G_{ji}^{<}\big(t_1, t_2\big) = \mp \mathrm{i}\hbar \sum_m Q_m(j) Q_m^*(i) \exp\left(\frac{1}{\mathrm{i}\hbar}\big(E_m^{(N-1)} - E_0^{(N)}\big)\big(t_2 - t_1\big)\right), \quad (140)$$

$$G_{ji}^{>}\big(t_1, t_2\big) = \mathrm{i}\hbar \sum_m P_m(j) P_m^*(i) \exp\left(\frac{1}{\mathrm{i}\hbar}\big(E_m^{(N+1)} - E_0^{(N)}\big)\big(t_1 - t_2\big)\right), \quad (141)$$

where the amplitudes $Q_m(i)$, $P_m(i)$ are defined as

$$P_m(i) = \left\langle \Psi_0^{(N)} \big| \hat{c}_i \big| \Psi_m^{(N+1)} \right\rangle, \tag{142}$$

$$Q_m(i) = \left\langle \Psi_m^{(N-1)} \big| \hat{c}_i \big| \Psi_0^{(N)} \right\rangle. \tag{143}$$

They are the overlap matrix elements of the $(N-1)$- and $(N+1)$-particle states of energy $E_m$ with the state which originates from removing/adding one particle in the $i$-th basis state from/to the $N$-particle state. By Fourier transforming and via the knowledge of $E_0^{(N)}$, the $(N-1)$- and $(N+1)$-particle energies can be determined from the propagation of the two-time Green function of the $N$-particle system. In other words, the correlation function $G^{<}$ contains information about the occupied states of the $N$-particle system and the transition energies to the $N-1$-particle system. In contrast, $G^{>}$ contains information about the unoccupied states ("holes") of the $N$-particle system and the transition energies to the $N+1$-particle system when one of the unoccupied states is being filled.

The combination of both functions directly yields the *spectral function*,

$$A_{ji}(t_1, t_2) = \mathrm{i}\hbar \left\{ G_{ji}^{>}\big(t_1, t_2\big) - G_{ji}^{<}\big(t_1, t_2\big) \right\}. \tag{144}$$

Aside from the two-particle energy, cf. Eq. (139), the single-particle Green function gives also access to another two-particle quantity—the *local two-particle density*. Indeed, via the collision integral, the time-dependent double-occupations, $n_i^{(2)}\big(t\big) = \langle \hat{n}_{i\uparrow}(t) \hat{n}_{i\downarrow}(t) \rangle$, i.e., the probability that one electron with spin up and another with spin down simultaneously occupy the same spatial orbital on a Hubbard site "i", can be computed as

$$n_i^{(2)}\big(t\big) = \frac{1}{2U} \sum_l \int_{t_0}^t \mathrm{d}\bar{t} \left\{ \Sigma_{il}^{>}\big(t, \bar{t}\big) G_{li}^{<}(\bar{t}, t) - \Sigma_{il}^{<}\big(t, \bar{t}\big) G_{li}^{>}(\bar{t}, t) \right\}, \tag{145}$$

which becomes obvious when taking into account the relation between the selfenergy and the two-particle Green function, cf. Eqs. (64) and (65). Further, the *two-particle local density correlation* (pair-correlation function) can be computed by subtracting the uncorrelated (mean-field) expression of the two-particle density which (for the Hubbard model) is nothing but the product of two single-particle densities,

$$\delta n_i^{(2)}(t) = n_i^{(2)}(t) - n_{i\uparrow}(t) n_{i\downarrow}(t) \,. \tag{146}$$

This quantity is identical to zero if the system is uncorrelated and thus directly measures effects beyond Hartree–Fock. A quantity that also measures the space-resolved correlations in a many-body system is the local entanglement entropy [26],

$$\begin{aligned}
S_i(t) = & -2 \left( \frac{n_i}{2} - n_i^{(2)} \right) \log_2 \left( \frac{n_i}{2} - n_i^{(2)} \right) - n_i^{(2)} \log_2 n_i^{(2)} \\
& - \left( 1 - n_i + n_i^{(2)} \right) \log_2 \left( 1 - n_i + n_i^{(2)} \right) \,.
\end{aligned} \tag{147}$$

We now turn to a survey of recent computational results that were obtained by (part of) the authors. Our focus is on comparison of different selfenergy approximations and on tests of their accuracy.

### 3.3. Numerical results for the correlated ground state

In this section, numerical results for the interacting ground-state energies are presented. As an example, we consider a 6-site Hubbard model for which exact results can be obtained. Our main interest is to test the performance of the different selfenergy approximation schemes introduced in Section 4 and 5 with respect to filling level $n$ (i.e. density) and interaction strength, $U/J$, of the system, by comparison with exact calculations.

Following the adiabatic-switching algorithm of Section 2.11.2, the system is initially prepared in the non-interacting ground state for $N_\uparrow = N_\downarrow = 1, 2, 3$ particles, i.e., filling levels $n = 1/6$, $n = 1/3$ and $n = 0.5$. Using the switching function (128) with parameters $\tau = 19.0 \, J^{-1}$ and $t_{\rm H} = 25.0 \, J^{-1}$, ensuing, the interaction $U$ is ramped up to the final values $U = 0.1, 0.5, 1.0$ and $2.0$. In Table 1, we list the selfenergy approximations that are being used and compared.

### 3.3.1. Results for the ground-state energy.
We begin with a detailed analysis of the ground-state energy to understand the quality of the different selfenergy approximation, in dependence on the coupling strength and the filling. Starting with $U = 0.1$, which is very close to an ideal system, the results for $N_\uparrow = N_\downarrow = 1, 2, 3$ are listed in the second columns of Tables (2), (3) and (4), respectively.

For all three filling factors, the Hartree results differ from the exact results in the third decimal place, while all methods beyond Hartree agree up to the fourth decimal place. This can be explained by the fact that the Hartree approximation is only correct up to the first order in the interaction strength, while all other methods agree up to the second order in the interaction with each other and the exact solution. For both smaller-than-half-filling factors, the results of the non-Hartree methods behave similarly.

| $\Sigma$ | $U/J = 0.1$ | $U/J = 0.5$ | $U/J = 1.0$ | $U/J = 2.0$ |
|---|---|---|---|---|
| H | $-0.533604$ | $-0.44956$ | $-0.34718$ | $-0.14939$ |
| TPP | $\mathbf{-0.534067}$ | $-0.45981$ | $-0.38291$ | $-0.26319$ |
| **Exact** | $\mathbf{-0.534073}$ | $\mathbf{-0.46065}$ | $\mathbf{-0.38810}$ | $\mathbf{-0.28820}$ |
| TOA | $\mathbf{-0.534075}$ | $\mathbf{-0.46048}$ | $\mathbf{-0.38545}$ | $-0.26738$ |
| FLEX | $-0.534076$ | $\mathbf{-0.46129}$ | $-0.39685$ | $-0.36059$ |
| SOA | $-0.534084$ | $-0.46164$ | $\mathbf{-0.39472}$ | $\mathbf{-0.32514}$ |
| GWA | $-0.534084$ | $-0.46181$ | $-0.39723$ | $-0.35068$ |
| TEH | $-0.534093$ | $-0.46293$ | $-0.40535$ | $-0.39065$ |

**Table 2.** Ground-state energies, $E_{gs}/J$, for $N_s = 6$ sites and 1/6 filling, i.e. $N_\uparrow = N_\downarrow = 1$, for different couplings and for different selfenergy approximations. Approximations are ordered by $E_{gs}$ for the smallest $U$. In each column, the two results that are closest to the exact one are typed bold.

| $\Sigma$ | $U/J = 0.1$ | $U/J = 0.5$ | $U/J = 1.0$ | $U/J = 2.0$ |
|---|---|---|---|---|
| H | $-2.642487$ | $-2.35948$ | $-2.00967$ | $-1.31891$ |
| TPP | $\mathbf{-2.643354}$ | $-2.37963$ | $-2.08369$ | $-1.57367$ |
| **Exact** | $\mathbf{-2.643367}$ | $\mathbf{-2.38107}$ | $\mathbf{-2.09367}$ | $\mathbf{-1.63342}$ |
| TOA | $\mathbf{-2.643367}$ | $\mathbf{-2.38093}$ | $\mathbf{-2.09182}$ | $\mathbf{-1.61455}$ |
| FLEX | $-2.643368$ | $\mathbf{-2.38174}$ | $-2.10424$ | $-1.61455$ |
| SOA | $-2.643376$ | $-2.38208$ | $\mathbf{-2.10058}$ | $\mathbf{-1.67092}$ |
| GWA | $-2.643376$ | $-2.38226$ | $-2.10357$ | $-1.71303$ |
| TEH | $-2.643390$ | $-2.38400$ | $-2.11746$ | $-1.80334$ |

**Table 3.** Ground-state energies, $E_{gs}/J$, for $N_s = 6$ sites and 1/3 filling, i.e. $N_\uparrow = N_\downarrow = 2$, for different couplings and for different selfenergy approximations. Approximations are ordered by $E_{gs}$ for the smallest $U$. In each column, the two results that are closest to the exact one are typed bold.

The best results, which agree up to at least 5 decimal places with the exact result, are achieved by the third-order and the FLEX approximations, which are both correct up the third order in the interaction strength.

On the other hand, when only one of the two third-order selfenergy contributions is included, the particle–particle and electron–hole $T$-matrix approximations (TPP, TEH), do not improve the result significantly compared to the second-order approximation (SOA) for $n = 1/3$, although, for $n = 1/6$, the particle–particle $T$ matrix performs better than the electron–hole $T$ matrix, since the latter contributes less as the electronic density is comparatively small. For $U = 0.1J$, the $GW$ approximation shows no difference to the SOA results, since both approximations only differ in the fourth order, as the GWA has no third-order diagram, cf. Eq. (315). For half filling, $n = 0.5$, all correlated methods (i.e. all methods except Hartree), apart from the $T$ matrices, agree with each other and with the exact solution up to 5 decimal places. This is explained by the so-called particle–hole symmetry in the Hubbard model [138], which only occurs at half filling. In this particular case there is an exact cancellation of all electron–hole and particle–particle $T$-matrix terms of odd orders in the interaction and equality of all even order terms. Therefore, for half filling, SOA and TOA yield exactly the same results and agree with GWA and FLEX up to the fourth order in the interaction.

| $\Sigma$ | $U/J = 0.1$ | $U/J = 0.5$ | $U/J = 1.0$ | $U/J = 2.0$ |
|---|---|---|---|---|
| H | $-6.837918$ | $-6.23792$ | $-5.48791$ | $-3.98791$ |
| TPP | $\mathbf{-6.839299}$ | $-6.26949$ | $-5.60173$ | $-4.36896$ |
| **Exact** | $\mathbf{-6.839331}$ | $\mathbf{-6.27322}$ | $\mathbf{-5.62889}$ | $\mathbf{-4.54631}$ |
| TOA | $\mathbf{-6.839331}$ | $-6.27285$ | $-5.62356$ | $-4.48569$ |
| SOA | $-6.839331$ | $\mathbf{-6.27285}$ | $\mathbf{-5.62356}$ | $\mathbf{-4.48569}$ |
| GWA | $-6.839331$ | $\mathbf{-6.27347}$ | $\mathbf{-5.63259}$ | $\mathbf{-4.58604}$ |
| FLEX | $-6.839334$ | $-6.27471$ | $-5.64923$ | $-4.71630$ |
| TEH | $-6.839363$ | $-6.27741$ | $-5.66104$ | $-4.73547$ |

**Table 4.** Ground-state energies, $E_{gs}/J$, for $N_s = 6$ sites, and half filling, i.e. $N_\uparrow = N_\downarrow = 3$, for different couplings and for different selfenergy approximations. Approximations are ordered by $E_{gs}$ for the smallest $U$. In each column, the two results that are closest to the exact one are typed bold.

For the increased but still small interaction strength $U = 0.5J$, the results for $N_\uparrow = N_\downarrow = 1, 2, 3$ ($n = 1/6, 1/3, 0.5$) are shown in the third columns of Tables (2), (3) and (4), respectively. Here, the Hartree results differ from the exact result already in the second decimal place. In contrast, all correlated methods agree with the exact result in the second decimal place. Compared to the $U = 0.1J$ results, only the TOA remains close to the exact solution whereas all other methods show more pronounced deviations. For $n = 1/6$, FLEX and TPP yield comparably good results. Further, TEH is worse compared to its particle–particle counterpart, and GWA is worse than the SOA. For $n = 1/3$, FLEX becomes better than TPP, showing the increasing importance of the third-order electron–hole contribution. For half filling, the GWA result is the closest to the exact solution, closely followed by the equal results of the SOA and TOA. This gives a first hint that, at $U = 0.5J$, the fourth- and possibly higher-order terms gain influence, although among the approximations containing fourth-order and higher terms, only the GWA gives improved energies compared to second-/third-order perturbative results.

For the higher interaction strengths $U/J = 1.0$ and $U/J = 2.0$, the results are shown in the fourth and fifth columns of Tables (2), (3) and (4), respectively. Compared to the results for $U = 0.5J$, the relative order of the performance of the different selfenergy approximations remains the same, but the absolute differences to the exact results increase. Summarizing the results for the ground-state energies, it is evident that, away from half filling, the TOA outperforms all other perturbative and non-perturbative selfenergy approximations, at least in the covered range of coupling parameters $U/J = 0.1 \dots 2.0$. For half filling, on the other hand, the best results are obtained from the $GW$ approximation.

*3.3.2. Results for the spectral function in the ground state.* Let us now consider another important quantity of the correlated ground state, which goes beyond the description of the lowest energy level: the single-particle spectral function, that was introduced in Section 3.2. For an $N$-particle system, it shows the transition energies into the $(N-1)$-particle as well as the $(N+1)$-particle system, i.e., the single-particle removal and addition energies. The removal energies are carried by the off-timediagonal values of $G^<$, while the addition energies are similarly encoded in $G^>$. They can be made visible by transforming into relative and center-of-mass time and, afterwards, Fourier transforming with respect to the relative time. Comparing with Eqs. (140) and (141), one expects peaks at the energy levels of $(N \pm 1)$-particle systems shifted by the ground-state energy

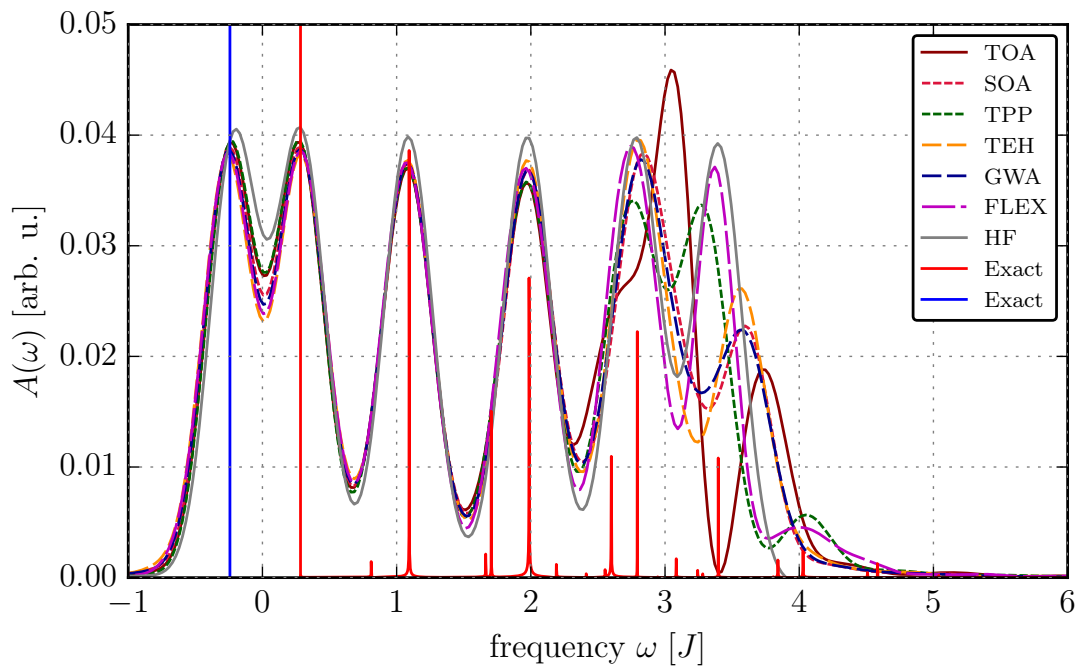

**Figure 6.** 6-site Hubbard cluster with $U = 1$ at one-sixth filling. Spectral function generated from different selfenergy approximations: brown full line: third-order approximation (TOA), crimson dashed line: second-order approximation (SOA), green dashed line: particle–particle $T$ matrix (TPP), yellow dashed line: electron–hole $T$ matrix (TEH), blue dashed line: $GW$ approximation (GWA), purple semi-dashed line: fluctuating-exchange approximation (FLEX), gray full line: Hartree–Fock approximation (HF). For comparison: Exact excitation spectra of the systems with $N^\uparrow = 1$, $N^\downarrow = 0$ (blue) and $N^\uparrow = 2$, $N^\downarrow = 1$ (red). The spectra are shifted such that $E = 0$ lies in the center between the highest removal energy and the lowest addition energy.

of the $N$-particle system.

For the six-site Hubbard system of interaction strength $U = 1J$, the performance of the different selfenergy approximations is studied for the three filling factors $n = 1/6, 1/3$ and $1/2$. The results are compared with exact excitation spectra of the relevant $(N \pm 1)$-particle systems, which in turn are generated by excitation with a $\delta$-kick and subsequent Fourier transform of the time-dependent density evolution.

For $n = 1/6$, the results are shown in Fig. 6. The frequency axis of the spectrum is shifted such that all removal energies, corresponding to $G^<$, have negative values while the addition energies, corresponding to $G^>$, have positive energies. Since the spectrum of the $(N-1)$-particle system only contains one spin-up or spin-down particle[xiii] it has no interaction effects and, thus, is ideal. This corresponds to the less-part of the

---

[xiii] As the fermionic Hubbard Hamiltonian contains no terms which are different for up or down spin-orientation, a system with $N$ spin-up and $M$ spin-down particles behaves like the system with $M$ spin-up particles and $N$ spin-down particles.

spectrum having only one spectral line which matches the exact result (blue line) for every approximation including Hartree(–Fock). Analyzing the greater part of the spectrum which belongs to the system with two particles of one spin-direction and one particle of the other, one can separate two sets of spectral lines. The three spectral lines belonging to the lowest addition energies are in exact agreement throughout all approximations including Hartree and, thus, indicate mostly uncorrelated states. The position of the peak for the next higher addition energy begins to differ between the approximations. The best agreement with the exact solution is reached by the GWA followed by SOA, TEH and HF. The positions for TPP and FLEX are slightly shifted to lower energies. For TOA, the peak position cannot be easily distinguished with the shown spectral resolution, but the knee-structure within the left slope of the peak for the next higher energy suggests that the accuracy is comparable to that of the other approximations. Unlike any other tested approximation, though, the TOA is able to show the energy level just above $\omega = 3$. The next-higher energy level at $\omega \approx 3.3$ is best captured by the FLEX approximation followed by TPP, which slightly shifts to lower energies. The SOA, GWA and the TEH show this peak shifted to higher energies. For the TOA, it remains questionable if the peak at $\omega \approx 3.8$ is to be attributed to exact energy level at $\omega \approx 3.3$ or if it shows the energy level at $\omega \approx 3.9$. The level just above $\omega = 4$ is only shown by the TPP and FLEX approximations, which indicates that these states embody a high degree of correlation. Summing up the findings for one-sixth filling, the best overall results are achieved by the TPP and FLEX approximation, with the latter performing slightly better. In addition, the TOA shows energy levels which are not captured by any other approximation.

For $n = 0.33$, the results are shown in Fig. 7. Two removal energy levels with a large amplitude and one with a small amplitude are visible. It is noteworthy that the removal energies of the one-third filled Hubbard cluster connect to, i.e., that of two and one particle of both spin-directions, respectively, is the same the addition energies of the one-sixth filled cluster connect to. Comparing with the right part of Fig. 6, one immediately recognizes that the shown energy levels are not the same. For $n = 1/3$, the destination energy levels are levels 1, 2 and 6, while, for $n = 1/6$, the levels $1, 3, 6$ and higher levels are reached. Thus, the combined information of both fillings can be used to gain insight into the energy spectrum of the intermediate system that differs by one particle to both. Concerning the quality of the approximations for the removal energy levels, all approximation agree with each other and with the exact solution, indicating that the states belonging to the removal energies are mainly uncorrelated. For the addition energies, the same is true for the first two levels. Starting from just above $\omega = 2$, there are many close-lying energy levels in the range up to $\omega \approx 4$, which renders an attribution to the different approximations difficult. In general, confirming the trend found for $n = 1/6$, the FLEX approximation yields results which very well agree with the exact energy levels, while the TOA reveals correct energy levels not found with the other approximations.

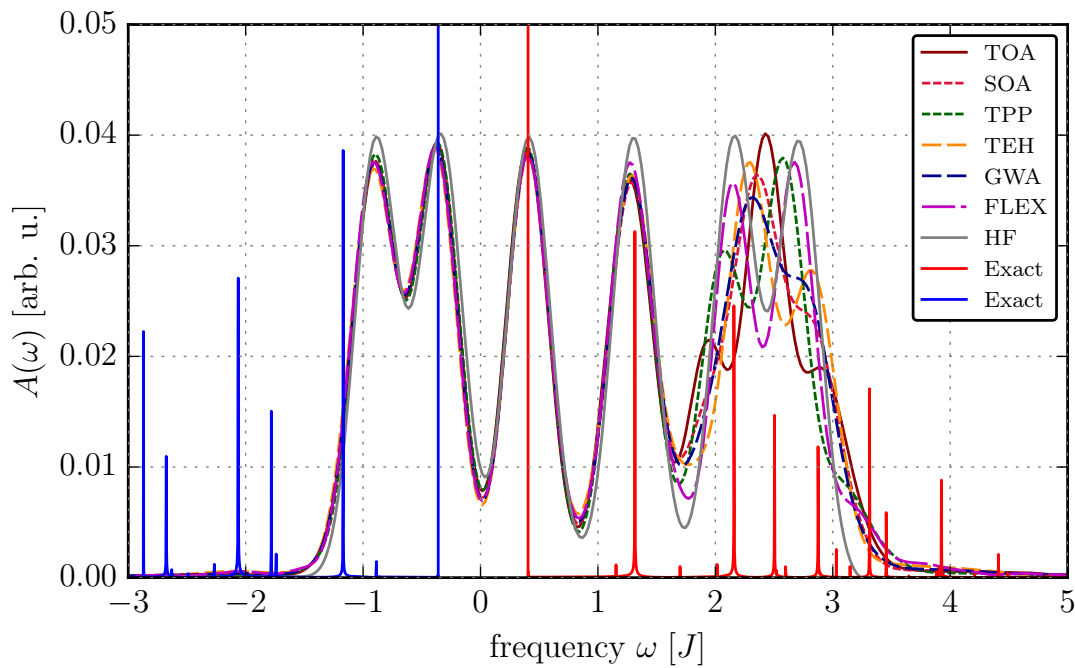

**Figure 7.** 6-site Hubbard cluster with $U = 1$ at one-third filling. Spectral function generated from different selfenergy approximations: brown full line: third-order approximation (TOA), crimson dashed line: second-order approximation (SOA), green dashed line: particle–particle $T$ matrix (TPP), yellow dashed line: electron–hole $T$ matrix (TEH), blue dashed line: $GW$ approximation (GWA), purple semi-dashed line: fluctuating-exchange approximation (FLEX), gray full line: Hartree–Fock approximation (HF). For comparison: Excitation spectra of the systems with $N^{\uparrow} = 2$, $N^{\downarrow} = 1$ (blue) and $N^{\uparrow} = 3$, $N^{\downarrow} = 2$ (red). The spectra are shifted such that $E = 0$ lies in the center between the highest removal energy and the lowest addition energy.

Turning to the results for half filling, shown in Fig. 8, one immediately recognizes two peculiarities of this setup. First, the removal and addition part of the spectrum is symmetric with respect to $E = 0$. This is, again, due to the occurrence of particle–hole symmetry [138]. Second, comparing the quality of the different approximations, one can discern only minor differences at energy levels farther away from $E = 0$, which are most pronounced for the FLEX approximation. As a special note, the good performance of the Hartree approximation for the spectral function indicates that the use of the GKBA with Hartree propagators is justified for half filling and explains the excellent results that could be achieved [76, 124, 139].

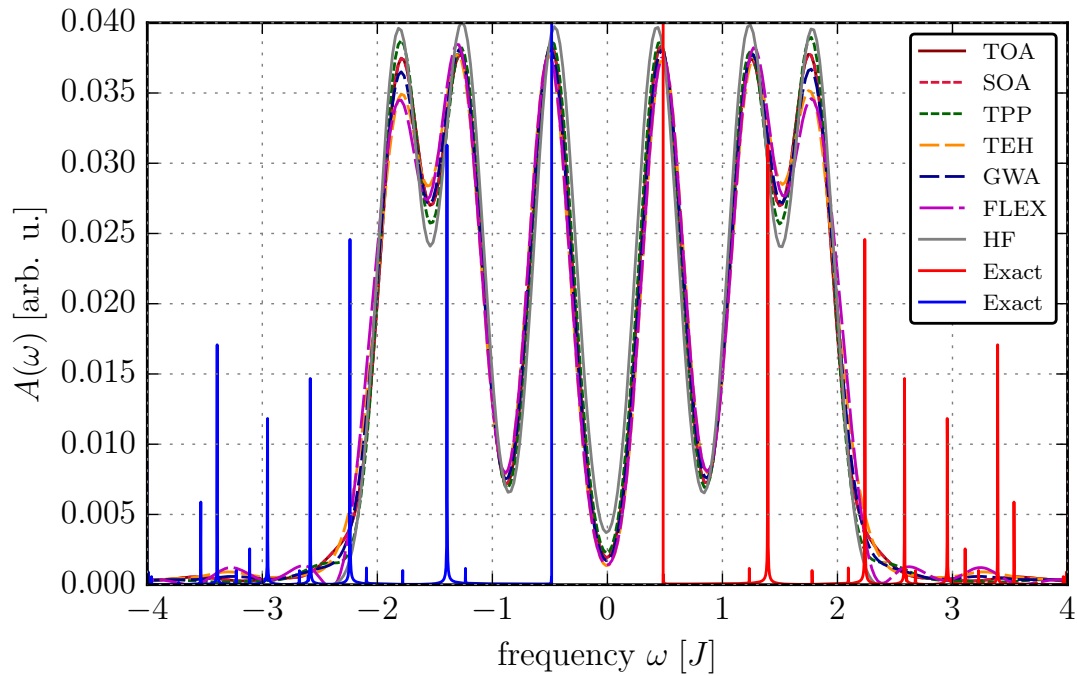

**Figure 8.** 6-site Hubbard cluster with $U = J$ at half filling. Spectral function generated from different selfenergy approximations: brown full line: third-order approximation (TOA), crimson dashed line: second-order approximation (SOA), green dashed line: particle–particle $T$ matrix (TPP), yellow dashed line: electron–hole $T$ matrix (TEH), blue dashed line: $GW$ approximation (GWA), purple semi-dashed line: fluctuating-exchange approximation (FLEX), gray full line: Hartree–Fock approximation (HF). For comparison: Excitation spectra of the systems with $N^{\uparrow} = 3$, $N^{\downarrow} = 2$ (blue) and $N^{\uparrow} = 4$, $N^{\downarrow} = 3$ (red). The spectra are shifted such that $E = 0$ lies in the center between the highest removal energy and the lowest addition energy.

Summarizing the ability of NEGF methods to describe the spectral function of Hubbard clusters, one can state that the overall agreement for small to medium interaction strength is good and especially via combination of different approximation methods as well as probing from both systems with adjacent number of particles, one can gain a large part of the spectral information.

### 3.4. Time evolution following an external excitation

After analyzing simulation results for the correlated ground state and the quality of different selfenergy approximations, we now turn to time-dependent simulations. Thereby the ground-state data serve as the initial condition of the system prior to the excitation.

The dynamics of the system which is driven out of equilibrium by an external excitation are studied numerically using the correlated selfenergy approximations that were described in Section 4 and 5. Thereby we focus on separate dynamics studies following different types of excitations.

Again, the motivation here is to analyze the accuracy of different approximations by performing tests against benchmark data. These include exact-diagonalization (CI) calculations that are possible for small systems; examples are given in Secs. 3.4.4 and 3.4.5. For larger systems, comparisons with time-dependent density-matrix renormalization group (DMRG) simulations can be performed, [cf. Secs. 3.4.1, 3.4.2] which, however, are restricted to one-dimensional systems, due to the present limitations of DMRG. Finally, comparisons can be made to experiments with ultracold atoms, cf. Sec. 3.4.1.

*3.4.1. Time evolution following a confinement quench.* A rather simple excitation of a finite system is to start with a spatially localized configuration that is achieved by a strong confinement potential and then to rapidly remove the confinement at some time $t = 0$. This resembles classical diffusion experiments where a localized particle density expands into vacuum. Such a configuration of a quantum system is straightforwardly realized with ultracold atoms in a trap or an optical lattice. An example of a rapid expansion of cold fermionic atoms was presented by Schneider *et al.* in Ref. [140].

With the NEGF approach it is fairly straightforward to simulate such an expansion scenario of atoms on an optical lattice because the latter accurately reproduces the Hubbard Hamiltonian (27) with onsite interaction. In the present experiment a two-dimensional geometry was used. Before presenting NEGF results for this setup we focus on benchmark calculations where comparison with DMRG simulations were performed. In Ref. [50], expansion simulations for $N = 34$ fermions in a one-dimensional configuration have been carried out. Initially, the 17 central sites were doubly occupied whereas the outer sites were empty, cf. the bottom row (left column) of Fig. 9. The expansion dynamics are quite interesting and differ significantly from the classical case due to the rectangular shape of the initial density profile. Furthermore, the expansion of fermions is constrained by the Pauli principle, i.e., the innermost particles cannot move until the fermions at the edge have (partially) emptied their sites.

In Figure 9 three sets of simulations are presented. The full lines are time-dependent DMRG results, cf. Ref. [50] for details. In addition, the authors show NEGF results with particle–particle $T$-matrix selfenergies (here TMA) in order to accurately simulate strong-correlation effects. One set of results is from a full two-time simulation (TMA, green dashes), the other one, from a single-time approximation using the GKBA with Hartree–Fock propagators (HF-GKBA+TMA, yellow dots). In the present simulations the coupling was moderate, $U/J = 1$ and, not surprisingly, the agreement of the NEGF results for the density profile with DMRG is excellent. An interesting observation is that the two-time simulations show a faster dynamics than the one-time approximation (HF-GKBA).

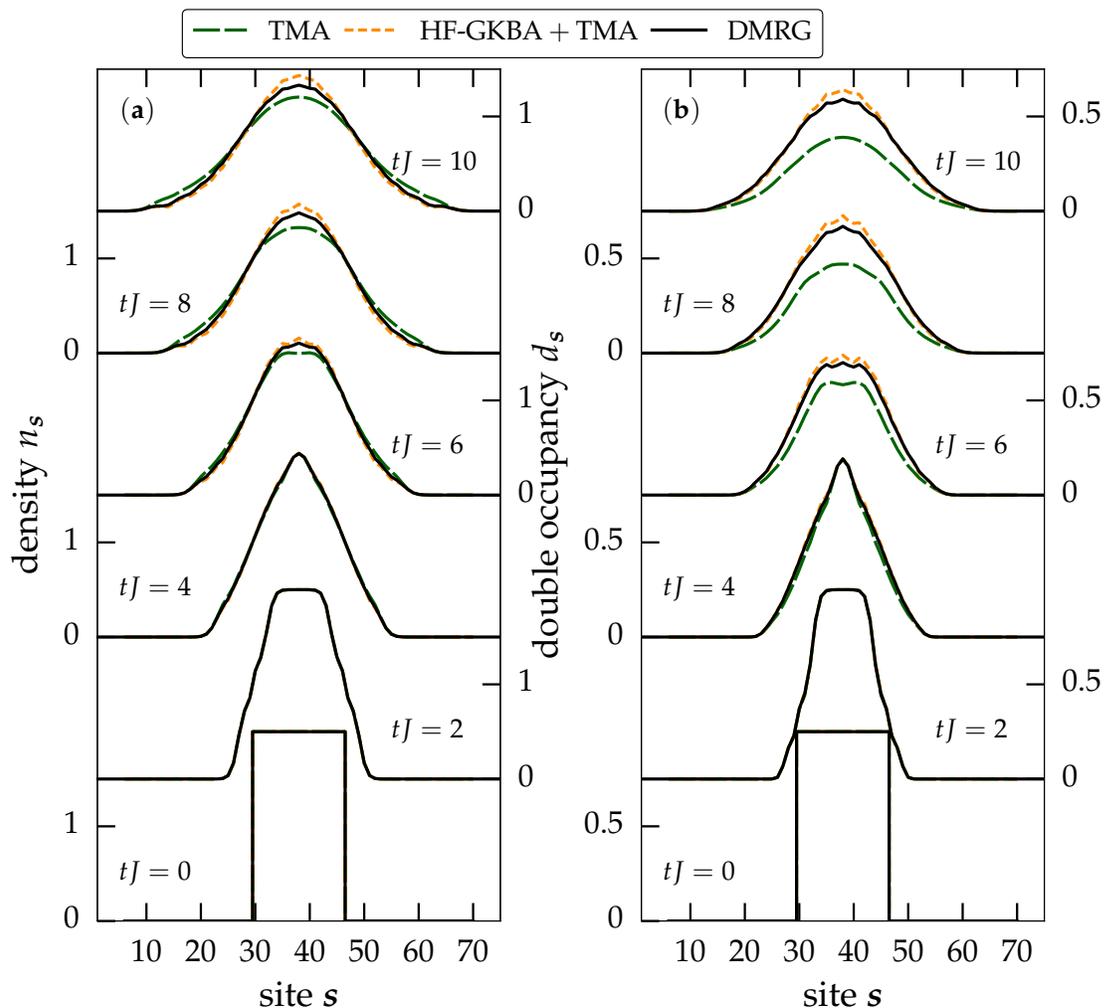

**Figure 9.** Symmetric 1D sudden expansion of a Hubbard chain of $N = 34$ fermions at $U = J$. Time evolution of (a) density $n_s$ and (b) double occupancy $d_s$ for six times (from bottom to top): $tJ = 0, 2, 4, 6, 8, 10$. Solid lines: DMRG, long dashes: TMA (two-time NEGF with $T$-matrix selfenergy), dashed lines: $T$ matrix with HF-GKBA. From Ref. [50].

A more sensitive quantity than the density is the local doublon number, Eq. (145), which is plotted in the right column of Fig. 9. Here the agreement with the DMRG data is similar. While the two-time result for $d_s(t)$ shows stronger deviations than the single-particle density, the HF-GKBA exhibits the same high accuracy for both quantities. A very interesting observation is that the exact (DMRG) result is enclosed by the two-time and GKBA results. This behavior was confirmed for a broad range of coupling parameters, selfenergy approximations and in many other setups as well, e.g. Ref. [50]. This has important implications for the use of NEGF simulations for more complex systems where independent benchmark data are not available.

Returning to the physics of the expansion of fermions—one of the most interesting questions is how the expansion speed depends on the interaction strength $U/J$ (i.e., on the correlations in the system). A particularly interesting theoretical prediction was [91] that fermions that are in doubly occupied lattice sites ("doublons") should expand slower compared to singly occupied sites ("singlons") giving rise to a spatial separation of the two components ("quantum distillation"). A first test can be made by comparing the two columns of Fig. 9. There, indeed, the doublon expansion is slightly slower than that of the total density, in particular, for the initial time frames (cf. the lowest three rows). Here, however, the interaction is comparably weak and the effect is small.

Such a peculiar separated expansion of doublons and singlons was, in fact, observed experimentally for strongly correlated fermionic atoms in Ref. [140]. They demonstrated that, when the system is initially in a fully doubly occupied configuration, after removal of the confinement, doublons remain dominantly in the trap center. Moreover, the expansion speed of this central part ("core") decreases when the coupling strength $U/J$ increases. The experimental results for the "core expansion velocity" $C_{\exp}$ are reproduced in Fig. 10 by the full black line. Interestingly, $C_{\exp}$ even becomes negative what means that the "core shrinks". Furthermore, the result is exactly the same for attractive and repulsive interaction (negative and positive $U$, respectively), what is an exact property of the Hubbard model.

Figure 10 also contains results from a semiclassical kinetic simulation in relaxation-time approximation (grey dashed curve [140]) which reproduces the overall trend but exhibits very strong deviations, for most values of $U$. It is, therefore, of high interest to apply the NEGF approach to this problem since, due to the 2D geometry, DMRG simulations are not possible. Such NEGF simulations were developed by three of the present authors and published in Ref. [26]. To correctly describe strong-coupling effects, the second-Born approximation cannot be applied. Instead, $T$-matrix selfenergies were used. An advantage in these simulations is that the initial state is uncorrelated (it is an Hartree–Fock state), so no adiabatic switching needs to be done. This is particularly important since the simulations have to be sufficiently long to reach the regime where the expansion velocity approaches a constant value ("hydrodynamic stage").

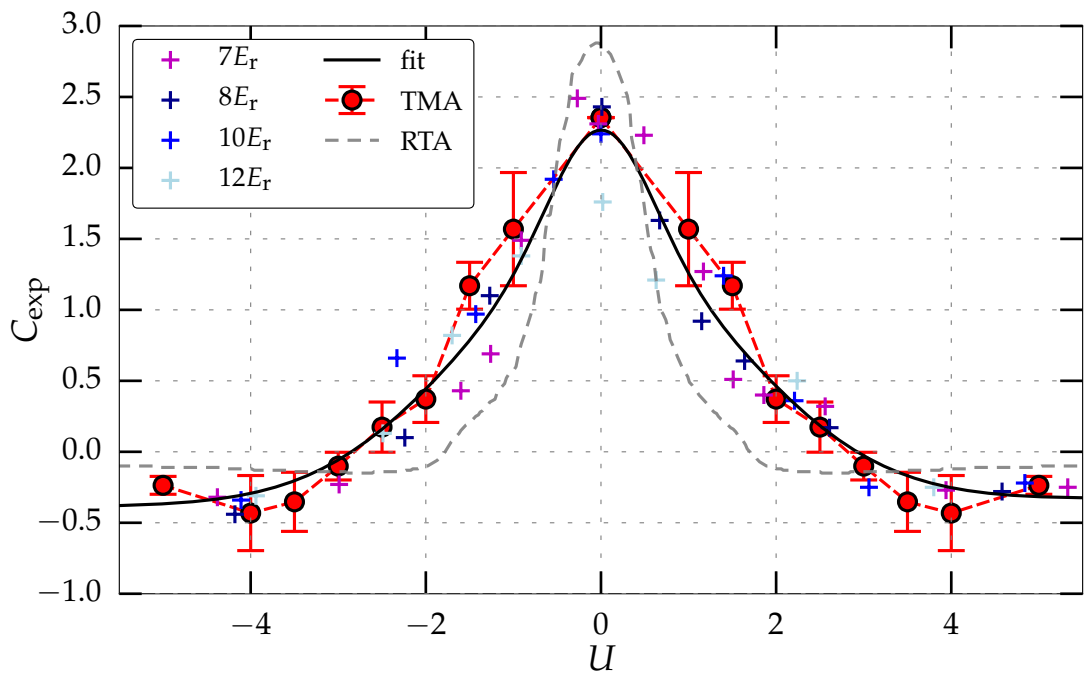

**Figure 10.** Asymptotic core (half width at half maximum of the density) expansion velocity $C_{\text{exp}}$. Plus signs: experimental results for different lattice depths in units of the recoil energy $E_r$; gray dashed line: relaxation-time approximation model of Ref. [140]; red circles: two-time NEGF results with $T$-matrix selfenergies, the error bars denote the statistical uncertainty due to the extrapolation with respect to time and particle number (see text). The black line is a fit through the experimental points to guide the eye. From Ref. [26].

The present NEGF simulations with particle–particle $T$-matrix selfenergies were carried out for a broad range of particle numbers, up to about $N = 100$ particles. It turned out that the expansion velocity shows a simple scaling with $N$, so an extrapolation to the thermodynamics limit, $N \to \infty$ (the experiments used several hundred thousands of atoms) was possible. At the same time, the statistical error of the extrapolation provides a measure of the numerical uncertainty of the macroscopic results [77]. These results are also included in Fig. 10 by the red dots and the associated error bars. Obviously, the agreement with the experiment, over the whole range of coupling parameters, is impressive. These have been the first and so far the only quantum-dynamics simulations that allow for a direct comparison with cold-atom experiments in two dimensions. Moreover, in Ref. [26] results for three-dimensional lattice configurations were presented and the dependence of the expansion velocity on the dimensionality was analyzed.

The time-dependent NEGF simulations provide extensive additional information on the expansion dynamics. From the nonequilibrium Green functions it is straightforward to obtain full space (site)-resolved information. An example is shown in Fig. 11 where the time evolution of various quantities with single-site resolution is presented [26]. This includes the density (first three rows), the doublon density [row 4, Eq. (145)], the doublon correlation function, [Eq. (146)] and the entanglement entropy, [Eq. (147)]. The latter two quantities are of particular interest, as they allow to separate the effect of the buildup of correlations in the system: as mentioned above, the initial state is uncorrelated, and with removal of the confinement correlations start to from at the cluster edge and then spread in and outward. These simulation results can be directly compared to experiments with ultracold atoms where single-site resolution has been achieved with quantum-gas microscopes [19–21]. More details on the present simulations can be found in Ref. [77].

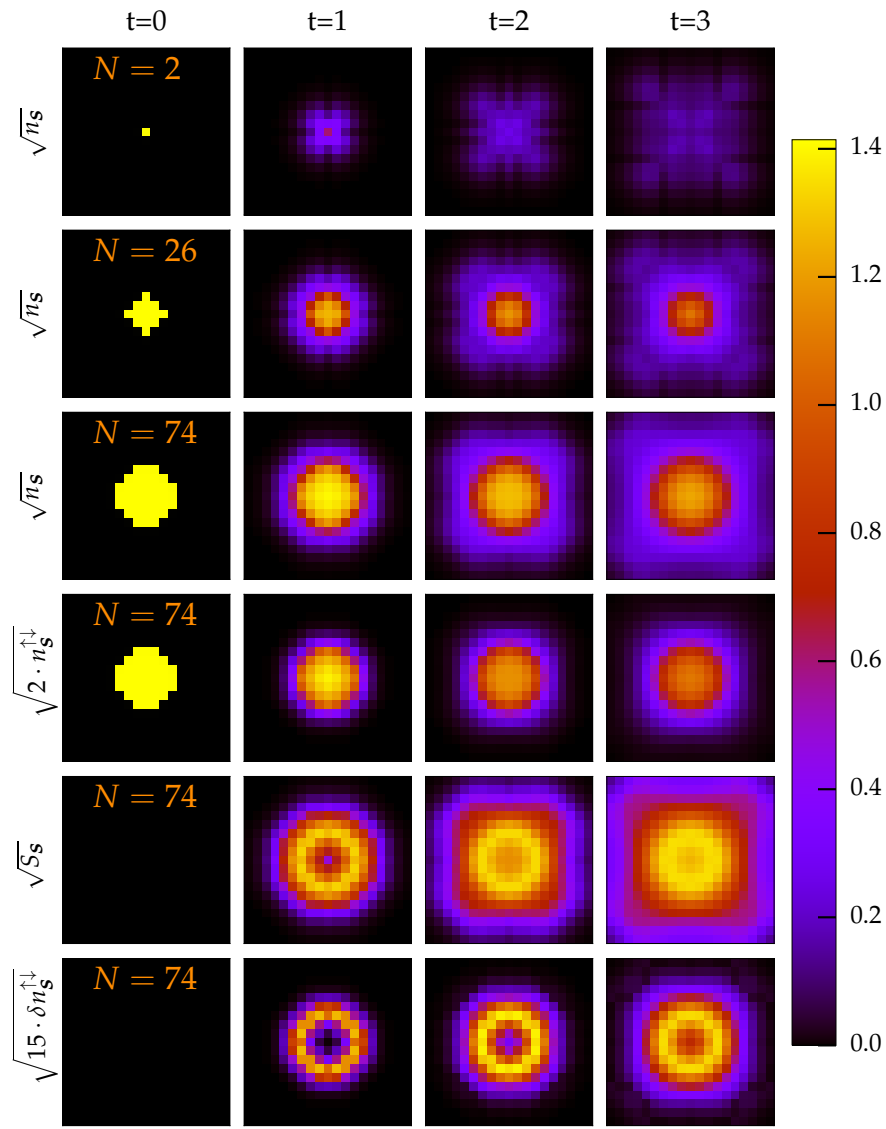

**Figure 11.** Site-resolved expansion dynamics in a 2D $19 \times 19$ Hubbard lattice at $U/J = 4$ for four times (in units of $J^{-1}$). Top three rows: square root of density $n_s$ for $N = 2$, $N = 26$, and $N = 74$, respectively. Rows 4–6: square root of double occupation, local entanglement entropy $S_s$, Eq. (147), and the pair-correlation function $\delta n_s^{\uparrow\downarrow} = \delta n_s^{(2)}$, Eq. (146). From Ref. [26].

Concluding this section, we note that the present NEGF simulations with $T$-matrix selfenergies are rather costly and require large computer resources. In particular, simulations for large $U$ become increasingly difficult which explains the choice of the maximum $U$-values displayed in Fig. 10. To reach large coupling values, improved computational approaches and, possibly, further improved selfenergy approximations, have to be developed.

*3.4.2. Time evolution starting from a charge-density-wave state.* Next, we focus on the time evolution of correlated electrons after a confinement quench starting from a different initial state than before. In Ref. [50] the authors considered a state of alternating doubly occupied and empty sites which will be called "charge-density-wave state" (CDW). After removal of the confinement density can spread to the originally empty sites. To compare with benchmark data from DMRG simulations the simulations are limited to 1D geometry. In Ref. [50] extensive NEGF–DMRG comparisons have been carried out. Here we show some typical results, cf. Fig. 12.

The first question to answer for the NEGF simulations is again the proper choice of the selfenergy. Since, again, correlation effects and, in particular, large values of $U/J$ are of interest, the second-Born approximation is not appropriate. The particle–particle $T$ matrix that showed impressive results in the diffusion setup of Sec. 3.4.1 is not expected to work well here. The reason is that the $T$-matrix selfenergy treats interaction effects accurately on the two-particle level but neglects three-particle effects. It is, therefore, expected to be adequate for low densities. In application to Hubbard systems, this corresponds to low (or, due to particle–hole symmetry, high) filling factors. In the diffusion setup, the filling factor is low, except for the initial dynamics in the core region.

In contrast, in the present CDW setup the entire system is initially at half filling and remains at half filling. There, the particle–particle $T$ matrix is inaccurate. At half filling we found in Sec. 3.3 that the third-order approximation (TOA) provides much more accurate data for the ground-state energy. It is, therefore, expected that also the dynamics will be treated more accurately within TOA selfenergies. Consequently, this selfenergy is being used, in addition to applying the HF-GKBA. The results are shown in Fig. 12 for two values of the coupling strength and 5 different chain lengths, in the range of $L = 6$ and $L = 36$. Note that we do not show the dynamics of the densities—there the agreement is excellent—but we focus on the more sensitive double occupations, Eq. (145), summed over the entire cluster.

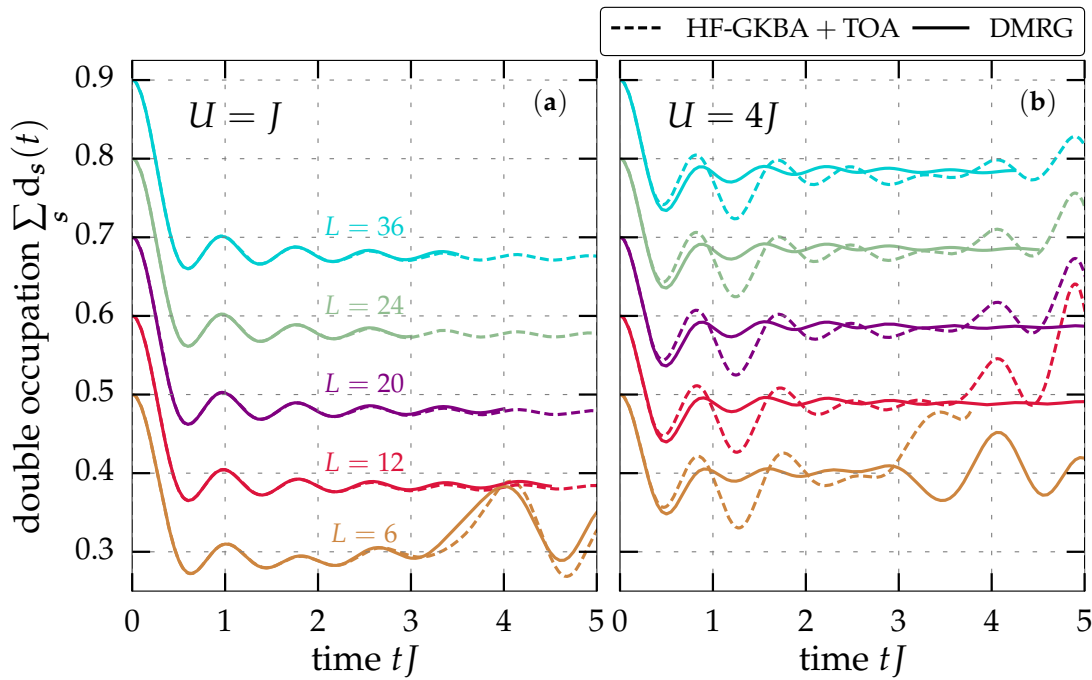

**Figure 12.** Charge-density wave excitation (the initial state consits of alternating doubly occupied and empty lattice sites): Time evolution of the total double occupation, cf. Eq. (145) for two coupling strengths and 5 different systems sizes, $L = 6, \ldots 36$. Comparison of DMRG (full lines) and NEGF with third-order selfenergies within the single-time HF-GKBA scheme. From Ref. [50].

For moderate coupling, $U/J = 1$, left column, the NEGF results are practically indistinguishable from the DMRG data. Only for the smallest system and for long times, small deviations are visible. For the case of larger coupling $U/J = 4$ significantly larger deviations are observed. While the overall trends, such as the mean value of the total double occupation, $d(t)$, is well reproduced for the initial time interval, the oscillations of $d(t)$ occur with a slightly modified frequency, and the amplitude of the HF-GKBA results is substantially larger than in the DMRG data. Also, the "density revival" seen in small systems at weak coupling (bottom curve of left column) which seems to be present also at larger coupling (bottom right DMRG curve) seems to be amplified by the HF-GKBA. Most interestingly, the agreement of the NEGF data with the DMRG systematically improves with increasing system size. For more details, the reader is referred to Ref. [50].

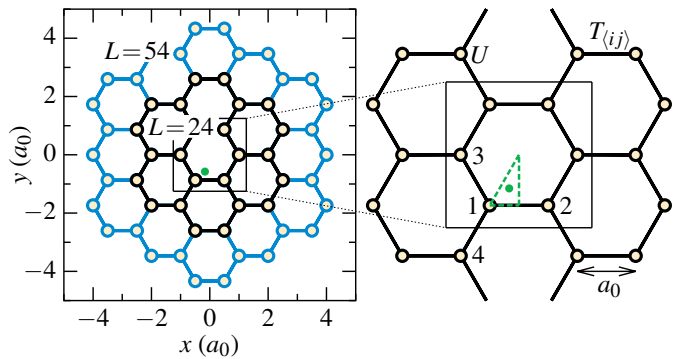

**Figure 13.** Sketch of the lattice structure of circular honeycomb clusters with $L = 24$ (black) and $L = 54$ sites (blue). The green point indicates the position where the projectile hits the lattice plain. From Ref. [23]

### 3.4.3. Time evolution following a charged-particle impact.

Let us now consider a very different type of excitation that is caused by the impact of an energetic projectile in a correlated Hubbard cluster. The interaction of the projectile with the electrons of the cluster is particularly strong for a charged particle. This type of excitation differs from the quenches above by its strongly localized character: typically only the few nearest neighbors of the impact point will be strongly affected. Depending on the velocity of the projectile, the interaction is also localized in time where the interaction duration is controlled by the initial velocity of the projectile.

The associated energy loss of the projectile ("stopping power") has been studied experimentally and theoretically for many years, for an overview see e.g. Ref. [141], and broad purpose numerical simulation tools (e.g. SRIM) were developed, e.g. Ref. [142]. These models are based on extensive averages and experimental input. Moreover, they assume that the solid can be treated as following the excitation adiabatically. More recently, time-dependent simulations have provided very detailed information on the complex physical processes. This mostly concerns time-dependent DFT simulations, e.g. Refs. [143, 144]. At the same time, TD-DFT does not allow one to reliably describe electronic correlation effects and the dynamics of correlations. Therefore, NEGF simulations are of high interest.

Time-dependent NEGF simulations of ion stopping were first presented by Balzer *et al.* in Ref. [23]. Here we summarize a few representative results. To investigate electronic correlation effects one can apply an NEGF–Ehrenfest approach where the dynamics of the projectile are treated classically. The corresponding electronic Hamiltonian is given by

$$\hat{H}_e = - J \sum_{\langle i,j \rangle, \sigma} c_{i\sigma}^\dagger c_{j\sigma} + U \sum_i \left( n_{i\uparrow} - \frac{1}{2} \right) \left( n_{i\downarrow} - \frac{1}{2} \right)$$
$$- \frac{Z_p e^2}{4\pi\epsilon_0} \sum_{i,\sigma} \frac{c_{i\sigma}^\dagger c_{i\sigma}}{|\vec{r}_p(t) - \vec{R}_i|} + \sum_{\langle i,j \rangle, \sigma} W_{ij}(t) c_{i\sigma}^\dagger c_{j\sigma} \,, \tag{148}$$

which, in addition to the standard Hubbard Hamiltonian, now includes the Coulomb interaction between a $Z_p$-fold charged ion and all electrons of the target. The trajectory of the projectile, $\vec{r}_p(t)$, is obtained by solving Newton's equation where the force is given by the total force from the interaction with all electrons. By solving the coupled Keldysh–Kadanoff–Baym equations for the electrons and Newton's equation for the projectile one obtains the time-dependent energy exchange between the ion and the electrons. The energy loss of the projectile follows from the asymptotic values for long times after and before the impact,

$$S_e[E_{\text{kin}}] = m_p \frac{\dot{r}_p^2(t \to -\infty)}{2} - m_p \frac{\dot{r}_p^2(t \to +\infty)}{2} \,. \tag{149}$$

Before starting the time-dependent simulations, again, the initial state has to be generated. Here two approximations are considered: (A) the initial state is a Hartree–Fock state and (B) the initial state is correlated and generated by adiabatic switching. From a computational stand point, (A) is advantageous, whereas (B) is more accurate but requires a substantially increased total computation duration.

The dependence of the energy loss on the initial kinetic energy, $E_{\text{kin}}$, of the projectile is plotted in Fig. 14 and show the characteristic single-peak behavior. For very fast projectiles, the interaction duration vanishes, and so does the energy exchange. On the other hand, for very slow projectiles, the initial kinetic energy is small. Therefore, obviously, a maximum exists, for an optimal choice of $E_{\text{kin}}$. For the present honeycomb clusters (see Fig. 13) this peak is in the range of 10keV.

Since no spatial homogeneity is assumed, the energy loss can be compared in finite clusters of different size. In Fig. 14 Balzer *et al.* [23] compare the energy loss of a proton of the same energy in two clusters [cf. the sketch in Fig. 13] of size $L = 54$ (top figure) and $L = 24$ (bottom figure). Clearly, the energy loss increases with cluster size because more electrons are being excited by the projectile.

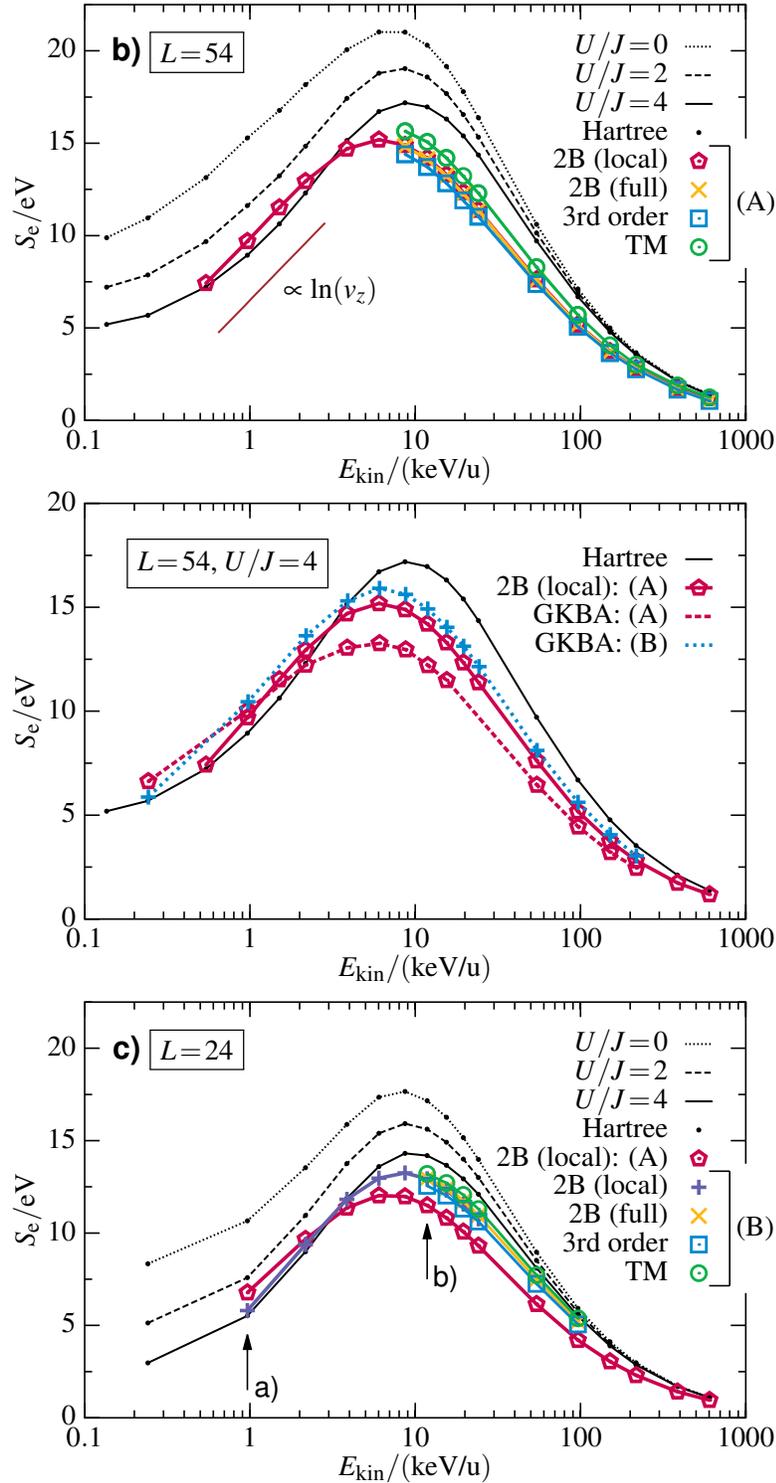

**Figure 14.** Energy loss $S_e$, Eq. (149), for protons passing through the honeycomb clusters of Fig. 13 of size $L = 54$ [panels (a) and (b)] and $L = 24$ [panel (c)]. In all panels, the value of the onsite interaction $U/J$ is encoded in the line style, the symbols correspond to NEGF results, and the black lines indicate the results of the Hartree approximation. The shortcuts stand for 2B (3rd order): second-order Born (third-order) approximation, TM: particle–particle $T$ matrix, local: local approximation of the selfenergy, $\Sigma_{ij} \to \Sigma_i \delta_{ij}$. From Ref. [23].

Next, the role of interaction effects in the substrate on the stopping power is analyzed. This can be achieved by comparing mean-field (Hartree) simulations with different coupling parameters $U/J$ as well as Hartree simulations to correlated NEGF results. Obviously, one should expect that interaction effects will be of minor importance at large projectile energies because, in this case, kinetic energy exceeds interaction energy. This is indeed observed in all simulations shown in Fig. 14. At energies exceeding 200keV the differences between different approximations quickly vanish. In contrast, for lower impact energies and, in particular, for energies below the peak energy, strong deviations are observed. The general trend is that, with increasing $U/J$ the stopping power decreases, regardless of the chosen selfenergy approximation. A comparison of Hartree and correlated simulations reveals that correlations, at large energies, tend to make the system "more rigid" what reduces the stopping power. Interestingly, at low impact energies, $E_{\text{kin}} \lesssim 3\text{keV}$ the situation changes and correlation effects lead to an increase of the stopping power. The explanation is that the fast and strong impact of the projectile excites a large number of electronic transitions in the system, including double excitations that are missing in a mean-field approach, see also Sec. 3.4.4.

Let us now discuss the influence of different selfenergy approximations. Figure 14 contains a large variety of approximations. Note, however, that these calculations are quite expensive and become increasingly more difficult when the impact energy decreases because then the required simulation duration grows. Therefore, most results were obtained for large impact energies where the differences between different approximations are small. In Ref. [23] the authors performed simulations using second-order Born (2B), third-order and particle–particle $T$-matrix (TM) selfenergies. The differences are small with the $T$ matrix yielding the largest stopping power. Comparing two-time simulations and single-time simulations within the HF-GKBA (top two frames of Fig. 14), the GKBA leads to a slight reduction of the stopping power. However, this comparison has only been done for a few cases, and more simulations are required to obtain a systematic picture. Also, the effect of initial correlations has not yet been fully clarified.

More recently, a special correlation effect has been investigated: the formation of local electronic double occupations, Eq. (145), due to the ion impact [145]. There it was shown that, at low projectile energies, electrons may be efficiently excited across the Hubbard gap (e.g. at half filling) which gives rise to an enhancement of the stopping power. This effect has been explored in more detail in Refs. [24, 146].

To summarize this section: NEGF simulations coupled to an Ehrenfest description of a classical ion have been shown to be a powerful tool to model the energy transfer between a projectile and correlated target. They allow for a fully time-resolved analysis and deep insight into the electronic transitions that can be triggered by the ion impact. At the same time, these simulations are very expensive at low impact energies and further optimization is needed. Interesting future questions include the quantum treatment of the projectile and of the electronic excitations in the ion as well as possible charge transfer processes between projectile and substrate [146, 147].

*3.4.4. Time evolution following a short enhancement ("kick") of the single-particle potential* We now consider another excitation scenario where an external single-particle potential is turned on for a very short time only,

$$f_i(t) = f_{i0}\,\delta(t - t_0). \tag{150}$$

Such a very short excitation, is spectrally (energetically) broad which means that a broad range of energetic transitions will be excited. Following the time-dependent dynamics $B(t)$ of a suitable observable one can easily reconstruct the spectral information contained in it, via Fourier transformation. It turns out that this is a very efficient way to obtain high-quality spectral information, provided the propagation can be extended to sufficiently long times, to avoid windowing effects in the Fourier transform. If, furthermore, the excitation is weak, i.e. linear response applies, then one accurately probes the properties of the unperturbed system, e.g. the ground state or the equilibrium properties.

This approach was first used in NEGF simulations by Kwong *et al.* to compute the optical absorption of a semiconductor [148]. There, the frequency-resolved absorption coefficient was obtained, after applying a short optical laser pulse, from Fourier transforming the interband polarization $P(t)$, for details, see Ref. [131]. Similarly, it was shown that one can obtain the dynamical structure factor and dielectric function of a correlated system (e.g. electron gas or plasma) by applying a short monochromatic electric field with wavenumber $q$, i.e. in Eq. (150) we replace $f_i \to U_0 \cos{(qr)}$, that excites a density modulation $\delta n(r, t)$ of wavelength $2\pi/q$ which yields the linear dynamic density-response function $\chi(q, \omega)$ [149]. As a consequence one obtains results for the dynamic dielectric function and for the dynamic structure factor that selfconsistently include correlation effects thereby obeying the relevant sum rules.

After successful applications to macroscopic systems this method was also used for finite systems. Van Leeuwen *et al.* computed the optical absorption of atoms by Fourier transforming the time-dependent dipole transition signal [69]. The method can also be applied to compute, via time-dependent NEGF simulations, the spectrum of electronic excitations of finite correlated systems. Balzer *et al.* considered a four electron model quantum well and showed the second-order Born selfenergies yield accurate results for the electronic double excitations [120]. Similar results were obtained by Säkinen et al. [150].

Here we illustrate this approach for a small Hubbard cluster of 8 sites and coupling strength $U/J = 0.1$. In the simulations of Hermanns *et al.* [76] the excitation was local on one site, i.e. in Eq. (150) the excitation amplitude was $f_{i0} \to f_{ij\alpha\beta} = w_0 \delta_{i,1} \delta_{ij} \delta_{\alpha,\beta}$, where $\alpha$ and $\beta$ denote the spin projections. Choosing a very small amplitude, $w_0 = 0.01J$, the system remained well inside the linear-response regime. Performing a very long simulation of duration $T = 1000J^{-1}$ provided an accurate excitation spectrum. The selfenergy was used on the second-order Born level, which is adequate for the present weak-coupling case. To achieve the desired long simulation duration a single-time simulation (with the Hartree–Fock GKBA) was carried out. The result is shown in Fig. 15 where the NEGF result is compared to an exact-diagonalization calculation. Obviously, the agreement is excellent. The NEGF simulations accurately reproduce the exact spectrum over a broad range of energies and practically capture all peaks. This agreement extends over an impressive seven orders of magnitude (note the logarithmic scale). Small deviations are visible for increasing frequency where peak shapes exhibit slight deviations. This calculation also allows to clearly single out correlation effects and explore the limitations of the mean-field result (cf. the blue curves labeled HF). While HF well captures the low-frequency peaks, it entirely misses peaks occuring above $\omega \sim 3.8J$. But also at lower energy, many peaks are missing, e.g. at $\omega/J \approx 2.7, 3.2$ and $3.6$ that are associated to double excitations.

This result confirms the power of this approach and the capability of NEGF simulations to obtain accurate ground-state (or equilibrium) spectra via time propagation. These results can, of course, be directly compared to independent pure ground-state (or equilibrium) simulations, e.g. within the framework of the Bethe–Salpeter equation for the Matsubara Green function. The comparison of the two approaches reveals that [149] time-dependent NEGF simulations with a selfenergy $\Sigma$ correspond to Bethe–Salpeter results with a two-particle kernel $K_{\mathrm{BSE}} = \delta\Sigma/\delta G$. For text-book discussions, see Refs. [72, 131].

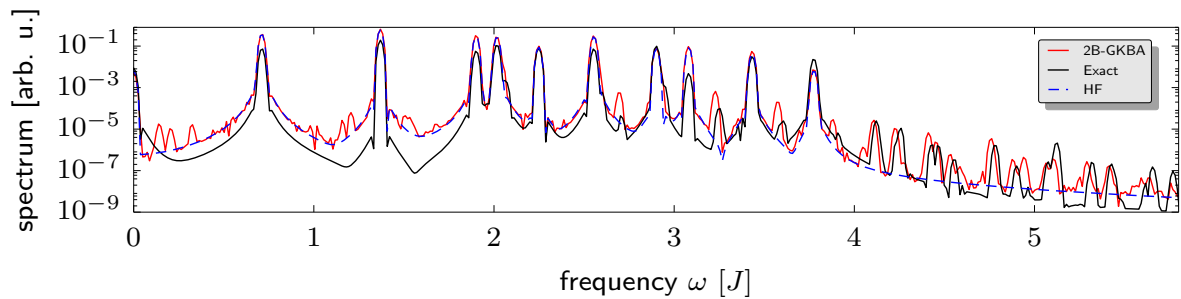

**Figure 15.** Spectral function of an 8-site Hubbard cluster at $U/J = 0.1$ computed via Fourier transform after a single-particle potential kick, Eq. (150). NEGF simulations with second-order Born selfenergies (Hartree–Fock-GKBA) are compared to Hartree and exact diagonalization results. From Ref. [76]

*3.4.5. Time evolution following a strong rapid quench of the onsite potential.* We now turn to the final example of time-dependent excitation that is very different from the previous cases. This section investigates the dynamics of small Hubbard clusters that are driven out of equilibrium by a very strong sudden quench of the on-site potential of the form

$$f_{ij}(t) = \Theta(t, t_0)\delta_{ij}\delta_{i1}w_0, \tag{151}$$

where throughout this section the value $w_0 = 5J$ is used. Thus at $t = 0$ site $i = 1$ is very strongly excited by a constant potential. This excitation, initially, drives a depletion of this site which is followed by a subsequent oscillation of the electronic density throughout the system. Such strong excitations of this form [Eq. (151)] of very small Hubbard clusters were studied in detail by Verdozzi and co-workers [117, 118] using selfconsistent two-time solutions of the KBE. They made a surprising observation: in contrast to the exact solution (which is easily found for these small systems), the NEGF dynamics of the density oscillations are strongly damped. The authors of these papers explained this artifact by the selfconsistency of the solution of the Dyson equation (or the KBE) which contains selfenergy contribution to arbitrary orders (powers). This leads to a series of peaks in the spectral function that are not present in the exact result. They also observed that the effect is particularly strong for small clusters.

In the following we illustrate this effect for a few examples. In particular, we are interested in 1) how this damping behavior depends on the chosen selfenergy approximation, 2) on the filling and 3) how results from the HF-GKBA behave. To this end we focus on very small systems containing just two electrons on two and four sites, corresponding to half and quarter filling, respectively. The case of a half-filled Hubbard dimer, with an interaction of $U = J$ and excitation strength of $w_0 = 5.0J$, is shown in Fig. 16 [76]. There, three of us analyzed the time-evolution of the density on the first (excited) site, comparing the exact result to selfconsistent two-time NEGF simulations and also single-time GKBA calculations with HF propagators. As expected, the exact solution exhibits undamped oscillations of the density because electrons periodically move between the two sites after an initial depletion of the first site. Consider now the result of fully selfconsistent solutions of the KBE with second-order selfenergy (top figure). Here, reasonably good agreement with the exact solution is only observed for the first 1.5 oscillation periods. For later times the oscillation period becomes smaller and the oscillations quickly damp, reaching an artificial steady state that is not present in the exact result. It is now of interest to reduce the level of selfconsistency for which different approximations can be considered [151]. Here we followed a different and more systematic strategy: we applied the Hartree–Fock-GKBA propagators as described in Sec. 2.10. The results for the case of second-order Born selfenergies are also included in the top part of Fig. 16. Evidently, application of the HF-GKBA indeed "cures" the artificial damping and qualitatively agrees with the exact solution. The quantitative agreement, though, can only be considered satisfactory for the first five oscillation periods up to 6 inverse hopping amplitudes. While the exact oscillation period is roughly reproduced, the GKBA exhibits several phase changes that are not present in the exact dynamics.

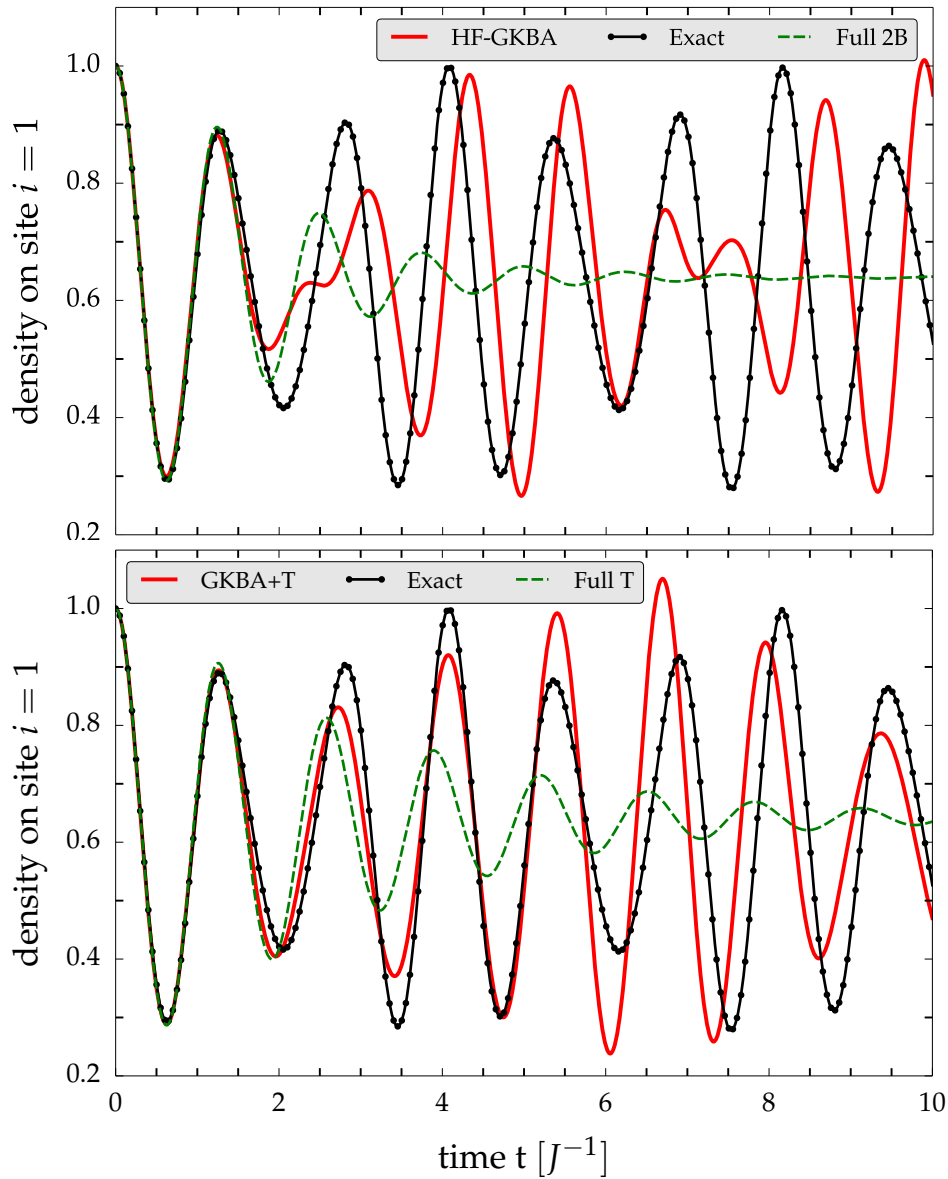

**Figure 16.** Half-filled Hubbard dimer for $U = J$. Density evolution on the first site following a sudden switch-on of the onsite potential [Eq. (151)] of strength $w_0 = 5.0J$ on site $i = 1$. Black line with dots: exact results; green dashed line: two-time NEGF result, and red full line: HF-GKBA. Top figure: NEGF and GKBA with second-order Born selfenergy (Full 2B). Bottom figure: NEGF and GKBA with $T$-matrix selfenergy (Full T). From Ref. [76].

To further improve the agreement with the exact results we then combined the HF-GKBA with a higher-order non-perturbative selfenergy. Results for the particle–particle $T$-matrix approximation, both, for two-time solutions and in combinations with the HF-GKBA, are shown in the bottom part of Fig. 16. In this case, the GKBA achieves very good agreement with the exact result. The first four periods are reproduced very accurately, wheres for later times a slight dephasing is observed. In contrast, the full two-time solution of the KBE with the particle–particle $T$ matrix is quickly damped, as in the case of the second-order Born approximation (Full 2B, SOA, top figure). At the same time, the $T$-matrix result is better than the second-Born approximation which is explained by the moderate coupling strength. Based on these results we conclude that the HF-GKBA, if used with the proper selfenergy, provides an excellent method to solve the problem of artificial damping of two-time simulations in the case of very strong excitation.

We now turn to the case of a four-site fermionic Hubbard model with one quarter filling, and an interaction strength of $U = 1.5J$. The excitation is the same as before. For this setup, *Friesen et. al.* [151] reported the particle–particle $T$ matrix to show very good agreement with the exact solution, while the $GW$ approximation performed much worse. Here, the same setup is re-examined using all selfenergy approximations that were introduced in Section 4 and 5. We observe that the selfenergies can be categorized into three groups based on the amount of artificial damping they exhibit. The three most strongly damped methods, which quickly start to deviate from the exact result are the

electron–hole $T$ matrix, the $GW$ approximation and the second-order Born selfenergy and are shown in the top part of Fig. 17.

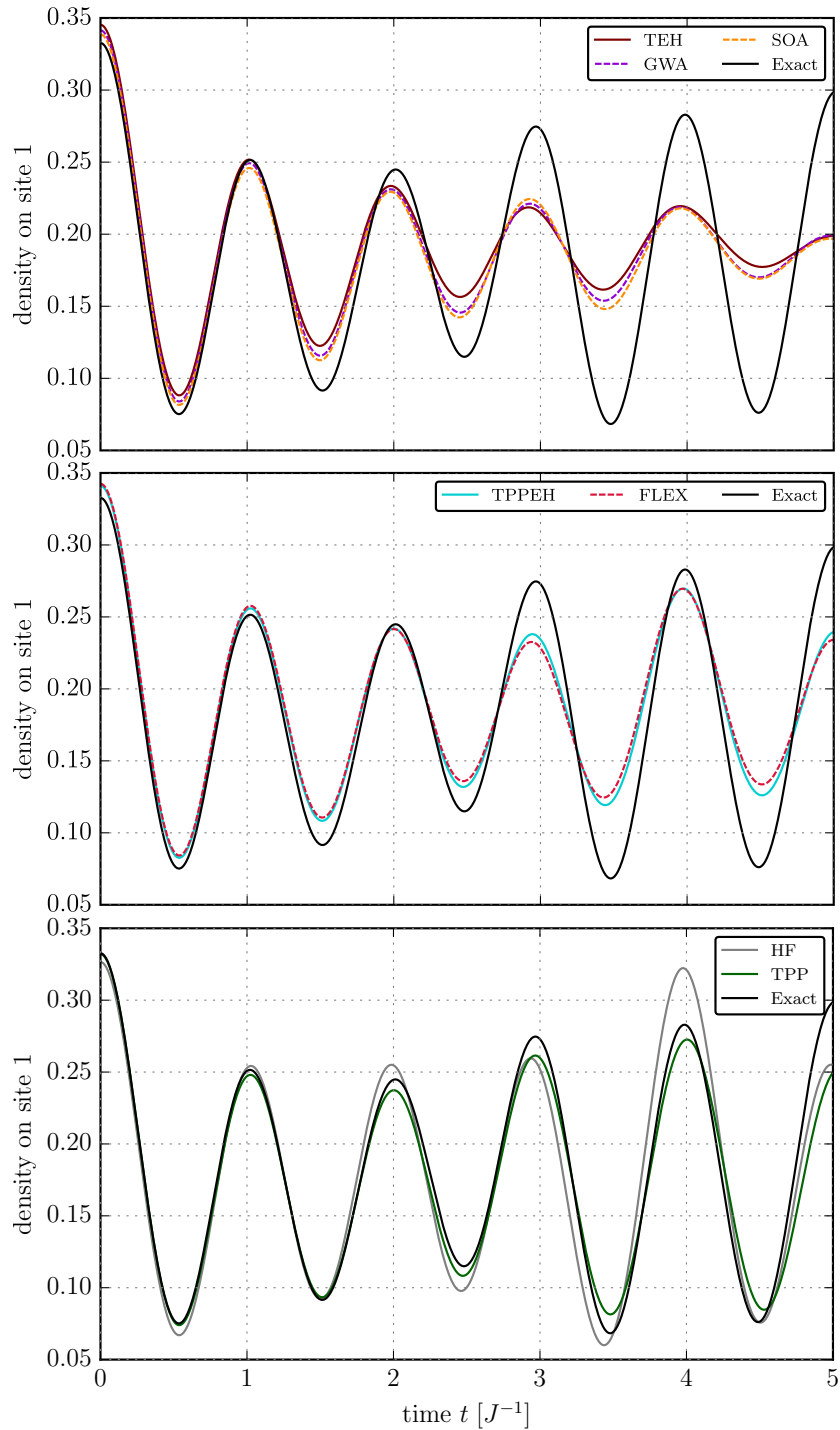

**Figure 17.** One quarter-filled four-site Hubbard cluster for $U = 1.5J$. Density evolution of the first (driven) site following a sudden switch-on of an onsite potential of strength $w_0 = 5.0J$. The selfenergy approximations are given in the inset. Top (middle): strongly (moderately) damped methods, bottom: undamped results.

For all three methods, the damping sets in after the second oscillation period, and the amplitude quickly drops to roughly one third of the exact amplitude. For the fluctuating-exchange approximation (FLEX) and the similarly constructed approximation combining only the electron–hole and the particle–particle $T$ matrix (TPPEH), the results are shown in the middle part of Fig. 17. Here, one notices only a slight damping, and the overall agreement with the exact result is significantly improved, compared to the first group of selfenergies approximations. The third group of methods that don't exhibit conceivable damping, in the present situation, are the particle–particle $T$ matrix and, by construction, the uncorrelated Hartree(–Fock) approximation. The corresponding results are shown in the bottom part of Fig. 17. Both methods are in good agreement with the exact result, but the quality of the $T$-matrix approximation is much better than that of the uncorrelated Hartree approximation. The superior quality of the particle–particle $T$ matrix, in the present case, is of course arising from the low density in the system as was noted also in the previous sections. For higher density the situation changes. Most importantly, we conclude that the artificial damping in strongly excited finite systems is not a generic feature of all two-time simulations but is observed for selfenergy approximations that most strongly deviate from the exact Hamiltonian. In addition, we have seen before that the artificial damping is removed almost completely by invoking the GKBA with Hartree–Fock propagators.

### 3.5. Discussion of the numerical results and outlook

We conclude the overview of numerical solutions of the KBE in equilibrium and nonequilibrium with a brief summary of the main findings and an outlook. In Sec. 3.3 it has been shown that, for the interacting ground state, the order-by-order expansion with respect to the interaction strength yields good results, already with the second-order approximation (SOA). The results significantly improve further when also the third-order terms are into account (TOA). On the other hand, the non-perturbative expansions, such as the particle–particle and particle–hole $T$-matrix approximations and the $GW$ approximation, as well as combinations of several non-perturbative approximations, do, in general, not reach the same accuracy. Concerning the single-particle spectrum, both, perturbative and non-perturbative methods, yield good results and especially with the combination of the results for all available methods, most of the spectral information can be computed.

For time-dependent processes involving strong excitations, the GKBA has been found to be a valuable tool to mitigate correlation-induced damping. Combined with the particle–particle $T$ matrix, which has shown to produce excellent results in full calculations for lower filling factors, it yields a density evolution which is close to the exact solution. With the great variety of available selfenergy approximations, covered in Section 4 and 5 both, with and without application of the GKBA, a powerful toolset is available to study systems of arbitrary dimension and large particle numbers, where currently no exact methods are available.

In section 3.4 we considered the dynamics of small- and intermediate-size Hubbard clusters in response to various kinds of excitations ranging from short to long and weak to strong. Where comparisons with benchmark data were available NEGF calculations have obtained very good agreement, provided the proper selfenergy approximation has been chosen. This means that $\Sigma$ has to be chosen such that is matches both the coupling strength and filling (density). While, at weak coupling, the second-order Born approximation is adequate, at moderate coupling, $U \gtrsim J$, this approximation shows significant deviations and there is even no guarantee that it captures the dominant trends. Therefore, it is crucial to have a sufficiently large arsenal of selfenergy approximations available that can be used in a flexible manner. Thus, we can conclude that NEGF simulations have, indeed, reached a level of accuracy where reliable predictions can be made. For an efficient use of the proper selfenergies, it is crucial to have explicit results for each approximation available that can be rapidly implemented. This will be discussed in great detail in the next sections.

Before doing this we note that time-dependent NEGF simulations have seen a dramatic surge in activity in many areas. Even though the most accurate and best tested results were obtained for lattice models, as discussed above, there exist many further applications that are outside the scope of this article but should be briefly mentioned, together with a few relevant references. Indeed, second-order Born simulations were reported for electrons in quantum dots [152, 153], the laser excitation of small atoms and molecules [69–71]. Interestingly, second-order Born calculations were shown to be applicable also to the photoionization of larger atoms such as krypton [154] and to well reproduce two-electron processes such as Auger ionization [155]. Other finite systems the relaxation dynamics of which were recently studied include graphene-type clusters such as graphene nanoribbons [156]. Correlation effects of particular interest here are, e.g., carrier multiplication effects.

While in the applications listed above the electronic system was typically treated as isolated, in many cases the coupling to the environment (bath) has to be included. Examples are transport problems. Since the bath is typically much larger than the physical system of interest with time scales often well separated from those of the system, suitable procedures to eliminate the bath degrees of freedom are of high interest. An important example is transport through nanoscale systems where it is often advantageous to eliminate the effect of the leads. Here a highly efficient solution within the NEGF scheme is provided by an embedding approach [157], for a text-book discussion see Ref. [72]. This approach has been extended to the photoionization of atoms [154] and charge transfer processes between atoms and a solid surface [24, 147].

Due to the success of these NEGF simulations we expect that the number and scope of applications will continue to increase over the next years. Further progress requires new developments in several directions. One is certainly the use of improved more realistic models. Here we mention the idea to combine NEGF simulations with an ab initio basis set that is provided by a Kohn–Sham simulation. This concept is realized, e.g., within the Yambo code of Marini *et al.*, e.g. Refs. [158, 159]. A major problem for these approaches is the large basis size which leads to very large requirements of CPU time and computer memory.

One approach that allows to mitigate these problems, at least partially, is the generalized Kadanoff–Baym ansatz (GKBA) that was introduced in Sec. 5.2. We have seen throughout the present section that the GKBA, combined with Hartree–Fock propagators, indeed provides the expected major savings of resource. Moreover, in many cases it yields excellent results that may be not worse than the full two-time simulation, e.g. Ref. [76]. This, however, does not mean that two-time simulations become obsolete. In contrast, as we have seen from the comparison to DMRG results in Secs. 3.4.1 and 3.4.2, in many cases, the exact result is enclosed between two-time and single-time NEGF results. Thus, both types of simulations should be developed in parallel.

Future developments in this field should also aim at improving the GKBA simulations. Part of the problems of the GKBA will be overcome if, instead of Hartree–Fock propagators, correlated propagators are being used. Here, we mention recent promising proposals of Refs. [24, 160]. Furthermore, it will also be important to include correlated initial states into GKBA simulations and to develop efficient schemes that reduce the associated computational overhead, e.g. Refs. [24, 161–163].

Finally, all of the NEGF applications discussed above crucially depend on the availability of a large arsenal of selfenergy approximations and their optimization for special basis sets. In the remainder of this paper, we present a detailed overview of practical formulas that are ready to use in NEGF simulations.

## 4. Selfenergy approximations I: Perturbation expansions

To study time-dependent observables of the system of interest by solving the Keldysh–Kadanoff–Baym equations, one has to know the less and greater components of the two-time Green function, $G^{\lessgtr}(t_1, t_2)$. These components can be generated from the solution of the contour Dyson equation, which is the first equation of Hedin's equations, cf. Eq. (91). Its main ingredient is the selfenergy $\Sigma$. It is of great importance, both, from a physical point of view—since it incorporates all different classes of inter-particle effects and processes—as well as from the computational view—since a large portion of the numerical resources is consumed for its determination. $\Sigma$ is the solution of the second Hedin equation, Eq. (95), which, in turn, is dependent on the third to fifth Hedin equation, Eqs. (102), (105) and (106), and, in turn, also on the Dyson equation via the Green function entering it. Since the selfenergy, by iteration of Hedin's equations, consists of an infinite number of terms, a strategy has to be used to reach a good approximation with only a small finite subset. The two most common approaches are detailed in the following. The first applies a perturbative approach with respect to the interaction strength, i.e., with respect to the powers of the potential $w$, whereas the second uses a resummation idea. The resummation involves (infinitely many) diagrams in all orders of the interaction strength belonging to certain topological classes, namely the particle–particle, the particle–hole $T$-matrix approximation (TPP, TEH), the $GW$ approximation (GWA), or a combination of all of them or some subsets. The present Section 4 deals with the expansion in orders of the interaction strength where we systematically study selfenergies of first, second and third order. After this, in Section 5, the resummation approaches, are introduced and discussed for all relevant special cases. In these two sections we introduce all selfenergies that were applied in the calculations of Section 3.

In the remainder of this section we focus on the expansion of the selfenergy with respect to the number of interaction factors $w$ involved. This means that the $n$-th order approximation contains only terms with no more than $n$ interactions. This procedure has two sources of reasoning behind it. First, for small interaction strength (in units of the single-particle energy), higher-order processes with more interaction factors usually have small amplitudes. Second, even for larger coupling strength, higher-order terms also contain several Green functions corresponding to the correlated creation and annihilation of several particles. The strength of these correlations is not directly coupled to the interaction strength and, therefore, a lower-order approximation may give good results even for stronger interactions.

Regarding the treatment of the Green functions occuring in the expansion of the selfenergy, two different approaches are common. In the *self-consistent approach*, one starts from the non-interacting Green function $G^{(0)}$ and computes the selfenergy according to Eq. (95), within the chosen approximation. Then, the Dyson equation is evaluated, taking $G = G^{(0)}$ on the right-hand side. With the resulting $G$, the selfenergy is reevaluated. This procedure is continued iteratively until convergence is reached. In contrast, the *free-particle approach*—expands $G$ with respect to the number of occurrences of the interaction, as it is done with the other quantities in Hedin's equations. That way, it is ensured that the $n$-th order approximation contains no terms of higher order, which is in contrast to the self-consistent approach (through the iteration procedure terms of all orders are produced). Nonetheless, these terms are valid terms of order higher than $n$, so it cannot be answered beforehand which method is superior.

In the remainder of this section, both these approaches will be analyzed in detail by considering selfenergies up to the third order in the interaction, i.e. Hartree–Fock (HF), second-order Born approximation (SOA) and third-order approximation (TOA). These approximations will be systematically deduced from Hedin's equation first, for a general basis. After this, each result will be specified to two important cases: a basis where the interaction is diagonal ("diagonal basis") and the Hubbard basis. For all three basis representations we present the quantities first on the Keldysh contour and then we derive the greater/less and retarded/advanced components. In addition, we separately present the results for bosons and fermions. In cases when there are differences for different spin projections, the different cases will be specified separately. In addition to the formulas we present the graphical representation in terms of Feynman diagrams that will be introduced in Fig. 18.

### 4.1. First-order terms. Hartree and Fock selfenergies

To determine the first-order contributions to $\Sigma$, one starts from the second Hedin equation, Eq. (95). The first term, the Hartree term $\Sigma^{\mathrm{H}}$, is of first order, since it contains one $w$. It was already given in Eq. (96) and is repeated here for consistency,

$$\Sigma_{ij}^{\mathrm{H}}\big(z_1, z_2\big) = \pm i\hbar \delta_{\mathcal{C}}\big(z_1, z_2\big) \sum_{mn} w_{mijn}\big(z_1\big) G_{nm}\big(z_1, z_{1^+}\big) . \tag{152}$$

Since this expression contains a contour delta function, $\delta_{\mathcal{C}}$, the only non-vanishing Keldysh component is

$$\begin{aligned}
\Sigma_{ij}^{\mathrm{H},\delta}\big(t_1\big) = \Sigma_{ij}^{\mathrm{H}}\big(z_1, z_1\big) = && (153) \\
= \pm i\hbar \sum_{mn} w_{mijn}\big(z_1\big) G_{nm}\big(z_1, z_{1^+}\big) = \pm i\hbar \sum_{mn} w_{mijn}\big(t_1\big) G_{nm}^{<}\big(t_1, t_1\big) .
\end{aligned}$$

The second first-order term belongs to $\Sigma^{\mathrm{xc}}$, cf. Eq. (95), and is generated by the first-order term $W^{(1)}$ of Eq. (102),

$$W_{ijkl}^{(1)}\big(z_1, z_2\big) = W_{ijkl}^{\mathrm{bare}}\big(z_1, z_2\big) = \delta_{\mathcal{C}}\big(z_1, z_2\big) w_{ijkl}\big(z_1\big) , \tag{154}$$

and the zeroth order vertex $\Gamma^{(0)}$, cf. Eq. (106),

$$\Gamma_{ijkl}^{(0)}\big(z_1, z_2, z_3\big) = \delta_{\mathcal{C}}\big(z_1, z_{2^+}\big) \delta_{\mathcal{C}}\big(z_3, z_2\big) \delta_{ik} \delta_{jl} . \tag{155}$$

This yields the *Fock term*, $\Sigma^{\mathrm{F}}$, which is of time-diagonal structure as the Hartree term and is given by

$$\begin{aligned}
\Sigma_{ij}^{\mathrm{F}}\big(z_1, z_2\big) = \Sigma_{ij}^{\mathrm{xc}}\big(W^{(1)} \equiv W^{\mathrm{bare}}, \Gamma^{(0)}\big) = && (156) \\
= i\hbar \delta_{\mathcal{C}}\big(z_1, z_2\big) \sum_{mn} w_{injm}\big(z_1\big) G_{mn}\big(z_1, z_{1^+}\big) .
\end{aligned}$$

The $\delta$-component (prefactor of the delta function) is given by

$$\Sigma_{ij}^{\mathrm{F},\delta}\big(t_1\big) = \Sigma_{ij}^{\mathrm{F}}\big(z_1, z_1\big) = \sum_{mn} w_{injm}\big(t_1\big) G_{mn}^{<}\big(t_1, t_1\big) . \tag{157}$$

Because there is no further term stemming from $W^{(0)}$ and $\Gamma^{(1)}$, since $W^{(0)} \equiv 0$, the final result for the first-order selfenergy $\Sigma^{(1)}$ is given by

$$\Sigma^{(1)} = \Sigma^{\mathrm{H}} + \Sigma^{\mathrm{F}} , \tag{158}$$

which both are time-diagonal. For the *non-selfconsistent treatment*, the Green functions appearing in $\Sigma^{\mathrm{H}}$ and $\Sigma^{\mathrm{F}}$ are taken as free Green functions, i.e., $G \longrightarrow G^{(0)}$. Otherwise, all expressions remain the same. There are no additional terms containing higher-order Green functions in first order.

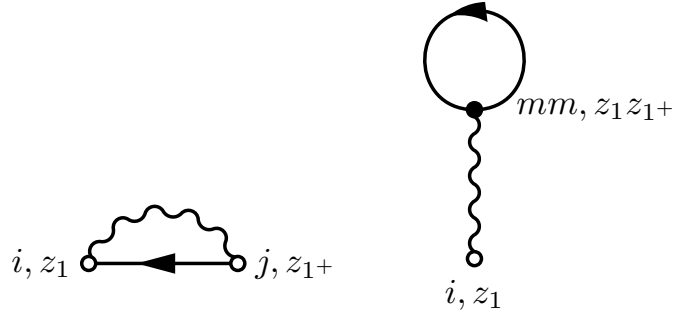

**Figure 18.** First-order diagrams in the diagonal basis. **Left:** Fock diagram, $\Sigma^{\mathrm{F,diagonal}}$. **Right:** Hartree diagram, $\Sigma^{\mathrm{H,diagonal}}$. Both diagrams are time-diagonal.

***Diagonal basis***. In a basis where the interactions are diagonal, $w_{ijkl} = \delta_{il}\delta_{jk}w_{ij}$, the first-order terms read

$$\Sigma_{ij}^{\mathrm{H,diagonal}}\big(z_1,z_2\big) = \pm i\hbar\delta_{\mathcal{C}}\big(z_1,z_2\big)\delta_{ij}\sum_m w_{mi}\big(z_1\big)G_{mm}\big(z_1,z_{1+}\big)\,, \tag{159}$$

$$\Sigma_{ij}^{\mathrm{F,diagonal}}\big(z_1,z_2\big) = i\hbar\delta_{\mathcal{C}}\big(z_1,z_2\big)w_{ij}\big(z_1\big)G_{ij}\big(z_1,z_{1+}\big)\,, \tag{160}$$

with the $\delta$-components

$$\Sigma_{ij}^{\mathrm{H,diagonal},\delta}\big(t_1\big) = \pm i\hbar\delta_{ij}\sum_m w_{mi}\big(t_1\big)G_{mm}^<\big(t_1,t_1\big)\,, \tag{161}$$

$$\Sigma_{ij}^{\mathrm{F,diagonal},\delta}\big(t_1\big) = i\hbar w_{ij}\big(t_1\big)G_{ij}^<\big(t_1,t_1\big)\,. \tag{162}$$

**Feynman diagrams.** The structure of the selfenergy contributions can be suitably visualized by using Feynman diagrams [164]. In this diagrammatic representation, Green functions are depicted as solid lines with an arrow pointing from the second argument to the first argument (since the creation operator in $G$ has the second argument and the annihilation operator has the first argument). The interaction is represented by a wiggly line which has two endpoints (in the diagonal basis). Employing the Feynman diagram technique, the two first-order contributions to the selfenergy are depicted in Fig. 18.

***Hubbard basis***. For the Hubbard basis, introduced in Section 2.3, the first-order selfenergy terms can be directly worked out. Here we separately consider the cases of bosons and fermions (superscripts $\mathfrak{b}$ and $\mathfrak{f}$, respectively), cf. Eqs. (28) and (35). For **bosons** we obtain

$$\Sigma_{i\alpha j\alpha}^{\mathrm{H,\mathfrak{b}}}\big(z_1,z_2\big) = i\hbar\delta_{\mathcal{C}}\big(z_1,z_2\big)\delta_{ij}\sum_\epsilon U\big(z_1\big)G_{i\epsilon i\epsilon}\big(z_1,z_{1+}\big)\,, \tag{163}$$

$$\Sigma_{i\alpha j\alpha}^{\mathrm{F,\mathfrak{b}}}\big(z_1,z_2\big) = i\hbar\delta_{\mathcal{C}}\big(z_1,z_2\big)\delta_{ij}U\big(z_1\big)G_{i\alpha i\alpha}\big(z_1,z_{1+}\big)\,, \tag{164}$$

with the $\delta$-components

$$\Sigma_{i\alpha j\alpha}^{\mathsf{H},\mathfrak{b},\delta}\big(t_1\big) = \mathrm{i}\hbar\delta_{ij}\sum_{\epsilon} U\big(t_1\big) G_{i\epsilon i\epsilon}^{<}\big(t_1, t_1\big),\qquad(165)$$

$$\Sigma_{i\alpha j\alpha}^{\mathsf{F},\mathfrak{b},\delta}\big(t_1\big) = \mathrm{i}\hbar\delta_{ij} U\big(t_1\big) G_{i\alpha i\alpha}^{<}\big(t_1, t_1\big).\qquad(166)$$

On the other hand, for **fermions** one obtains the following Hartree terms

$$\Sigma_{i\alpha j\alpha}^{\mathsf{H},\mathfrak{f}}\big(z_1, z_2\big) = -\mathrm{i}\hbar\delta_{\mathcal{C}}\big(z_1, z_2\big)\delta_{ij}\sum_{\epsilon\neq\alpha} U\big(z_1\big) G_{i\epsilon i\epsilon}\big(z_1, z_{1^+}\big),\qquad(167)$$

$$\Sigma_{i\alpha j\alpha}^{\mathsf{H},\mathfrak{f},\delta}\big(t_1\big) \quad= -\mathrm{i}\hbar\delta_{ij}\sum_{\epsilon\neq\alpha} U\big(t_1\big) G_{i\epsilon i\epsilon}^{<}\big(t_1, t_1\big),\qquad(168)$$

whereas the fermionic Fock terms vanish exactly in the Hubbard basis, $\Sigma_{i\alpha j\alpha}^{\mathsf{F},\mathfrak{f}}\big(z_1, z_2\big) \equiv \Sigma_{i\alpha j\alpha}^{\mathsf{F},\mathfrak{f},\delta}\big(t_1\big) \equiv 0$.

We also consider the important special cases of spin-0 bosons and spin-1/2 fermions, respectively (the spin is indicated by an additional superscript). For spin-0 bosons the terms attain the form

$$\Sigma_{ij}^{\mathsf{H},\mathfrak{b},0}\big(z_1\big) \quad= \mathrm{i}\hbar\delta_{\mathcal{C}}\big(z_1, z_2\big)\delta_{ij} U\big(z_1\big) G_{ii}\big(z_1, z_{1^+}\big),$$
$$\Sigma_{ij}^{\mathsf{F},\mathfrak{b},0}\big(z_1\big) \quad= \mathrm{i}\hbar\delta_{\mathcal{C}}\big(z_1, z_2\big)\delta_{ij} U\big(z_1\big) G_{ii}\big(z_1, z_{1^+}\big),$$
$$\Sigma_{ij}^{\mathsf{H},\mathfrak{b},0,\delta}\big(t_1\big) = \mathrm{i}\hbar\delta_{ij} U\big(t_1\big) G_{ii}^{<}\big(t_1, t_1\big),$$
$$\Sigma_{ij}^{\mathsf{F},\mathfrak{b},0,\delta}\big(t_1\big) = \mathrm{i}\hbar\delta_{ij} U\big(t_1\big) G_{ii}^{<}\big(t_1, t_1\big),$$

whereas for spin-1/2 fermions the results are

$$\Sigma_{i\uparrow j\uparrow}^{\mathsf{H},\mathfrak{f},1/2}\big(z_1\big) \quad= -\mathrm{i}\hbar\delta_{\mathcal{C}}\big(z_1, z_2\big)\delta_{ij} U\big(z_1\big) G_{i\downarrow i\downarrow}\big(z_1, z_{1^+}\big),$$
$$\Sigma_{i\downarrow j\downarrow}^{\mathsf{H},\mathfrak{f},1/2}\big(z_1\big) \quad= -\mathrm{i}\hbar\delta_{\mathcal{C}}\big(z_1, z_2\big)\delta_{ij} U\big(z_1\big) G_{i\uparrow i\uparrow}\big(z_1, z_{1^+}\big),$$
$$\Sigma_{i\uparrow j\uparrow}^{\mathsf{H},\mathfrak{f},1/2,\delta}\big(t_1\big) \quad= -\mathrm{i}\hbar\delta_{ij} U\big(t_1\big) G_{i\downarrow i\downarrow}^{<}\big(t_1, t_1\big),$$
$$\Sigma_{i\downarrow j\downarrow}^{\mathsf{H},\mathfrak{f},1/2,\delta}\big(t_1\big) \quad= -\mathrm{i}\hbar\delta_{ij} U\big(t_1\big) G_{i\uparrow i\uparrow}^{<}\big(t_1, t_1\big),$$

where the fermionic Fock terms are again zero. The corresponding diagrams for spin-0 bosons and spin-1/2 fermions are shown in Figs. 19 and 20, respectively.

## 4.2. Second-order terms. Second-Born approximation (SOA)

We now return to Eqs. (95) and (101), and investigate the selfconsistent second-order contribution,

$$\Sigma^{(2)} = \Sigma^{\mathrm{xc},(2)}.\qquad(169)$$

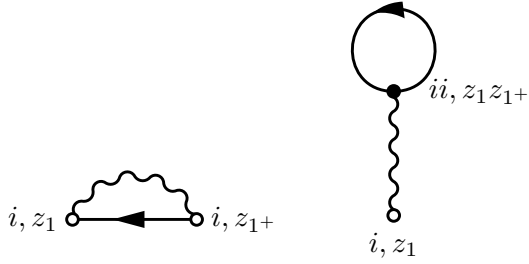

**Figure 19.** First-order diagrams in Hubbard basis for spin-0 bosons. **Left:** Fock diagram, $\Sigma^{\mathrm{F,b},0}$. **Right:** Hartree diagram, $\Sigma^{\mathrm{H,b},0}$. Note that both diagrams coincide for the Hubbard basis.

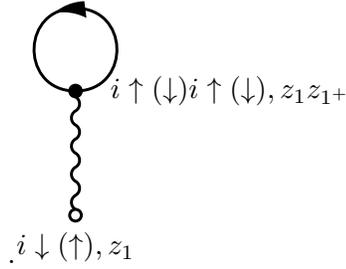

**Figure 20.** First-order (Hartree) diagram in Hubbard basis for spin-1/2 fermions, $\Sigma^{\mathrm{H,f},1/2}$.

The second-order terms of $\Sigma^{\mathrm{xc},(2)}$ can only be of either one of two forms

$$\Sigma^{(2),2,0} = \Sigma^{\mathrm{xc},(2)}\left(W^{(2)}, \Gamma^{(0)}\right), \qquad \Sigma^{(2),1,1} = \Sigma^{\mathrm{xc},(2)}\left(W^{(1)}, \Gamma^{(1)}\right), \qquad (170)$$

where the superscripts refer to the orders of $W$ and $\Gamma$. The first term involves $W^{(2)}$, the structure of which is determined from Eq. (102),

$$W^{(2)} = W^{\mathrm{ns}}\left(P^{(0)}, W^{(1)}\right), \qquad (171)$$

where the zeroth order polarization is

$$P^{(0)} = P\left(\Gamma^{(0)}\right), \qquad (172)$$

which is explicitly given by[xiv]

$$
\begin{aligned}
P^{(0)}\left(1, 2\right) &= \pm \mathrm{i}\hbar G\left(1, 3\right) G\left(4, 1\right) \\
&\quad \delta\left(3, 4^+\right) \delta\left(2, 4\right) \\
&= \pm \mathrm{i}\hbar G\left(1, 2\right) G\left(2, 1\right)
\end{aligned}
\qquad (173)
$$

Inserting this result into Eq. (171) and, employing Eq. (104), one arrives at

---

[xiv]For the purpose of better understanding, the following derivations are given in the simplified notation, as introduced in Section 2.8. The way to the first second-order selfenergy term in the full notation is presented in Appendix A.1.1.

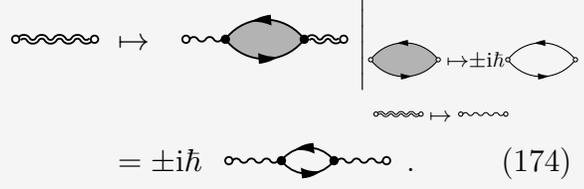

$$W^{(2)}\left(1,2\right) = w\left(1,3\right)P^{(0)}\left(3,4\right)$$
$$W^{(1)}\left(4,2\right)$$
$$= \pm i\hbar w\left(1,3\right)G\left(3,4\right)$$
$$G\left(4,3\right)w\left(4,2\right) \tag{174}$$

With this, $\Sigma^{(2),2,0}$ can be calculated as, [cf. Eq. (101)],

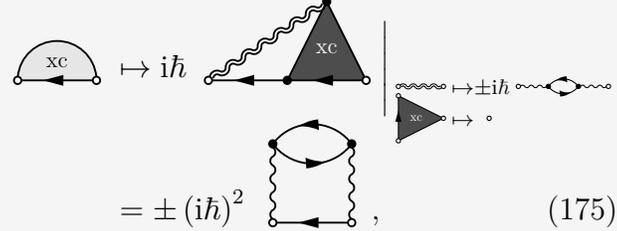

$$\Sigma^{(2),2,0}\left(1,2\right)$$
$$= i\hbar W^{(2)}\left(1,3\right)G\left(1,4\right)$$
$$\Gamma^{(0)}\left(4,2,3\right)$$
$$= \pm\left(i\hbar\right)^2 w\left(1,3\right)G\left(3,4\right)$$
$$G\left(4,3\right)w\left(4,2\right)G\left(1,2\right) \tag{175}$$

or, in the full extended notation,

$$\Sigma_{ij}^{(2),2,0}\left(z_1,z_2\right) = \pm\left(i\hbar\right)^2 \sum_{mn} G_{mn}\left(z_1,z_2\right) \sum_{st} G_{st}\left(z_1,z_2\right) \times \tag{176}$$
$$\times \sum_r w_{irsm}\left(z_1\right) \sum_u w_{tnju}\left(z_2\right) G_{ur}\left(z_2,z_1\right).$$

The other second-order selfenergy term, $\Sigma^{(2),1,1}$, requires the first-order term of the vertex $\Gamma$, the structure of which is

$$\Gamma^{(1)} = \Gamma\left(\delta\Sigma^{\mathrm{xc},(1)}/\delta G, \Gamma^{(0)}\right), \tag{177}$$

and involves the functional derivative of $\Sigma^{\mathrm{xc},(1)}$ with respect to $G$,

$$\frac{\delta\Sigma^{\mathrm{xc},(1)}\left(1,2\right)}{\delta G\left(5,6\right)} = \frac{\delta\Sigma^{\mathrm{xc},(1),\mathrm{F}}\left(1,2\right)}{\delta G\left(5,6\right)}. \tag{178}$$

Employing Eq. (156), one finds[xv]

---

[xv]For the full-notation derivation, see Appendix A.1.2.

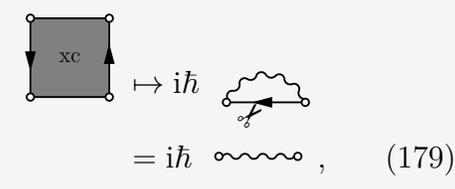

$$\frac{\delta \Sigma^{\mathrm{xc},(1)}\big(1,2\big)}{\delta G\big(5,6\big)} = \mathrm{i}\hbar w\big(1,2\big)\frac{\delta G\big(1,2\big)}{\delta G\big(5,6\big)}$$

$$= \mathrm{i}\hbar \delta\big(1,5\big)\delta\big(2,6\big)w\big(1,2\big) \qquad (179)$$

where the functional derivative with respect to $G$, in the diagrams, corresponds to cutting the $G$-line (or, more generally, all $G$-lines one by one) which is symbolized by the scissors. In case of different arguments of the Green functions, the result is

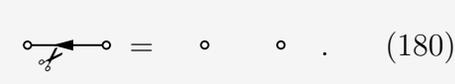

$$\frac{\delta G\big(1,2\big)}{\delta G\big(5,6\big)} = \delta\big(1,5\big)\delta\big(2,6\big) \qquad (180)$$

Using Eqs. (179) and (155), one arrives at

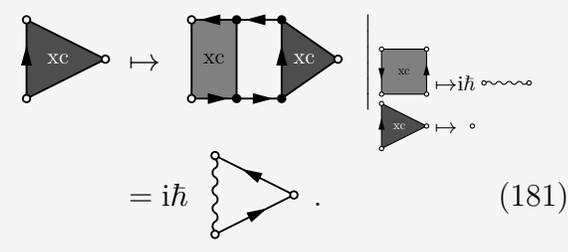

$$\Gamma^{(1)}\big(1,2,3\big)$$
$$= \frac{\delta \Sigma^{\mathrm{xc},(1)}\big(1,2\big)}{\delta G\big(4,5\big)}G\big(4,6\big)$$
$$G\big(7,5\big)\Gamma^{(0)}\big(6,7,3\big)$$
$$= \mathrm{i}\hbar w\big(1,2\big)G\big(1,3\big)G\big(3,2\big) \qquad (181)$$

Inserting this result, together with Eq. (154), yields

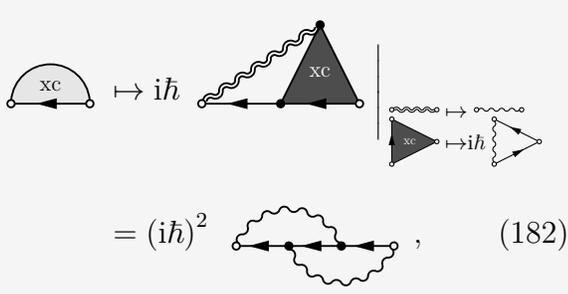

$$\Sigma^{(2),1,1}\big(1,2\big)$$
$$= \mathrm{i}\hbar W^{(1)}\big(1,3\big)G\big(1,4\big)$$
$$\Gamma^{(1)}\big(4,2,3\big)$$
$$= \big(\mathrm{i}\hbar\big)^2 w\big(1,3\big)G\big(1,4\big)$$
$$w\big(4,2\big)G\big(4,3\big)G\big(3,2\big) \qquad (182)$$

or, in the full notation,

$$\Sigma_{ij}^{(2),1,1}\big(z_1, z_2\big) = \big(\mathrm{i}\hbar\big)^2 \sum_{mpq} w_{ipqm}\big(z_1\big) \sum_n G_{mn}\big(z_1, z_2\big) \times \tag{183}$$
$$\times \sum_{rs} w_{nsjr}\big(z_2\big) G_{rp}\big(z_2, z_1\big) G_{qs}\big(z_1, z_2\big).$$

This is the final result that is still written on the Keldysh contour, i.e. is an equation for Keldysh matrices. The corresponding matrix elements (greater, less, retarded and advanced components) of the selfconsistent second-order selfenergy contributions $\Sigma_{ij}^{(2),2,0}$ and $\Sigma_{ij}^{(2),1,1}$ are straightforwardly extracted, applying the Langreth rules:

$$\Sigma_{ij}^{(2),2,0,\gtrless}\big(t_1, t_2\big) = \pm \big(\mathrm{i}\hbar\big)^2 \sum_{mn} G_{mn}^{\gtrless}\big(t_1, t_2\big) \sum_{st} G_{st}^{\gtrless}\big(t_1, t_2\big) \times \tag{184}$$
$$\times \sum_r w_{irsm}\big(t_1\big) \sum_u w_{tnju}\big(t_2\big) G_{ur}^{\lessgtr}\big(t_2, t_1\big),$$

$$\Sigma_{ij}^{(2),1,1,\gtrless}\big(t_1, t_2\big) = \big(\mathrm{i}\hbar\big)^2 \sum_{mpq} w_{ipqm}\big(t_1\big) \sum_n G_{mn}^{\gtrless}\big(t_1, t_2\big) \times \tag{185}$$
$$\times \sum_{rs} w_{nsjr}\big(t_2\big) G_{rp}^{\lessgtr}\big(t_2, t_1\big) G_{qs}^{\gtrless}\big(t_1, t_2\big).$$

All the above results where for the selfconsistent approach where all expressions contain full Green functions. As we noted in the beginning of this section, alternatively one can perform a **non-selfconsistent treatment**, where all Green functions are replaced by non-interacting functions. In that case, the possible second-order classes are

$$\Sigma^{(2),(2),2,0,0} = \Sigma^{\mathrm{xc},(2)}\big(W^{(2)}, G^{(0)}, \Gamma^{(0)}\big) \equiv \Sigma^{(2),2,0}\big(G \to G^{(0)}\big), \tag{186}$$

$$\Sigma^{(2),(2),1,0,1} = \Sigma^{\mathrm{xc},(2)}\big(W^{(1)}, G^{(0)}, \Gamma^{(1)}\big) \equiv \Sigma^{(2),1,1}\big(G \to G^{(0)}\big) \tag{187}$$

and

$$\Sigma^{(2),\{\mathrm{H},0\},1} = \Sigma^{\mathrm{H},0}\big(G^{(1)}\big), \qquad \Sigma^{(2),\{\mathrm{F},0\},1} = \Sigma^{\mathrm{F},0}\big(G^{(1)}\big). \tag{188}$$

A detailed list of these contributions, including all Keldysh matrix components for all considered basis sets, is given in Appendix B.

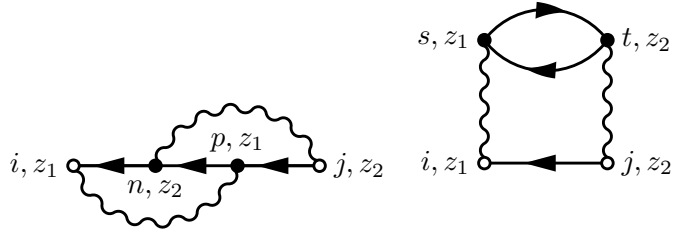

**Figure 21.** Selfconsistent second-order diagrams in the diagonal basis. **Left:** Exchange diagram $\Sigma^{(2),1,1,\text{diagonal}}$. **Right:** Direct diagram, $\Sigma^{(2),2,0,\text{diagonal}}$.

***Diagonal basis***. For a basis where the interaction is diagonal, $w_{ijkl} = \delta_{il}\delta_{jk}w_{ij}$, the selfconsistent second-order selfenergy terms, [cf. Eqs. (176) and (183)], simplify to

$$\Sigma_{ij}^{(2),2,0,\text{diagonal}}(z_1, z_2) = \tag{189}$$
$$= \pm(i\hbar)^2 G_{ij}(z_1, z_2) \sum_s w_{is}(z_1) \sum_t G_{st}(z_1, z_2) G_{ts}(z_2, z_1) w_{tj}(z_2),$$

$$\Sigma_{ij}^{(2),1,1,\text{diagonal}}(z_1, z_2) = \tag{190}$$
$$= (i\hbar)^2 \sum_p w_{ip}(z_1) \sum_n G_{in}(z_1, z_2) w_{nj}(z_2) G_{np}(z_2, z_1) G_{pj}(z_1, z_2),$$

with the corresponding Keldysh matrix components

$$\Sigma_{ij}^{(2),2,0,\text{diagonal},\gtrless}(t_1, t_2) = \tag{191}$$
$$= \pm(i\hbar)^2 G_{ij}^{\gtrless}(t_1, t_2) \sum_s w_{is}(t_1) \sum_t G_{st}^{\gtrless}(t_1, t_2) G_{ts}^{\lessgtr}(t_2, t_1) w_{tj}(t_2),$$

$$\Sigma_{ij}^{(2),1,1,\text{diagonal},\gtrless}(t_1, t_2) = \tag{192}$$
$$= (i\hbar)^2 \sum_p w_{ip}(t_1) \sum_n G_{in}^{\gtrless}(t_1, t_2) w_{nj}(t_2) G_{np}^{\lessgtr}(t_2, t_1) G_{pj}^{\gtrless}(t_1, t_2).$$

The Feynman diagrams for these expressions are shown in Fig. 21.

***Hubbard basis***. For the Hubbard basis, we give the selfconsistent second-order selfenergy contributions first for **bosons**

$$\Sigma_{i\alpha j\alpha}^{(2),2,0,\text{Hubbard},\mathfrak{b}}(z_1, z_2) = \tag{193}$$
$$= (i\hbar)^2 G_{i\alpha j\alpha}(z_1, z_2) U(z_1) \sum_\epsilon G_{i\epsilon j\epsilon}(z_1, z_2) G_{j\epsilon i\epsilon}(z_2, z_1) U(z_2),$$

$$\Sigma_{i\alpha j\alpha}^{(2),1,1,\text{Hubbard},\mathfrak{b}}(z_1, z_2) = \tag{194}$$
$$= (i\hbar)^2 U(z_1) G_{i\alpha j\alpha}(z_1, z_2) U(z_2) G_{j\alpha i\alpha}(z_2, z_1) G_{i\alpha j\alpha}(z_1, z_2).$$

Similarly, we obtain for **fermions**:

$$\Sigma_{i\alpha j\alpha}^{(2),2,0,\mathrm{Hubbard,f}}(z_1, z_2) = \tag{195}$$
$$= -(\mathrm{i}\hbar)^2 G_{i\alpha j\alpha}(z_1, z_2) U(z_1) \sum_{\epsilon \neq \alpha} G_{i\epsilon j\epsilon}(z_1, z_2) G_{j\epsilon i\epsilon}(z_2, z_1) U(z_2),$$

whereas the second expression vanishes, $\Sigma_{i\alpha j\alpha}^{(2),1,1,\mathrm{Hubbard,f}}(z_1, z_2) \equiv 0$.

The corresponding greater/less Keldysh matrix components read, for **bosons**,

$$\Sigma_{i\alpha j\alpha}^{(2),2,0,\mathrm{Hubbard,b},\gtrless}(t_1, t_2) =$$
$$= (\mathrm{i}\hbar)^2 G_{i\alpha j\alpha}^{\gtrless}(t_1, t_2) U(t_1) \sum_{\epsilon} G_{i\epsilon j\epsilon}^{\gtrless}(t_1, t_2) G_{j\epsilon i\epsilon}^{\lessgtr}(t_2, t_1) U(t_2),$$

$$\Sigma_{i\alpha j\alpha}^{(2),1,1,\mathrm{Hubbard,b},\gtrless}(t_1, t_2) =$$
$$= (\mathrm{i}\hbar)^2 U(t_1) G_{i\alpha j\alpha}^{\gtrless}(t_1, t_2) U(t_2) G_{j\alpha i\alpha}^{\lessgtr}(t_2, t_1) G_{i\alpha j\alpha}^{\gtrless}(t_1, t_2).$$

Analogously, we have, for **fermions**,

$$\Sigma_{i\alpha j\alpha}^{(2),2,0,\mathrm{Hubbard,f},\gtrless}(t_1, t_2) =$$
$$= -(\mathrm{i}\hbar)^2 G_{i\alpha j\alpha}^{\gtrless}(t_1, t_2) U(t_1) \sum_{\epsilon \neq \alpha} G_{i\epsilon j\epsilon}^{\gtrless}(t_1, t_2) G_{j\epsilon i\epsilon}^{\lessgtr}(t_2, t_1) U(t_2)$$

and, as before, $\Sigma_{i\alpha j\alpha}^{(2),1,1,\mathrm{Hubbard,f},\gtrless}(t_1, t_2) \equiv 0$.

We again consider the special cases of **spin-0 bosons and spin-1/2 fermions**, respectively. For spin-0 bosons, the selfconsistent second-order contributions are given by

$$\Sigma_{ij}^{(2),2,0,\mathrm{Hubbard,b},0}(z_1, z_2) = \tag{196}$$
$$= (\mathrm{i}\hbar)^2 G_{ij}(z_1, z_2) U(z_1) G_{ij}(z_1, z_2) G_{ji}(z_2, z_1) U(z_2),$$

$$\Sigma_{ij}^{(2),1,1,\mathrm{Hubbard,b},0}(z_1, z_2) = \tag{197}$$
$$= (\mathrm{i}\hbar)^2 U(z_1) G_{ij}(z_1, z_2) U(z_2) G_{ji}(z_2, z_1) G_{ij}(z_1, z_2).$$

Similarly, for spin-1/2 fermions we obtain

$$\Sigma_{i\downarrow(\uparrow)j\downarrow(\uparrow)}^{(2),2,0,\mathrm{Hubbard,f},1/2}(z_1, z_2) = \tag{198}$$
$$= -(\mathrm{i}\hbar)^2 G_{i\downarrow(\uparrow)j\downarrow(\uparrow)}(z_1, z_2) U(z_1) G_{i\uparrow(\downarrow)j\uparrow(\downarrow)}(z_1, z_2) G_{j\uparrow(\downarrow)i\uparrow(\downarrow)}(z_2, z_1) U(z_2).$$

The Feynman diagrams of the self-consistent second-order selfenergy contributions for spin-0 bosons and spin-1/2 fermions are depicted in Figs. 22 and 23, respectively.

Consider again the corresponding greater and less Keldysh matrix components. For

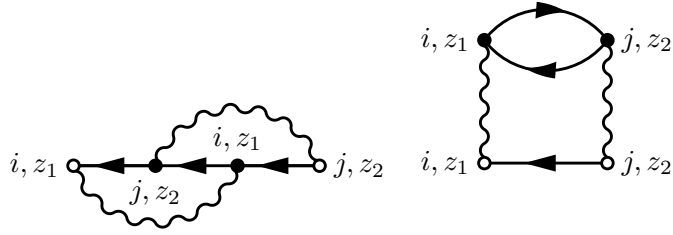

**Figure 22.** Selfconsistent second-order diagrams for spin-0 bosons in the Hubbard basis. **Left:** Exchange diagram $\Sigma^{(2),1,1,\mathrm{Hubbard},\mathrm{b},0}$. **Right:** Direct diagram, $\Sigma^{(2),2,0,\mathrm{Hubbard},\mathrm{b},0}$. Note that both diagrams coincide for the Hubbard basis.

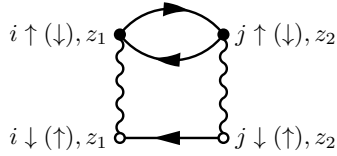

**Figure 23.** Selfconsistent second-order diagram for spin-1/2 fermions in the Hubbard basis, $\Sigma^{(2),2,0,\mathrm{Hubbard},\mathrm{f},1/2}$.

spin-0 bosons, we have

$$\Sigma_{ij}^{(2),2,0,\mathrm{Hubbard},\mathrm{b},0,\gtrless}\big(t_1,t_2\big)$$
$$= \big(\mathrm{i}\hbar\big)^2 G_{ij}^{\gtrless}\big(t_1,t_2\big) U\big(t_1\big) G_{ij}^{\gtrless}\big(t_1,t_2\big) G_{ji}^{\lessgtr}\big(t_2,t_1\big) U\big(t_2\big),$$
$$\Sigma_{ij}^{(2),1,1,\mathrm{Hubbard},\mathrm{b},0,\gtrless}\big(t_1,t_2\big) =$$
$$= \big(\mathrm{i}\hbar\big)^2 U\big(t_1\big) G_{ij}^{\gtrless}\big(t_1,t_2\big) U\big(t_2\big) G_{ji}^{\lessgtr}\big(t_2,t_1\big) G_{ij}^{\gtrless}\big(t_1,t_2\big).$$

In similar manner, we find the correlation components for spin-1/2 fermions:

$$\Sigma_{i\downarrow(\uparrow)j\downarrow(\uparrow)}^{(2),2,0,\mathrm{Hubbard},\mathrm{f},1/2,\gtrless}\big(t_1,t_2\big) =$$
$$= -\big(\mathrm{i}\hbar\big)^2 G_{i\downarrow(\uparrow)j\downarrow(\uparrow)}^{\gtrless}\big(t_1,t_2\big) U\big(t_1\big) G_{i\uparrow(\downarrow)j\uparrow(\downarrow)}^{\gtrless}\big(t_1,t_2\big) G_{j\uparrow(\downarrow)i\uparrow(\downarrow)}^{\lessgtr}\big(t_2,t_1\big) U\big(t_2\big),$$

whereas the second contribution vanishes, as before.

### 4.3. Third-order selfenergy (TOA)

After discussing the frequently used first- and second-order contributions, we now turn to the selfenergy approximations that are of third order in the interaction. We have seen in the results section 3 that, in many cases, the third order provides surprisingly accurate results. On the other hand, this approximation has not been discussed in the literature before. Therefore, we discuss the third-order approximation and its different variants in detail below.

The structure of the selfconsistent third-order contributions to the selfenergy can again be deduced from Eq. (101). There is a total of three terms that contribute to the third-order selfenergy

$$\Sigma^{(3),3,0} = \Sigma^{\mathrm{xc}}\big(W^{(3)}, \Gamma^{(0)}\big), \tag{199}$$

$$\Sigma^{(3),2,1} = \Sigma^{\mathrm{xc}}\big(W^{(2)}, \Gamma^{(1)}\big), \tag{200}$$

$$\Sigma^{(3),1,2} = \Sigma^{\mathrm{xc}}\big(W^{(1)}, \Gamma^{(2)}\big). \tag{201}$$

For the first class, in turn, there exist two contributions to $W^{(3)}$:

$$W^{(3),0,2} = W^{\mathrm{ns}}\big(P^{(0)}, W^{(2)}\big), \qquad W^{(3),1,1} = W^{\mathrm{ns}}\big(P^{(1)}, W^{(1)}\big). \tag{202}$$

Using Eq. (104), together with Eq. (174), one finds[xvi]

$$
\begin{aligned}
&W^{(3),0,2}\big(1,2\big) \\
&= w\big(1,3\big)P^{(0)}\big(3,4\big) \\
&\quad W^{(2)}\big(4,2\big) \\
&= \big(\mathrm{i}\hbar\big)^2 w\big(1,3\big)G\big(3,4\big)G\big(4,3\big) \\
&\quad w\big(4,5\big)G\big(5,6\big)G\big(6,5\big)w\big(6,2\big)
\end{aligned}
\tag{203}
$$

Combining this with Eq. (155), the first term of the first third-order selfenergy class, $\Sigma^{(3),3,0}$, becomes

$$
\begin{aligned}
&\Sigma^{(3),\{3;0,2\},0}\big(1,2\big) \\
&= \mathrm{i}\hbar W^{(3),0,2}\big(1,3\big)G\big(1,4\big) \\
&\quad \Gamma^{(0)}\big(4,2,3\big) \\
&= \big(\mathrm{i}\hbar\big)^3 G\big(1,2\big)w\big(1,3\big)G\big(3,4\big)G\big(4,3\big) \\
&\quad w\big(4,5\big)G\big(5,6\big)G\big(6,5\big)w\big(6,2\big)
\end{aligned}
\tag{204}
$$

---

[xvi]The corresponding equations in the full notation are given in Appendix A.2.1.

or, in the full notation,

$$\Sigma_{ij}^{(3),\{3;0,2\},0}\big(z_1, z_2\big) = \tag{205}$$

$$= \big(\mathrm{i}\hbar\big)^3 \sum_{mn} G_{mn}\big(z_1, z_2\big) \sum_{rs} w_{irsm}\big(z_1\big) \int_{\mathcal{C}} \mathrm{d}z_3 \sum_{tu} G_{st}\big(z_1, z_3\big) G_{ur}\big(z_3, z_1\big) \times$$

$$\times \sum_{vw} w_{tvwu}\big(z_3\big) \sum_{xy} G_{wx}\big(z_3, z_2\big) G_{yv}\big(z_2, z_3\big) w_{xnjy}\big(z_2\big)\,.$$

For the second class of the interaction, $W^{(3),1,1}$, contributing to Eq. (199), the first-order contribution to the polarization is needed, which is given by[xvii][cf. Eqs. (105) and (181)],

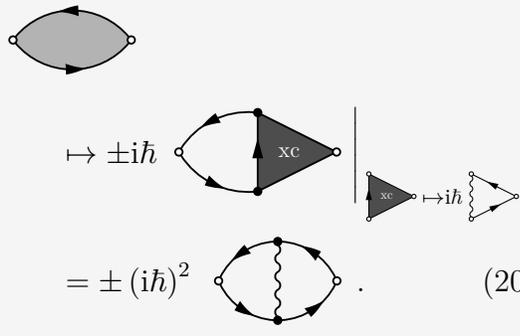

$$P^{(1)}\big(1, 2\big)$$
$$= \pm\mathrm{i}\hbar G\big(1, 3\big) G\big(4, 1\big) \Gamma^{(1)}\big(3, 4, 2\big)$$
$$= \pm\big(\mathrm{i}\hbar\big)^2 G\big(1, 3\big) G\big(4, 1\big)$$
$$w\big(3, 4\big) G\big(3, 2\big) G\big(2, 4\big) \tag{206}$$

Inserting this result back, one finds, using Eq. (104),

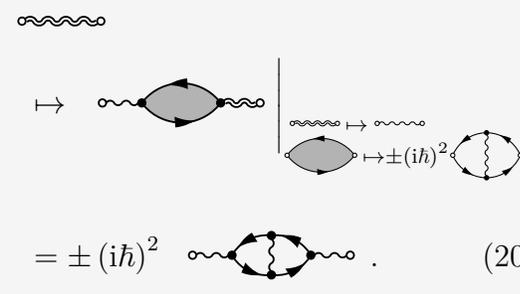

$$W^{(3),1,1}\big(1, 2\big)$$
$$= w\big(1, 3\big) P^{(1)}\big(3, 4\big) W^{(1)}\big(4, 2\big)$$
$$= \pm\big(\mathrm{i}\hbar\big)^2 w\big(1, 3\big) G\big(3, 5\big) G\big(6, 3\big)$$
$$w\big(5, 6\big) G\big(5, 4\big) G\big(4, 6\big) w\big(4, 2\big) \tag{207}$$

With these results, the second term of the class $\Sigma^{(3),3,0}$ is found, using Eqs. (101) and (155),

---

[xvii]The corresponding equations in the full notation are given in Appendix A.2.2.

$$\Sigma^{(3),\{3;1,1\},0}(1,2)$$
$$= i\hbar W^{(3),1,1}(1,3)G(1,4)$$
$$\Gamma^{(0)}(4,2,3)$$
$$= \pm(i\hbar)^3 w(1,3)G(3,5)G(6,3)$$
$$w(5,6)G(5,4)G(4,6)w(4,2)G(1,2)$$

$$= \pm(i\hbar)^3 \qquad , \qquad (208)$$

or, in the full notation,

$$\Sigma_{ij}^{(3),\{3;1,1\},0}(z_1,z_2) = \qquad\qquad (209)$$
$$= \pm(i\hbar)^3 \sum_{mn} G_{mn}(z_1,z_2) \sum_{rs} w_{irsm}(z_1) \int_{\mathcal{C}} \mathrm{d}z_3 \sum_t G_{st}(z_1,z_3) \sum_u G_{ur}(z_3,z_1)$$
$$\times \sum_{vw} w_{twuv}(z_3) \sum_{xy} G_{vx}(z_3,z_2)G_{yw}(z_2,z_3)w_{xnjy}(z_2).$$

We continue with the second class, $\Sigma^{(3),2,1}$, which is straightforwardly worked out[xviii] by combining Eqs. (174) and (181),

$$\Sigma^{(3),2,1}(1,2) = i\hbar W^{(2)}(1,3)$$
$$G(1,4)\Gamma^{(1)}(4,2,3)$$
$$= \pm(i\hbar)^3 w(1,5)G(5,6)G(6,5)$$
$$w(6,3)G(1,4)w(4,2)G(4,3)G(3,2)$$

$$= \pm(i\hbar)^3 \qquad , \qquad (210)$$

or, in the full notation,

$$\Sigma_{ij}^{(3),2,1}(z_1,z_2) = \qquad\qquad (211)$$
$$= \pm(i\hbar)^3 \int_{\mathcal{C}} \mathrm{d}z_3 \sum_{mrs} w_{irsm}(z_1) \sum_{tu} G_{st}(z_1,z_3)G_{ur}(z_3,z_1) \sum_{pq} w_{tpqu}(z_3) \times$$
$$\times \sum_n G_{mn}(z_1,z_2) \sum_{vw} w_{nwjv}(z_2)G_{vp}(z_2,z_3)G_{qw}(z_3,z_2).$$

---

[xviii]The corresponding equations in the full notation are given in Appendix A.2.3.

For the third class $\Sigma^{(3),1,2}$, the second-order contributions to the vertex, $\Gamma^{(2)}$, have to be computed. There are two structural classes to consider:

$$\Gamma^{(2),1,1} = \Gamma\left(\delta\Sigma^{\mathrm{xc},(1)}/\delta G, \Gamma^{(1)}\right), \tag{212}$$

$$\Gamma^{(2),2,0} = \Gamma\left(\delta\Sigma^{\mathrm{xc},(2)}/\delta G, \Gamma^{(0)}\right). \tag{213}$$

For the class $\Gamma^{(2),1,1}$, there exists a single contribution[xix] which is found by employing Eqs. (179) and (181),

$$\Gamma^{(2),1,1}\left(1,2,3\right) = \frac{\delta\Sigma^{\mathrm{xc},(1)}\left(1,2\right)}{\delta G\left(4,5\right)}$$

$$G\left(4,6\right)G\left(7,5\right)\Gamma^{(1)}\left(6,7,3\right)$$

$$= \left(\mathrm{i}\hbar\right)^2 w\left(1,2\right)G\left(1,6\right)$$

$$G\left(7,2\right)w\left(6,7\right)G\left(6,3\right)G\left(3,7\right) \tag{214}$$

This enables the computation of $\Sigma^{(3),1,\{2;1,1\}}$ with Eqs. (101) and (154),

$$\Sigma^{(3),1,\{2;1,1\}}\left(1,2\right) = \mathrm{i}\hbar W^{(1)}\left(1,3\right)$$

$$G\left(1,4\right)\Gamma^{(2),1,1}\left(4,2,3\right)$$

$$= \left(\mathrm{i}\hbar\right)^3 w\left(1,3\right)G\left(1,4\right)w\left(4,2\right)G\left(4,5\right)$$

$$G\left(6,2\right)w\left(5,6\right)G\left(5,3\right)G\left(3,6\right) \tag{215}$$

i.e., in the full notation,

$$\Sigma_{ij}^{(3),1,\{2;1,1\}}\left(z_1, z_2\right) = \tag{216}$$

$$= \left(\mathrm{i}\hbar\right)^3 \sum_{mpq} w_{ipqm}\left(z_1\right) \sum_n G_{mn}\left(z_1, z_2\right) \sum_{rs} w_{nsjr}\left(z_2\right) \int_{\mathcal{C}} \mathrm{d}z_3 \sum_t G_{rt}\left(z_2, z_3\right) \times$$

$$\times \sum_u G_{us}\left(z_3, z_2\right) \sum_{vw} w_{twuv}\left(z_3\right) G_{vp}\left(z_3, z_1\right) G_{qw}\left(z_1, z_3\right).$$

The vertex class $\Gamma^{(2),2,0}$ has six members stemming from the derivatives with respect to each of the three Green functions in both second-order contributions to $\Sigma^{(2)}$, cf.

---

[xix]The corresponding equations in the full notation are given in Appendix A.2.4.

Eqs. (176) and (183),

$$\Gamma^{(2),\{2;2,0\},0} = \Gamma\left(\delta\Sigma^{(2),2,0}/\delta G, \Gamma^{(0)}\right), \tag{217}$$

$$\Gamma^{(2),\{2;1,1\},0} = \Gamma\left(\delta\Sigma^{(2),1,1}/\delta G, \Gamma^{(0)}\right). \tag{218}$$

For the first terms, one finds[xx]

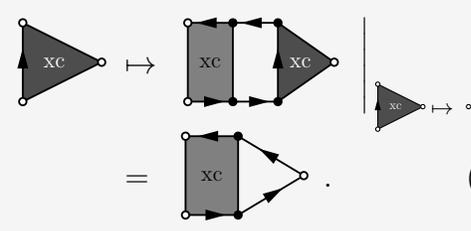

$$\Gamma^{(2),\{2;2,0\},0}\left(1,2,3\right) = \frac{\delta\Sigma^{(2),2,0}\left(1,2\right)}{\delta G\left(4,5\right)}$$
$$G\left(4,6\right)G\left(7,5\right)\Gamma^{(0)}\left(6,7,3\right)$$
$$= \frac{\delta\Sigma^{(2),2,0}\left(1,2\right)}{\delta G\left(4,5\right)}G\left(4,3\right)G\left(3,5\right) \tag{219}$$

Inserting Eq. (176), the occurring derivative evaluates to

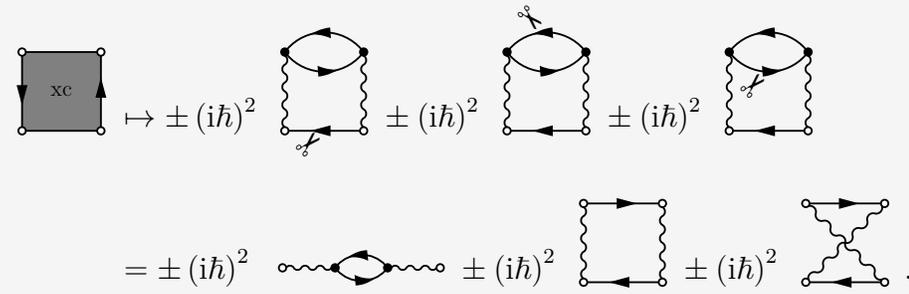

$$\frac{\delta\Sigma^{(2),2,0}\left(1,2\right)}{\delta G\left(4,5\right)} = \pm\left(\mathrm{i}\hbar\right)^2 \frac{\delta\left(w\left(1,6\right)G\left(6,7\right)G\left(7,6\right)w\left(7,2\right)G\left(1,2\right)\right)}{\delta G\left(4,5\right)}$$
$$= \pm\left(\mathrm{i}\hbar\right)^2 \delta\left(1,4\right)\delta\left(2,5\right)w\left(1,6\right)G\left(6,7\right)G\left(7,6\right)w\left(7,2\right)$$
$$\pm\left(\mathrm{i}\hbar\right)^2 w\left(1,4\right)G\left(5,4\right)w\left(5,2\right)G\left(1,2\right)$$
$$\pm\left(\mathrm{i}\hbar\right)^2 w\left(1,5\right)G\left(5,4\right)w\left(4,2\right)G\left(1,2\right) \tag{220}$$

With that, the resulting vertex splits up into three parts labeled "A", "B" and "C":

[xx]The corresponding equations in the full notation are given in Appendix A.2.5.

$$\Gamma^{(2),\{2;2,0\},0,\text{A}}(1,2,3)$$
$$= \frac{\delta\Sigma^{(2),2,0,\text{A}}(1,2)}{\delta G(4,5)}G(4,3)G(3,5)$$
$$= \pm(\mathrm{i}\hbar)^2 w(1,6)G(6,7)$$
$$G(7,6)w(7,2)G(1,3)G(3,2)$$

$$= \pm(\mathrm{i}\hbar)^2 \qquad , \qquad (221)$$

$$\Gamma^{(2),\{2;2,0\},0,\text{B}}(1,2,3)$$
$$= \frac{\delta\Sigma^{(2),2,0,\text{B}}(1,2)}{\delta G(4,5)}G(4,3)G(3,5)$$
$$= \pm(\mathrm{i}\hbar)^2 w(1,4)G(5,4)$$
$$w(5,2)G(1,2)G(4,3)G(3,5)$$

$$= \pm(\mathrm{i}\hbar)^2 \qquad , \qquad (222)$$

$$\Gamma^{(2),\{2;2,0\},0,\text{C}}(1,2,3)$$
$$= \frac{\delta\Sigma^{(2),2,0,\text{C}}(1,2)}{\delta G(4,5)}G(4,3)G(3,5)$$
$$= \pm(\mathrm{i}\hbar)^2 w(1,5)G(5,4)$$
$$w(4,2)G(1,2)G(4,3)G(3,5)$$

$$= \pm(\mathrm{i}\hbar)^2 \qquad . \qquad (223)$$

Similarly, one finds

$$\Gamma^{(2),\{2;1,1\},0}(1,2,3) =$$
$$\frac{\delta\Sigma^{(2),1,1}(1,2)}{\delta G(4,5)}G(4,3)G(3,5)$$

$$\qquad . \qquad (224)$$

Inserting Eq. (183), the occurring derivative evaluates to

$$\frac{\delta\Sigma^{(2),1,1}\big(1,2\big)}{\delta G\big(4,5\big)} = \big(\mathrm{i}\hbar\big)^2\frac{\delta\big(w\big(1,6\big)G\big(1,7\big)w\big(7,2\big)G\big(7,6\big)G\big(6,2\big)\big)}{\delta G\big(4,5\big)}$$

$$= \big(\mathrm{i}\hbar\big)^2\delta\big(1,4\big)w\big(1,6\big)w\big(5,2\big)G\big(5,6\big)G\big(6,2\big)$$

$$+ \big(\mathrm{i}\hbar\big)^2 w\big(1,5\big)G\big(1,4\big)w\big(4,2\big)G\big(5,2\big)$$

$$+ \big(\mathrm{i}\hbar\big)^2\delta\big(2,5\big)w\big(1,4\big)G\big(1,7\big)w\big(7,2\big)G\big(7,4\big)$$

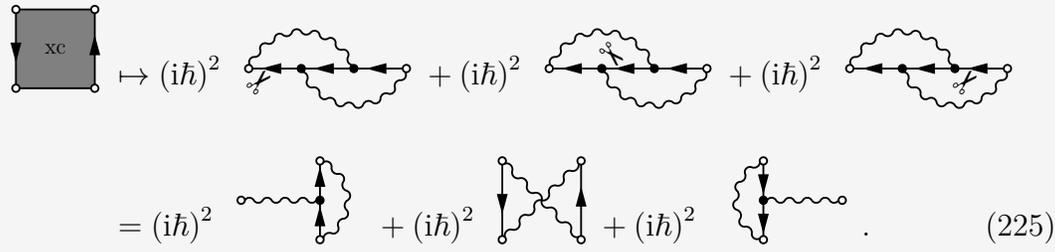

$$(225)$$

Again, three vertex contributions A, B and C are generated:

$$\Gamma^{(2),\{2;1,1\},0,\mathrm{A}}\big(1,2,3\big)$$

$$= \frac{\delta\Sigma^{(2),1,1,\mathrm{A}}\big(1,2\big)}{\delta G\big(4,5\big)}G\big(4,3\big)G\big(3,5\big)$$

$$= \big(\mathrm{i}\hbar\big)^2 w\big(1,6\big)w\big(5,2\big)G\big(5,6\big)$$

$$G\big(6,2\big)G\big(1,3\big)G\big(3,5\big)$$

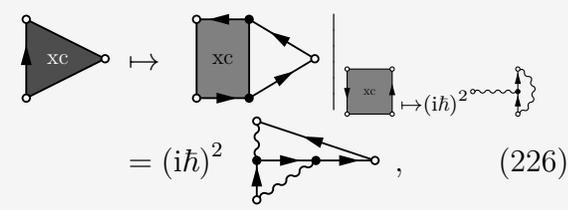

$$(226)$$

$$\Gamma^{(2),\{2;1,1\},0,\mathrm{B}}\big(1,2,3\big)$$

$$= \frac{\delta\Sigma^{(2),1,1,\mathrm{B}}\big(1,2\big)}{\delta G\big(4,5\big)}G\big(4,3\big)G\big(3,5\big)$$

$$= \big(\mathrm{i}\hbar\big)^2 w\big(1,5\big)G\big(1,4\big)w\big(4,2\big)$$

$$G\big(5,2\big)G\big(4,3\big)G\big(3,5\big)$$

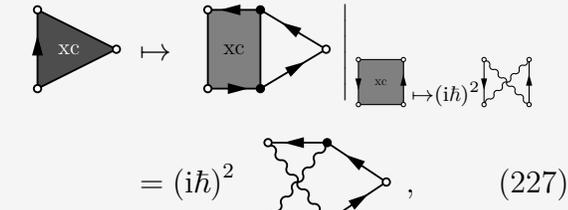

$$(227)$$

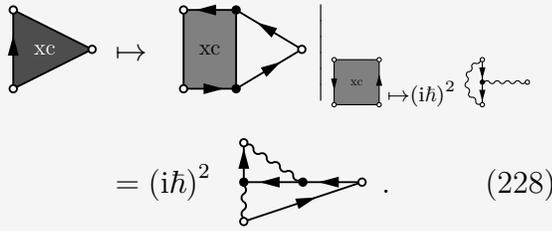

$$\Gamma^{(2),\{2;1,1\},0,C}(1,2,3)$$
$$= \frac{\delta \Sigma^{(2),1,1,C}(1,2)}{\delta G(4,5)} G(4,3) G(3,5)$$
$$= (i\hbar)^2 w(1,4) G(1,7) w(7,2)$$
$$G(7,4) G(4,3) G(3,2)$$

$$= (i\hbar)^2 \qquad . \qquad (228)$$

With this result, the corresponding selfenergy terms$^{xxi}$ can be computed by combining Eqs. (101) and (154) with Eqs. (221-228),

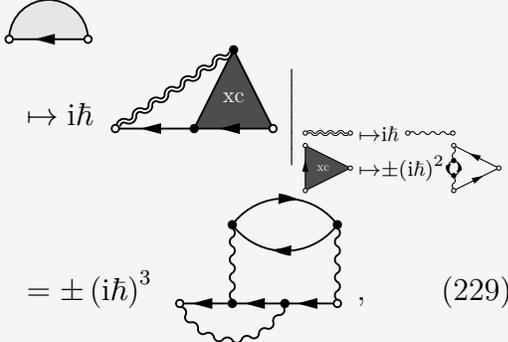

$$\Sigma^{(3),1,\{2;\{2;2,0\},0,A\}}(1,2) = i\hbar W^{(1)}(1,3)$$
$$G(1,4) \Gamma^{(2),\{2;2,0\},0,A}(4,2,3)$$
$$= \pm(i\hbar)^3 w(1,3) G(1,4) w(4,6)$$
$$G(6,7) G(7,6) w(7,2) G(4,3) G(3,2)$$

$$= \pm(i\hbar)^3 \qquad , \qquad (229)$$

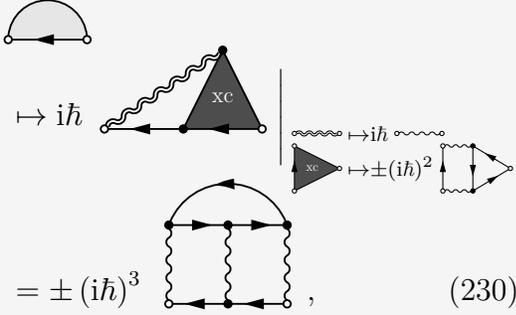

$$\Sigma^{(3),1,\{2;\{2;2,0\},0,B\}}(1,2) = i\hbar W^{(1)}(1,3)$$
$$G(1,4) \Gamma^{(2),\{2;2,0\},0,B}(4,2,3)$$
$$= \pm(i\hbar)^3 w(1,3) G(1,4) w(4,5)$$
$$G(6,5) w(6,2) G(4,2) G(5,3) G(3,6)$$

$$= \pm(i\hbar)^3 \qquad , \qquad (230)$$

---

$^{xxi}$The corresponding equations in the full notation are given in Appendix A.2.6.

$$\Sigma^{(3),1,\{2;\{2;2,0\},0,C\}}(1,2) = i\hbar W^{(1)}(1,3)$$
$$G(1,4)\Gamma^{(2),\{2;2,0\},0,C}(4,2,3)$$
$$= \pm(i\hbar)^3 w(1,3)G(1,4)w(4,6)$$
$$G(6,5)w(5,2)G(4,2)G(5,3)G(3,6)$$

$$\mapsto i\hbar$$

$$= \pm(i\hbar)^3 \qquad , \qquad (231)$$

$$\Sigma^{(3),1,\{2;\{2;1,1\},0,A\}}(1,2) = i\hbar W^{(1)}(1,3)$$
$$G(1,4)\Gamma^{(2),\{2;1,1\},0,A}(4,2,3)$$
$$= (i\hbar)^3 w(1,3)G(1,4)w(4,6)w(5,2)$$
$$G(5,6)G(6,2)G(4,3)G(3,5)$$

$$\mapsto i\hbar$$

$$= \pm(i\hbar)^3 \qquad , \qquad (232)$$

$$\Sigma^{(3),1,\{2;\{2;1,1\},0,B\}}(1,2) = i\hbar W^{(1)}(1,3)$$
$$G(1,4)\Gamma^{(2),\{2;1,1\},0,B}(4,2,3)$$
$$= (i\hbar)^3 w(1,3)G(1,4)w(4,6)G(4,5)$$
$$w(5,2)G(6,2)G(5,3)G(3,6)$$

$$\mapsto i\hbar$$

$$= \pm(i\hbar)^3 \qquad , \qquad (233)$$

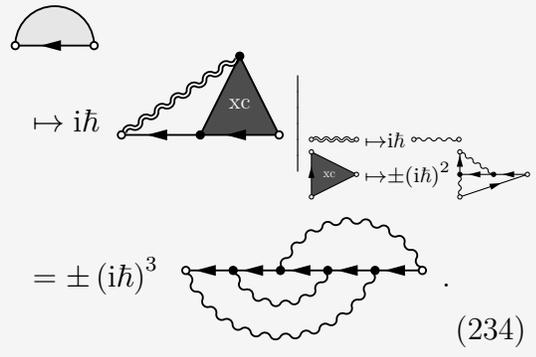

$$\Sigma^{(3),1,\{2;\{2;1,1\},0,C\}}\big(1,2\big) = \mathrm{i}\hbar W^{(1)}\big(1,3\big)$$
$$G\big(1,4\big)\Gamma^{(2),\{2;1,1\},0,C}\big(4,2,3\big)$$
$$= \big(\mathrm{i}\hbar\big)^3 w\big(1,3\big)G\big(1,4\big)w\big(4,5\big)G\big(4,6\big)$$
$$w\big(6,2\big)G\big(6,5\big)G\big(5,3\big)G\big(3,2\big)$$

$$= \pm\,(\mathrm{i}\hbar)^3 \qquad (234)$$

In the full notation, these contributions read,

$$\Sigma_{ij}^{(3),1,\{2;\{2;2,0\},0,A\}}\big(z_1,z_2\big) = \pm\big(\mathrm{i}\hbar\big)^3 \sum_{mpq} w_{ipqm}\big(z_1\big) \times \tag{235}$$
$$\times \int_{\mathcal{C}} \mathrm{d}z_4 \sum_n G_{mn}\big(z_1,z_4\big) \sum_{rsuv} G_{uv}\big(z_4,z_2\big) \sum_t w_{ntur}\big(z_4\big) \times$$
$$\times \sum_w w_{vsjw}\big(z_2\big) G_{wt}\big(z_2,z_4\big) G_{rp}\big(z_4,z_1\big) G_{qs}\big(z_1,z_2\big),$$

$$\Sigma_{ij}^{(3),1,\{2;\{2;2,0\},0,B\}}\big(z_1,z_2\big) = \tag{236}$$
$$= \pm\big(\mathrm{i}\hbar\big)^3 \sum_{mpq} w_{ipqm}\big(z_1\big) \int_{\mathcal{C}} \mathrm{d}z_4 \sum_n G_{mn}\big(z_1,z_4\big) \sum_{rs}\sum_{tu} G_{tu}\big(z_4,z_2\big) \times$$
$$\times \sum_v w_{nvrt}\big(z_4\big) \sum_w w_{sujw}\big(z_2\big) G_{wv}\big(z_2,z_4\big) G_{rp}\big(z_4,z_1\big) G_{qs}\big(z_1,z_2\big),$$

$$\Sigma_{ij}^{(3),1,\{2;\{2;2,0\},0,C\}}\big(z_1,z_2\big) = \tag{237}$$
$$= \pm\big(\mathrm{i}\hbar\big)^3 \sum_{mpq} w_{ipqm}\big(z_1\big) \int_{\mathcal{C}} \mathrm{d}z_4 \sum_n G_{mn}\big(z_1,z_4\big) \sum_{rs}\sum_{tu} G_{tu}\big(z_4,z_2\big) \times$$
$$\times \sum_{vw} G_{vw}\big(z_4,z_2\big) w_{nsvt}\big(z_4\big) w_{wujr}\big(z_2\big) G_{rp}\big(z_2,z_1\big) G_{qs}\big(z_1,z_4\big),$$

$$\Sigma_{ij}^{(3),1,\{2;\{2;1,1\},0,A\}}\big(z_1,z_2\big) = \tag{238}$$
$$= \big(\mathrm{i}\hbar\big)^3 \sum_{mpq} w_{ipqm}\big(z_1\big) \int_{\mathcal{C}} \mathrm{d}z_4 \sum_n G_{mn}\big(z_1,z_4\big) \sum_{rs}\sum_{tu} w_{ntur}\big(z_4\big) \times$$
$$\times \sum_{vw} w_{swjv}\big(z_2\big) G_{vt}\big(z_2,z_4\big) G_{uw}\big(z_4,z_2\big) G_{rp}\big(z_4,z_1\big) G_{qs}\big(z_1,z_2\big),$$

$$\Sigma_{ij}^{(3),1,\{2;\{2;1,1\},0,\text{B}\}}\big(z_1,z_2\big) = \tag{239}$$

$$= \big(\text{i}\hbar\big)^3 \sum_{mpq} w_{ipqm}\big(z_1\big) \int_{\mathcal{C}} \text{d}z_4 \sum_n G_{mn}\big(z_1,z_4\big) \sum_{rs} \sum_{tv} w_{nsvt}\big(z_4\big) \times$$

$$\times \sum_u G_{tu}\big(z_4,z_2\big) \sum_w w_{uwjr}\big(z_2\big) G_{vw}\big(z_4,z_2\big) G_{rp}\big(z_2,z_1\big) G_{qs}\big(z_1,z_4\big)\,,$$

$$\Sigma_{ij}^{(3),1,\{2;\{2;1,1\},0,\text{C}\}}\big(z_1,z_2\big) = \tag{240}$$

$$= \big(\text{i}\hbar\big)^3 \sum_{mpq} w_{ipqm}\big(z_1\big) \int_{\mathcal{C}} \text{d}z_4 \sum_n G_{mn}\big(z_1,z_4\big) \sum_{rs} \sum_{tv} w_{nvrt}\big(z_4\big) \times$$

$$\times \sum_u G_{tu}\big(z_4,z_2\big) \sum_w w_{usjw}\big(z_2\big) G_{wv}\big(z_2,z_4\big) G_{rp}\big(z_4,z_1\big) G_{qs}\big(z_1,z_2\big)\,.$$

**Greater and less Keldysh coomponents:** Let us now turn to the $\gtrless$-components which are the central ingredient for the numerical implementation. For all third-order selfenergy terms these components can be computed in a generic fashion, splitting all integrals at the points where the arguments of each Green function change their relative ordering on the contour. Consider first the **"less" component**, to $\Sigma_{ij}^{(3),1,\{2;\{2;2,0\},0,\text{B}\}}\big(z_1,z_2\big)$. The result consists of three terms, [cf. Eq. (74)],

$$\Sigma_{ij}^{(3),1,\{2;\{2;2,0\},0,\text{B}\},<}\big(t_1,t_2\big) = \Sigma_{ij}^{(3),1,\{2;\{2;2,0\},0,\text{B}\},<}\big(z_{1-},z_{2+}\big) = \tag{241}$$

$$= \pm\big(\text{i}\hbar\big)^3 \sum_{mpq} w_{ipqm}\big(z_{1-}\big) \times$$

$$\times \left\{ I_{mpq}^{1,<}\big(z_{1-},z_{2+}\big) + I_{mpq}^{2,<}\big(z_{1-},z_{2+}\big) + I_{mpq}^{3,<}\big(z_{1-},z_{2+}\big) \right\}\,,$$

with the three terms given by

$$I^{1,<}_{mpq}\big(z_{1-}, z_{2+}\big) = \int_{z_{0-}}^{z_{1-}} \mathrm{d}z_4 \sum_n G^>_{mn}\big(z_{1-}, z_4\big) \times \tag{242}$$
$$\times \sum_{rs} \sum_{tu} G^<_{tu}\big(z_4, z_{2+}\big) \sum_v w_{nvrt}\big(z_4\big) \times$$
$$\times \sum_w w_{sujw}\big(z_{2+}\big) G^>_{wv}\big(z_{2+}, z_4\big) G^<_{rp}\big(z_4, z_{1-}\big) G^<_{qs}\big(z_{1-}, z_{2+}\big),$$

$$I^{2,<}_{mpq}\big(z_{1-}, z_{2+}\big) = \int_{z_{1-}}^{z_{2+}} \mathrm{d}z_4 \sum_n G^<_{mn}\big(z_{1-}, z_4\big) \times \tag{243}$$
$$\times \sum_{rs} \sum_{tu} G^<_{tu}\big(z_4, z_{2+}\big) \sum_v w_{nvrt}\big(z_4\big) \times$$
$$\times \sum_w w_{sujw}\big(z_{2+}\big) G^>_{wv}\big(z_{2+}, z_4\big) G^>_{rp}\big(z_4, z_{1-}\big) G^<_{qs}\big(z_{1-}, z_{2+}\big),$$

$$I^{3,<}_{mpq}\big(z_{1-}, z_{2+}\big) = \int_{z_{2+}}^{z_{0+}} \mathrm{d}z_4 \sum_n G^<_{mn}\big(z_{1-}, z_4\big) \tag{244}$$
$$\sum_{rs} \sum_{tu} G^>_{tu}\big(z_4, z_{2+}\big) \sum_v w_{nvrt}\big(z_4\big)$$
$$\sum_w w_{sujw}\big(z_{2+}\big) G^<_{wv}\big(z_{2+}, z_4\big) G^>_{rp}\big(z_4, z_{1-}\big) G^<_{qs}\big(z_{1-}, z_{2+}\big).$$

We now transform these expression to real-time integrals:

$$\Sigma^{(3),1,\{2;\{2;2,0\},0,\mathrm{B}\},<}_{ij}\big(t_1, t_2\big) = \pm\big(\mathrm{i}\hbar\big)^3 \sum_{mpq} w_{ipqm}\big(t_1\big) \times \tag{245}$$
$$\times \Big\{ I^{1,<}_{mpq}\big(t_1, t_2\big) + I^{2,<}_{mpq}\big(t_1, t_2\big) + I^{3,<}_{mpq}\big(t_1, t_2\big) \Big\},$$

with the corresponding results for the three contributions

$$
\begin{aligned}
I_{mpq}^{1,<}\left(t_1, t_2\right) = \int_{t_0}^{t_1} \mathrm{d}t_4 \sum_n G_{mn}^{>}\left(t_1, t_4\right) \times \\
\times \sum_{rs} \sum_{tu} G_{tu}^{<}\left(t_4, t_2\right) \sum_v w_{nvrt}\left(t_4\right) \times \\
\times \sum_w w_{sujw}\left(t_2\right) G_{wv}^{>}\left(t_2, t_4\right) G_{rp}^{<}\left(t_4, t_1\right) G_{qs}^{<}\left(t_1, t_2\right),
\end{aligned}
\tag{246}
$$

$$
\begin{aligned}
I_{mpq}^{2,<}\left(t_1, t_2\right) = \int_{t_1}^{t_2} \mathrm{d}t_4 \sum_n G_{mn}^{<}\left(t_1, t_4\right) \times \\
\times \sum_{rs} \sum_{tu} G_{tu}^{<}\left(t_4, t_2\right) \sum_v w_{nvrt}\left(t_4\right) \times \\
\times \sum_w w_{sujw}\left(t_2\right) G_{wv}^{>}\left(t_2, t_4\right) G_{rp}^{>}\left(t_4, t_1\right) G_{qs}^{<}\left(t_1, t_2\right),
\end{aligned}
\tag{247}
$$

$$
\begin{aligned}
I_{mpq}^{3,<}\left(t_1, t_2\right) = \int_{t_2}^{t_0} \mathrm{d}t_4 \sum_n G_{mn}^{<}\left(t_1, t_4\right) \times \\
\times \sum_{rs} \sum_{tu} G_{tu}^{>}\left(t_4, t_2\right) \sum_v w_{nvrt}\left(t_4\right) \times \\
\times \sum_w w_{sujw}\left(t_2\right) G_{wv}^{<}\left(t_2, t_4\right) G_{rp}^{>}\left(t_4, t_1\right) G_{qs}^{<}\left(t_1, t_2\right).
\end{aligned}
\tag{248}
$$

For the **greater component**, we find analogously

$$
\begin{aligned}
\Sigma_{ij}^{(3),1,\{2;\{2;2,0\},0,\mathrm{B}\},>}\left(t_1, t_2\right) = \Sigma_{ij}^{(3),1,\{2;\{2;2,0\},0,\mathrm{B}\},>}\left(z_{1+}, z_{2-}\right) = \\
= \pm\left(\mathrm{i}\hbar\right)^3 \sum_{mpq} w_{ipqm}\left(z_{1+}\right) \times \\
\times \left\{ I_{mpq}^{1,>}\left(z_{1+}, z_{2-}\right) + I_{mpq}^{2,>}\left(z_{1+}, z_{2-}\right) + I_{mpq}^{3,>}\left(z_{1+}, z_{2-}\right) \right\},
\end{aligned}
\tag{249}
$$

which, again, consists of three terms:

$$I_{mpq}^{1,>}\big(z_{1+}, z_{2-}\big) = \int_{z_{0-}}^{z_{2-}} \mathrm{d}z_4 \sum_n G_{mn}^>\big(z_{1+}, z_4\big) \times \tag{250}$$
$$\times \sum_{rs} \sum_{tu} G_{tu}^<\big(z_4, z_{2-}\big) \sum_v w_{nvrt}\big(z_4\big) \times$$
$$\times \sum_w w_{sujw}\big(z_{2-}\big) G_{wv}^>\big(z_{2-}, z_4\big) G_{rp}^<\big(z_4, z_{1+}\big) G_{qs}^>\big(z_{1+}, z_{2-}\big),$$

$$I_{mpq}^{2,>}\big(z_{1+}, z_{2-}\big) = \int_{z_{2-}}^{z_{1+}} \mathrm{d}z_4 \sum_n G_{mn}^>\big(z_{1+}, z_4\big) \times \tag{251}$$
$$\times \sum_{rs} \sum_{tu} G_{tu}^>\big(z_4, z_{2-}\big) \sum_v w_{nvrt}\big(z_4\big) \times$$
$$\times \sum_w w_{sujw}\big(z_{2-}\big) G_{wv}^<\big(z_{2-}, z_4\big) G_{rp}^<\big(z_4, z_{1+}\big) G_{qs}^>\big(z_{1+}, z_{2-}\big),$$

$$I_{mpq}^{3,>}\big(z_{1+}, z_{2-}\big) = \int_{z_{1+}}^{z_{0^+}} \mathrm{d}z_4 \sum_n G_{mn}^<\big(z_{1+}, z_4\big) \times \tag{252}$$
$$\times \sum_{rs} \sum_{tu} G_{tu}^>\big(z_4, z_{2-}\big) \sum_v w_{nvrt}\big(z_4\big) \times$$
$$\times \sum_w w_{sujw}\big(z_{2-}\big) G_{wv}^<\big(z_{2-}, z_4\big) G_{rp}^>\big(z_4, z_{1+}\big) G_{qs}^>\big(z_{1+}, z_{2-}\big).$$

Transforming, again, to real-time integrals, we obtain

$$\Sigma_{ij}^{(3),1,\{2;\{2;2,0\},0,B\},>}\big(t_1, t_2\big) = \pm\big(\mathrm{i}\hbar\big)^3 \sum_{mpq} w_{ipqm}\big(t_1\big) \times \tag{253}$$
$$\times \left\{ I_{mpq}^{1,>}\big(t_1, t_2\big) + I_{mpq}^{2,>}\big(t_1, t_2\big) + I_{mpq}^{3,>}\big(t_1, t_2\big) \right\},$$

with the three contributions becoming

$$I^{1,>}_{mpq}\big(t_1,t_2\big) = \int_{t_0}^{t_2} \mathrm{d}t_4 \sum_n G^{>}_{mn}\big(t_1,t_4\big) \times \tag{254}$$
$$\times \sum_{rs}\sum_{tu} G^{<}_{tu}\big(t_4,t_2\big) \sum_v w_{nvrt}\big(t_4\big) \times$$
$$\times \sum_w w_{sujw}\big(t_2\big) G^{>}_{wv}\big(t_2,t_4\big) G^{<}_{rp}\big(t_4,t_1\big) G^{>}_{qs}\big(t_1,t_2\big),$$

$$I^{2,>}_{mpq}\big(t_1,t_2\big) = \int_{t_2}^{t_1} \mathrm{d}t_4 \sum_n G^{>}_{mn}\big(t_1,t_4\big) \times \tag{255}$$
$$\times \sum_{rs}\sum_{tu} G^{>}_{tu}\big(t_4,t_2\big) \sum_v w_{nvrt}\big(t_4\big) \times$$
$$\times \sum_w w_{sujw}\big(t_2\big) G^{<}_{wv}\big(t_2,t_4\big) G^{<}_{rp}\big(t_4,t_1\big) G^{>}_{qs}\big(t_1,t_2\big),$$

$$I^{3,>}_{mpq}\big(t_1,t_2\big) = \int_{t_1}^{t_0} \mathrm{d}t_4 \sum_n G^{<}_{mn}\big(t_1,t_4\big) \times \tag{256}$$
$$\times \sum_{rs}\sum_{tu} G^{>}_{tu}\big(t_4,t_2\big) \sum_v w_{nvrt}\big(t_4\big) \times$$
$$\times \sum_w w_{sujw}\big(t_2\big) G^{<}_{wv}\big(t_2,t_4\big) G^{>}_{rp}\big(t_4,t_1\big) G^{>}_{qs}\big(t_1,t_2\big).$$

**Non-selfconsistent expansion:** We now briefly discuss how the above results change in the case that all expressions are expanded in terms of noninteracting Green functions. The additional non-selfconsistent diagrams are of either of the structures

$$\Sigma^{(3)}\big(G \longrightarrow G^{(0)}\big), \qquad\qquad \text{(10 terms)}$$
$$\Sigma^{(2)}\big(G^{(0)} \longrightarrow G^{(0)}\Sigma^{(1)}G^{(0)}\big), \qquad\qquad (6\cdot 2 = 12 \text{ terms})$$
$$\Sigma^{\mathrm{H}}/\Sigma^{\mathrm{F}}\big(G^{(0)} \longrightarrow G^{(0)}\Sigma^{(2)}G^{(0)}\big), \qquad\qquad (2\cdot 6 = 12 \text{ terms})$$
$$\Sigma^{\mathrm{H}}/\Sigma^{\mathrm{F}}\big(G^{(0)} \longrightarrow G^{(0)}\Sigma^{(1)}G^{(0)}\Sigma^{(1)}G^{(0)}\big), \quad (2\cdot 2\cdot 2 = 8 \text{ terms})$$

This makes a total of 42 non-selfconsistent third-order terms.

***Diagonal basis***. In a diagonal basis, the selfconsistent third-order selfenergy contributions become

$$\Sigma^{(3),\{3;0,2\},0,\mathrm{diag}}_{ij}\big(z_1,z_2\big) = \tag{257}$$
$$= \big(\mathrm{i}\hbar\big)^3 G_{ij}\big(z_1,z_2\big) \sum_r w_{ir}\big(z_1\big) \int_{\mathcal{C}} \mathrm{d}z_3 \sum_t G_{rt}\big(z_1,z_3\big) G_{tr}\big(z_3,z_1\big) \times$$
$$\times \sum_v w_{tv}\big(z_3\big) \sum_x G_{vx}\big(z_3,z_2\big) G_{xv}\big(z_2,z_3\big) w_{xj}\big(z_2\big),$$

$$\Sigma_{ij}^{(3),\{3;1,1\},0,\text{diag}}\left(z_1,z_2\right) = \tag{258}$$

$$= \pm\left(\mathrm{i}\hbar\right)^3 G_{ij}\left(z_1,z_2\right) \sum_r w_{ir}\left(z_1\right) \int_{\mathcal{C}} \mathrm{d}z_3 \sum_t G_{rt}\left(z_1,z_3\right) \sum_u G_{ur}\left(z_3,z_1\right) \times$$

$$\times\, w_{tu}\left(z_3\right) \sum_x G_{tx}\left(z_3,z_2\right) G_{xu}\left(z_2,z_3\right) w_{xj}\left(z_2\right),$$

$$\Sigma_{ij}^{(3),2,1,\text{diag}}\left(z_1,z_2\right) = \tag{259}$$

$$= \pm\left(\mathrm{i}\hbar\right)^3 \int_{\mathcal{C}} \mathrm{d}z_3 \sum_r w_{ir}\left(z_1\right) \sum_t G_{rt}\left(z_1,z_3\right) G_{tr}\left(z_3,z_1\right) \sum_p w_{tp}\left(z_3\right) \times$$

$$\times \sum_n G_{in}\left(z_1,z_2\right) w_{nj}\left(z_2\right) G_{np}\left(z_2,z_3\right) G_{pj}\left(z_3,z_2\right),$$

$$\Sigma_{ij}^{(3),1,\{2;1,1\},\text{diag}}\left(z_1,z_2\right) = \tag{260}$$

$$= \left(\mathrm{i}\hbar\right)^3 \sum_p w_{ip}\left(z_1\right) \sum_n G_{in}\left(z_1,z_2\right) w_{nj}\left(z_2\right) \int_{\mathcal{C}} \mathrm{d}z_3 \sum_t G_{nt}\left(z_2,z_3\right) \times$$

$$\times \sum_u G_{uj}\left(z_3,z_2\right) w_{tu}\left(z_3\right) G_{tp}\left(z_3,z_1\right) G_{pu}\left(z_1,z_3\right),$$

Now we again provide the three contributions labeled "A, B, C", respectively:

$$\Sigma_{ij}^{(3),1,\{2;\{2;2,0\},0,\text{A}\},\text{diag}}\left(z_1,z_2\right) = \tag{261}$$

$$= \pm\left(\mathrm{i}\hbar\right)^3 \sum_p w_{ip}\left(z_1\right) \int_{\mathcal{C}} \mathrm{d}z_4 \sum_n G_{in}\left(z_1,z_4\right) \sum_{tv} G_{tv}\left(z_4,z_2\right) \times$$

$$\times\, w_{nt}\left(z_4\right) w_{vj}\left(z_2\right) G_{vt}\left(z_2,z_4\right) G_{np}\left(z_4,z_1\right) G_{pj}\left(z_1,z_2\right),$$

$$\Sigma_{ij}^{(3),1,\{2;\{2;2,0\},0,\text{B}\},\text{diag}}\left(z_1,z_2\right) = \tag{262}$$

$$= \pm\left(\mathrm{i}\hbar\right)^3 \sum_p w_{ip}\left(z_1\right) \int_{\mathcal{C}} \mathrm{d}z_4 \sum_n G_{in}\left(z_1,z_4\right) \sum_{rs} G_{nj}\left(z_4,z_2\right) \times$$

$$\times\, w_{nr}\left(z_4\right) w_{sj}\left(z_2\right) G_{sr}\left(z_2,z_4\right) G_{rp}\left(z_4,z_1\right) G_{ps}\left(z_1,z_2\right),$$

$$\Sigma_{ij}^{(3),1,\{2;\{2;2,0\},0,\text{C}\},\text{diag}}\left(z_1,z_2\right) = \tag{263}$$

$$= \pm\left(\mathrm{i}\hbar\right)^3 \sum_p w_{ip}\left(z_1\right) \int_{\mathcal{C}} \mathrm{d}z_4 \sum_n G_{in}\left(z_1,z_4\right) \sum_{rs} G_{nj}\left(z_4,z_2\right) \times$$

$$\times\, G_{sr}\left(z_4,z_2\right) w_{ns}\left(z_4\right) w_{rj}\left(z_2\right) G_{rp}\left(z_2,z_1\right) G_{ps}\left(z_1,z_4\right)$$

and, similarly, for the second class of selfenergy contributions:

$$\Sigma_{ij}^{(3),1,\{2;\{2;1,1\},0,\text{A}\},\text{diag}}\left(z_1,z_2\right) = \tag{264}$$

$$= \left(\mathrm{i}\hbar\right)^3 \sum_p w_{ip}\left(z_1\right) \int_{\mathcal{C}} \mathrm{d}z_4 \sum_n G_{in}\left(z_1,z_4\right) \sum_{st} w_{nt}\left(z_4\right) \times$$

$$\times\, w_{sj}\left(z_2\right) G_{st}\left(z_2,z_4\right) G_{tj}\left(z_4,z_2\right) G_{np}\left(z_4,z_1\right) G_{ps}\left(z_1,z_2\right),$$

$$\Sigma_{ij}^{(3),1,\{2;\{2;1,1\},0,\mathrm{B}\},\mathrm{diag}}(z_1,z_2) = \tag{265}$$

$$= (\mathrm{i}\hbar)^3 \sum_p w_{ip}(z_1) \int_{\mathcal{C}} \mathrm{d}z_4 \sum_n G_{in}(z_1,z_4) \sum_{rs} w_{ns}(z_4) \times$$
$$\times\, G_{nr}(z_4,z_2) w_{rj}(z_2) G_{sj}(z_4,z_2) G_{rp}(z_2,z_1) G_{ps}(z_1,z_4)\,,$$

$$\Sigma_{ij}^{(3),1,\{2;\{2;1,1\},0,\mathrm{C}\},\mathrm{diag}}(z_1,z_2) = \tag{266}$$

$$= (\mathrm{i}\hbar)^3 \sum_p w_{ip}(z_1) \int_{\mathcal{C}} \mathrm{d}z_4 \sum_n G_{in}(z_1,z_4) \sum_r w_{nr}(z_4) \times$$
$$\times \sum_u G_{nu}(z_4,z_2) w_{uj}(z_2) G_{ur}(z_2,z_4) G_{rp}(z_4,z_1) G_{pj}(z_1,z_2)\,.$$

The corresponding Keldysh matrix components as well as the non-selfconsistent selfenergy contributions can be worked out in analogy to those in the non-diagonal basis. The diagrams of the selfconsistent third-order selfenergy contributions in a diagonal basis are shown in Fig. 24.

***Hubbard basis. Spin-0 bosons.*** For the Hubbard basis we separately consider spin-0 bosons and spin-1/2 fermions. For the case of spin-0 bosons, the third-order selfenergy contributions separate into two classes. The first is given by

$$\Sigma_{ij}^{(3),\{3;0,2\},0,\flat,0}(z_1,z_2) = \Sigma_{ij}^{(3),\{3;1,1\},0,\flat,0}(z_1,z_2) = \tag{267}$$
$$= \Sigma_{ij}^{(3),2,1,\flat,0}(z_1,z_2) = \Sigma_{ij}^{(3),1,\{2;1,1\},\flat,0}(z_1,z_2) =$$
$$= \Sigma_{ij}^{(3),1,\{2;\{2;2,0\},0,\mathrm{A}\},\flat,0}(z_1,z_2) = \Sigma_{ij}^{(3),1,\{2;\{2;2,0\},0,\mathrm{B}\},\flat,0}(z_1,z_2) =$$
$$= \Sigma_{ij}^{(3),1,\{2;\{2;1,1\},0,\mathrm{A}\},\flat,0}(z_1,z_2) = \Sigma_{ij}^{(3),1,\{2;\{2;1,1\},0,\mathrm{C}\},\flat,0}(z_1,z_2) =$$
$$= (\mathrm{i}\hbar)^3 G_{ij}(z_1,z_2) U(z_1) \int_{\mathcal{C}} \mathrm{d}z_3 \sum_t G_{it}(z_1,z_3) G_{ti}(z_3,z_1) \times$$
$$\times\, U(z_3) G_{tj}(z_3,z_2) G_{jt}(z_2,z_3) U(z_2)\,,$$

and the second is given by

$$\Sigma_{ij}^{(3),1,\{2;\{2;2,0\},0,\mathrm{C}\},\flat,0}(z_1,z_2) = \Sigma_{ij}^{(3),1,\{2;\{2;1,1\},0,\mathrm{B}\},\flat,0}(z_1,z_2) = \tag{268}$$
$$= (\mathrm{i}\hbar)^3 U(z_1) \int_{\mathcal{C}} \mathrm{d}z_4 \sum_n G_{in}(z_1,z_4) G_{nj}(z_4,z_2) \times$$
$$\times\, G_{nj}(z_4,z_2) U(z_4) U(z_2) G_{ji}(z_2,z_1) G_{in}(z_1,z_4)\,.$$

The corresponding diagrams are shown in Fig. 25.

Let us now turn to the Hubbard result For **spin-1/2 fermions**. In this case only the terms with the superscripts "B" and "C" exist which are denoted by $\Sigma^{(3),1,\{2;\{2;2,0\},0,\mathrm{B}\},\mathrm{f},1/2}$

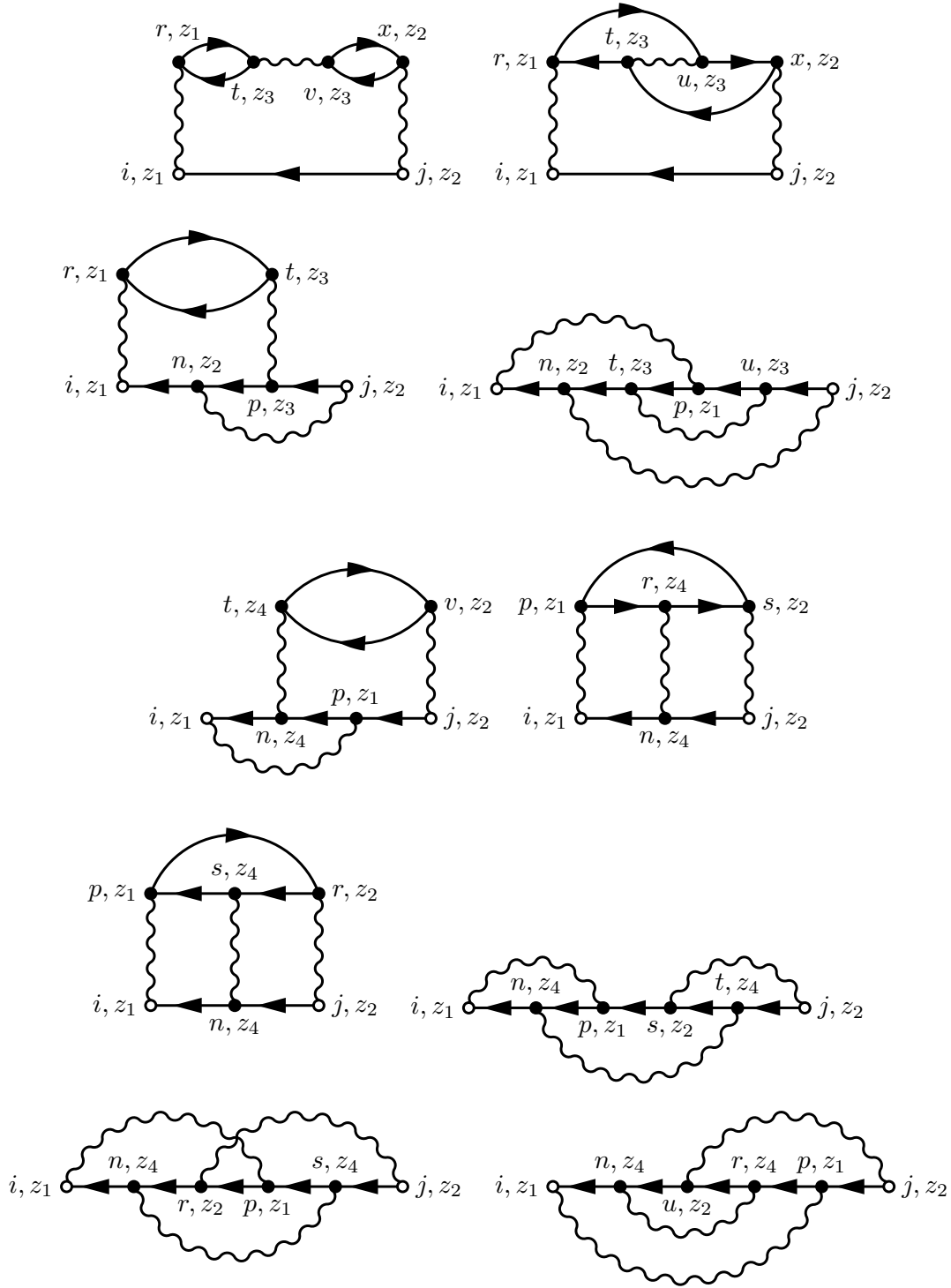

**Figure 24.** Third-order diagrams in diagonal basis from left to right. **First row:** $\Sigma^{(3),\{3;0,2\},0,\mathrm{diag}}$, $\Sigma^{(3),\{3;1,1\},0,\mathrm{diag}}$ **Second row:** $\Sigma^{(3),2,1,\mathrm{diag}}$, $\Sigma^{(3),1,\{2;1,1\},\mathrm{diag}}$ **Third row:** $\Sigma^{(3),1,\{2;\{2;2,0\},0,\mathrm{A}\},\mathrm{diag}}$, $\Sigma^{(3),1,\{2;\{2;2,0\},0,\mathrm{B}\},\mathrm{diag}}$ **Fourth row:** $\Sigma^{(3),1,\{2;\{2;2,0\},0,\mathrm{C}\},\mathrm{diag}}$, $\Sigma^{(3),1,\{2;\{2;1,1\},0,\mathrm{A}\},\mathrm{diag}}$ **Fifth row:** $\Sigma^{(3),1,\{2;\{2;1,1\},0,\mathrm{B}\},\mathrm{diag}}$, $\Sigma^{(3),1,\{2;\{2;1,1\},0,\mathrm{C}\},\mathrm{diag}}$.

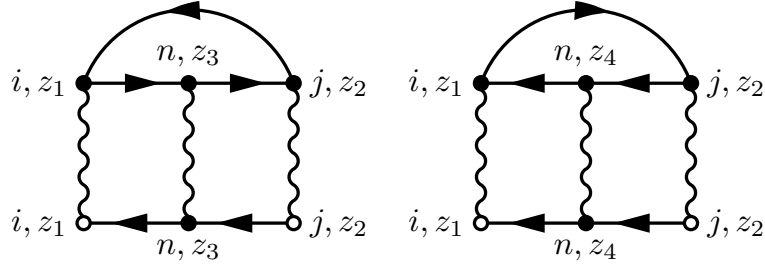

**Figure 25.** Third-order diagrams in the Hubbard basis for spin-0 bosons. **Left:** First equivalence class of $\Sigma^{(3),\{3;0,2\},0,\mathrm{diag}}$, $\Sigma^{(3),\{3;1,1\},0,\mathrm{diag}}$, $\Sigma^{(3),2,1,\mathrm{diag}}$, $\Sigma^{(3),1,\{2;1,1\},\mathrm{diag}}$, $\Sigma^{(3),1,\{2;\{2;2,0\},0,\mathrm{A}\},\mathrm{diag}}$, $\Sigma^{(3),1,\{2;\{2;2,0\},0,\mathrm{B}\},\mathrm{diag}}$, $\Sigma^{(3),1,\{2;\{2;1,1\},0,\mathrm{A}\},\mathrm{diag}}$ and $\Sigma^{(3),1,\{2;\{2;1,1\},0,\mathrm{C}\},\mathrm{diag}}$. **Right:** Second equivalence class of $\Sigma^{(3),1,\{2;2,0,\mathrm{C}\},\mathrm{diag}}$ and $\Sigma^{(3),1,\{2;1,1,\mathrm{B}\},\mathrm{diag}}$.

and $\Sigma^{(3),1,\{2;\{2;2,0\},0,\mathrm{C}\},\mathrm{f},1/2}$, respectively. The "B"-terms are given by

$$\Sigma^{(3),1,\{2;\{2;2,0\},0,\mathrm{B}\},\mathrm{f},1/2}_{i\uparrow j\uparrow}\big(z_1,z_2\big) = \tag{269}$$
$$= -\big(\mathrm{i}\hbar\big)^3 U\big(z_1\big)\int_{\mathcal{C}}\mathrm{d}z_4 \sum_n G_{i\uparrow n\uparrow}\big(z_1,z_4\big)G_{n\uparrow j\uparrow}\big(z_4,z_2\big)\times$$
$$\times U\big(z_4\big)U\big(z_2\big)G_{j\downarrow n\downarrow}\big(z_2,z_4\big)G_{n\downarrow i\downarrow}\big(z_4,z_1\big)G_{i\downarrow j\downarrow}\big(z_1,z_2\big),$$

$$\Sigma^{(3),1,\{2;\{2;2,0\},0,\mathrm{B}\},\mathrm{f},1/2}_{i\downarrow j\downarrow}\big(z_1,z_2\big) = \tag{270}$$
$$= -\big(\mathrm{i}\hbar\big)^3 U\big(z_1\big)\int_{\mathcal{C}}\mathrm{d}z_4 \sum_n G_{i\downarrow n\downarrow}\big(z_1,z_4\big)G_{n\downarrow j\downarrow}\big(z_4,z_2\big)\times$$
$$\times U\big(z_4\big)U\big(z_2\big)G_{j\uparrow n\uparrow}\big(z_2,z_4\big)G_{n\uparrow i\uparrow}\big(z_4,z_1\big)G_{i\uparrow j\uparrow}\big(z_1,z_2\big),$$

and, for the "C"-terms, we find

$$\Sigma^{(3),1,\{2;\{2;2,0\},0,\mathrm{C}\},\mathrm{f},1/2}_{i\uparrow j\uparrow}\big(z_1,z_2\big) = \tag{271}$$
$$= -\big(\mathrm{i}\hbar\big)^3 U\big(z_1\big)\int_{\mathcal{C}}\mathrm{d}z_4 \sum_n G_{i\uparrow n\uparrow}\big(z_1,z_4\big)G_{n\uparrow j\uparrow}\big(z_4,z_2\big)\times$$
$$\times U\big(z_4\big)U\big(z_2\big)G_{n\downarrow j\downarrow}\big(z_4,z_2\big)G_{i\downarrow n\downarrow}\big(z_1,z_4\big)G_{j\downarrow i\downarrow}\big(z_2,z_1\big),$$

$$\Sigma^{(3),1,\{2;\{2;2,0\},0,\mathrm{C}\},\mathrm{f},1/2}_{i\downarrow j\downarrow}\big(z_1,z_2\big) = \tag{272}$$
$$= -\big(\mathrm{i}\hbar\big)^3 U\big(z_1\big)\int_{\mathcal{C}}\mathrm{d}z_4 \sum_n G_{i\downarrow n\downarrow}\big(z_1,z_4\big)G_{n\downarrow j\downarrow}\big(z_4,z_2\big)\times$$
$$\times U\big(z_4\big)U\big(z_2\big)G_{n\uparrow j\uparrow}\big(z_2,z_4\big)G_{i\uparrow n\uparrow}\big(z_4,z_1\big)G_{j\uparrow i\uparrow}\big(z_1,z_2\big),$$

whereas the "A" terms vanish, as well as all other contributions in third order. The corresponding Feynman diagrams are shown in Fig. 26.

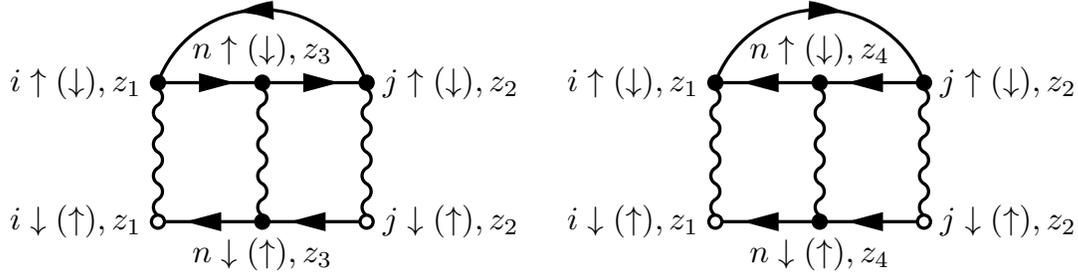

**Figure 26.** Third-order diagrams in Hubbard basis for spin-1/2 fermions. **Left:** $\Sigma^{(3),1,\{2;\{2;2,0\},0,B\},f,1/2}$. **Right:** $\Sigma^{(3),1,\{2;\{2;2,0\},0,C\},f,1/2}$.

### 4.4. Selfenergies of orders higher than three.

All terms of orders higher than three can be generated similarly as was demonstrated above for the lower orders. So we only outline the main steps. As before, one computes all possible permutations of the quantities involved in Hedin's equations that lead to the desired total order. An approach that is suitable for a systematic recursive algorithm, starts by eliminating the polarizability from Hedin's equations, yielding, for the selfenergy and the screened potential [cf. Eq. (95) to Eq. (106)],

$$\Sigma_{ij}(z_1, z_2) = \Sigma_{ij}^{H}(z_1, z_2) + i\hbar \int_{\mathcal{C}} dz_3 \sum_{mpq} W_{ipqm}(z_1, z_3) \times \tag{273}$$
$$\times \int_{\mathcal{C}} dz_4 \sum_n G_{mn}(z_1, z_4) \Gamma_{nqpj}(z_4, z_2, z_3),$$

$$W_{ijkl}(z_1, z_2) = \delta_{\mathcal{C}}(z_1, z_2) w_{ijkl}(z_1) \pm \tag{274}$$
$$\pm i\hbar \sum_{mn} w_{imnl}(z_1) \int_{\mathcal{C}} dz_3 \sum_{pq} \int_{\mathcal{C}} dz_4 \sum_r G_{nr}(z_1, z_4) \times$$
$$\times \int_{\mathcal{C}} dz_5 \sum_s G_{sm}(z_5, z_1) \Gamma_{rqps}(z_4, z_5, z_3) W_{pjkq}(z_3, z_2)$$

and, for the vertex function,

$$\Gamma_{ijkl}(z_1, z_2, z_3) = \delta_{\mathcal{C}}(z_1, z_{2^+}) \delta_{\mathcal{C}}(z_3, z_2) \delta_{ik} \delta_{jl} + \tag{275}$$
$$+ \int_{\mathcal{C}} dz_4 dz_5 \sum_{mn} \frac{\delta \Sigma_{il}^{xc}(z_1, z_2)}{\delta G_{mn}(z_4, z_5)} \int_{\mathcal{C}} dz_6 \sum_p G_{mp}(z_4, z_6) \times$$
$$\times \int_{\mathcal{C}} dz_7 \sum_q G_{qn}(z_7, z_5) \Gamma_{pjkq}(z_6, z_7, z_3).$$

With this set of three coupled equations, the following recursive algorithm can be applied to calculate the $N$-th order selfenergy contributions:

(i) Initialize $\Sigma^{(1)} = \Sigma^{H}$, cf. Eq. (273),

(ii) Initialize $W^{(1)} = w$, cf. Eq. (274),

(iii) Initialize $\Gamma = \Gamma^{(0)}$, cf. Eq. (275),

(iv) Loop over $n = 1 \ldots N$:

    (a) If $n > 1$: Loop over all orders $m = 1 \ldots (n-1)$:

        – Loop over all selfenergy contributions $\Sigma^{(m)}$ of order $m$:

          – Loop over all vertex contributions $\Gamma^{(n-1-m)}$ of order $n - 1 - m$:

          – Calculate the new vertex contribution $\Gamma^{(n-1)}$ of order $m + n - 1 - m = n - 1$, from $\Sigma^{(m)}$ and $\Gamma^{(n-1-m)}$, via Eq. (275),

    (b) If $n > 1$: Loop over all orders $m = 1 \ldots (n-1)$:

        – Loop over all contributions to the screened interaction $W^{(m)}$ of order $m$:

          – Loop over all vertex contributions $\Gamma^{(n-1-m)}$ of order $n - 1 - m$:

          – Calculate the new contribution to the screened interaction $W^{(n)}$ of order $1 + m + n - 1 - m = n$, from $w$, $W^{(m)}$ and $\Gamma^{(n-1-m)}$, via Eq. (274)

    (c) Loop over all orders $m = 1 \ldots n$:

        – Loop over all contributions to the screened interaction $W^{(m)}$ of order $m$:

          – Loop over all vertex contributions $\Gamma^{(n-m)}$ of order $n - m$:

          – Calculate the new selfenergy contribution $\Sigma^{(n)}$ from $W^{(m)}$ and $\Gamma^{(n-m)}$ of order $m + n - m = n$ via Eq. (273)

A similar algorithm yielding the diagrams with respect to the bare interaction, $w$, can be deduced, replacing the full vertex $\Gamma$ by the bare vertex $\Lambda$, cf. Eqs. (97) and (98). Further, the generation of the non-selfconsistent diagrams is straightforward by inclusion of the Dyson equation and, additionally, taking into account the order of the Green functions in the respective equations.

    With this we conclude the discussion of the perturbative approaches to the selfenergy.

## 5. Selfenergy approximations II: Diagram resummation. $GW$, $T$ matrix, FLEX

In this section we discuss an alternative to the perturbative expansion of the selfenergy in terms of the interaction strength that was presented in Section 4. The perturbation expansion is expected to become questionable or, at least, inefficient with increasing interaction strength. This was confirmed in the Sec. 3 where we demonstrated that a number of non-perturbative approaches, such as the $GW$ approximation, the $T$-matrix approximation or the FLEX approach are significantly more accurate, in many cases.

Thus, when perturbation expansions fail, a more appropriate approach consists in diagram resummation techniques that sum an entire infinite perturbation series and which is in the focus of the present section. The underlying idea is to take into consideration one or several classes of terms with a recursive structure which occur in all orders of the interaction, based on physical intuition about their importance. In fact such resummation approaches have a long history. For example the $T$-matrix approximation has been successfully applied in scattering theory and in nuclear physics. On the other hand, the concept of dynamical screening ($GW$ approximation) has been employed for electrolytes and plasmas. The major advance provided by Green functions theory is the extension of the concept to arbitrary nonequilibrium situations.

**$GW$ approximation.** Starting from the notion of the screened interaction, $W$, the simplest choice is the $GW$ approximation [81] which centers around treating $W$ exactly according to Eq. (102) while taking the screened vertex $\Gamma$ only in zeroth-order approximation. This leads to the familiar concept of dynamical screening and plasmon dynamics which is of particular importance for long-range Coulomb interaction in dense plasmas [131] or molecules. The resulting structure of the selfenergy approximation is discussed in the ensuing Section 5.2.

**$T$-matrix approximation.** In contrast to $GW$, the $T$-matrix approximation, treats the interaction only at the level of the bare interaction, but focuses instead on a good representation of the bare vertex functions $\Lambda$. This approximation sums the entire Born series and is, thus expected to be more accurate than the second-Born approximation, at strong coupling. The $T$-matrix approximation exists in two flavors—the particle–particle $T$-matrix approximation (TPP) and the particle–hole $T$-matrix approximation (TPH).

**Combination of strong coupling and dynamical screening.** Furthermore, several other approaches have been introduced that mix screened and bare interaction [165]. An example for this group is the **second-order screened-exchange** (SOSEX) approximation [101, 165, 166], which takes the second-order exchange diagram, cf. Eq. (183), and replaces one of the bare interactions $w$ by the screened interaction $W$. Doing that, the total complexity is still of $\mathcal{O}\left(N_{\mathrm{t}}^3\right)$, since the determination of the selfenergy, the computation of the screened interaction according to Eq. (104) and the solution of the KBE, [cf. Eqs. (66) and (67)], all scale as $\mathcal{O}\left(N_{\mathrm{t}}^3\right)$.

Another possible way is to combine several existing approximations. Thereby one has to correct for possible double counting. This strategy is pursued in the so-called **fluctuating-exchange (FLEX) approximation** which adds the diagrams of third and higher order of the $GW$ approximation and both $T$ matrices to the second-order diagrams, which are taken only once. The FLEX approximation which will be detailed in Section 5.5 can be seen as a the starting term of the more sophisticated plaquet theory [167–173], where one uses coupled equations for the vertex functions in the particle–particle and particle–hole channels.

Before discussing in detail the $GW$ approximation, the $T$-matrix approximation and the FLEX approximation, in Sections 5.2, 5.3 and 5.5, respectively, we investigate in some detail the two-particle Green function $G^{(2)}$. We start by introducing the Hartree and the Fock approximation for $G^{(2)}$ in Section 5.1, since they will be used later.

### 5.1. Mean field. Hartree and Fock approximations for $G^{(2)}$

In the following, the two simplest approximations for the two-particle Green function are defined which will are the starting point for simplifying the expressions occuring in the resummation approaches. Those are the Hartree Green function, $G^{(2),\mathrm{H}}$, and the Fock Green function, $G^{(2),\mathrm{F}}$.

$G^{(2),\mathrm{H}}$ corresponds to the approximation that two particles are uncorrelated. Then the two-particle Green function $G^{(2)}$ is approximated as

$$G^{(2)}_{ijkl}\big(z_1, z_2, z_3, z_4\big) \equiv G^{(2),\mathrm{H}}_{ijkl}\big(z_1, z_2, z_3, z_4\big) := G_{ik}\big(z_1, z_3\big) G_{jl}\big(z_2, z_4\big). \quad (276)$$

This approximation applies to classical and quantum many-particle systems alike. In contrast, an additional approximation that exists only in the case of quantum systems and reflects the indistinguishability of quantum particles (exchange effects of bosons or fermions) is the Fock approximations which, is denoted as $G^{(2),\mathrm{F}}$ and reads

$$G^{(2)}_{ijkl}\big(z_1, z_2, z_3, z_4\big) \equiv G^{(2),\mathrm{F}}_{ijkl}\big(z_1, z_2, z_3, z_4\big) := \pm G_{il}\big(z_1, z_4\big) G_{jk}\big(z_2, z_3\big). \quad (277)$$

For better illustration both two-particle quantities are repeated in the compact notation:

$$G^{(2)}\big(1,2,3,4\big) \approx G^{(2),\mathrm{H}}\big(1,2,3,4\big)$$
$$:= G\big(1,3\big)G\big(2,4\big)$$

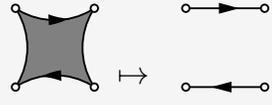

(278)

$$G^{(2)}\big(1,2,3,4\big) \approx G^{(2),\mathrm{F}}\big(1,2,3,4\big)$$
$$:= \pm G\big(1,4\big)G\big(2,3\big)$$

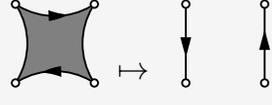

(279)

Following the above considerations, in a quantum system the mean-field approximation for $G^{(2)}$ is the sum of the Hartree and Fock contributions.

In the resummation expansions presented in this section, both quantities play a central role. However, in most cases, we will only need simpler two-time versions of the two-particle functions, $G^{\mathrm{H}}$ and $G^{\mathrm{F}}$, respectively These expressions follow by setting equal two pairs of time-arguments and adding a dimensionality factor $\mathrm{i}\hbar$. This leads to (we skip the superscript "(2)" since it is clear that a function with four basis indices refers to a two-particle Green function)

$$G^{\mathrm{H}}_{ijkl}\big(z_1,z_2\big) := \mathrm{i}\hbar G^{(2),\mathrm{H}}_{ijkl}\big(z_1,z_1,z_2,z_2\big) \equiv \mathrm{i}\hbar G_{ik}\big(z_1,z_2\big) G_{jl}\big(z_1,z_2\big),\quad (280)$$

whereas, for the Fock approximation, one has

$$G^{\mathrm{F}}_{ijkl}\big(z_1,z_2\big) = \mathrm{i}\hbar G^{(2),\mathrm{F}}_{ijkl}\big(z_1,z_2,z_1,z_2\big) \equiv \pm\mathrm{i}\hbar G_{il}\big(z_1,z_2\big) G_{jk}\big(z_2,z_1\big).\quad (281)$$

We now provide the Keldysh matrix components of Eqs. (280) and (281). For the Hartree function, we have the following greater/less and advanced/retarded components:

$$G^{\mathrm{H},\gtrless}_{ijkl}\big(t_1,t_2\big) = \mathrm{i}\hbar G^{\gtrless}_{ik}\big(t_1,t_2\big) G^{\gtrless}_{jl}\big(t_1,t_2\big),$$
$$G^{\mathrm{H},\mathcal{A}/\mathcal{R}}_{ijkl}\big(t_1,t_2\big) = \mp\mathrm{i}\hbar\Theta[\pm(t_2-t_1)]\big[G^{>}_{ik}\big(t_1,t_2\big) G^{>}_{jl}\big(t_1,t_2\big) - G^{<}_{ik}\big(t_1,t_2\big) G^{<}_{jl}\big(t_1,t_2\big)\big],$$

Similarly, for the Fock two-particle function follows

$$G^{\mathrm{F},\gtrless}_{ijkl}\big(t_1,t_2\big) = \pm\mathrm{i}\hbar G^{\gtrless}_{il}\big(t_1,t_2\big) G^{\lessgtr}_{jk}\big(t_2,t_1\big),$$
$$G^{\mathrm{F},\mathcal{A}/\mathcal{R}}_{ijkl}\big(t_1,t_2\big) = \mp\mathrm{i}\hbar\Theta[\pm(t_2-t_1)]\big[G^{>}_{il}\big(t_1,t_2\big) G^{<}_{jk}\big(t_2,t_1\big) - G^{<}_{il}\big(t_1,t_2\big) G^{>}_{jk}\big(t_2,t_1\big)\big],$$

## 5.2. Polarization bubble resummation. GW approximation

The *GW* approximation aims at treating long-range interaction effects that lead to dynamical screening and collective excitations (plasmons). These effects are of particular relevance for charged many-particle systems, including ionized gases (plasmas), the electron gas in metals, electron-hole plasmas in semiconductors and so on. In fact, many-body approximations that go beyond the static second-Born approximation, on one hand, and beyond the statical-screening concept of Debye and Hückel, on the other hand, have a long tradition in plasma physics. In fact kinetic equations with collision integrals the include a complete resummation of all polarization diagrams have been derived in the 1960s by Lenard and Balescu [174, 175] and analyzed in detail by Klimontovich [176] and many others. A quantum derivation within density-operator theory (BBGKY-hierarchy) and a discussion of its relation to Green functions can be found in Ref. [131].

The Green functions approach to dynamical screening and collective excitation is based on Hedin's equations for the screened interaction $W$ according to Eq. (104) with the zeroth order vertex $\Gamma^{(0)}$. The set of equations is given by the Dyson equation[xxii][cf. Eq. (91)]

$$
\begin{aligned}
G\big(1,2\big) &= G^{(0)}\big(1,2\big) \\
&+ G^{(0)}\big(1,3\big)\Sigma\big(3,4\big)G\big(4,2\big)
\end{aligned}
\qquad
\Bigg|
\qquad
\text{}
$$

(282)

the equation for the selfenergy [cf. Eq. (95)]

$$
\Sigma\big(1,2\big) = \Sigma^{\mathrm{H}}\big(1,2\big) + \Sigma^{\mathrm{xc}}\big(1,2\big)
\qquad
\Bigg|
\qquad
\text{}
$$

(283)

with

[xxii]The corresponding equations in the full notation are given in Appendix A.3.

$$\Sigma^{\mathrm{xc}}\big(1,2\big)$$
$$= \mathrm{i}\hbar W\big(1,3\big)G\big(1,4\big)\Gamma^{(0)}\big(4,2,3\big)$$
$$= \mathrm{i}\hbar W\big(1,2\big)G\big(1,2\big)$$

$$\tag{284}$$

the zeroth-order polarizability [cf. Eq. (173)]

$$P^{(0)}\big(1,2\big) = \pm\mathrm{i}\hbar G\big(1,3\big)G\big(4,1\big)$$
$$\Gamma^{(0)}\big(3,4,2\big)$$
$$= \pm\mathrm{i}\hbar G\big(1,2\big)G\big(2,1\big)$$

$$\tag{285}$$

and the screened interaction [cf. Eqs. (102) and (104)]

$$W\big(1,2\big)$$
$$= w\big(1,2\big)$$
$$+ w\big(1,3\big)P^{(0)}\big(3,4\big)W\big(4,2\big)$$
$$= w\big(1,2\big)$$
$$\pm \mathrm{i}\hbar w\big(1,3\big)G\big(3,4\big)G\big(4,3\big)W\big(4,2\big)$$

$$\tag{286}$$

To solve this set of equations, one has to determine the selfconsistent solution of Eq. (286). Due to the time-diagonal structure (i.e. due to the time delta function) of the bare interaction [cf. Eq. (60)] the computational solution of Eq. (286) in the displayed form becomes ill-defined. Therefore, it is advantageous to eliminate the singular bare interaction by defining the "non-singular" part of the interaction (or the induced potential which will be labeled with the superscript "ns")

$$W^{\mathrm{ns}}\big(1,2\big) := W\big(1,2\big) - W^{\mathrm{bare}}\big(1,2\big)$$
$$= W\big(1,2\big) - w\big(1,2\big)$$

$$\overset{\mathrm{ns}}{\sim\!\!\sim\!\!\sim} \;=\; \sim\!\!\sim\!\!\sim \;-\; \sim\!\!\sim\!\!\sim \qquad (287)$$

Using this and, by comparison with Eq. (156), one arrives at

$$\Sigma^{GW}\big(1,2\big) = \Sigma^{\mathrm{H}}\big(1,2\big) + \mathrm{i}\hbar W\big(1,2\big) G\big(1,2\big)$$
$$= \Sigma^{\mathrm{H}}\big(1,2\big) + \mathrm{i}\hbar w\big(1,2\big) G\big(1,2\big) + \mathrm{i}\hbar W^{\mathrm{ns}}\big(1,2\big) G\big(1,2\big)$$
$$=: \Sigma^{\mathrm{H}}\big(1,2\big) + \Sigma^{\mathrm{F}}\big(1,2\big) + \Sigma^{GW,\mathrm{corr}}\big(1,2\big)$$

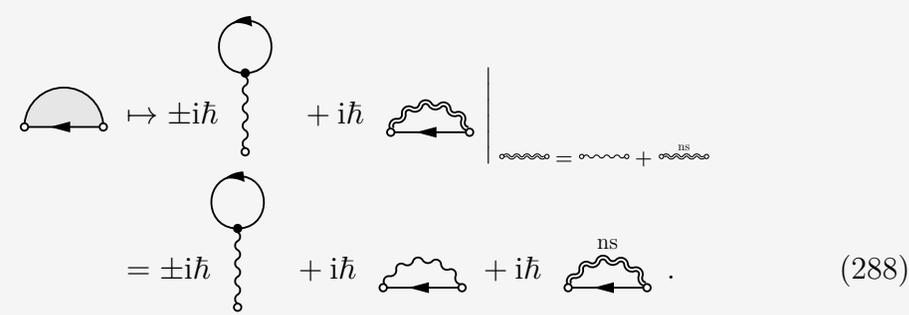

$$(288)$$

Thus the full selfenergy in the $GW$ approximation contains, in addition to the Hartree–Fock selfenergy, a correlation contribution which is denoted $\Sigma^{GW,\mathrm{corr}}$ and which will be in the focus of the subsequent analysis.

For the non-singular part of the screened interaction, we have

$$W^{\mathrm{ns}}\big(1,2\big)$$
$$= \pm \mathrm{i}\hbar w\big(1,3\big) G\big(3,4\big) G\big(4,3\big) w\big(4,2\big)$$
$$\pm \mathrm{i}\hbar w\big(1,3\big) G\big(3,4\big) G\big(4,3\big) W^{\mathrm{ns}}\big(4,2\big)$$

$$\overset{\mathrm{ns}}{\sim\!\!\sim\!\!\sim} \;=\; \pm\,\mathrm{i}\hbar \;\sim\!\!\sim\!\!\bigcirc\!\!\sim\!\!\sim$$
$$\pm\,\mathrm{i}\hbar \;\sim\!\!\sim\!\!\bigcirc\!\!\overset{\mathrm{ns}}{\sim\!\!\sim\!\!\sim}\;.$$

$$(289)$$

Returning to the full notation and the single-time interaction one can use the definition of $G^{\mathrm{F}}$ in Eq. (281) to simplify Eq. (289) to

$$W_{ijkl}^{\mathrm{ns}}\big(z_1,z_2\big) = \Phi_{ijkl}^{GW}\big(z_1,z_2\big) + \sum_{mn} w_{imnl}\big(z_1\big) \int_{\mathcal{C}} \mathrm{d}z_3 \sum_{pq} G_{nqmp}^{\mathrm{F}}\big(z_1,z_3\big) W_{pjkq}^{\mathrm{ns}}\big(z_3,z_2\big) ,$$
(290)

$$\Phi_{ijkl}^{GW}\big(z_1,z_2\big) =: \sum_{mn} w_{imnl}\big(z_1\big) \sum_{pq} G_{nqmp}^{\mathrm{F}}\big(z_1,z_2\big) w_{pjkq}\big(z_2\big) ,$$
(291)

where we introduced the short notation $\Phi$ for the second-Born contribution to the screened potential.

Finally, we provide the correlation components of Eq. (288)

$$\Sigma_{ij}^{GW,\mathrm{corr},\gtrless}\big(t_1,t_2\big) = \mathrm{i}\hbar \sum_{mp} W_{ipjm}^{\mathrm{ns},\gtrless}\big(t_1,t_2\big) G_{mp}^{\gtrless}\big(t_1,t_2\big) ,$$
(292)

which require knowledge of the correlation components of the non-singular screened potential Eq. (290), that are given by

$$W_{ijkl}^{\mathrm{ns},\gtrless}\big(t_1,t_2\big) = \Phi_{ijkl}^{GW,\gtrless}\big(t_1,t_2\big) + \sum_{mn} w_{imnl}\big(t_1\big) \sum_{pq}$$
(293)

$$\left( \int_{t_0}^{t_1} \mathrm{d}t_3\, G_{nqmp}^{\mathrm{F},\mathcal{R}}\big(t_1,t_3\big) W_{pjkq}^{\mathrm{ns},\gtrless}\big(t_3,t_2\big) + \int_{t_0}^{t_2} \mathrm{d}t_3\, G_{nqmp}^{\mathrm{F},\gtrless}\big(t_1,t_3\big) W_{pjkq}^{\mathrm{ns},\mathcal{A}}\big(t_3,t_2\big) \right) ,$$

as well as of the advanced/retarded components,

$$W_{ijkl}^{\mathrm{ns},\mathcal{A}/\mathcal{R}}\big(t_1,t_2\big) = \Phi_{ijkl}^{GW,\mathcal{A}/\mathcal{R}}\big(t_1,t_2\big) +$$
(294)

$$+ \sum_{mn} w_{imnl}\big(t_1\big) \int_{t_{1/2}}^{t_{2/1}} \mathrm{d}t_3 \sum_{pq} G_{nqmp}^{\mathrm{F},\mathcal{A}/\mathcal{R}}\big(t_1,t_3\big) W_{pjkq}^{\mathrm{ns},\mathcal{A}/\mathcal{R}}\big(t_3,t_2\big) .$$

In the integration limits, in the notation $t_{1/2}$, the first (second) subscript refers to the advanced (retarded) function.

**Diagonal basis**.   In a diagonal basis set, Eq. (290) simplifies to

$$W_{ijkl}^{\mathrm{ns},\mathrm{diag}}\big(z_1,z_2\big) = \delta_{il}\delta_{jk}\Phi_{ijji}^{GW,\mathrm{diag}}\big(z_1,z_2\big) +$$
(295)

$$+ \delta_{il} \sum_m w_{im}\big(z_1\big) \int_{\mathcal{C}} \mathrm{d}z_3 \sum_{pq} G_{mpmq}^{\mathrm{F}}\big(z_1,z_3\big) W_{pjkq}^{\mathrm{ns},\mathrm{diag}}\big(z_3,z_2\big) .$$

$$\Phi_{ijji}^{GW,\mathrm{diag}}\big(z_1,z_2\big) =: \sum_m w_{im}\big(z_1\big) \sum_p G_{mpmp}^{\mathrm{F}}\big(z_1,z_2\big) w_{pj}\big(z_2\big) ,$$
(296)

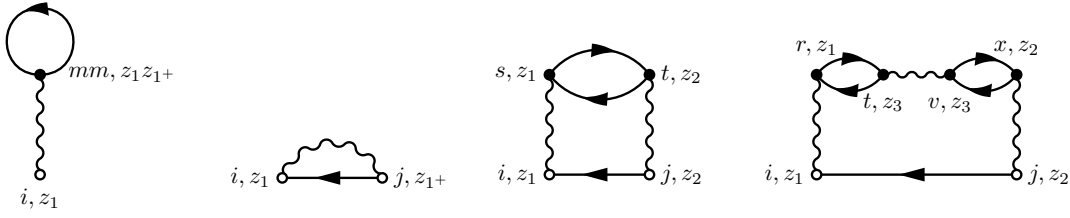

**Figure 27.** The first four terms of the *GW* selfenergy, including Hartree and Fock terms, in a diagonal basis.

where we again used the function $\Phi$, Eq. (291). By iteration, it becomes evident that $W^{\mathrm{ns,diag}}$ and $\Phi_{ijji}^{GW,\mathrm{diag}}$ are always of the form

$$W_{ijkl}^{\mathrm{ns,diag}}\big(z_1,z_2\big) = \delta_{il}\delta_{jk}W_{ijji}^{\mathrm{ns,diag}}\big(z_1,z_2\big) =: \delta_{il}\delta_{jk}W_{ij}^{\mathrm{ns,diag}}\big(z_1,z_2\big), \tag{297}$$

$$\Phi_{ijkl}^{GW,\mathrm{diag}}\big(z_1,z_2\big) = \delta_{il}\delta_{jk}\Phi_{ijji}^{GW,\mathrm{diag}}\big(z_1,z_2\big) =: \delta_{il}\delta_{jk}\Phi_{ij}^{GW,\mathrm{diag}}\big(z_1,z_2\big). \tag{298}$$

With this, for a diagonal basis, Eq. (290) attains the form

$$W_{ij}^{\mathrm{ns,diag}}\big(z_1,z_2\big) = \Phi_{ij}^{GW,\mathrm{diag}}\big(z_1,z_2\big) + \tag{299}$$
$$+ \sum_m w_{im}(z_1)\int_{\mathcal{C}}\mathrm{d}z_3\sum_p G_{mpmp}^{\mathrm{F}}\big(z_1,z_3\big)W_{pj}^{\mathrm{ns,diag}}\big(z_3,z_2\big).$$

The correlation part of the selfenergy, Eq. (288), reads

$$\Sigma_{ij}^{GW,\mathrm{corr,diag}}\big(z_1,z_2\big) = +\mathrm{i}\hbar W_{ij}^{\mathrm{ns,diag}}\big(z_1,z_2\big)G_{ij}\big(z_1,z_2\big). \tag{300}$$

The first four terms of the *GW* selfenergy (mean field plus correlation selfenergy) are shown diagrammatically in Fig. 27. The sum continues to infinite order (infinite number of polarization bubbles).

The relevant components of the Keldysh matrix read [cf. Eqs. (293) and (294)],

$$W_{ij}^{\mathrm{ns,diag},\gtrless}\big(t_1,t_2\big) = \Phi_{ij}^{GW,\mathrm{diag},\gtrless}\big(t_1,t_2\big) + \sum_{mp}w_{im}\big(t_1\big) \tag{301}$$
$$\left(\int_{t_0}^{t_1}\mathrm{d}t_3\,G_{mpmp}^{\mathrm{F,\mathcal{R}}}\big(t_1,t_3\big)W_{pj}^{\mathrm{ns,diag},\gtrless}\big(t_3,t_2\big) + \right.$$
$$\left. + \int_{t_0}^{t_2}\mathrm{d}t_3\,G_{mpmp}^{\mathrm{F,\gtrless}}\big(t_1,t_3\big)W_{pj}^{\mathrm{ns,diag},\mathcal{A}}\big(t_3,t_2\big)\right),$$

$$W_{ij}^{\mathrm{ns,diag},\mathcal{A}/\mathcal{R}}\big(t_1,t_2\big) = \Phi_{ij}^{GW,\mathrm{diag},\mathcal{A}/\mathcal{R}}\big(t_1,t_2\big) + \tag{302}$$
$$+ \sum_{mp}w_{im}\big(t_1\big)\int_{t_{1/2}}^{t_{2/1}}\mathrm{d}t_3\,G_{mpmp}^{\mathrm{F},\mathcal{A}/\mathcal{R}}\big(t_1,t_3\big)W_{pj}^{\mathrm{ns,diag},\mathcal{A}/\mathcal{R}}\big(t_3,t_2\big),$$

and [cf. Eq. (292)],

$$\Sigma_{ij}^{GW,\mathrm{corr,diag},\gtrless}\big(t_1,t_2\big) = \mathrm{i}\hbar W_{ij}^{\mathrm{ns,diag},\gtrless}\big(t_1,t_2\big) G_{ij}^{\gtrless}\big(t_1,t_2\big). \tag{303}$$

**Hubbard basis.** In the Hubbard basis, cf. Section 2.3, the $GW$ approximation simplifies considerably. We start by presenting the equations for **bosons**:

$$W_{i\alpha j\beta}^{\mathrm{ns,b}}\big(z_1,z_2\big) = \Phi_{ij}^{GW,\mathrm{b}}\big(z_1,z_2\big) + U\big(z_1\big)\int_{\mathcal{C}}\mathrm{d}z_3\sum_{p\epsilon}G_{i\epsilon p\epsilon i\epsilon p\epsilon}^{\mathrm{F}}\big(z_1,z_3\big)W_{p\epsilon j\beta}^{\mathrm{ns,b}}\big(z_3,z_2\big), \tag{304}$$

$$\Phi_{ij}^{GW,\mathrm{b}}\big(z_1,z_2\big) =: U\big(z_1\big)\sum_{\epsilon}G_{i\epsilon j\epsilon i\epsilon j\epsilon}^{\mathrm{F}}\big(z_1,z_2\big)U\big(z_2\big), \tag{305}$$

and the correlation selfenergy on the Keldysh contour is

$$\Sigma_{i\alpha j\alpha}^{GW,\mathrm{corr,b}}\big(z_1,z_2\big) = \mathrm{i}\hbar W_{i\alpha j\alpha}^{\mathrm{ns,b}}\big(z_1,z_2\big)G_{i\alpha j\alpha}\big(z_1,z_2\big). \tag{306}$$

The Keldysh matrix components of the screened potential in the Hubbard basis become

$$\begin{aligned}
W_{i\alpha j\beta}^{\mathrm{ns,b},\gtrless}\big(t_1,t_2\big) \quad &= \Phi_{ij}^{GW,\mathrm{b},\gtrless}\big(t_1,t_2\big) + \\
&\quad + U\big(t_1\big)\int_{t_0}^{t_1}\mathrm{d}t_3\sum_{p\epsilon}G_{i\epsilon p\epsilon i\epsilon p\epsilon}^{\mathrm{F},\mathcal{R}}\big(t_1,t_3\big)W_{p\epsilon j\beta}^{\mathrm{ns,b},\gtrless}\big(t_3,t_2\big) + \\
&\quad + U\big(t_1\big)\int_{t_0}^{t_2}\mathrm{d}t_3\sum_{p\epsilon}G_{i\epsilon p\epsilon i\epsilon p\epsilon}^{\mathrm{F},\gtrless}\big(t_1,t_3\big)W_{p\epsilon j\beta}^{\mathrm{ns,b},\mathcal{A}}\big(t_3,t_2\big), \\
W_{i\alpha j\beta}^{\mathrm{ns,b},\mathcal{A}/\mathcal{R}}\big(t_1,t_2\big) &= \Phi_{ij}^{GW,\mathrm{b},\mathcal{A}/\mathcal{R}}\big(t_1,t_2\big) + \\
&\quad + U\big(t_1\big)\int_{t_{1/2}}^{t_{2/1}}\mathrm{d}t_3\sum_{p\epsilon}G_{i\epsilon p\epsilon i\epsilon p\epsilon}^{\mathrm{F},\mathcal{A}/\mathcal{R}}\big(t_1,t_3\big)W_{p\epsilon j\beta}^{\mathrm{ns,b},\mathcal{A}/\mathcal{R}}\big(t_3,t_2\big),
\end{aligned}$$

and the greater/less components of the correlation selfenergy (306) are

$$\Sigma_{i\alpha j\alpha}^{GW,\mathrm{corr,b},\gtrless}\big(t_1,t_2\big) = \mathrm{i}\hbar W_{i\alpha j\alpha}^{\mathrm{ns,b},\gtrless}\big(t_1,t_2\big)G_{i\alpha j\alpha}^{\gtrless}\big(t_1,t_2\big). \tag{307}$$

For the special case of **spin-0 bosons**, the screened potential on the Keldysh contour has the form

$$W_{ij}^{\mathrm{ns,b,0}}\big(z_1,z_2\big) = \Phi_{ij}^{GW,\mathrm{b,0}}\big(z_1,z_2\big) + U\big(z_1\big)\int_{\mathcal{C}}\mathrm{d}z_3\sum_{p}G_{ipip}^{\mathrm{F}}\big(z_1,z_3\big)W_{pj}^{\mathrm{ns,b,0}}\big(z_3,z_2\big),$$

$$\Phi_{ij}^{GW,\mathrm{b,0}}\big(z_1,z_2\big) =: U\big(z_1\big)G_{ijij}^{\mathrm{F}}\big(z_1,z_2\big)U\big(z_2\big), \tag{308}$$

and the correlation selfenergy in $GW$ approximation becomes

$$\Sigma_{ij}^{GW,\mathrm{corr,b,0}}\big(z_1,z_2\big) = \mathrm{i}\hbar W_{ij}^{\mathrm{ns,b,0}}\big(z_1,z_2\big)G_{ij}\big(z_1,z_2\big).$$

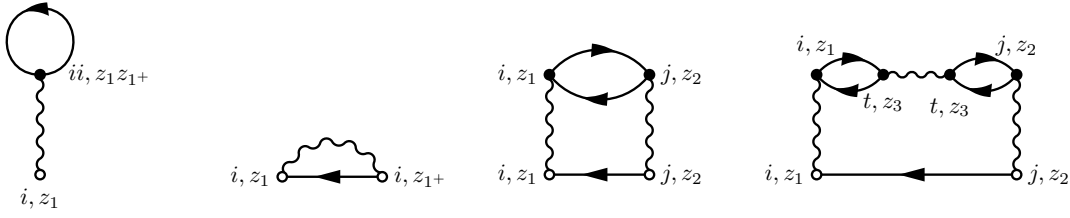

**Figure 28.** The first four terms of the *GW* selfenergy (Hartree, Fock and correlation part) in the Hubbard basis for spin-0 bosons.

The greater, less and advanced/retarded matrix components become

$$W_{ij}^{\mathrm{ns,b},0,\gtrless}\left(t_1,t_2\right) = \Phi_{ij}^{GW,\mathrm{b},0,\gtrless}\left(t_1,t_2\right) +$$
$$+ U\left(t_1\right) \int_{t_0}^{t_1} \mathrm{d}t_3 \sum_p G_{ipip}^{\mathrm{F},\mathcal{R}}\left(t_1,t_3\right) W_{pj}^{\mathrm{ns,b},0,\gtrless}\left(t_3,t_2\right) +$$
$$+ U\left(t_1\right) \int_{t_0}^{t_2} \mathrm{d}t_3 \sum_p G_{ipip}^{\mathrm{F},\gtrless}\left(t_1,t_3\right) W_{pj}^{\mathrm{ns,b},0,\mathcal{A}}\left(t_3,t_2\right),$$

$$W_{ij}^{\mathrm{ns,b},0,\mathcal{A/R}}\left(t_1,t_2\right) = \Phi_{ij}^{GW,\mathrm{b},0,\mathcal{A/R}}\left(t_1,t_2\right) +$$
$$+ U\left(t_1\right) \int_{t_{1/2}}^{t_{2/1}} \mathrm{d}t_3 \sum_p G_{ipip}^{\mathrm{F},\mathcal{A/R}}\left(t_1,t_3\right) W_{pj}^{\mathrm{ns,b},0,\mathcal{A/R}}\left(t_3,t_2\right)$$

and, for the greater/less components of the correlation selfenergy, we have

$$\Sigma_{ij}^{GW,\mathrm{corr,b},0,\gtrless}\left(t_1,t_2\right) = \mathrm{i}\hbar W_{ij}^{\mathrm{ns,b},0,\gtrless}\left(t_1,t_2\right) G_{ij}^{\gtrless}\left(t_1,t_2\right).$$

The diagrammatic representation of the leading terms for spin-0 bosons in the Hubbard basis is shown in Fig. 28.

Let us now turn to **fermions**. In that case the equations on the Keldysh contour attain the form

$$W_{i\alpha j\beta}^{\mathrm{ns,f}}\left(z_1,z_2\right) = \Phi_{ij\alpha\beta}^{GW,\mathrm{f}}\left(z_1,z_2\right) + \tag{309}$$
$$+ U\left(z_1\right) \int_{\mathcal{C}} \mathrm{d}z_3 \sum_p \sum_{\epsilon \neq \alpha} G_{i\epsilon p i \epsilon p \epsilon}^{\mathrm{F}}\left(z_1,z_3\right) W_{p\epsilon j\beta}^{\mathrm{ns,f}}\left(z_3,z_2\right)$$
$$\Phi_{ij\alpha\beta}^{GW,\mathrm{f}}\left(z_1,z_2\right) =: \sum_{\epsilon \neq \{\alpha,\beta\}} U\left(z_1\right) G_{i\epsilon j\epsilon i\epsilon j\epsilon}^{\mathrm{F}}\left(z_1,z_2\right) U\left(z_2\right), \tag{310}$$

whereas the correlation part of the selfenergy becomes

$$\Sigma_{i\alpha j\alpha}^{GW,\mathrm{corr,f}}\left(z_1,z_2\right) = \mathrm{i}\hbar W_{i\alpha j\alpha}^{\mathrm{ns,f}}\left(z_1,z_2\right) G_{i\alpha j\alpha}\left(z_1,z_2\right).$$

The Keldysh matrix components of the screened potential are now

$$W^{\mathrm{ns,f},\gtrless}_{i\alpha j\beta}\big(t_1,t_2\big) = \Phi^{GW,\mathrm{f},\gtrless}_{ij\alpha\beta}\big(t_1,t_2\big) +$$
$$+ U\big(t_1\big)\int_{t_0}^{t_1}\mathrm{d}t_3\sum_p\sum_{\epsilon\neq\alpha}G^{\mathrm{F},\mathcal{R}}_{i\epsilon p\epsilon i\epsilon p\epsilon}\big(t_1,t_3\big)W^{\mathrm{ns,f},\gtrless}_{p\epsilon j\beta}\big(t_3,t_2\big) +$$
$$+ U\big(t_1\big)\int_{t_0}^{t_2}\mathrm{d}t_3\sum_p\sum_{\epsilon\neq\alpha}G^{\mathrm{F},\gtrless}_{i\epsilon p\epsilon i\epsilon p\epsilon}\big(t_1,t_3\big)W^{\mathrm{ns,f},\mathcal{A}}_{p\epsilon j\beta}\big(t_3,t_2\big) ,$$

$$W^{\mathrm{ns,f},\mathcal{A}/\mathcal{R}}_{i\alpha j\beta}\big(t_1,t_2\big) = \Phi^{GW,\mathrm{f},\mathcal{A}/\mathcal{R}}_{ij\alpha\beta}\big(t_1,t_2\big) +$$
$$+ U\big(t_1\big)\int_{t_{1/2}}^{t_{2/1}}\mathrm{d}t_3\sum_p\sum_{\epsilon\neq\alpha}G^{\mathrm{F},\mathcal{A}/\mathcal{R}}_{i\epsilon p\epsilon i\epsilon p\epsilon}\big(t_1,t_3\big)W^{\mathrm{ns,f},\mathcal{A}/\mathcal{R}}_{p\epsilon j\beta}\big(t_3,t_2\big) .$$

The *GW* selfenergy has now the following correlation components

$$\Sigma^{GW,\mathrm{corr,f},\gtrless}_{i\alpha j\alpha}\big(t_1,t_2\big) = \mathrm{i}\hbar W^{\mathrm{ns,f},\gtrless}_{i\alpha j\alpha}\big(t_1,t_2\big)G^{\gtrless}_{i\alpha j\alpha}\big(t_1,t_2\big) .$$

For the special case of **spin-$\frac{1}{2}$-fermions**, the equations for the screened interaction require some care. Since the particles can have two spin projections, there are four different screened potentials each obeying its own equation which, in turn, are coupled. To underline these details we use different colors for the four potentials:

$$\textcolor{blue}{W^{\mathrm{ns,f},1/2}_{i\uparrow j\uparrow}\big(z_1,z_2\big)} = \Phi^{GW,\mathrm{f},1/2}_{ij\uparrow}\big(z_1,z_2\big) + U\big(z_1\big)\int_{\mathcal{C}}\mathrm{d}z_3\sum_m G^{\mathrm{F}}_{i\downarrow m\downarrow i\downarrow m\downarrow}\big(z_1,z_3\big)\textcolor{red}{W^{\mathrm{ns,f},1/2}_{m\downarrow j\uparrow}\big(z_3,z_2\big)} ,$$
$$\Phi^{GW,\mathrm{f},1/2}_{ij\uparrow}\big(z_1,z_2\big) =: U\big(z_1\big)G^{\mathrm{F}}_{i\downarrow j\downarrow i\downarrow j\downarrow}\big(z_1,z_2\big)U\big(z_2\big) ,$$
$$\textcolor{red}{W^{\mathrm{ns,f},1/2}_{i\downarrow j\uparrow}\big(z_1,z_2\big)} = U\big(z_1\big)\int_{\mathcal{C}}\mathrm{d}z_3\sum_m G^{\mathrm{F}}_{i\uparrow m\uparrow i\uparrow m\uparrow}\big(z_1,z_3\big)\textcolor{blue}{W^{\mathrm{ns,f},1/2}_{m\uparrow j\uparrow}\big(z_3,z_2\big)} , \tag{311}$$

$$\textcolor{green}{W^{\mathrm{ns,f},1/2}_{i\downarrow j\downarrow}\big(z_1,z_2\big)} = \Phi^{GW,\mathrm{f},1/2}_{ij\downarrow}\big(z_1,z_2\big) + U\big(z_1\big)\int_{\mathcal{C}}\mathrm{d}z_3\sum_m G^{\mathrm{F}}_{i\uparrow m\uparrow i\uparrow m\uparrow}\big(z_1,z_3\big)\textcolor{orange}{W^{\mathrm{ns,f},1/2}_{m\uparrow j\downarrow}\big(z_3,z_2\big)} ,$$
$$\Phi^{GW,\mathrm{f},1/2}_{ij\downarrow}\big(z_1,z_2\big) =: U\big(z_1\big)G^{\mathrm{F}}_{i\uparrow j\uparrow i\uparrow j\uparrow}\big(z_1,z_2\big)U\big(z_2\big) ,$$
$$\textcolor{orange}{W^{\mathrm{ns,f},1/2}_{i\uparrow j\downarrow}\big(z_1,z_2\big)} = U\big(z_1\big)\int_{\mathcal{C}}\mathrm{d}z_3\sum_m G^{\mathrm{F}}_{i\downarrow m\downarrow i\downarrow m\downarrow}\big(z_1,z_3\big)\textcolor{green}{W^{\mathrm{ns,f},1/2}_{m\downarrow j\downarrow}\big(z_3,z_2\big)} . \tag{312}$$

Interestingly, the equations for the screened potentials with different spin combinations, Eqs. (311) and (312), do not contain a contribution from the bare interaction.

The correlation selfenergies of fermions with spin up and down read, respectively,

$$\textcolor{blue}{\Sigma^{GW,\mathrm{corr,f},1/2}_{i\uparrow j\uparrow}\big(z_1,z_2\big)} = \mathrm{i}\hbar W^{\mathrm{ns,f},1/2}_{i\uparrow j\uparrow}\big(z_1,z_2\big)G_{i\uparrow j\uparrow}\big(z_1,z_2\big) , \tag{313}$$
$$\textcolor{green}{\Sigma^{GW,\mathrm{corr,f},1/2}_{i\downarrow j\downarrow}\big(z_1,z_2\big)} = \mathrm{i}\hbar W^{\mathrm{ns,f},1/2}_{i\downarrow j\downarrow}\big(z_1,z_2\big)G_{i\downarrow j\downarrow}\big(z_1,z_2\big) . \tag{314}$$

The diagrammatic representation of the leading terms for spin-1/2 fermions in the

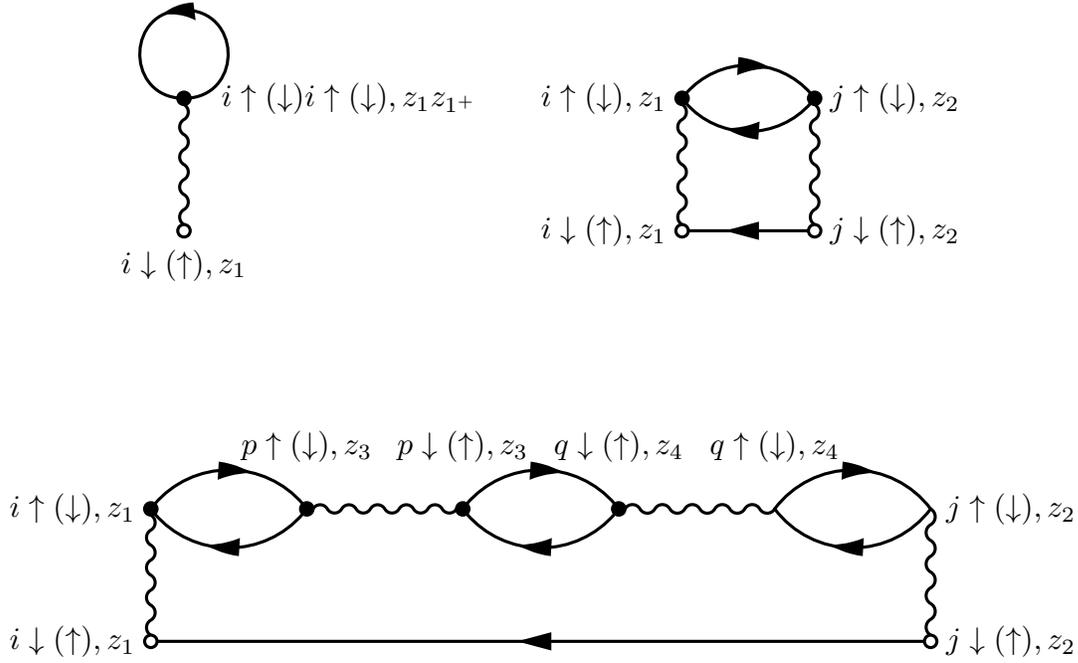

**Figure 29.** The first three terms of the $GW$ selfenergy (mean field plus correlation term) in the Hubbard basis for spin-1/2 fermions. Note that the Fock term equals zero.

Hubbard basis is shown in Fig. 29. One recognizes a special structure of these equations for spin-$\frac{1}{2}$-fermions. The selfenergy $\left(\Sigma_{i\uparrow j\uparrow}^{GW,\mathrm{f},1/2} \,/\, \Sigma_{i\downarrow j\downarrow}^{GW,\mathrm{f},1/2}\right)$, which couples only contributions of the same spin, also directly depends only on the same-spin parts of the screened interaction $\left(W_{i\uparrow j\uparrow}^{\mathrm{ns,f},1/2} G_{i\uparrow j\uparrow} \,/\, W_{i\downarrow j\downarrow}^{\mathrm{ns,f},1/2} G_{i\downarrow j\downarrow}\right)$, cf. Eqs. (313) and (314). The same-spin screened interaction, in turn, depends on the screened interaction with different spin orientations $\left(W_{m\downarrow j\uparrow}^{\mathrm{ns,f},1/2} \,/\, W_{m\uparrow j\downarrow}^{\mathrm{ns,f},1/2}\right)$, which itself couples back to the same-spin part.

Finally, we present the respective greater/less and retarded/advanced Keldysh components of the screened potential for all four spin combinations:

$$W_{i\uparrow j\uparrow}^{\mathrm{ns,f},1/2,\gtrless}\left(t_1,t_2\right) = \Phi_{ij\uparrow}^{GW,\mathrm{f},1/2,\gtrless}\left(t_1,t_2\right) +$$
$$+ U\left(t_1\right)\int_{t_0}^{t_1} \mathrm{d}t_3 \sum_m G_{i\downarrow m\downarrow i\downarrow m\downarrow}^{\mathrm{F},\mathcal{R}}\left(t_1,t_3\right) W_{m\downarrow j\uparrow}^{\mathrm{ns,f},1/2,\gtrless}\left(t_3,t_2\right) +$$
$$+ U\left(t_1\right)\int_{t_0}^{t_2} \mathrm{d}t_3 \sum_m G_{i\downarrow m\downarrow i\downarrow m\downarrow}^{\mathrm{F},\gtrless}\left(t_1,t_3\right) W_{m\downarrow j\uparrow}^{\mathrm{ns,f},1/2,\mathcal{A}}\left(t_3,t_2\right),$$

$$W_{i\uparrow j\uparrow}^{\mathrm{ns,f},1/2,\mathcal{A}/\mathcal{R}}\left(t_1,t_2\right) = \Phi_{ij\uparrow}^{GW,\mathrm{f},1/2,\mathcal{A}/\mathcal{R}}\left(t_1,t_2\right) +$$
$$+ U\left(t_1\right)\int_{t_{1/2}}^{t_{2/1}} \mathrm{d}t_3 \sum_m G_{i\downarrow m\downarrow i\downarrow m\downarrow}^{\mathrm{F},\mathcal{A}/\mathcal{R}}\left(t_1,t_3\right) W_{m\downarrow j\uparrow}^{\mathrm{ns,f},1/2,\mathcal{A}/\mathcal{R}}\left(t_3,t_2\right),$$

$$W_{i\downarrow j\uparrow}^{\mathrm{ns,f},1/2,\gtrless}\left(t_1,t_2\right) = U\left(t_1\right)\int_{t_0}^{t_1}\mathrm{d}t_3\sum_m G_{i\uparrow m\uparrow i\uparrow m\uparrow}^{\mathrm{F},\mathcal{R}}\left(t_1,t_3\right)W_{m\uparrow j\uparrow}^{\mathrm{ns,f},1/2,\gtrless}\left(t_3,t_2\right) +$$

$$+ U\left(t_1\right)\int_{t_0}^{t_2}\mathrm{d}t_3\sum_m G_{i\uparrow m\uparrow i\uparrow m\uparrow}^{\mathrm{F},\gtrless}\left(t_1,t_3\right)W_{m\uparrow j\uparrow}^{\mathrm{ns,f},1/2,\mathcal{A}}\left(t_3,t_2\right),$$

$$W_{i\downarrow j\uparrow}^{\mathrm{ns,f},1/2,\mathcal{A}/\mathcal{R}}\left(t_1,t_2\right) = U\left(t_1\right)\int_{t_{1/2}}^{t_{2/1}}\mathrm{d}t_3\sum_m G_{i\uparrow m\uparrow i\uparrow m\uparrow}^{\mathrm{F},\mathcal{A}/\mathcal{R}}\left(t_1,t_3\right)W_{m\uparrow j\uparrow}^{\mathrm{ns,f},1/2,\mathcal{A}/\mathcal{R}}\left(t_3,t_2\right),$$

$$W_{i\downarrow j\downarrow}^{\mathrm{ns,f},1/2,\gtrless}\left(t_1,t_2\right) = \Phi_{ij\downarrow}^{GW,\mathrm{f},1/2,\gtrless}\left(t_1,t_2\right) +$$

$$+ U\left(t_1\right)\int_{t_0}^{t_1}\mathrm{d}t_3\sum_m G_{i\uparrow m\uparrow i\uparrow m\uparrow}^{\mathrm{F},\mathcal{R}}\left(t_1,t_3\right)W_{m\uparrow j\downarrow}^{\mathrm{ns,f},1/2,\gtrless}\left(t_3,t_2\right) +$$

$$+ U\left(t_1\right)\int_{t_0}^{t_2}\mathrm{d}t_3\sum_m G_{i\uparrow m\uparrow i\uparrow m\uparrow}^{\mathrm{F},\gtrless}\left(t_1,t_3\right)W_{m\uparrow j\downarrow}^{\mathrm{ns,f},1/2,\mathcal{A}}\left(t_3,t_2\right),$$

$$W_{i\downarrow j\downarrow}^{\mathrm{ns,f},1/2,\mathcal{A}/\mathcal{R}}\left(t_1,t_2\right) = \Phi_{ij\downarrow}^{GW,\mathrm{f},1/2,\mathcal{A}/\mathcal{R}}\left(t_1,t_2\right) +$$

$$+ U\left(t_1\right)\int_{t_{1/2}}^{t_{2/1}}\mathrm{d}t_3\sum_m G_{i\uparrow m\uparrow i\uparrow m\uparrow}^{\mathrm{F},\mathcal{A}/\mathcal{R}}\left(t_1,t_3\right)W_{m\uparrow j\downarrow}^{\mathrm{ns,f},1/2,\mathcal{A}/\mathcal{R}}\left(t_3,t_2\right),$$

$$W_{i\uparrow j\downarrow}^{\mathrm{ns,f},1/2,\gtrless}\left(t_1,t_2\right) = U\left(t_1\right)\int_{t_0}^{t_1}\mathrm{d}t_3\sum_m G_{i\downarrow m\downarrow i\downarrow m\downarrow}^{\mathrm{F},\mathcal{R}}\left(t_1,t_3\right)W_{m\downarrow j\downarrow}^{\mathrm{ns,f},1/2,\gtrless}\left(t_3,t_2\right) +$$

$$+ U\left(t_1\right)\int_{t_0}^{t_2}\mathrm{d}t_3\sum_m G_{i\downarrow m\downarrow i\downarrow m\downarrow}^{\mathrm{F},\gtrless}\left(t_1,t_3\right)W_{m\downarrow j\downarrow}^{\mathrm{ns,f},1/2,\mathcal{A}}\left(t_3,t_2\right)$$

$$W_{i\uparrow j\downarrow}^{\mathrm{ns,f},1/2,\mathcal{A}/\mathcal{R}}\left(t_1,t_2\right) = U_1\int_{t_{1/2}}^{t_{2/1}}\mathrm{d}t_3\sum_m G_{i\downarrow m\downarrow i\downarrow m\downarrow}^{\mathrm{F},\mathcal{A}/\mathcal{R}}\left(t_1,t_3\right)W_{m\downarrow j\downarrow}^{\mathrm{ns,f},1/2,\mathcal{A}/\mathcal{R}}\left(t_3,t_2\right).$$

The correlation components of the selfenergy read

$$\Sigma_{i\uparrow j\uparrow}^{GW,\mathrm{corr},\mathrm{f},\gtrless}\left(t_1,t_2\right) = \mathrm{i}\hbar W_{i\uparrow j\uparrow}^{\mathrm{ns,f},\gtrless}\left(t_1,t_2\right)G_{i\uparrow j\uparrow}^{\gtrless}\left(t_1,t_2\right), \tag{315}$$

$$\Sigma_{i\downarrow j\downarrow}^{GW,\mathrm{corr},\mathrm{f},\gtrless}\left(t_1,t_2\right) = \mathrm{i}\hbar W_{i\downarrow j\downarrow}^{\mathrm{ns,f},\gtrless}\left(t_1,t_2\right)G_{i\downarrow j\downarrow}^{\gtrless}\left(t_1,t_2\right). \tag{316}$$

With this we conclude the discussion of the *GW* approximation. As mentioned before the strength of this approximation is the account of long-range screening effects and collective excitations (plasmons) which is of particular relevance for systems with Coulomb interaction. As we have seen in Sec. 3, the *GW* approximation has an impressively high accuracy, in many cases. At the same time, the *GW* approximation does not take into account strong-coupling effects since it includes only terms of second order in the (screened) interaction. Effects of multiple scattering are, thus, missing.

The inclusion of these effects for the case of a static pair interaction is the goal of the *T*-matrix approximation that is studied in Sec. 5.3. On the other hand, to account for multiple scattering and dynamical screening simultaneously, is the goal of the FLEX scheme that is discussed in Sec. 5.5.

## 5.3. Strong coupling. $T$-matrix approximation. Particle–particle and particle–hole $T$ matrices

The goal of the $T$-matrix approximation is to capture effects of multiple scattering that are important in strongly coupled systems, but are missing in the second-Born approximation. The solution is to sum all higher-order Born terms up what leads to an effective interaction. In contrast to the $GW$ approximation, the $T$-matrix approximation takes only the bare interaction, $w$, into account (at least in its standard versions), neglecting dynamical screening, and aims, instead, at a good approximation of the bare vertex function $\Lambda$. The goal here is to accurately capture multiple scattering effects. Thus, its constitutive equations are the Dyson equation[xxiii]

$$G\big(1,2\big) = G^{(0)}\big(1,2\big) + G^{(0)}\big(1,3\big)\Sigma\big(3,4\big)G\big(4,2\big)$$

$$(317)$$

the equation for the selfenergy, cf. Eqs. (95) and (97),

$$\Sigma\big(1,2\big) = \Sigma^{\mathrm{H}}\big(1,2\big) + \Sigma^{\mathrm{xc}}\big(1,2\big)$$

$$(318)$$

with the exchange–correlation selfenergy (all contributions beyond Hartree) given by

$$\Sigma^{\mathrm{xc}}\big(1,2\big) = \mathrm{i}\hbar w\big(1,3\big)G\big(1,4\big)\Lambda\big(4,2,3\big)$$

$$(319)$$

The bare vertex $\Lambda$ is self-consistently given as the solution of the integral equation

---

[xxiii]The corresponding equations in the full notation are given in Appendix A.4.

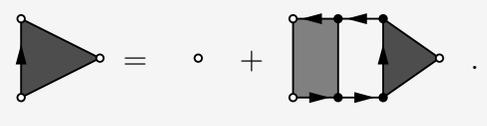

$$\Lambda\big(1,2,3\big) = \delta\big(1,2^+\big)\delta\big(3,2\big)$$
$$+ \frac{\delta\Sigma\big(1,2\big)}{\delta G\big(4,5\big)} G\big(4,6\big) G\big(7,5\big) \Lambda\big(6,7,3\big)$$

(320)

If these equations are iterated *ad infinitum*, all selfenergy terms will be generated. To break the circular dependence between Eqs. (319) and (320), the *T*-matrix approximation starts by taking the bare vertex on the right-hand side of Eq. (320) only in zeroth order, transforming it into

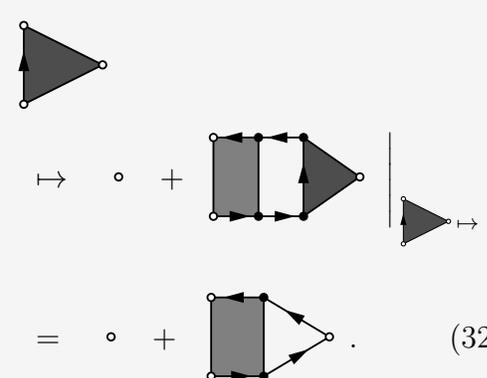

$$\Lambda^{\mathrm{cl}}\big(1,2,3\big) = \delta\big(1,2^+\big)\delta\big(3,2\big)$$
$$+ \frac{\delta\Sigma^{\mathrm{cl}}\big(1,2\big)}{\delta G\big(4,5\big)} G\big(4,6\big) G\big(7,5\big) \Lambda^{(0)}\big(6,7,3\big)$$
$$= \delta\big(1,2^+\big)\delta\big(3,2\big)$$
$$+ \frac{\delta\Sigma^{\mathrm{cl}}\big(1,2\big)}{\delta G\big(4,5\big)} G\big(4,3\big) G\big(3,5\big)$$

(321)

This closure of the equation for $\Lambda$ allows to systematically generate results for the selfenergy which we will denote by the superscript "cl". Using this result in Eq. (318), we obtain [cf. Eq. (156)]

$$\Sigma^{\text{cl}}(1,2)$$
$$= \Sigma^{\text{H}}(1,2)$$
$$+ \mathrm{i}\hbar w(1,3) G(1,4) \Lambda^{\text{cl}}(4,2,3)$$
$$= \Sigma^{\text{H}}(1,2) + \Sigma^{\text{F}}(1,2)$$
$$+ \mathrm{i}\hbar w(1,3) G(1,4) \frac{\delta \Sigma^{\text{cl}}(4,2)}{\delta G(5,6)}$$
$$G(5,3) G(3,6)$$

$$\tag{322}$$

To solve for $\delta\Sigma^{\text{cl}}/\delta G$, we differentiate the whole equation with respect to $G$:

$$\frac{\delta \Sigma^{\mathrm{cl}}(1,2)}{\delta G(7,8)} = \pm \mathrm{i}\hbar \delta(1,2)\delta(7,8) w(1,7) + \mathrm{i}\hbar \delta(1,7)\delta(2,8) w(1,2)$$

$$+ \mathrm{i}\hbar \delta(1,7) w(1,3) \frac{\delta \Sigma^{\mathrm{cl}}(8,2)}{\delta G(5,6)} G(5,3) G(3,6)$$

$$+ \mathrm{i}\hbar w(1,8) G(1,4) \frac{\delta \Sigma^{\mathrm{cl}}(4,2)}{\delta G(7,6)} G(8,6)$$

$$+ \mathrm{i}\hbar w(1,7) G(1,4) \frac{\delta \Sigma^{\mathrm{cl}}(4,2)}{\delta G(5,8)} G(5,7)$$

$$+ \mathrm{i}\hbar w(1,3) G(1,4) \frac{\delta \Sigma^{\mathrm{cl}}(4,2)}{\delta G(5,6)\delta G(7,8)} G(5,3) G(3,6)$$

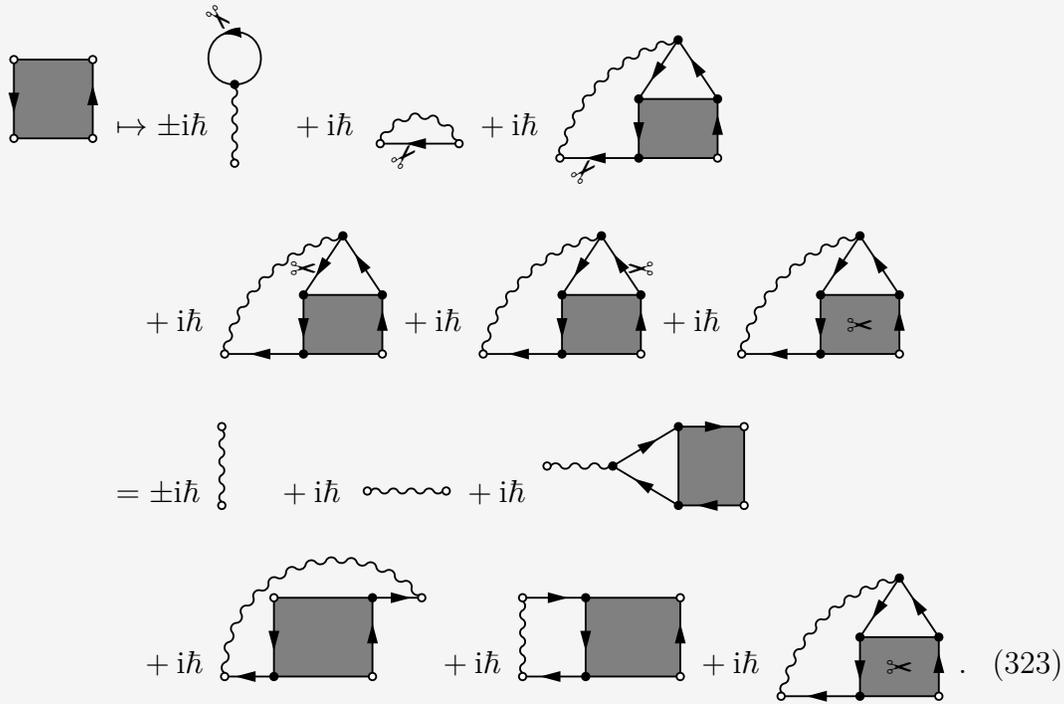

$(323)$

This equation is still very complicated. Therefore, to make further progress, we transform Eq. (323) into a closed equation for $\frac{\delta \Sigma^{\mathrm{cl}}}{\delta G}$ by neglecting the term $\frac{\delta \Sigma^{\mathrm{cl}}(4,2)}{\delta G(5,6)\delta G(7,8)}$. The first iteration yields the second-order terms

$$\frac{\delta\Sigma^{\mathrm{cl},(2)}\left(1,2\right)}{\delta G\left(7,8\right)} \approx i\hbar\delta\left(1,7\right)w\left(1,3\right)\frac{\delta\Sigma^{\mathrm{cl},(1)}\left(8,2\right)}{\delta G\left(5,6\right)}G\left(5,3\right)G\left(3,6\right)$$

$$+ i\hbar w\left(1,8\right)G\left(1,4\right)\frac{\delta\Sigma^{\mathrm{cl},(1)}\left(4,2\right)}{\delta G\left(7,6\right)}G\left(8,6\right)$$

$$+ i\hbar w\left(1,7\right)G\left(1,4\right)\frac{\delta\Sigma^{\mathrm{cl},(1)}\left(4,2\right)}{\delta G\left(5,8\right)}G\left(5,7\right)$$

$$= \pm\left(i\hbar\right)^{2}\delta\left(1,7\right)\delta\left(8,2\right)w\left(1,3\right)w\left(2,5\right)G\left(5,3\right)G\left(3,5\right)$$

$$+ \left(i\hbar\right)^{2}\delta\left(1,7\right)w\left(1,3\right)w\left(8,2\right)G\left(8,3\right)G\left(3,2\right)$$

$$\pm\left(i\hbar\right)^{2}w\left(1,8\right)G\left(1,2\right)w\left(2,7\right)G\left(8,7\right)$$

$$+ \left(i\hbar\right)^{2}w\left(1,8\right)G\left(1,7\right)w\left(7,2\right)G\left(8,2\right)$$

$$\pm\left(i\hbar\right)^{2}w\left(1,7\right)G\left(1,2\right)w\left(2,8\right)G\left(8,7\right)$$

$$+ \left(i\hbar\right)^{2}\delta\left(2,8\right)w\left(1,7\right)G\left(1,4\right)w\left(4,2\right)G\left(4,7\right)$$

$$\tag{324}$$

recovering Eqs. (220) and (225) from the derivation of the third-order selfenergy guided by Hedin's equations. Note that, for the first iteration, $\frac{\delta \Sigma^{\mathrm{cl}}(4,2)}{\delta G(5,6)\delta G(7,8)}$ is exactly equal to zero, thus Eq. (324) is also exact up to second order in $w$.

Considering Eq. (323), in the following, each of the three leading higher-order terms will be treated separately, starting off its particular diagrammatic series. Looking back at Eqs. (323) and (324), it is convenient to choose a common starting point for all series, leading to the same first- and second-order selfenergy contributions,

$$\Sigma^{\mathrm{cl},(1)}(1,2) = \Sigma^{\mathrm{H}}(1,2) + \Sigma^{\mathrm{F}}(1,2)$$

$$\Sigma^{\mathrm{cl},(2)}(1,2)$$

$$= \mathrm{i}\hbar w(1,3)G(1,4)\frac{\delta \Sigma^{\mathrm{cl},(1)}(4,2)}{\delta G(5,6)}$$

$$G(5,3)G(3,6)$$

$$= \pm(\mathrm{i}\hbar)^2 w(1,3)G(1,2)w(2,5)$$

$$G(5,3)G(3,5)$$

$$+ (\mathrm{i}\hbar)^2 w(1,3)G(1,4)w(4,2)$$

$$G(4,3)G(3,2)$$

(325)

(326)

which agree with the exact first and second-order terms, already encountered in Eqs. (158), (176) and (183).

The first **diagram series (A)** is generated by decoupling the first higher-order contribution of the derivation of the selfenergy in Eq. (323), leading to

$$\frac{\delta \Sigma^{\mathrm{cl},\mathrm{A}}(1,2)}{\delta G(7,8)}$$

$$= \pm \mathrm{i}\hbar \delta(1,2)\delta(7,8)w(1,7)$$

$$+ \mathrm{i}\hbar \delta(1,7)\delta(2,8)w(1,2)$$

$$+ \mathrm{i}\hbar \delta(1,7)w(1,3)\frac{\delta \Sigma^{\mathrm{cl},\mathrm{A}}(8,2)}{\delta G(5,6)}$$

$$G(5,3)G(3,6)$$

(327)

The corresponding third-order selfenergy terms of series A follow by insertion of the respective second-order derivative terms in Eq. (324) as

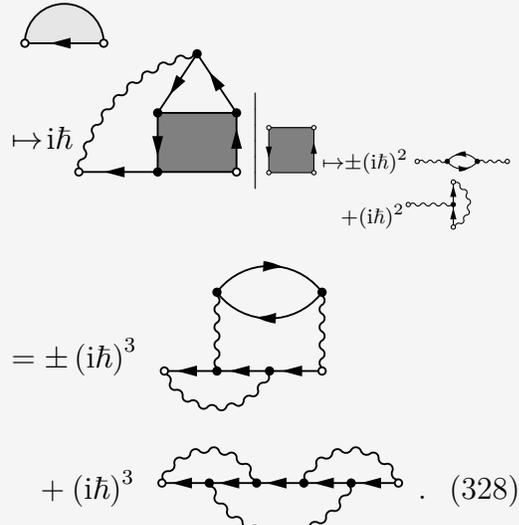

$$
\begin{aligned}
&\Sigma^{\mathrm{cl,A(3)}}\big(1,2\big)\\
&= \mathrm{i}\hbar w\big(1,3\big)G\big(1,4\big)\frac{\delta\Sigma^{\mathrm{cl,A(2)}}\big(4,2\big)}{\delta G\big(5,6\big)}\\
&\quad G\big(5,3\big)G\big(3,6\big)\\
&= \pm\big(\mathrm{i}\hbar\big)^{3}w\big(1,3\big)G\big(1,4\big)w\big(4,5\big)w\big(2,6\big)\\
&\quad G\big(6,5\big)G\big(5,6\big)G\big(4,3\big)G\big(3,2\big)\\
&\quad +\big(\mathrm{i}\hbar\big)^{3}w\big(1,3\big)G\big(1,4\big)w\big(4,5\big)w\big(6,2\big)\\
&\quad G\big(6,5\big)G\big(5,2\big)G\big(4,3\big)G\big(3,6\big)
\end{aligned}
\tag{328}
$$

At this point it is convenient to go back to the full notation and, particularly, the single-time interaction, to estimate the numerical effort of computationally solving the first diagrammatic series. By introducing the kernel $K$ in the following way,

$$
K^{\mathrm{A}}_{ijkl}\big(z_1,z_2,z_3,z_4\big):=\pm\frac{\delta\Sigma^{\mathrm{cl,A}}_{ik}\big(z_1,z_3\big)}{\delta G_{lj}\big(z_4,z_2\big)}\,,
\tag{329}
$$

we arrive at a closed equation for $K^{\mathrm{A}}_{ijkl}\big(z_1,z_2,z_3,z_4\big)$,

$$
\begin{aligned}
K^{\mathrm{A}}_{ijkl}\big(z_1,z_2,z_3,z_4\big)&=\pm\mathrm{i}\hbar\delta_{\mathcal{C}}\big(z_1,z_3\big)\delta_{\mathcal{C}}\big(z_1,z_4\big)\delta_{\mathcal{C}}\big(z_{1^+},z_2\big)w^{\pm}_{ijkl}\big(z_1\big)+\\
&\quad+\mathrm{i}\hbar\delta_{\mathcal{C}}\big(z_1,z_4\big)\sum_{pq}w_{ipql}\big(z_1\big)\times\\
&\quad\times\int_{\mathcal{C}}\mathrm{d}z_5\mathrm{d}z_6\sum_{rs}K^{\mathrm{A}}_{jskr}\big(z_2,z_6,z_3,z_5\big)G_{rp}\big(z_5,z_1\big)G_{qs}\big(z_1,z_6\big)\,,
\end{aligned}
\tag{330}
$$

where we introduced the (anti-)symmetrized interaction, $w^{\pm}_{ijkl}\big(z_1\big):=w_{ijkl}\big(z_1\big)\pm w_{jikl}\big(z_1\big)$. Iterating this equation, starting from

$$
K^{\mathrm{A,(1)}}_{ijkl}\big(z_1,z_2,z_3,z_4\big)=\pm\mathrm{i}\hbar\delta_{\mathcal{C}}\big(z_1,z_3\big)\delta_{\mathcal{C}}\big(z_1,z_4\big)\delta_{\mathcal{C}}\big(z_{1^+},z_2\big)w^{\pm}_{ijkl}\big(z_1\big)\,,
$$

we have, for the second iteration,

$$K_{ijkl}^{\mathrm{A},(2)}\big(z_1, z_2, z_3, z_4\big) = \mathrm{i}\hbar\delta_{\mathcal{C}}\big(z_1, z_4\big) \sum_{pq} w_{ipql}\big(z_1\big) \times$$

$$\times \int_{\mathcal{C}} \mathrm{d}z_5 \mathrm{d}z_6 \sum_{rs} K_{jskr}^{\mathrm{A},(1)}\big(z_2, z_6, z_3, z_5\big) G_{rp}\big(z_5, z_1\big) G_{qs}\big(z_1, z_6\big)$$

$$= \pm\big(\mathrm{i}\hbar\big)^2 \delta_{\mathcal{C}}\big(z_1, z_4\big) \delta_{\mathcal{C}}\big(z_2, z_3\big) \sum_{pq} w_{ipql}\big(z_1\big) \sum_{rs} w_{jskr}^{\pm}\big(z_2\big) G_{rp}\big(z_2, z_1\big) G_{qs}\big(z_1, z_2\big),$$

Similarly, we obtain the third and fourth iterations,

$$K_{ijkl}^{\mathrm{A},(3)}\big(z_1, z_2, z_3, z_4\big) = \mathrm{i}\hbar\delta_{\mathcal{C}}\big(z_1, z_4\big) \sum_{pq} w_{ipql}\big(z_1\big) \times$$

$$\times \int_{\mathcal{C}} \mathrm{d}z_5 \mathrm{d}z_6 \sum_{rs} K_{jskr}^{\mathrm{A},(2)}\big(z_2, z_6, z_3, z_5\big) G_{rp}\big(z_5, z_1\big) G_{qs}\big(z_1, z_6\big)$$

$$= \pm\mathrm{i}\hbar\delta_{\mathcal{C}}\big(z_1, z_4\big) \sum_{pq} w_{ipql}\big(z_1\big) \sum_{rs} K_{jskr}^{\mathrm{A},(2)}\big(z_2, z_3, z_3, z_2\big) G_{rp}\big(z_2, z_1\big) G_{qs}\big(z_1, z_3\big),$$

$$K_{ijkl}^{\mathrm{A},(4)}\big(z_1, z_2, z_3, z_4\big) = \mathrm{i}\hbar\delta_{\mathcal{C}}\big(z_1, z_4\big) \sum_{pq} w_{ipql}\big(z_1\big) \times$$

$$\int_{\mathcal{C}} \mathrm{d}z_5 \mathrm{d}z_6 \sum_{rs} K_{jskr}^{\mathrm{A},(3)}\big(z_2, z_6, z_3, z_5\big) G_{rp}\big(z_5, z_1\big) G_{qs}\big(z_1, z_6\big)$$

$$= \mathrm{i}\hbar\delta_{\mathcal{C}}\big(z_1, z_4\big) \sum_{pq} w_{ipql}\big(z_1\big) \int_{\mathcal{C}} \mathrm{d}z_6 \sum_{rs} K_{jskr}^{\mathrm{A},(3)}\big(z_2, z_6, z_3, z_2\big) G_{rp}\big(z_2, z_1\big) G_{qs}\big(z_1, z_6\big),$$

and a similar structure for the higher orders. It is noteworthy that the computation of the fourth- and higher-order iterations are of complexity $\mathcal{O}\big(N_{\mathrm{t}}^4\big)$, since, due to the appearance of the delta function, $\delta_{\mathcal{C}}\big(z_1, z_4\big)$,

$$K_{ijkl}^{\mathrm{A},(N)}\big(z_1, z_2, z_3, z_4\big) = \delta_{\mathcal{C}}\big(z_1, z_4\big) K_{ijkl}^{\mathrm{A},(N)}\big(z_1, z_2, z_3, z_1\big), \tag{331}$$

and the right-hand side contains one integral over an intermediate time, which typically limits the applicability of this approximation to very short time scales. Therefore, series A is usually omitted in efficient quantum-many-body frameworks.

**The second approximation (B)** is generated by the second higher-order contribution of $\frac{\delta\Sigma^{\mathrm{cl}}}{\delta G}$ in Eq. (323), leading to

$$\frac{\delta \Sigma^{\mathrm{cl,B}}(1,2)}{\delta G(7,8)}$$

$$= \pm i\hbar \delta(1,2)\delta(7,8) w(1,7)$$

$$+ i\hbar \delta(1,7)\delta(2,8) w(1,2)$$

$$+ i\hbar w(1,8) G(1,4) \frac{\delta \Sigma^{\mathrm{cl,B}}(4,2)}{\delta G(7,6)} G(8,6)$$

$$\mapsto \pm i\hbar \qquad + i\hbar \qquad \qquad + i\hbar \qquad \qquad . \tag{332}$$

Using the respective second-order derivative terms in Eq. (324) one obtains the third-order selfenergy contributions of series B,

$$\Sigma^{\mathrm{cl,B(3)}}(1,2)$$

$$= i\hbar w(1,3) G(1,4) \frac{\delta \Sigma^{\mathrm{cl,B(2)}}(4,2)}{\delta G(5,6)}$$

$$\quad G(5,3) G(3,6)$$

$$= \pm (i\hbar)^3 w(1,3) G(1,4) w(4,6) G(4,2)$$

$$\quad w(2,5) G(6,5) G(5,3) G(3,6)$$

$$+ (i\hbar)^3 w(1,3) G(1,4) w(4,6) G(4,5)$$

$$\quad w(5,2) G(6,2) G(5,3) G(3,6)$$

$$\mapsto i\hbar \qquad \qquad \mapsto \pm (i\hbar)^2$$

$$\qquad \qquad \qquad \qquad + (i\hbar)^2$$

$$= \pm (i\hbar)^3$$

$$+ (i\hbar)^3 \qquad \qquad . \tag{333}$$

In the full notation with the single-time interaction we again introduce a kernel $K$ for the series B:

$$K^{\mathrm{B}}_{ijkl}(z_1, z_2, z_3, z_4) := \pm \frac{\delta \Sigma^{\mathrm{cl,B}}_{ik}(z_1, z_3)}{\delta G_{lj}(z_4, z_2)}, \tag{334}$$

which obeys the equation of motion

$$K^{\mathrm{B}}_{ijkl}(z_1, z_2, z_3, z_4) = \pm i\hbar \delta_{\mathcal{C}}(z_1, z_3) \delta_{\mathcal{C}}(z_1, z_4) \delta_{\mathcal{C}}(z_{1^+}, z_2) w^{\pm}_{ijkl}(z_1) +$$

$$+ i\hbar \delta_{\mathcal{C}}(z_1, z_2) \sum_{mq} w_{ijqm}(z_1) \int_{\mathcal{C}} \mathrm{d}z_5 \sum_n G_{mn}(z_1, z_5) \times$$

$$\int_{\mathcal{C}} \mathrm{d}z_6 \sum_s K^{\mathrm{B}}_{nskl}(z_5, z_6, z_3, z_4) G_{qs}(z_1, z_6).$$

Iterating, starting again from the first iteration,

$$K_{ijkl}^{\mathrm{B},(1)}\big(z_1,z_2,z_3,z_4\big) = \pm\mathrm{i}\hbar\delta_{\mathcal{C}}\big(z_1,z_3\big)\delta_{\mathcal{C}}\big(z_1,z_4\big)\delta_{\mathcal{C}}\big(z_{1^+},z_2\big)w_{ijkl}^{\pm}\big(z_1\big),$$

we find, for the second iteration,

$$
\begin{aligned}
K_{ijkl}^{\mathrm{B},(2)}\big(z_1,z_2,z_3,z_4\big) &= \mathrm{i}\hbar\delta_{\mathcal{C}}\big(z_1,z_2\big)\sum_{mq}w_{ijqm}\big(z_1\big)\int_{\mathcal{C}}\mathrm{d}z_5\sum_n G_{mn}\big(z_1,z_5\big)\times\\
&\quad\times\int_{\mathcal{C}}\mathrm{d}z_6\sum_s K_{nskl}^{\mathrm{B},(1)}\big(z_5,z_6,z_3,z_4\big)G_{qs}\big(z_1,z_6\big) =\\
&= \mathrm{i}\hbar\delta_{\mathcal{C}}\big(z_1,z_2\big)\delta_{\mathcal{C}}\big(z_3,z_4\big)\sum_{mq}w_{ijqm}\big(z_1\big)\sum_n G_{mn}\big(z_1,z_3\big)\sum_s K_{nskl}^{\mathrm{B},(1)}\big(z_3,z_3,z_3,z_3\big)G_{qs}\big(z_1,z_3\big)
\end{aligned}
\tag{335}
$$

and, for the third iteration,

$$
\begin{aligned}
K_{ijkl}^{\mathrm{B},(3)}\big(z_1,z_2,z_3,z_4\big) &= \mathrm{i}\hbar\delta_{\mathcal{C}}\big(z_1,z_2\big)\sum_{mq}w_{ijqm}\big(z_1\big)\int_{\mathcal{C}}\mathrm{d}z_5\sum_n G_{mn}\big(z_1,z_5\big)\times\\
&\quad\int_{\mathcal{C}}\mathrm{d}z_6\sum_s K_{nskl}^{\mathrm{B},(2)}\big(z_5,z_6,z_3,z_4\big)G_{qs}\big(z_1,z_6\big) =\\
&= \mathrm{i}\hbar\delta_{\mathcal{C}}\big(z_1,z_2\big)\delta_{\mathcal{C}}\big(z_3,z_4\big)\sum_{mq}w_{ijqm}\big(z_1\big)\int_{\mathcal{C}}\mathrm{d}z_5\sum_n G_{mn}\big(z_1,z_5\big)\times\\
&\quad\sum_s K_{nskl}^{\mathrm{B},(2)}\big(z_5,z_5,z_3,z_3\big)G_{qs}\big(z_1,z_5\big).
\end{aligned}
\tag{336}
$$

In similar manner higher orders are derived. The structure of this approximation is such that the computation of all terms scales as $\mathcal{O}\big(N_{\mathrm{t}}^3\big)$ since, due to the two delta functions, we have, for each iteration order $(N)$,

$$K_{ijkl}^{\mathrm{B},(N)}\big(z_1,z_2,z_3,z_4\big) = \delta_{\mathcal{C}}\big(z_1,z_2\big)\delta_{\mathcal{C}}\big(z_3,z_4\big)K_{ijkl}^{\mathrm{B},(N)}\big(z_1,z_2\big),$$

and one integration on the right-hand side. With this, one finds [cf. Eq. (336)],

$$
\begin{aligned}
K_{ijkl}^{\mathrm{B}}\big(z_1,z_3\big) &= \delta_{\mathcal{C}}\big(z_1,z_3\big)K_{ijkl}^{\mathrm{B},(1)}\big(z_1,z_1\big) +\\
&\quad+ \sum_{mq}w_{ijqm}\big(z_1\big)\int_{\mathcal{C}}\mathrm{d}z_5\sum_{ns}G_{mqns}^{\mathrm{H}}\big(z_1,z_5\big)K_{nskl}^{\mathrm{B}}\big(z_5,z_3\big),
\end{aligned}
\tag{337}
$$

where Eq. (280) has been used, in the last line. To simplify the following expressions further, it is useful to eliminate the contributions with the delta function by introducing the non-singular kernel (superscript "ns")[xxiv] [cf. Eqs. (337) and (335)], according to

$$
\begin{aligned}
K_{ijkl}^{\mathrm{B,ns}}\big(z_1,z_2\big) &:= K_{ijkl}^{\mathrm{B}}\big(z_1,z_2\big) - \delta_{\mathcal{C}}\big(z_1,z_2\big)K_{ijkl}^{\mathrm{B},(1)}\big(z_1,z_1\big) =\\
&= K_{ijkl}^{\mathrm{B}}\big(z_1,z_2\big) \mp \mathrm{i}\hbar\delta_{\mathcal{C}}\big(z_1,z_2\big)w_{ijkl}^{\pm}\big(z_1\big).
\end{aligned}
\tag{338}
$$

[xxiv]this follows the procedure applied for the GWA, Sec. 5.2

It obeys its own equation where no singular terms appear anymore [cf. Eq. (337)],

$$
\begin{aligned}
K_{ijkl}^{\mathrm{B,ns}}\big(z_1,z_2\big) &= \sum_{mq} w_{ijqm}\big(z_1\big) \int_{\mathcal{C}} \mathrm{d}z_5 \sum_{ns} G_{mqns}^{\mathrm{H}}\big(z_1,z_5\big) \times \\
&\quad \times \left\{ K_{nskl}^{\mathrm{B,ns}}\big(z_5,z_2\big) \pm \mathrm{i}\hbar \delta_{\mathcal{C}}\big(z_5,z_2\big) w_{nskl}^{\pm}\big(z_2\big) \right\} \\
&= \pm \mathrm{i}\hbar \sum_{mq} w_{ijqm}\big(z_1\big) \sum_{ns} G_{mqns}^{\mathrm{H}}\big(z_1,z_2\big) w_{nskl}^{\pm}\big(z_2\big) \\
&\quad + \sum_{mq} w_{ijqm}\big(z_1\big) \int_{\mathcal{C}} \mathrm{d}z_5 \sum_{ns} G_{mqns}^{\mathrm{H}}\big(z_1,z_5\big) K_{nskl}^{\mathrm{B,ns}}\big(z_5,z_2\big).
\end{aligned}
\tag{339}
$$

The corresponding selfenergy reads, [cf. Eq. (322)],

$$
\Sigma_{ij}^{\mathrm{cl,B}}\big(z_1,z_2\big) = \Sigma_{ij}^{\mathrm{H}}\big(z_1,z_2\big) + \Sigma_{ij}^{\mathrm{F}}\big(z_1,z_2\big) + \Sigma_{ij}^{\mathrm{cl,corr,B}}\big(z_1,z_2\big)
\tag{340}
$$

$$
\Sigma_{ij}^{\mathrm{cl,corr,B}}\big(z_1,z_2\big) =: \pm \mathrm{i}\hbar \sum_{mpq} w_{ipqm}\big(z_1\big) \int_{\mathcal{C}} \mathrm{d}z_3 \sum_n G_{mn}\big(z_1,z_3\big) \times
\tag{341}
$$

$$
\int_{\mathcal{C}} \mathrm{d}z_4 \mathrm{d}z_5 \sum_{rs} K_{nsjr}^{\mathrm{B}}\big(z_3,z_5,z_2,z_4\big) G_{rp}\big(z_4,z_1\big) G_{qs}\big(z_1,z_5\big),
$$

where we defined the correlation part of the selfenergy via the additional superscript "corr". Inserting the expression for $K^{\mathrm{B}}$, we find

$$
\begin{aligned}
\Sigma_{ij}^{\mathrm{cl,corr,B}}\big(z_1,z_2\big) &= \pm \mathrm{i}\hbar \sum_{mpq} w_{ipqm}\big(z_1\big) \int_{\mathcal{C}} \mathrm{d}z_3 \sum_n G_{mn}\big(z_1,z_3\big) \times \\
&\quad \times \sum_{rs} G_{qs}\big(z_1,z_3\big) K_{nsjr}^{\mathrm{B}}\big(z_3,z_2\big) G_{rp}\big(z_2,z_1\big),
\end{aligned}
$$

which, by insertion of Eq. (338), transforms to

$$
\begin{aligned}
\Sigma_{ij}^{\mathrm{cl,corr,B}}\big(z_1,z_2\big) &= \pm \sum_{mpq} w_{ipqm}\big(z_1\big) \int_{\mathcal{C}} \mathrm{d}z_3 \sum_{nrs} G_{mqns}^{\mathrm{H}}\big(z_1,z_3\big) \times \\
&\quad \times \left\{ K_{nsjr}^{\mathrm{B,ns}}\big(z_3,z_2\big) \pm \mathrm{i}\hbar \delta_{\mathcal{C}}\big(z_3,z_2\big) w_{nsjr}^{\pm}\big(z_2\big) \right\} G_{rp}\big(z_2,z_1\big).
\end{aligned}
$$

After restructuring, one has

$$
\begin{aligned}
\Sigma_{ij}^{\mathrm{cl,corr,B}}\big(z_1,z_2\big) &= \pm \bigg( \pm \mathrm{i}\hbar \sum_{mpq} w_{ipqm}\big(z_1\big) \sum_{nrs} G_{mqns}^{\mathrm{H}}\big(z_1,z_3\big) w_{nsjr}^{\pm}\big(z_2\big) + \\
&\quad + \sum_{mpq} w_{ipqm}\big(z_1\big) \int_{\mathcal{C}} \mathrm{d}z_3 \sum_{nrs} G_{mqns}^{\mathrm{H}}\big(z_1,z_3\big) K_{nsjr}^{\mathrm{B,ns}}\big(z_3,z_2\big) \bigg) G_{rp}\big(z_2,z_1\big).
\end{aligned}
$$

Looking at Eq. (339), it is obvious that the right-hand side already contains the first iteration of the recursion equation, so we simply have

$$
\Sigma_{ij}^{\mathrm{cl,corr,B}}\big(z_1,z_2\big) = \pm \sum_{pr} K_{ipjr}^{\mathrm{B,ns}}\big(z_1,z_2\big) G_{rp}\big(z_2,z_1\big).
\tag{342}
$$

With the definition

$$i\hbar T_{ijkl}^{\mathrm{pp}}\big(z_1, z_2\big) := \pm K_{ijkl}^{\mathrm{B,ns}}\big(z_1, z_2\big),\tag{343}$$

the coupled solution of, Eqs. (339) and (342) becomes,

$$T_{ijkl}^{\mathrm{pp}}\big(z_1, z_2\big) = \Phi_{ijkl}^{\mathrm{pp}}\big(z_1, z_2\big) + \sum_{mq} w_{ijqm}\big(z_1\big) \sum_{ns} \int_{\mathcal{C}} \mathrm{d}z_5 \, G_{mqns}^{\mathrm{H}}\big(z_1, z_5\big) T_{nskl}^{\mathrm{pp}}\big(z_5, z_2\big),\tag{344}$$

$$\Phi_{ijkl}^{\mathrm{pp}}\big(z_1, z_2\big) =: \sum_{mq} w_{ijqm}\big(z_1\big) \sum_{ns} G_{mqns}^{\mathrm{H}}\big(z_1, z_2\big) w_{nskl}^{\pm}\big(z_2\big),\tag{345}$$

where we introduced an abbreviation for the first term that contains the second-Born approximation [as we did before for the $GW$ approximation].

Thus, according to Eq. (342), the correlation part of the selfenergy in the so-called (particle–particle) $T$-matrix approximation is given by

$$\Sigma_{ij}^{\mathrm{pp}}\big(z_1, z_2\big) := \Sigma_{ij}^{T^{\mathrm{pp}},\mathrm{corr}}\big(z_1, z_2\big) = i\hbar \sum_{pr} T_{ipjr}^{\mathrm{pp}}\big(z_1, z_2\big) G_{rp}\big(z_2, z_1\big).\tag{346}$$

The *T matrix* relates to a similar quantity in scattering theory, which is called *transfer matrix* there. It describes an interacting scattering state of a system selfconsistently in terms of a free state of two particles which undergo multiple (in general, infinitely many) scattering events with each other, which can be resummed into the transfer matrix acting on the two particles. Looking at Eqs. (344) and (346), the same interpretation is possible, since $G^{\mathrm{H}}$ describes a particle pair, and their scattering is governed by $T^{\mathrm{pp}}$.

The Keldysh correlation components of the particle–particle $T$ matrix read

$$T_{ijkl}^{\mathrm{pp},\gtrless}\big(t_1, t_2\big) = \Phi_{ijkl}^{\mathrm{pp},\gtrless}\big(t_1, t_2\big) + \sum_{mq} w_{ijqm}\big(t_1\big) \sum_{ns} \times\tag{347}$$

$$\left( \int_{t_0}^{t_1} \mathrm{d}t_5 \, G_{mqns}^{\mathrm{H},\mathcal{R}}\big(t_1, t_5\big) T_{nskl}^{\mathrm{pp},\gtrless}\big(t_5, t_2\big) + \int_{t_0}^{t_2} \mathrm{d}t_5 \, G_{mqns}^{\mathrm{H},\gtrless}\big(t_1, t_5\big) T_{nskl}^{\mathrm{pp},\mathcal{A}}\big(t_5, t_2\big) \right),$$

where $\Phi^{\mathrm{pp},\gtrless}$ is the $\gtrless$-component of $\Phi^{\mathrm{pp}}$ that is obtained from Eq. (345) by replacing $G^{\mathrm{H}} \to G^{\mathrm{H},\gtrless}$. Analogously, we have for the advanced/retarded Keldysh component and

$$T_{ijkl}^{\mathrm{pp},\mathcal{A}/\mathcal{R}}\big(t_1, t_2\big) = \Phi_{ijkl}^{\mathrm{pp},\mathcal{A}/\mathcal{R}}\big(t_1, t_2\big) + \sum_{mq} w_{ijqm}\big(t_1\big) \sum_{ns} \int_{t_{1/2}}^{t_{2/1}} \mathrm{d}t_5 \, G_{mqns}^{\mathrm{H},\mathcal{A}/\mathcal{R}}\big(t_1, t_5\big) T_{nskl}^{\mathrm{pp},\mathcal{A}/\mathcal{R}}\big(t_5, t_2\big),$$

where $\Phi^{\mathrm{pp},\mathcal{A}/\mathcal{R}}$ again follows from $\Phi^{\mathrm{pp}}$ by replacing $G^{\mathrm{H}} \to G^{\mathrm{H},\mathcal{A}/\mathcal{R}}$, and in the notation $t_{1/2}$ the first (second) subscript refers to the advanced (retarded) function. For the

**Figure 30.** Leading terms of the particle–particle $T$-matrix selfenergy, $\Sigma^{\mathrm{pp,diag}}$ (including first-order terms), in a diagonal basis.

correlation components of the particle–particle $T$-matrix selfenergy (346), we have

$$\Sigma_{ij}^{\mathrm{pp},\gtrless}\big(t_1,t_2\big) = \mathrm{i}\hbar \sum_{pr} T_{ipjr}^{\mathrm{pp},\gtrless}\big(t_1,t_2\big) G_{rp}^{\lessgtr}\big(t_2,t_1\big) \, .$$

***Particle–particle $T$ matrix in a diagonal basis.*** In a basis with diagonal interaction, $w_{ijkl} = \delta_{il}\delta_{jk}w_{ij}$, Eq. (344) attains the form

$$T_{ijkl}^{\mathrm{pp,diag}}\big(z_1,z_2\big) =: \Phi_{ijkl}^{\mathrm{pp,diag}}\big(z_1,z_2\big) + \tag{348}$$
$$+ \, w_{ij}\big(z_1\big) \sum_{ns} \int_{\mathcal{C}} \mathrm{d}z_5 \, G_{ijns}^{\mathrm{H}}\big(z_1,z_5\big) T_{nskl}^{\mathrm{pp,diag}}\big(z_5,z_2\big) \, ,$$

where the diagonal basis version of Eq. (345) is

$$\Phi_{ijkl}^{\mathrm{pp,diag}}\big(z_1,z_2\big) = w_{ij}\big(z_1\big) \left\{ G_{ijlk}^{\mathrm{H}}\big(z_1,z_2\big) \pm G_{ijkl}^{\mathrm{H}}\big(z_1,z_2\big) \right\} w_{lk}\big(z_2\big) \, . \tag{349}$$

The correlation part of the selfenergy, corresponding to Eq. (346), reads

$$\Sigma_{ij}^{\mathrm{pp,diag}}\big(z_1,z_2\big) = \mathrm{i}\hbar \sum_{pr} T_{ipjr}^{\mathrm{pp,diag}}\big(z_1,z_2\big) G_{rp}\big(z_2,z_1\big) \, . \tag{350}$$

The leading contributions to the selfenergy for a diagonal basis are shown in Fig. 30. The respective Keldysh matrix components are

$$T^{\mathrm{pp,diag},\gtrless}_{ijkl}\big(t_1,t_2\big) = \Phi^{\mathrm{pp,diag},\gtrless}_{ijkl}\big(t_1,t_2\big) + \tag{351}$$

$$+ \, w_{ij}\big(t_1\big)\bigg(\int_{t_0}^{t_1}\mathrm{d}t_5 \sum_{ns} G^{\mathrm{H},\mathcal{R}}_{ijns}\big(t_1,t_5\big)T^{\mathrm{pp,diag},\gtrless}_{nskl}\big(t_5,t_2\big) +$$

$$+ \int_{t_0}^{t_2}\mathrm{d}t_5 \, G^{\mathrm{H},\gtrless}_{ijns}\big(t_1,t_5\big)T^{\mathrm{pp,diag},\mathcal{A}}_{nskl}\big(t_5,t_2\big)\bigg),$$

$$T^{\mathrm{pp,diag},\mathcal{A}/\mathcal{R}}_{ijkl}\big(t_1,t_2\big) = \Phi^{\mathrm{pp,diag},\mathcal{A}/\mathcal{R}}_{ijkl}\big(t_1,t_2\big) + \tag{352}$$

$$+ \, w_{ij}\big(t_1\big)\sum_{ns}\int_{t_{1/2}}^{t_{2/1}}\mathrm{d}t_5 \, G^{\mathrm{H},\mathcal{A}/\mathcal{R}}_{ijns}\big(t_1,t_5\big)T^{\mathrm{pp,diag},\mathcal{A}/\mathcal{R}}_{nskl}\big(t_5,t_2\big)$$

and, finally, the diagonal version of the correlation selfenergy (346) becomes

$$\Sigma^{\mathrm{pp,diag},\gtrless}_{ij}\big(t_1,t_2\big) = \sum_{pr} T^{\mathrm{pp,diag},\gtrless}_{ipjr}\big(t_1,t_2\big)G^{\lessgtr}_{rp}\big(t_2,t_1\big)\,.$$

***Particle–particle $T$ matrix in the Hubbard basis.*** In the Hubbard basis the expressions for the $T$ matrix simplify further. We start from the case of bosons. In the **bosonic Hubbard basis**, the particle–particle $T$ matrix reads

$$T^{\mathrm{pp},\mathfrak{b}}_{i\alpha j\beta k\alpha l\beta}\big(z_1,z_2\big) = \delta_{ij}\delta_{kl}\Phi^{\mathrm{pp},\mathfrak{b}}_{i\alpha i\beta k\alpha k\beta}\big(z_1,z_2\big) + \tag{353}$$

$$+ \, \delta_{ij}U\big(z_1\big)\sum_{ns}\int_{\mathcal{C}}\mathrm{d}z_5 \, G^{\mathrm{H}}_{i\alpha i\beta n\alpha s\beta}\big(z_1,z_5\big)T^{\mathrm{pp},\mathfrak{b}}_{n\alpha s\beta k\alpha l\beta}\big(z_5,z_2\big)\,,$$

where the bosonic Hubbard version of the function $\Phi^{\mathrm{pp}}$ is given by

$$\Phi^{\mathrm{pp},\mathfrak{b}}_{i\alpha i\beta k\alpha k\beta}\big(z_1,z_2\big) =: 2U\big(z_1\big)G^{\mathrm{H}}_{i\alpha i\beta k\alpha k\beta}\big(z_1,z_2\big)U\big(z_2\big)\,. \tag{354}$$

For the correlation part of the selfenergy, we now have

$$\Sigma^{\mathrm{pp},\mathfrak{b}}_{i\alpha j\alpha}\big(z_1,z_2\big) = \mathrm{i}\hbar \sum_{pr}\sum_{\epsilon} T^{\mathrm{pp},\mathfrak{b}}_{i\alpha p\epsilon j\alpha r\epsilon}\big(z_1,z_2\big)G_{r\epsilon p\epsilon}\big(z_2,z_1\big)\,.$$

From these equations it is evident that the equality

$$\Phi^{\mathrm{pp},\mathfrak{b}}_{i\alpha j\beta k\alpha l\beta}\big(z_1,z_2\big) = \delta_{ij}\delta_{kl}\Phi^{\mathrm{pp},\mathfrak{b}}_{i\alpha i\beta k\alpha k\beta}\big(z_1,z_2\big) =: \delta_{ij}\delta_{kl}\Phi^{\mathrm{pp},\mathfrak{b}}_{ik\alpha\beta}\big(z_1,z_2\big)\,, \tag{355}$$

implies $T^{\mathrm{pp},\mathfrak{b}}_{i\alpha j\beta k\alpha l\beta}\big(z_1,z_2\big) =: \delta_{ij}\delta_{kl}T^{\mathrm{pp},\mathfrak{b}}_{ik\alpha\beta}\big(z_1,z_2\big)\,,$ which is verified by iteration. Regarding the notation, one has to bear in mind that $T^{\mathrm{pp},\mathfrak{b}}_{ik\alpha\beta}$ and $G^{\mathrm{H}}_{ik\alpha\beta} := G^{\mathrm{H}}_{i\alpha i\beta k\alpha k\beta}$ are quantities of all four spin-space orbitals $\big|i\alpha\big\rangle, \big|i\beta\big\rangle, \big|k\alpha\big\rangle, \big|k\beta\big\rangle$. With this, the equations become

$$T^{\mathrm{pp},\mathfrak{b}}_{ik\alpha\beta}\big(z_1,z_2\big) = \Phi^{\mathrm{pp},\mathfrak{b}}_{ik\alpha\beta}\big(z_1,z_2\big) + U\big(z_1\big)\sum_{n}\int_{\mathcal{C}}\mathrm{d}z_5 \, G^{\mathrm{H}}_{in\alpha\beta}\big(z_1,z_5\big)T^{\mathrm{pp},\mathfrak{b}}_{nk\alpha\beta}\big(z_5,z_2\big)\,,$$

and the correlation selfenergy is

$$\Sigma_{i\alpha j\alpha}^{\mathrm{pp},\mathfrak{b}}\left(z_1,z_2\right) = \mathrm{i}\hbar \sum_\epsilon T_{ij\alpha\epsilon}^{\mathrm{pp},\mathfrak{b}}\left(z_1,z_2\right) G_{j\epsilon i\epsilon}\left(z_2,z_1\right).$$

The Keldysh greater/less matrix components of the selfenergy and the $T$ matrix read

$$\Sigma_{i\alpha j\alpha}^{\mathrm{pp},\mathfrak{b},\gtrless}\left(t_1,t_2\right) = \mathrm{i}\hbar \sum_\epsilon T_{ij\alpha\epsilon}^{\mathrm{pp},\mathfrak{b},\gtrless}\left(t_1,t_2\right) G_{j\epsilon i\epsilon}^{\lessgtr}\left(t_2,t_1\right),$$

$$T_{ik\alpha\beta}^{\mathrm{pp},\mathfrak{b},\gtrless}\left(t_1,t_2\right) = \Phi_{ik\alpha\beta}^{\mathrm{pp},\mathfrak{b},\gtrless}\left(t_1,t_2\right) + U\left(t_1\right)\sum_n \tag{356}$$

$$\left(\int_{t_0}^{t_1} \mathrm{d}t_5 \, G_{in\alpha\beta}^{\mathrm{H},\mathcal{R}}\left(t_1,t_5\right) T_{nk\alpha\beta}^{\mathrm{pp},\gtrless}\left(t_5,t_2\right) + \int_{t_0}^{t_2} \mathrm{d}t_5 \, G_{in\alpha\beta}^{\mathrm{H},\gtrless}\left(t_1,t_5\right) T_{nk\alpha\beta}^{\mathrm{pp},\mathcal{A}}\left(t_5,t_2\right)\right),$$

whereas the advanced/retarded component is given by

$$T_{ik\alpha\beta}^{\mathrm{pp},\mathfrak{b},\mathcal{A}/\mathcal{R}}\left(t_1,t_2\right) = \Phi_{ik\alpha\beta}^{\mathrm{pp},\mathfrak{b},\mathcal{A}/\mathcal{R}}\left(t_1,t_2\right) + \tag{357}$$

$$+ U\left(t_1\right)\sum_n \int_{t_{1/2}}^{t_{1/2}} \mathrm{d}t_5 \, G_{in\alpha\beta}^{\mathrm{H},\mathcal{A}/\mathcal{R}}\left(t_1,t_5\right) T_{nk\alpha\beta}^{\mathrm{pp},\mathcal{A}/\mathcal{R}}\left(t_5,t_2\right).$$

In the special case of **spin-0 bosons**, there is no spin index, and we have

$$T_{ik}^{\mathrm{pp},\mathfrak{b},0}\left(z_1,z_2\right) = \Phi_{ik}^{\mathrm{pp},\mathfrak{b},0}\left(z_1,z_2\right) + U\left(z_1\right)\sum_n \int_\mathcal{C} \mathrm{d}z_5 \, G_{in}^{\mathrm{H}}\left(z_1,z_5\right) T_{nk}^{\mathrm{pp},\mathfrak{b},0}\left(z_5,z_2\right),$$

and the correlation part of the selfenergy reduces to

$$\Sigma_{ij}^{\mathrm{pp},\mathfrak{b},0}\left(z_1,z_2\right) = \mathrm{i}\hbar T_{ij}^{\mathrm{pp},\mathfrak{b},0}\left(z_1,z_2\right) G_{ji}\left(z_2,z_1\right). \tag{358}$$

The greater/less components of the correlation selfenergy are

$$\Sigma_{ij}^{\mathrm{pp},\mathfrak{b},0,\gtrless}\left(t_1,t_2\right) = \mathrm{i}\hbar T_{ij}^{\mathrm{pp},\mathfrak{b},0,\gtrless}\left(t_1,t_2\right) G_{ji}^{\lessgtr}\left(t_2,t_1\right), \tag{359}$$

with the Keldysh components of the $T$ matrix given by

$$T_{ik}^{\mathrm{pp},\mathfrak{b},0,\gtrless}\left(t_1,t_2\right) = \Phi_{ik}^{\mathrm{pp},\mathfrak{b},0,\gtrless}\left(t_1,t_2\right) + U\left(t_1\right)\sum_n \tag{360}$$

$$\times \left(\int_{t_0}^{t_1} \mathrm{d}t_5 \, G_{in}^{\mathrm{H},\mathcal{R}}\left(t_1,t_5\right) T_{nk}^{\mathrm{pp},\mathfrak{b},0,\gtrless}\left(t_5,t_2\right) + \int_{t_0}^{t_2} \mathrm{d}t_5 \, G_{in}^{\mathrm{H},\gtrless}\left(t_1,t_5\right) T_{nk}^{\mathrm{pp},\mathfrak{b},0,\mathcal{A}}\left(t_5,t_2\right)\right),$$

$$T_{ik}^{\mathrm{pp},\mathfrak{b},0,\mathcal{A}/\mathcal{R}}\left(t_1,t_2\right) = \Phi_{ik}^{\mathrm{pp},\mathfrak{b},0,\mathcal{A}/\mathcal{R}}\left(t_1,t_2\right) + \tag{361}$$

$$+ U\left(t_1\right)\sum_n \int_{t_{1/2}}^{t_{2/1}} \mathrm{d}t_5 \, G_{in}^{\mathrm{H},\mathcal{A}/\mathcal{R}}\left(t_1,t_5\right) T_{nk}^{\mathrm{pp},\mathfrak{b},0,\mathcal{A}/\mathcal{R}}\left(t_5,t_2\right).$$

The corresponding diagrams are shown in Fig. 31.

**Fermions.** Let us now consider the results for the $T$-matrix approximation for the

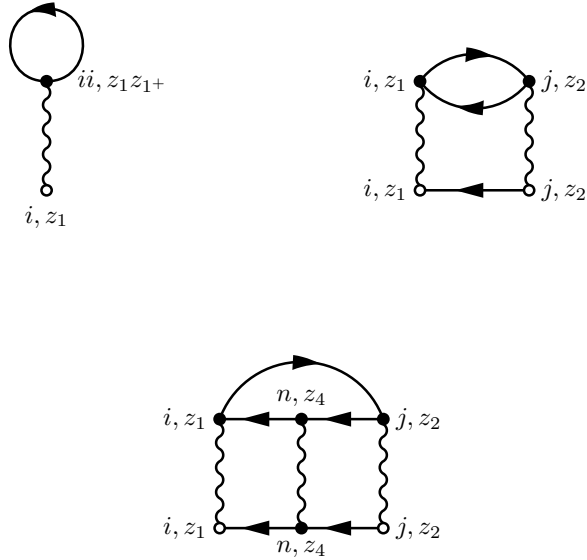

**Figure 31.** Leading terms of $\Sigma^{\mathrm{pp,b},0}$ (including first-order terms) in the Hubbard basis for spin-0 bosons. Each term carries a factor of two.

fermionic Hubbard model. The correlation part of the selfenergy[xxv] is

$$\Sigma^{\mathrm{pp,f}}_{i\alpha j\alpha}(z_1, z_2) = ih \sum_{pr} \sum_{\epsilon} T^{\mathrm{pp,f}}_{i\alpha pe j\alpha r\epsilon}(z_1, z_2) G_{r\epsilon p\epsilon}(z_2, z_1),$$

and the equation of motion for the $T$ matrix reads

$$
\begin{aligned}
T^{\mathrm{pp,f}}_{i\alpha j\beta k\alpha l\beta}(z_1, z_2) = &\, U(z_1)\delta_{ij}\delta_{kl}\bar{\delta}_{\alpha\beta} \times \\
&\left\{ G^{\mathrm{H}}_{i\alpha i\beta k\beta k\alpha}(z_1, z_2)\delta_{\alpha\beta} - G^{\mathrm{H}}_{i\alpha i\beta k\alpha k\beta}(z_1, z_2)\delta_{\alpha\alpha}\delta_{\beta\beta} \right\} U(z_2) + \\
&+ U(z_1)\bar{\delta}_{\alpha\beta}\delta_{ij} \sum_{ns} \int_{\mathcal{C}} \mathrm{d}z_5\, G^{\mathrm{H}}_{i\alpha i\beta n\alpha s\beta}(z_1, z_5)T^{\mathrm{pp,f}}_{n\alpha s\beta k\alpha l\beta}(z_5, z_2).
\end{aligned}
\tag{362}
$$

After cancellations, the equation becomes

$$
\begin{aligned}
T^{\mathrm{pp,f}}_{i\alpha j\beta k\alpha l\beta}(z_1, z_2) = &\, \bar{\delta}_{\alpha\beta}\delta_{ij}\delta_{kl}\Phi^{\mathrm{pp,f}}_{iikk\alpha\beta}(z_1, z_2) + \\
&+ U(z_1)\bar{\delta}_{\alpha\beta}\delta_{ij} \sum_{ns} \int_{\mathcal{C}} \mathrm{d}z_5\, G^{\mathrm{H}}_{i\alpha i\beta n\alpha s\beta}(z_1, z_5)T^{\mathrm{pp,f}}_{n\alpha s\beta k\alpha l\beta}(z_5, z_2),
\end{aligned}
\tag{363}
$$

$$\Phi^{\mathrm{pp,f}}_{iikk\alpha\beta}(z_1, z_2) =: -U(z_1)G^{\mathrm{H}}_{i\alpha i\beta k\alpha k\beta}(z_1, z_2)U(z_2),\tag{364}$$

where we again introduced the function $\Phi^{\mathrm{pp}}$. Similar as for bosons, by iteration starting with

$$\Phi^{\mathrm{pp,f}}_{i\alpha j\beta k\alpha l\beta}(z_1, z_2) = \bar{\delta}_{\alpha\beta}\delta_{ij}\delta_{kl}\Phi^{\mathrm{pp,f}}_{iikk\alpha\beta}(z_1, z_2) =: \bar{\delta}_{\alpha\beta}\delta_{ij}\delta_{kl}\Phi^{\mathrm{pp,f}}_{ik\alpha\neq\beta}(z_1, z_2),$$

[xxv]The total selfenergy contains the Hartree selfenergy. There is no Fock term.

the particle–particle $T$ matrix is also of the structure

$$T_{i\alpha j\beta k\alpha l\beta}^{\mathrm{pp,f}}(z_1, z_2) := \bar{\bar{\delta}}_{\alpha\beta}\delta_{ij}\delta_{kl}T_{iikk\alpha\beta}^{\mathrm{pp,f}}(z_1, z_2) =: \delta_{ij}\delta_{kl}\bar{\bar{\delta}}_{\alpha\beta}T_{ik\alpha\neq\beta}^{\mathrm{pp,f}}(z_1, z_2).$$

The resulting equations for the $T$ matrix and the correlation selfenergy read

$$T_{ik\alpha\neq\beta}^{\mathrm{pp,f}}(z_1, z_2) = \Phi_{ik\alpha\neq\beta}^{\mathrm{pp,f}}(z_1, z_2) + \tag{365}$$
$$+ U(z_1)\sum_n\int_{\mathcal{C}}\mathrm{d}z_5\, G_{in\alpha\neq\beta}^{\mathrm{H}}(z_1, z_5)T_{nk\alpha\neq\beta}^{\mathrm{pp,f}}(z_5, z_2),$$

$$\Sigma_{i\alpha j\alpha}^{\mathrm{pp,f}}(z_1, z_2) = \mathrm{i}\hbar\sum_{\epsilon\neq\alpha}T_{ij\alpha\neq\epsilon}^{\mathrm{pp,f}}(z_1, z_2)G_{j\epsilon i\epsilon}(z_2, z_1). \tag{366}$$

The greater/less components of the selfenergy are given by

$$\Sigma_{i\alpha j\alpha}^{\mathrm{pp,f},\gtrless}(t_1, t_2) = \mathrm{i}\hbar\sum_{\epsilon\neq\alpha}T_{ij\alpha\neq\epsilon}^{\mathrm{pp,f},\gtrless}(t_1, t_2)G_{j\epsilon i\epsilon}^{\lessgtr}(t_2, t_1),$$

whereas the Keldysh components of the $T$ matrix are

$$T_{ik\alpha\neq\beta}^{\mathrm{pp,f},\gtrless}(t_1, t_2) = \Phi_{ik\alpha\neq\beta}^{\mathrm{pp,f},\gtrless}(t_1, t_2) + \tag{367}$$
$$U(t_1)\sum_n\left(\int_{t_0}^{t_1}\mathrm{d}t_5\, G_{in\alpha\neq\beta}^{\mathrm{H},\mathcal{R}}(t_1, t_5)T_{nk\alpha\neq\beta}^{\mathrm{pp,f},\gtrless}(t_5, t_2) + \right.$$
$$\left. + \int_{t_0}^{t_2}\mathrm{d}t_5\, G_{in\alpha\neq\beta}^{\mathrm{H},\gtrless}(t_1, t_5)T_{nk\alpha\neq\beta}^{\mathrm{pp,f},\mathcal{A}}(t_5, t_2)\right),$$

and

$$T_{ik\alpha\neq\beta}^{\mathrm{pp,f},\mathcal{A}/\mathcal{R}}(t_1, t_2) = \Phi_{ik\alpha\neq\beta}^{\mathrm{pp,f},\mathcal{A}/\mathcal{R}}(t_1, t_2) + U(t_1)\sum_n\int_{t_{1/2}}^{t_{2/1}}\mathrm{d}t_5\, G_{in\alpha\neq\beta}^{\mathrm{H},\mathcal{A}/\mathcal{R}}(t_1, t_5)T_{nk\alpha\neq\beta}^{\mathrm{pp,f},\mathcal{A}/\mathcal{R}}(t_5, t_2).$$

In the special case of **spin-$\frac{1}{2}$ fermions**, we have

$$\Phi_{ik\alpha\neq\beta}^{\mathrm{pp,f},1/2}(z_1, z_2) =: \Phi_{ik}^{\mathrm{pp,f},1/2}(z_1, z_2),$$
$$G_{ik\alpha\neq\beta}^{\mathrm{H}}(z_1, z_2) =: G_{ik}^{\mathrm{H}}(z_1, z_2),$$

and, consequently, $T_{ik\alpha\neq\beta}^{\mathrm{pp,f},1/2}(z_1, z_2) =: T_{ik}^{\mathrm{pp,f},1/2}(z_1, z_2)$, holds. With this, one arrives at

$$T_{ik}^{\mathrm{pp,f},1/2}(z_1, z_2) = \Phi_{ik}^{\mathrm{pp,f},1/2}(z_1, z_2) + \tag{368}$$
$$+ U(z_1)\sum_n\int_{\mathcal{C}}\mathrm{d}z_5\, G_{in}^{\mathrm{H}}(z_1, z_5)T_{nk}^{\mathrm{pp,f},1/2}(z_5, z_2),$$

and obtains for the spin up and spin down selfenergies

$$\Sigma_{i\uparrow j\uparrow}^{\mathrm{pp,f},1/2}(z_1, z_2) = \mathrm{i}\hbar T_{ij}^{\mathrm{pp,f},1/2}(z_1, z_2)G_{j\downarrow i\downarrow}(z_2, z_1),$$
$$\Sigma_{i\downarrow j\downarrow}^{\mathrm{pp,f},1/2}(z_1, z_2) = \mathrm{i}\hbar T_{ij}^{\mathrm{pp,f},1/2}(z_1, z_2)G_{j\uparrow i\uparrow}(z_2, z_1).$$

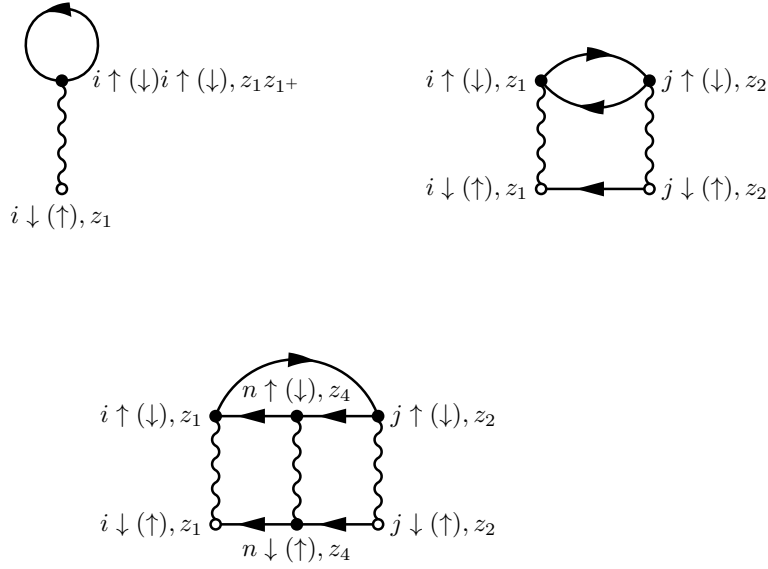

**Figure 32.** Leading terms of $\Sigma^{\mathrm{pp,f},1/2}$ (including first-order terms) in the Hubbard basis for spin-$1/2$ fermions.

The greater/less components of the selfenergy are given by

$$\Sigma^{\mathrm{pp,f},1/2,\gtrless}_{i\uparrow j\uparrow}\big(t_1, t_2\big) = \mathrm{i}\hbar T^{\mathrm{pp,f},1/2,\gtrless}_{ij}\big(t_1, t_2\big) G^{\lessgtr}_{j\downarrow i\downarrow}\big(t_2, t_1\big),$$

$$\Sigma^{\mathrm{pp,f},1/2,\gtrless}_{i\downarrow j\downarrow}\big(t_1, t_2\big) = \mathrm{i}\hbar T^{\mathrm{pp,f},1/2,\gtrless}_{ij}\big(t_1, t_2\big) G^{\lessgtr}_{j\uparrow i\uparrow}\big(t_2, t_1\big),$$

and the corresponding greater/less Keldysh components of the $T$ matrix are

$$\begin{aligned}
T^{\mathrm{pp,f},1/2,\gtrless}_{ik}\big(t_1, t_2\big) &= \Phi^{\mathrm{pp,f},1/2,\gtrless}_{ik}\big(t_1, t_2\big) + \\
&\quad + U\big(t_1\big) \sum_n \int_{t_0}^{t_1} \mathrm{d}t_5\, G^{\mathrm{H},\mathcal{R}}_{in}\big(t_1, t_5\big) T^{\mathrm{pp,f},1/2,\gtrless}_{nk}\big(t_5, t_2\big) + \\
&\quad + U\big(t_1\big) \sum_n \int_{t_0}^{t_2} \mathrm{d}t_5\, G^{\mathrm{H},\gtrless}_{in}\big(t_1, t_5\big) T^{\mathrm{pp,f},1/2,\mathcal{A}}_{nk}\big(t_5, t_2\big),
\end{aligned} \tag{369}$$

whereas the advanced and retarded components become

$$\begin{aligned}
T^{\mathrm{pp,f},1/2,\mathcal{A/R}}_{ik}\big(t_1, t_2\big) &= \Phi^{\mathrm{pp,f},1/2,\mathcal{A/R}}_{ik}\big(t_1, t_2\big) \\
&\quad + U\big(t_1\big) \sum_n \int_{t_{1/2}}^{t_{2/1}} \mathrm{d}t_5\, G^{\mathrm{H},\mathcal{A/R}}_{in}\big(t_1, t_5\big) T^{\mathrm{pp,f},1/2,\mathcal{A/R}}_{nk}\big(t_5, t_2\big).
\end{aligned} \tag{370}$$

Finally, the corresponding Feynman diagrams are shown in Fig. 32.

## 5.4. Particle–hole $T$-matrix approximation

Returning to Eq. (323) and taking the third approximation (C), we have

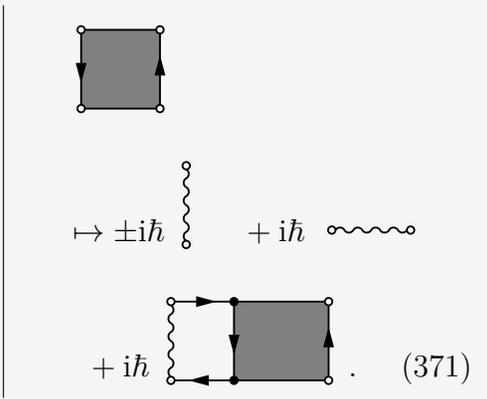

$$\frac{\delta \Sigma^{\mathrm{cl,C}}(1,2)}{\delta G(7,8)}$$

$$= \pm i\hbar \delta(1,2)\delta(7,8)w(1,7)$$

$$+ i\hbar \delta(1,7)\delta(2,8)w(1,2)$$

$$+ i\hbar w(1,7)G(1,4)\frac{\delta \Sigma^{\mathrm{cl,C}}(4,2)}{\delta G(5,8)}G(5,7) \qquad (371)$$

The corresponding third-order selfenergy terms follow as [cf. Eq. (324)]

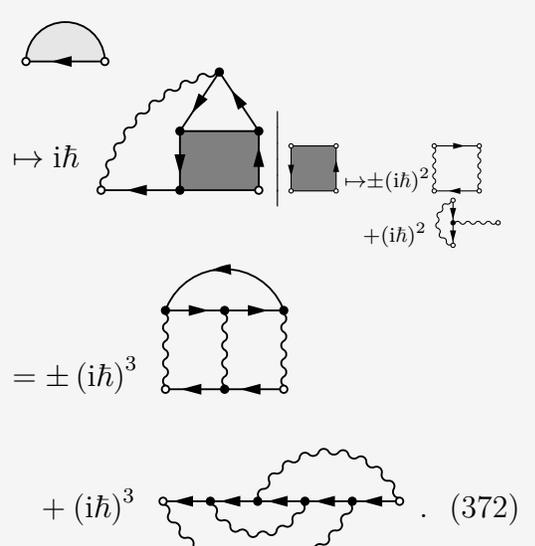

$$\Sigma^{\mathrm{cl,C(3)}}(1,2)$$

$$= i\hbar w(1,3)G(1,4)\frac{\delta \Sigma^{\mathrm{cl,C(2)}}(4,2)}{\delta G(5,6)}$$

$$G(5,3)G(3,6)$$

$$= \pm(i\hbar)^3 w(1,3)G(1,4)G(4,2)w(4,5)$$

$$w(2,6)G(6,5)G(5,3)G(3,6)$$

$$+ (i\hbar)^3 w(1,3)G(1,4)w(4,5)G(4,6)$$

$$w(6,2)G(6,5)G(5,3)G(3,2) \qquad (372)$$

The Kernel of series C is again introduced in the full notation with the single-time interaction,

$$K^{\mathrm{C}}_{ijkl}(z_1,z_2,z_3,z_4) := \pm\frac{\delta \Sigma^{\mathrm{cl,C}}_{ik}(z_1,z_3)}{\delta G_{lj}(z_4,z_2)},$$

and obeys the equation

$$K^{\mathrm{C}}_{ijkl}(z_1,z_2,z_3,z_4) = \pm i\hbar \delta_{\mathcal{C}}(z_1,z_3)\delta_{\mathcal{C}}(z_1,z_4)\delta_{\mathcal{C}}(z_{1^+},z_2)w^{\pm}_{ijkl}(z_1)$$

$$+ i\hbar \delta_{\mathcal{C}}(z_1,z_4)\sum_{mp} w_{iplm}(z_1)\int_{\mathcal{C}} \mathrm{d}z_5 \sum_n G_{mn}(z_1,z_5) \times$$

$$\times \int_{\mathcal{C}} \mathrm{d}z_6 \sum_r K^{\mathrm{C}}_{njkr}(z_5,z_2,z_3,z_6)G_{rp}(z_6,z_1).$$

Iterating as before, starting with

$$K_{ijkl}^{\mathrm{C},(1)}\big(z_1, z_2, z_3, z_4\big) = \pm \mathrm{i}\hbar \delta_{\mathcal{C}}\big(z_1, z_3\big)\delta_{\mathcal{C}}\big(z_1, z_4\big)\delta_{\mathcal{C}}\big(z_{1^+}, z_2\big)w_{ijkl}^{\pm}\big(z_1\big), \quad (373)$$

we arrive at

$$K_{ijkl}^{\mathrm{C},(2)}\big(z_1, z_2, z_3, z_4\big) = \mathrm{i}\hbar \delta_{\mathcal{C}}\big(z_1, z_4\big) \sum_{mp} w_{iplm}\big(z_1\big) \int_{\mathcal{C}} \mathrm{d}z_5 \sum_n G_{mn}\big(z_1, z_5\big) \times$$
$$\int_{\mathcal{C}} \mathrm{d}z_6 \sum_r K_{njkr}^{\mathrm{C},(1)}\big(z_5, z_2, z_3, z_6\big)G_{rp}\big(z_6, z_1\big). \tag{374}$$

Use of the $\delta$-structure of $K^{\mathrm{C},(1)}$, leads to

$$K_{ijkl}^{\mathrm{C},(2)}\big(z_1, z_2, z_3, z_4\big) = \mathrm{i}\hbar \delta_{\mathcal{C}}\big(z_1, z_4\big)\delta_{\mathcal{C}}\big(z_2, z_3\big) \sum_{mp} w_{iplm}\big(z_1\big) \sum_n G_{mn}\big(z_1, z_2\big)$$
$$\times \sum_r K_{njkr}^{\mathrm{C},(1)}\big(z_2, z_2, z_2, z_2\big)G_{rp}\big(z_2, z_1\big), \tag{375}$$

$$K_{ijkl}^{\mathrm{C},(3)}\big(z_1, z_2, z_3, z_4\big) = \mathrm{i}\hbar \delta_{\mathcal{C}}\big(z_1, z_4\big) \sum_{mp} w_{iplm}\big(z_1\big) \int_{\mathcal{C}} \mathrm{d}z_5 \sum_n G_{mn}\big(z_1, z_5\big)$$
$$\times \int_{\mathcal{C}} \mathrm{d}z_6 \sum_r K_{njkr}^{\mathrm{C},(2)}\big(z_5, z_2, z_3, z_6\big)G_{rp}\big(z_6, z_1\big) \tag{376}$$
$$= \mathrm{i}\hbar \delta_{\mathcal{C}}\big(z_1, z_4\big)\delta_{\mathcal{C}}\big(z_2, z_3\big) \sum_{mp} w_{iplm}\big(z_1\big) \int_{\mathcal{C}} \mathrm{d}z_5 \sum_n G_{mn}\big(z_1, z_5\big)$$
$$\times \sum_r K_{njkr}^{\mathrm{C},(2)}\big(z_5, z_2, z_2, z_5\big)G_{rp}\big(z_5, z_1\big).$$

The structure again remains the same for higher orders, so that this approximation lies within $\mathcal{O}\big(N_{\mathrm{t}}^3\big)$ due to

$$K_{ijkl}^{\mathrm{C},(N)}\big(z_1, z_2, z_3, z_4\big) =: \delta_{\mathcal{C}}\big(z_1, z_4\big)\delta_{\mathcal{C}}\big(z_2, z_3\big)K_{ijkl}^{\mathrm{C},(N)}\big(z_1, z_2\big),$$

and one integration on the right-hand side. With this, we arrive at [cf. Eq. (373)]

$$K_{ijkl}^{\mathrm{C}}\big(z_1, z_2\big) = \delta_{\mathcal{C}}\big(z_1, z_2\big)K_{ijkl}^{\mathrm{C},(1)}\big(z_1, z_1\big) + \mathrm{i}\hbar \sum_{mp} w_{iplm}\big(z_1\big)$$
$$\times \int_{\mathcal{C}} \mathrm{d}z_5 \sum_n G_{mn}\big(z_1, z_5\big) \sum_r K_{njkr}^{\mathrm{C}}\big(z_5, z_2\big)G_{rp}\big(z_5, z_1\big). \tag{377}$$

To simplify, we again go over to a non-singular kernel

$$K_{ijkl}^{\mathrm{C,ns}}\big(z_1, z_2\big) := K_{ijkl}^{\mathrm{C}}\big(z_1, z_2\big) - \delta_{\mathcal{C}}\big(z_1, z_2\big)K_{ijkl}^{\mathrm{C},(1)}\big(z_1, z_1\big)$$
$$= K_{ijkl}^{\mathrm{C}}\big(z_1, z_2\big) \mp \mathrm{i}\hbar \delta_{\mathcal{C}}\big(z_1, z_2\big)w_{ijkl}^{\pm}\big(z_1\big). \tag{378}$$

This quantity obeys the equation [cf. Eq. (376)]

$$K_{ijkl}^{\mathrm{C,ns}}\big(z_1, z_2\big) = \mathrm{i}\hbar \sum_{mp} w_{iplm}\big(z_1\big) \int_{\mathcal{C}} \mathrm{d}z_5 \sum_n G_{mn}\big(z_1, z_5\big) \sum_r G_{rp}\big(z_5, z_1\big)$$
$$\times \left\{ K_{njkr}^{\mathrm{C,ns}}\big(z_5, z_2\big) \pm \mathrm{i}\hbar \delta_{\mathcal{C}}\big(z_5, z_2\big) w_{njkr}^{\pm}\big(z_2\big) \right\}. \tag{379}$$

Restructuring, this implies

$$K_{ijkl}^{\mathrm{C,ns}}\big(z_1, z_2\big) = \pm \big(\mathrm{i}\hbar\big)^2 \sum_{mp} w_{iplm}\big(z_1\big) \sum_n G_{mn}\big(z_1, z_2\big) \sum_r G_{rp}\big(z_2, z_1\big) w_{njkr}^{\pm}\big(z_2\big)$$
$$+ \mathrm{i}\hbar \sum_{mp} w_{iplm}\big(z_1\big) \int_{\mathcal{C}} \mathrm{d}z_5 \sum_n G_{mn}\big(z_1, z_5\big) \sum_r G_{rp}\big(z_5, z_1\big) K_{njkr}^{\mathrm{C,ns}}\big(z_5, z_2\big)$$

and, finally,

$$K_{ijkl}^{\mathrm{C,ns}}\big(z_1, z_2\big) = \mathrm{i}\hbar \sum_{mp} w_{iplm}\big(z_1\big) \sum_{nr} G_{mrpn}^{\mathrm{F}}\big(z_1, z_2\big) w_{njkr}^{\pm}\big(z_2\big) \tag{380}$$
$$\pm \sum_{mp} w_{iplm}\big(z_1\big) \int_{\mathcal{C}} \mathrm{d}z_5 \sum_{nr} G_{mrpn}^{\mathrm{F}}\big(z_1, z_5\big) K_{njkr}^{\mathrm{C,ns}}\big(z_5, z_2\big),$$

where in the last line, Eq. (281) has been used.

With this, the correlation selfenergy reads[xxvi], [cf. Eq. (322)]

$$\Sigma_{ij}^{\mathrm{cl,corr,C}}\big(z_1, z_2\big) = \pm \mathrm{i}\hbar \sum_{mpq} w_{ipqm}\big(z_1\big) \int_{\mathcal{C}} \mathrm{d}z_3 \sum_n G_{mn}\big(z_1, z_3\big) \times$$
$$\int_{\mathcal{C}} \mathrm{d}z_4 \mathrm{d}z_5 \sum_{rs} K_{nsjr}^{\mathrm{C}}\big(z_3, z_5, z_2, z_4\big) G_{rp}\big(z_4, z_1\big) G_{qs}\big(z_1, z_5\big).$$

Using Eq. (378), we arrive at

$$\Sigma_{ij}^{\mathrm{cl,corr,C}}\big(z_1, z_2\big) = \pm \mathrm{i}\hbar \sum_{mpqs} w_{ipqm}\big(z_1\big) G_{qs}\big(z_1, z_2\big) \sum_{nr} \int_{\mathcal{C}} \mathrm{d}z_3 \, G_{mn}\big(z_1, z_3\big) G_{rp}\big(z_3, z_1\big)$$
$$\times \left\{ K_{nsjr}^{\mathrm{C,ns}}\big(z_3, z_2\big) \pm \mathrm{i}\hbar \delta_{\mathcal{C}}\big(z_3, z_2\big) w_{nsjr}^{\pm}\big(z_2\big) \right\}.$$

[xxvi]the total selfenergy contains, in addition the mean-field terms $\Sigma_{ij}^{\mathrm{H}}\big(z_1, z_2\big) + \Sigma_{ij}^{\mathrm{F}}\big(z_1, z_2\big)$

Evaluating this expression further, we find

$$\Sigma_{ij}^{\text{cl,corr,C}}\big(z_1, z_2\big) = \pm \sum_{qs} G_{qs}\big(z_1, z_2\big) \times$$
$$\times \bigg( \mathrm{i}\hbar \sum_{pm} w_{ipqm}\big(z_1\big) \sum_{rn} G_{mrpn}^{\text{F}}\big(z_1, z_2\big) w_{nsjr}^{\pm}\big(z_2\big) \pm$$
$$\pm \sum_{pm} w_{ipqm}\big(z_1\big) \int_{\mathcal{C}} \mathrm{d}z_3 \sum_{rn} G_{mrpn}^{\text{F}}\big(z_1, z_3\big) K_{nsjr}^{\text{C,ns}}\big(z_3, z_2\big) \bigg).$$

As for the particle–particle $T$ matrix, the right-hand side already contains the first iteration of Eq. (380) and, thus, we can simplify

$$\Sigma_{ij}^{\text{cl,corr,C}}\big(z_1, z_2\big) = \pm \sum_{qs} G_{qs}\big(z_1, z_2\big) K_{isjq}^{\text{C,ns}}\big(z_1, z_2\big).$$

Defining $\mathrm{i}\hbar T_{ijkl}^{\text{ph}}\big(z_1, z_2\big) := \pm K_{ijkl}^{\text{C,ns}}\big(z_1, z_2\big)$, and solving Eqs. (380) and (381), we have

$$T_{ijkl}^{\text{ph}}\big(z_1, z_2\big) = \Phi_{ijkl}^{\text{ph}}\big(z_1, z_2\big) \pm$$
$$\pm \sum_{mp} w_{iplm}\big(z_1\big) \int_{\mathcal{C}} \mathrm{d}z_5 \sum_{nr} G_{mrpn}^{\text{F}}\big(z_1, z_5\big) T_{njkr}^{\text{ph}}\big(z_5, z_2\big), \quad (381)$$
$$\Phi_{ijkl}^{\text{ph}}\big(z_1, z_2\big) = \pm \sum_{mp} w_{iplm}\big(z_1\big) \sum_{nr} G_{mrpn}^{\text{F}}\big(z_1, z_2\big) w_{njkr}^{\pm}\big(z_2\big), \quad (382)$$

where we again defined the proper function $\Phi^{\text{ph}}$ corresponding to the first iteration (second-Born approximation).

With these definitions, we obtain the correlation selfenergy

$$\Sigma_{ij}^{\text{ph}}\big(z_1, z_2\big) := \Sigma_{ij}^{T^{\text{ph}},\text{corr}}\big(z_1, z_2\big) = \mathrm{i}\hbar \sum_{qs} G_{qs}\big(z_1, z_2\big) T_{isjq}^{\text{ph}}\big(z_1, z_2\big),$$

which is the so-called *particle–hole $T$-matrix approximation*[xxvii] (TPH). In contrast to the particle–particle $T$ matrix, which describes the recurrent scattering of a pair of particles, the particle–hole $T$ matrix describes the (multiple) scattering of a particle–hole pair. The Keldysh components of the particle–hole $T$ matrix read

$$T_{ijkl}^{\text{ph},\gtrless}\big(t_1, t_2\big) = \Phi_{ijkl}^{\text{ph},\gtrless}\big(t_1, t_2\big) \pm \sum_{mp} w_{iplm}\big(t_1\big) \sum_{nr} \qquad (383)$$
$$\bigg( \int_{t_0}^{t_1} \mathrm{d}t_5 \, G_{mrpn}^{\text{F},\mathcal{R}}\big(t_1, t_5\big) T_{njkr}^{\text{ph},\gtrless}\big(t_5, t_2\big) + \int_{t_0}^{t_2} \mathrm{d}t_5 \, G_{mrpn}^{\text{F},\gtrless}\big(t_1, t_5\big) T_{njkr}^{\text{ph},\mathcal{A}}\big(t_5, t_2\big) \bigg),$$

---

[xxvii]In the context of the Fermi–Hubbard model for electrons, the particle–hole $T$ matrix will later be called *electron–hole $T$ matrix* (TEH).

and

$$T^{\text{ph},\mathcal{A}/\mathcal{R}}_{ijkl}\big(t_1,t_2\big) = \Phi^{\text{ph},\mathcal{A}/\mathcal{R}}_{ijkl}\big(t_1,t_2\big) \pm \tag{384}$$

$$\pm \sum_{mp} w_{iplm}\big(t_1\big) \sum_{nr} \int_{t_{1/2}}^{t_{2/1}} \mathrm{d}t_5\, G^{\text{H},\mathcal{A}/\mathcal{R}}_{mrpn}\big(t_1,t_5\big) T^{\text{ph},\mathcal{A}/\mathcal{R}}_{njkr}\big(t_5,t_2\big)\,.$$

For the greater/less components of the selfenergy, we have

$$\Sigma^{\text{ph},\gtrless}_{ij}\big(t_1,t_2\big) = \sum_{qs} G^{\gtrless}_{qs}\big(t_1,t_2\big) T^{\text{ph},\gtrless}_{isjq}\big(t_1,t_2\big)\,.$$

***Particle–hole $T$ matrix in a diagonal basis.*** For diagonal basis sets with $w_{ijkl} = \delta_{il}\delta_{jk}w_{ij}$, Eqs. (381) and (383) become

$$T^{\text{ph,diag}}_{ijkl}\big(z_1,z_2\big) = \Phi^{\text{ph,diag}}_{ijkl}\big(z_1,z_2\big) \pm w_{il}\big(z_1\big) \int_{\mathcal{C}} \mathrm{d}z_5 \sum_{nr} G^{\text{F}}_{irln}\big(z_1,z_5\big) T^{\text{ph,diag}}_{njkr}\big(z_5,z_2\big)\,,$$

$$\Phi^{\text{ph,diag}}_{ijkl}\big(z_1,z_2\big) =: \pm w_{il}\big(z_1\big) \left\{ \delta_{jk} \sum_n G^{\text{F}}_{inln}\big(z_1,z_2\big) w_{nj}\big(z_2\big) \pm G^{\text{F}}_{ijlk}\big(z_1,z_2\big) w_{jk}\big(z_2\big) \right\}\,,$$

where we defined the diagonal version of the particle–hole function $\Phi^{\text{ph}}$. Then the selfenergy in the diagonal basis becomes

$$\Sigma^{\text{ph,diag}}_{ij}\big(z_1,z_2\big) = \mathrm{i}\hbar \sum_{qs} G_{qs}\big(z_1,z_2\big) T^{\text{ph,diag}}_{isjq}\big(z_1,z_2\big)\,.$$

The greater/less Keldysh components are

$$T^{\text{ph,diag},\gtrless}_{ijkl}\big(t_1,t_2\big) = \Phi^{\text{ph,diag},\gtrless}_{ijkl}\big(t_1,t_2\big) \pm w_{il}\big(t_1\big) \sum_{nr} \tag{385}$$

$$\times \left( \int_{t_0}^{t_1} \mathrm{d}t_5\, G^{\text{F},\mathcal{R}}_{irln}\big(t_1,t_5\big) T^{\text{ph,diag},\gtrless}_{njkr}\big(t_5,t_2\big) + \int_{t_0}^{t_2} \mathrm{d}t_5\, G^{\text{F},\gtrless}_{irln}\big(t_1,t_5\big) T^{\text{ph,diag},\mathcal{A}}_{njkr}\big(t_5,t_2\big) \right),$$

whereas, for the advanced/retarded $T$ matrices we obtain

$$T^{\text{ph,diag},\mathcal{A}/\mathcal{R}}_{ijkl}\big(t_1,t_2\big) = \Phi^{\text{ph,diag},\mathcal{A}/\mathcal{R}}_{ijkl}\big(t_1,t_2\big) \pm \tag{386}$$

$$\pm w_{il}\big(t_1\big) \sum_{nr} \int_{t_{1/2}}^{t_{2/1}} \mathrm{d}t_5\, G^{\text{F},\mathcal{A}/\mathcal{R}}_{irln}\big(t_1,t_5\big) T^{\text{ph,diag},\mathcal{A}/\mathcal{R}}_{njkr}\big(t_5,t_2\big)\,.$$

For the greater/less components of the selfenergy, we have

$$\Sigma^{\text{ph,diag},\gtrless}_{ij}\big(t_1,t_2\big) = \mathrm{i}\hbar \sum_{qs} G^{\gtrless}_{qs}\big(t_1,t_2\big) T^{\text{ph,diag},\gtrless}_{isjq}\big(t_1,t_2\big)\,. \tag{387}$$

The diagrammatic representation of the first terms of the particle–hole $T$ matrix is given in Fig. 33.

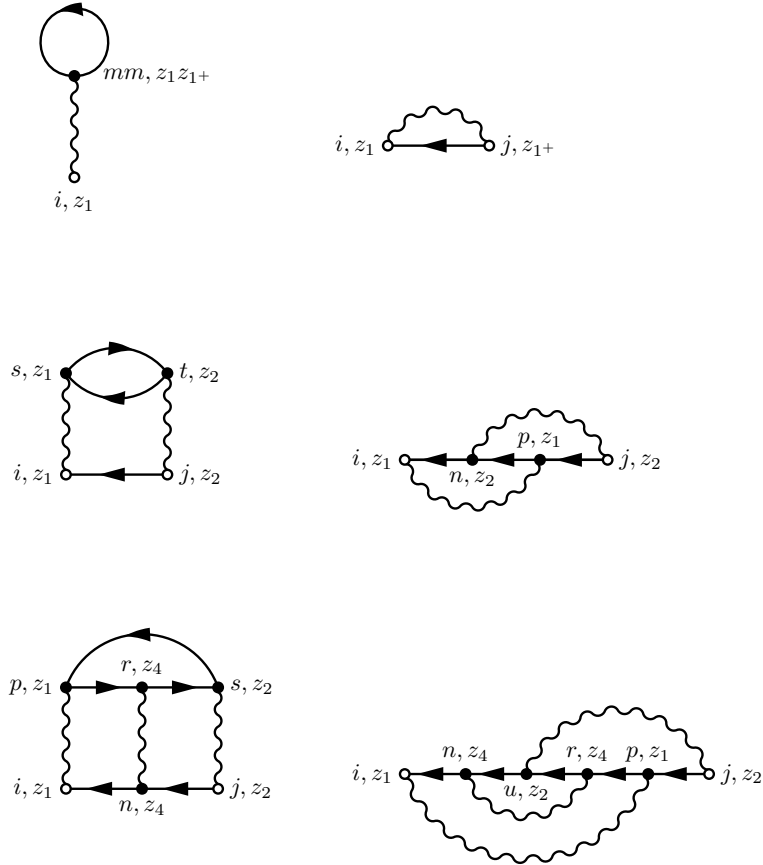

**Figure 33.** The first six terms of the particle–hole $T$-matrix selfenergy (including first-order terms) in a diagonal basis.

***Particle–hole $T$ matrix in the Hubbard basis.*** For the **bosonic Hubbard basis**, the particle–hole $T$ matrix reads

$$T^{\mathrm{ph},\mathfrak{b}}_{i\alpha j\beta k\alpha l\beta}\big(z_1, z_2\big) = \delta_{il}\delta_{jk}U\big(z_1\big)\left\{\delta_{\alpha\beta}G^{\mathrm{F}}_{i\alpha j\alpha i\alpha j\alpha}\big(z_1, z_2\big) + G^{\mathrm{F}}_{i\alpha j\beta i\beta j\alpha}\big(z_1, z_2\big)\right\}U\big(z_2\big) +$$
$$+ \,\delta_{il}U\big(z_1\big)\int_{\mathcal{C}}\mathrm{d}z_5\sum_{nr}G^{\mathrm{F}}_{i\alpha r\beta i\beta n\alpha}\big(z_1, z_5\big)T^{\mathrm{ph},\mathfrak{b}}_{n\alpha j\beta k\alpha r\beta}\big(z_5, z_2\big)\,.$$

Introducing the function $\Phi^{\mathrm{ph}}$, as before, this expression becomes

$$T^{\mathrm{ph},\mathfrak{b}}_{i\alpha j\beta k\alpha l\beta}\big(z_1, z_2\big) =: \delta_{il}\delta_{jk}\Phi^{\mathrm{ph},\mathfrak{b}}_{i\alpha j\beta i\alpha j\beta}\big(z_1, z_2\big) +$$
$$+ \,\delta_{il}U\big(z_1\big)\int_{\mathcal{C}}\mathrm{d}z_5\sum_{nr}G^{\mathrm{F}}_{i\alpha r\beta i\beta n\alpha}\big(z_1, z_5\big)T^{\mathrm{ph},\mathfrak{b}}_{n\alpha j\beta k\alpha r\beta}\big(z_5, z_2\big)\,.$$

By iteration, starting from

$$\Phi^{\mathrm{ph},\mathfrak{b}}_{i\alpha j\beta k\alpha l\beta}\big(z_1, z_2\big) = \delta_{il}\delta_{jk}\Phi^{\mathrm{ph},\mathfrak{b}}_{i\alpha j\beta j\alpha i\beta}\big(z_1, z_2\big) =: \delta_{il}\delta_{jk}\Phi^{\mathrm{ph},\mathfrak{b}}_{ij\alpha\beta}\big(z_1, z_2\big)\,,$$

it is evident that the particle–hole $T$ matrix is also of the structure

$$T^{\mathrm{ph},\mathfrak{b}}_{i\alpha j\beta k\alpha l\beta}\big(z_1,z_2\big) =: \delta_{il}\delta_{jk}T^{\mathrm{ph},\mathfrak{b}}_{ij\alpha\beta}\big(z_1,z_2\big)\,.$$

Thus the governing equation for the $T$ matrix becomes

$$T^{\mathrm{ph},\mathfrak{b}}_{ij\alpha\beta}\big(z_1,z_2\big) = \Phi^{\mathrm{ph},\mathfrak{b}}_{ij\alpha\beta}\big(z_1,z_2\big) + \tag{388}$$
$$+\, U\big(z_1\big)\int_{\mathcal{C}}\mathrm{d}z_5 \sum_n G^{\mathrm{F}}_{i\alpha n\beta i\beta n\alpha}\big(z_1,z_5\big)T^{\mathrm{ph},\mathfrak{b}}_{nj\alpha\beta}\big(z_5,z_2\big)\,,$$

and the resulting correlation selfenergy reads

$$\Sigma^{\mathrm{ph},\mathfrak{b}}_{i\alpha j\alpha}\big(z_1,z_2\big) = \mathrm{i}\hbar \sum_{\epsilon} G_{i\epsilon j\epsilon}\big(z_1,z_2\big)T^{\mathrm{ph}}_{ij\alpha\epsilon}\big(z_1,z_2\big)\,. \tag{389}$$

The greater/less components of the correlation selfenergy are given by

$$\Sigma^{\mathrm{ph},\mathfrak{b},\gtrless}_{i\alpha j\alpha}\big(t_1,t_2\big) = \mathrm{i}\hbar \sum_{\epsilon} G^{\gtrless}_{i\epsilon j\epsilon}\big(t_1,t_2\big)T^{\mathrm{ph},\gtrless}_{ij\alpha\epsilon}\big(t_1,t_2\big)\,, \tag{390}$$

with the greater/less components of the $T$ matrix:

$$T^{\mathrm{ph},\mathfrak{b},\gtrless}_{ij\alpha\beta}\big(t_1,t_2\big) = \Phi^{\mathrm{ph},\mathfrak{b},\gtrless}_{ij\alpha\beta}\big(t_1,t_2\big) + \tag{391}$$
$$+\, U\big(t_1\big)\Bigg(\int_{t_0}^{t_1}\mathrm{d}t_5 \sum_n G^{\mathrm{F},\mathcal{R}}_{i\alpha n\beta i\beta n\alpha}\big(t_1,t_5\big)T^{\mathrm{ph},\mathfrak{b},\gtrless}_{nj\alpha\beta}\big(t_5,t_2\big) +$$
$$+\int_{t_0}^{t_2}\mathrm{d}t_5 \sum_n G^{\mathrm{F},\gtrless}_{i\alpha n\beta i\beta n\alpha}\big(t_1,t_5\big)T^{\mathrm{ph},\mathfrak{b},\mathcal{A}}_{nj\alpha\beta}\big(t_5,t_2\big)\Bigg)\,,$$

and the advanced/retarded components,

$$T^{\mathrm{ph},\mathfrak{b},\mathcal{A}/\mathcal{R}}_{ij\alpha\beta}\big(t_1,t_2\big) = \Phi^{\mathrm{ph},\mathfrak{b},\mathcal{A}/\mathcal{R}}_{ij\alpha\beta}\big(t_1,t_2\big) \tag{392}$$
$$+\, U\big(t_1\big)\int_{t_{1/2}}^{t_{2/1}}\mathrm{d}t_5 \sum_n G^{\mathrm{F},\mathcal{A}/\mathcal{R}}_{i\alpha n\beta i\beta n\alpha}\big(t_1,t_5\big)T^{\mathrm{ph},\mathfrak{b},\mathcal{A}/\mathcal{R}}_{nj\alpha\beta}\big(t_5,t_2\big)\,.$$

For **spin-0 bosons**, the equations simplify to

$$T^{\mathrm{ph},\mathfrak{b},0}_{ij}\big(z_1,z_2\big) = \Phi^{\mathrm{ph},\mathfrak{b},0}_{ij}\big(z_1,z_2\big) + U\big(z_1\big)\int_{\mathcal{C}}\mathrm{d}z_5 \sum_n G^{\mathrm{F}}_{inin}\big(z_1,z_5\big)T^{\mathrm{ph},\mathfrak{b},0}_{nj}\big(z_5,z_2\big)$$

and the correlation selfenergy is

$$\Sigma^{\mathrm{ph},\mathfrak{b},0}_{ij}\big(z_1,z_2\big) = \mathrm{i}\hbar G_{ij}\big(z_1,z_2\big)T^{\mathrm{ph},\mathfrak{b},0}_{ij}\big(z_1,z_2\big)\,, \tag{393}$$

with the greater/less components

$$\Sigma^{\mathrm{ph},\mathfrak{b},0,\gtrless}_{ij}\big(t_1,t_2\big) = \mathrm{i}\hbar G^{\gtrless}_{ij}\big(t_1,t_2\big)T^{\mathrm{ph},\mathfrak{b},0,\gtrless}_{ij}\big(t_1,t_2\big)\,. \tag{394}$$

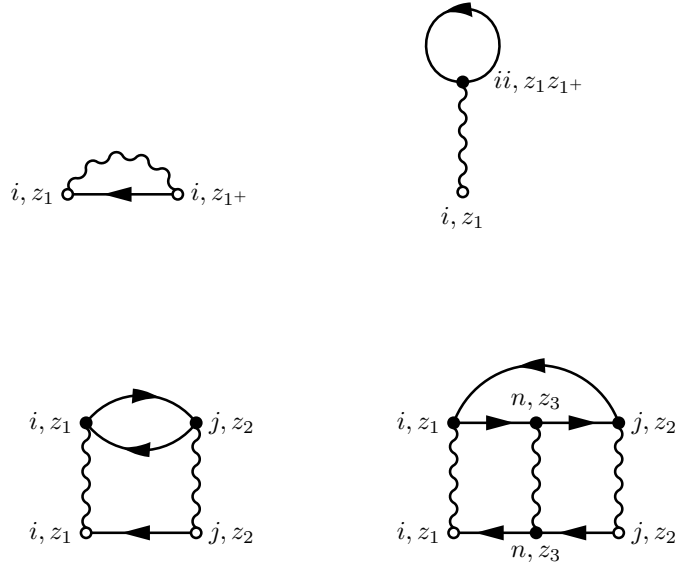

**Figure 34.** Leading terms of $\Sigma^{\mathrm{ph,b,0}}$ (including first-order terms) in the Hubbard basis for spin-0 bosons. Each term carries a factor of two.

The greater/less components of the $T$ matrix that enter this expression are given by

$$T_{ij}^{\mathrm{ph,b,0},\gtrless}(t_1, t_2) = \Phi_{ij}^{\mathrm{ph,b,0},\gtrless}(t_1, t_2) + U(t_1)\left(\int_{t_0}^{t_1} \mathrm{d}t_5 \sum_n G_{inin}^{\mathrm{F},\mathcal{R}}(t_1, t_5) T_{nj}^{\mathrm{ph,b,0},\gtrless}(t_5, t_2)\right.$$
$$\left. + \int_{t_0}^{t_2} \mathrm{d}t_5 \sum_n G_{inin}^{\mathrm{F},\gtrless}(t_1, t_5) T_{nj}^{\mathrm{ph,b,0},\mathcal{A}}(t_5, t_2)\right), \tag{395}$$

and the advanced/retarded components are

$$T_{ij}^{\mathrm{ph,b,0},\mathcal{A}/\mathcal{R}}(t_1, t_2) = \Phi_{ij}^{\mathrm{ph,b,0},\mathcal{A}/\mathcal{R}}(t_1, t_2) \tag{396}$$
$$+ U(t_1) \int_{t_{1/2}}^{t_{2/1}} \mathrm{d}t_5 \sum_n G_{inin}^{\mathrm{F},\mathcal{A}/\mathcal{R}}(t_1, t_5) T_{nj}^{\mathrm{ph,b,0},\mathcal{A}/\mathcal{R}}(t_5, t_2).$$

The leading terms of the corresponding diagrams are shown in Fig. 34.

For **fermionic particles and holes**, the particle–hole $T$ matrix satisfies

$$T_{i\alpha j\beta k\alpha l\beta}^{\mathrm{ph,f}}(z_1, z_2) = \delta_{il}\delta_{jk}\bar{\delta}_{\alpha\beta}U(z_1) \times \tag{397}$$
$$\times \left\{-\delta_{\alpha\beta}G_{i\alpha j\alpha i\alpha j\alpha}^{\mathrm{F}}(z_1, z_2) + G_{i\alpha j\beta i\beta j\alpha}^{\mathrm{F}}(z_1, z_2)\right\} U(z_2)$$
$$- \delta_{il}\bar{\delta}_{\alpha\beta}U(z_1) \int_{\mathcal{C}} \mathrm{d}z_5 \sum_{nr} G_{i\alpha r\beta i\beta n\alpha}^{\mathrm{F}}(z_1, z_5) T_{n\alpha j\beta k\alpha r\beta}^{\mathrm{ph,f}}(z_5, z_2).$$

Evaluating the terms, we have

$$T^{\text{ph,f}}_{i\alpha j\beta k\alpha l\beta}(z_1, z_2) = \delta_{il}\delta_{jk}\bar{\delta}_{\alpha\beta}U(z_1)G^{\text{F}}_{i\alpha j\beta i\beta j\alpha}(z_1, z_2)U(z_2) -$$
$$- \delta_{il}\bar{\delta}_{\alpha\beta}U(z_1)\int_{\mathcal{C}}\mathrm{d}z_5\sum_{nr}G^{\text{F}}_{i\alpha r\beta i\beta n\alpha}(z_1, z_5)T^{\text{ph,f}}_{n\alpha j\beta k\alpha r\beta}(z_5, z_2),$$

which, with the introduction of the function $\Phi^{\text{ph}}$, becomes

$$T^{\text{ph,f}}_{i\alpha j\beta k\alpha l\beta}(z_1, z_2) =: \delta_{il}\delta_{jk}\bar{\delta}_{\alpha\beta}\Phi^{\text{ph,f}}_{i\alpha j\beta i\alpha j\beta}(z_1, z_2) \tag{398}$$
$$- \delta_{il}\bar{\delta}_{\alpha\beta}U(z_1)\int_{\mathcal{C}}\mathrm{d}z_5\sum_{nr}G^{\text{F}}_{i\alpha r\beta i\beta n\alpha}(z_1, z_5)T^{\text{ph,f}}_{n\alpha j\beta k\alpha r\beta}(z_5, z_2).$$

Again by iteration, starting from $\Phi^{\text{ph,f}}_{i\alpha j\beta k\alpha l\beta}(z_1, z_2) =: \delta_{il}\delta_{jk}\bar{\delta}_{\alpha\beta}\Phi^{\text{ph,f}}_{ij\alpha\neq\beta}(z_1, z_2)$, it follows $T^{\text{ph,f}}_{i\alpha j\beta k\alpha l\beta}(z_1, z_2) =: \delta_{il}\delta_{jk}\bar{\delta}_{\alpha\beta}T^{\text{ph,f}}_{ij\alpha\neq\beta}(z_1, z_2)$. With this, Eq. (398) simplifies to

$$T^{\text{ph,f}}_{ij\alpha\neq\beta}(z_1, z_2) = \Phi^{\text{ph,f}}_{ij\alpha\neq\beta}(z_1, z_2) \tag{399}$$
$$- U(z_1)\int_{\mathcal{C}}\mathrm{d}z_5\sum_{n}G^{\text{F}}_{i\alpha n\beta i\beta n\alpha}(z_1, z_5)T^{\text{ph,f}}_{nj\alpha\neq\beta}(z_5, z_2).$$

The corresponding correlation selfenergy becomes[xxviii]

$$\Sigma^{\text{ph,f}}_{i\alpha j\alpha}(z_1, z_2) = \mathrm{i}\hbar\sum_{\epsilon\neq\alpha}G_{i\epsilon j\epsilon}(z_1, z_2)T^{\text{ph,f}}_{ij\alpha\neq\epsilon}(z_1, z_2). \tag{400}$$

The Keldysh components of the $T$ matrix read

$$T^{\text{ph,f},\gtrless}_{ij\alpha\neq\beta}(t_1, t_2) = \Phi^{\text{ph,f},\gtrless}_{ij\alpha\neq\beta}(t_1, t_2) \tag{401}$$
$$- U(t_1)\left(\int_{t_0}^{t_1}\mathrm{d}t_5\sum_{n}G^{\text{F},\mathcal{R}}_{i\alpha n\beta i\beta n\alpha}(t_1, t_5)T^{\text{ph,f},\gtrless}_{nj\alpha\neq\beta}(t_5, t_2) + \right.$$
$$\left. + \int_{t_0}^{t_2}\mathrm{d}t_5\sum_{n}G^{\text{F},\gtrless}_{i\alpha n\beta i\beta n\alpha}(t_1, t_5)T^{\text{ph,f},\mathcal{A}}_{nj\alpha\neq\beta}(t_5, t_2)\right),$$

$$T^{\text{ph,f},\mathcal{A}/\mathcal{R}}_{ij\alpha\neq\beta}(t_1, t_2) = \Phi^{\text{ph,f},\mathcal{A}/\mathcal{R}}_{ij\alpha\neq\beta}(t_1, t_2) \tag{402}$$
$$- U(t_1)\int_{t_{1/2}}^{t_{2/1}}\mathrm{d}t_5\sum_{n}G^{\text{F},\mathcal{A}/\mathcal{R}}_{i\alpha n\beta i\beta n\alpha}(t_1, t_5)T^{\text{ph,f},\mathcal{A}/\mathcal{R}}_{nj\alpha\neq\beta}(t_5, t_2).$$

For the correlation selfenergy, the greater/less components read

$$\Sigma^{\text{ph,f},\gtrless}_{i\alpha j\alpha}(t_1, t_2) = \mathrm{i}\hbar\sum_{\epsilon\neq\alpha}G^{\gtrless}_{i\epsilon j\epsilon}(t_1, t_2)T^{\text{ph,f},\gtrless}_{ij\alpha\neq\epsilon}(t_1, t_2).$$

For **spin-$\frac{1}{2}$ fermions**, we now switch to the name *electron–hole $T$ matrix* (TEH), since

[xxviii]the total selfenergy contains, in addition, the Hartree selfenergy, $\Sigma^{\text{H,f}}_{i\alpha j\alpha}(z_1, z_2)$

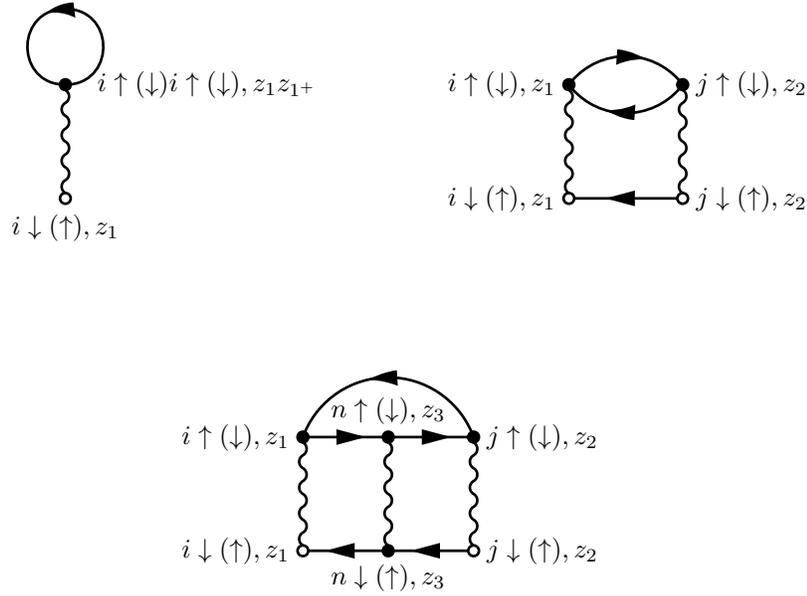

**Figure 35.** Leading terms of $\Sigma^{\text{eh,f},1/2}$ (including first-order terms) in the Hubbard basis for spin-1/2 fermions.

the quantity is predominantly used for (effective) electrons. We get

$$
\begin{aligned}
T^{\text{eh,f},1/2}_{ij\uparrow\downarrow}\big(z_1, z_2\big) &= T^{\text{eh,f},1/2}_{ij\downarrow\uparrow}\big(z_1, z_2\big) =: T^{\text{eh,f},1/2}_{ij}\big(z_1, z_2\big) \\
&= \Phi^{\text{eh,f},1/2}_{ij}\big(z_1, z_2\big) - U\big(z_1\big) \int_{\mathcal{C}} \mathrm{d}z_5 \sum_n G^{\text{F}}_{in}\big(z_1, z_5\big) T^{\text{eh,f},1/2}_{nj}\big(z_5, z_2\big),
\end{aligned}
$$

where we have defined

$$
\begin{aligned}
\Phi^{\text{eh,f},1/2}_{ij}\big(z_1, z_2\big) &:= \Phi^{\text{eh,f},1/2}_{ij\uparrow\downarrow}\big(z_1, z_2\big) = \Phi^{\text{eh,f},1/2}_{ij\downarrow\uparrow}\big(z_1, z_2\big), \\
G^{\text{F}}_{in}\big(z_1, z_2\big) &:= G^{\text{F}}_{i\uparrow n\downarrow i\downarrow n\uparrow}\big(z_1, z_2\big) = G^{\text{F}}_{i\downarrow n\uparrow i\uparrow n\downarrow}\big(z_1, z_2\big).
\end{aligned}
\tag{403}
$$

The correlation selfenergy reads

$$
\Sigma^{\text{eh,f},1/2}_{i\downarrow(\uparrow)j\downarrow(\uparrow)}\big(z_1, z_2\big) = \mathrm{i}\hbar G_{i\uparrow(\downarrow)j\uparrow(\downarrow)}\big(z_1, z_2\big) T^{\text{eh,f},1/2}_{ij}\big(z_1, z_2\big).
\tag{404}
$$

The first terms of the diagrammatic representation are shown in Fig. 35.

The greater/less components of the particle–hole $T$ matrix are

$$
\begin{aligned}
T^{\text{eh,f},1/2,\gtrless}_{ij}\big(t_1, t_2\big) &= \Phi^{\text{eh,f},1/2,\gtrless}_{ij}\big(t_1, t_2\big) - \\
&\quad - U\big(t_1\big)\bigg(\int_{t_0}^{t_1} \mathrm{d}t_5 \sum_n G^{\text{F},\mathcal{R}}_{in}\big(t_1, t_5\big) T^{\text{eh,f},1/2,\gtrless}_{nj}\big(t_5, t_2\big) + \\
&\quad \int_{t_0}^{t_2} \mathrm{d}t_5 \sum_n G^{\text{F},\gtrless}_{in}\big(t_1, t_5\big) T^{\text{eh,f},1/2,\mathcal{A}}_{nj}\big(t_5, t_2\big)\bigg),
\end{aligned}
\tag{405}
$$

and the advanced/retarded components become

$$T_{ij}^{\text{eh,f,1/2},\mathcal{A}/\mathcal{R}}\left(t_1, t_2\right) = \Phi_{ij}^{\text{eh,f,1/2},\mathcal{A}/\mathcal{R}}\left(t_1, t_2\right) - \tag{406}$$
$$- U\left(t_1\right) \int_{t_{1/2}}^{t_{2/1}} \mathrm{d}t_5 \sum_n G_{in}^{\text{F},\mathcal{A}/\mathcal{R}}\left(t_1, t_5\right) T_{nj}^{\text{eh,f,1/2},\mathcal{A}/\mathcal{R}}\left(t_5, t_2\right).$$

The greater/less components of the correlation selfenergy read

$$\Sigma_{i\downarrow(\uparrow)j\downarrow(\uparrow)}^{\text{eh,f,1/2},\gtrless}\left(t_1, t_2\right) = \mathrm{i}\hbar G_{i\uparrow(\downarrow)j\uparrow(\downarrow)}^{\gtrless}\left(t_1, t_2\right) T_{ij}^{\text{eh,f,1/2},\gtrless}\left(t_1, t_2\right).$$

With this we conclude the discussion of the $T$-matrix approximation. After considering separately the standard approximations—the particle–particle and particle–hole $T$ matrix, we briefly mention the limitations and possible extensions. The present approximations were based on the static pair interaction. While we took into account multiple scattering processes to all orders, on the other hand, dynamical-screening effects (as described by the $GW$ approximation in Sec. 5.2), have been neglected completely. An approximate combination of dynamical-screening and strong-coupling effects is, therefore, considered in the next section.

### 5.5. Fluctuating-exchange approximation (FLEX)

The idea to combine strong-coupling and dynamical-screening effects goes back several decades. A discussion in the frame of Green functions is summarized in Ref. [177]. An alternative approach has been presented within density-operator theory. The solution for the pair-correlation operator that includes both, ladder and polarization terms leads to the screened-ladder approximation, e.g. Ref. [131]. However, implementing these approximations for nonequilibrium situations is presently not computationally feasible. Therefore, it is reasonable to employ a simpler approach where contributions of both approximations are taken into account approximately. This idea was first realized for classical plasmas by Gould and DeWitt [178]. They had the idea to simultaneously include strong-coupling and dynamical-screening effects in a kinetic equation by simply adding the Boltzmann (B) and Lenard–Balescu (LB) collision integrals,

$$I^{\text{GDW}} = I^{\text{B}} + I^{\text{LB}} - I^{\text{L}}, \tag{407}$$

where the Boltzmann collision integral includes the entire Born series and Lenard–Balescu the entire ring-diagram sum [174, 175]. In the Green functions language the former corresponds to the $T$-matrix approximation and the latter to $GW$. Subtraction of the Landau integral, $I^{\mathrm{L}}$, is necessary to avoid double counting of terms. The Landau integral corresponds to the static second-Born approximation (collision integrals of second order in the pair potential) which are contained (as the lowest iteration orders) in both, the $T$ matrix and the dynamically screened potential, see Ref. [179] for a recent discussion and further references.

Extension of this idea to quantum systems directly leads to the fluctuating-exchange approximation (FLEX). The idea behind FLEX is to construct an approximation that includes both flavors of the $T$ matrix as well as the $GW$ approximation thereby neglecting cross-terms that mix the three different approximations. To avoid double counting, the common to all three second-order terms are subtracted twice in the correlation contribution. Thereby, the resulting FLEX selfenergy becomes

$$\begin{aligned}
\Sigma^{\mathrm{FLEX}} &= \Sigma^{\mathrm{H}} + \Sigma^{\mathrm{F}} + \Sigma^{\mathrm{FLEX,corr}} . \\
\Sigma^{\mathrm{FLEX,corr}} &= \Sigma^{GW,\mathrm{corr}} + \Sigma^{\mathrm{pp}} + \Sigma^{\mathrm{eh}} - 2\Sigma^{(2)} .
\end{aligned} \tag{408}$$

where Eq. (408) directly corresponds to the Gould–DeWitt approach, Eq. (407). This scheme can be applied in an arbitrary basis representation. The diagrammatic representation of the leading terms for a diagonal basis, and for spin-0 bosons as well as spin-1/2 fermions in the Hubbard basis are shown in Figs. 36 and 37 and Fig. 38, respectively.

We have implemented this scheme for a fermionic Hubbard basis and found excellent performance. Numerical results for the ground-state properties and for nonequilibrium dynamics were presented in Sec. 3 and confirm that this is a powerful and highly accurate approximation.

## 6. Discussion and outlook

***Summary of numerical results***.   In this article, an overview of recent progress in the dynamics of correlated fermions out of equilibrium has been given. The theoretical framework in the focus was nonequilibrium (real-time) Green functions that were introduced 55 years ago by Keldysh, in the Soviet Union, and Baym and Kadanoff, in the U.S. For more than two decades the method of NEGF was primarily a tool to systematically derive Boltzmann-type quantum-kinetic equations and improvements thereof. Only after the work of Danielewicz two decades later [180] it became a practical option to use the NEGF technique for numerical simulations. However, the computational effort that exceeds that of other many-body approaches by far, remained a major obstacle. The next step forward occured in the second half of the 1990s where important effects in semiconductor optics and transport, in nuclear matter and laser plasmas could be explained for the first time by using NEGF methods, for details see the text books [53, 54, 131, 181]. Not only were new approximations derived but also the number of groups that attempted numerical solutions increased rapidly.

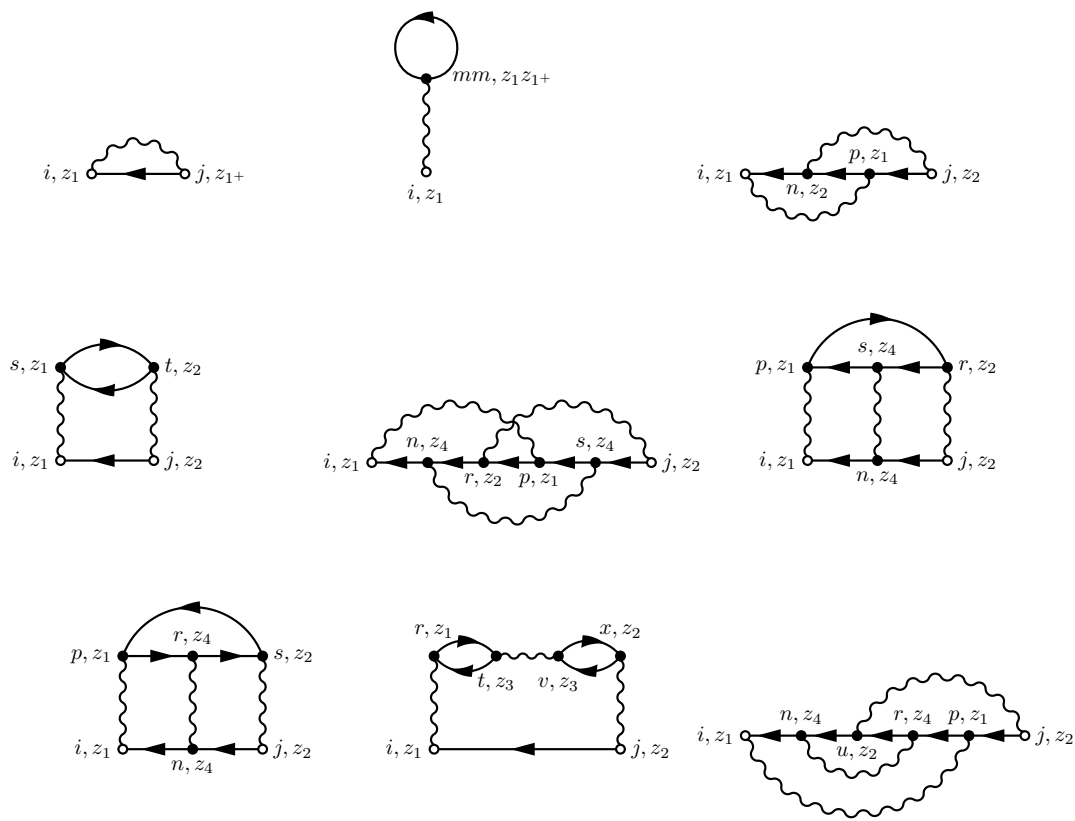

**Figure 36.** Leading terms of $\Sigma^{\text{FLEX,diag}}$ in a diagonal basis.

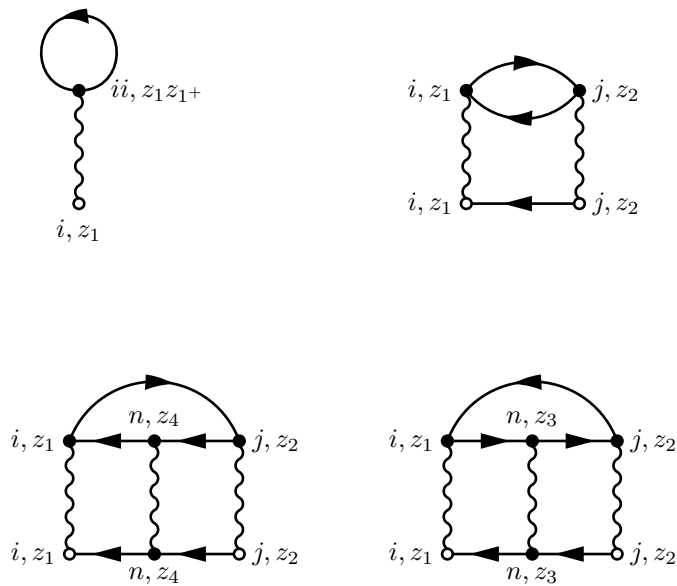

**Figure 37.** Leading terms of $\Sigma^{\text{FLEX,b},0}$ in the Hubbard basis for spin-0 bosons.

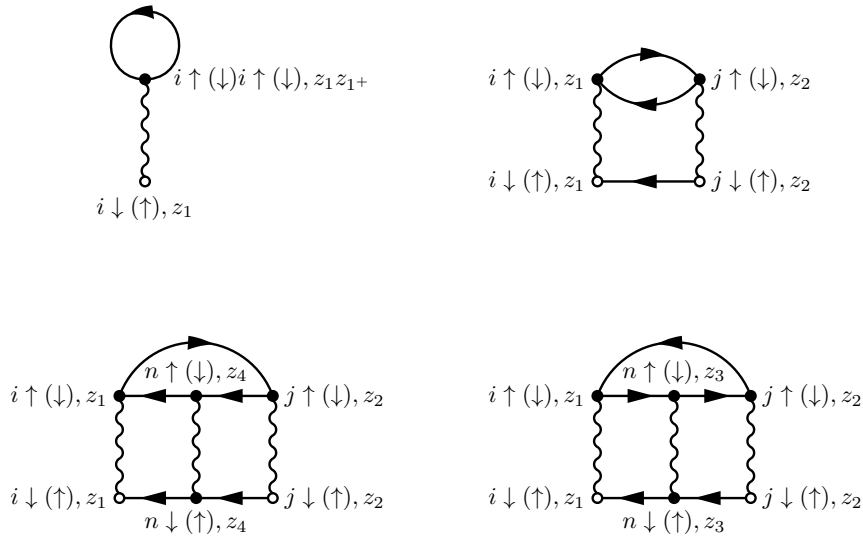

**Figure 38.** Leading terms of $\Sigma^{\mathrm{FLEX,f},1/2}$ in the Hubbard basis for spin-1/2 fermions.

The next spike of activity came 10 years later when NEGF methods were first applied to finite spatially inhomogenous systems including electrons in atoms, molecules or quantum dots [70, 72, 182]. NEGF simulations with second-order Born selfenergies (SOA) were able to reproduce the qualitative features of the excitation and ionization dynamics of optically excited few-electron systems. Another new application concerned finite Hubbard-type lattice models [76, 117, 118]. The simplicity of the basis allowed, for the first time, to systematically study strongly correlated systems in nonequilibrim with selfenergies beyond the simple second-Born approximation. This has allowed for NEGF applications in the fields of strongly correlated solids and cold atoms [26, 77, 145].

Still it remained unclear what level of accuracy NEGF simulations can provide and whether they are an approach that is competitive with other many-body methods. This question was answered in a series of papers where small Hubbard clusters were treated with NEGF simulations that could be compared to exact-diagonalization calculations [76, 151]. Recently the first systematic test of various selfenergy approximations was reported by two of the present authors by comparing to DMRG benchmark simulations for 1D

Hubbard systems [50]. The result was that, indeed, NEGF simulations are a highly accurate tool if, in each case, the proper selfenergy approximation is being used. The choice is dictated, primarily, by the coupling strength and the particle density (filling factor). With that NEGF simulations have reached the level of a predictive tool where the main observables can be accurately computed—with an error not exceeding the order of $10 \dots 20\%$. This allows extensions of the system size, system geometry and the simulation duration to situations that are out of the reach of alternative methods such as CI, DMRG or real-time quantum Monte Carlo.

Thus, NEGF simulations have the potential of becoming a broadly used tool in many fields of physics including atomic and molecular physics, condensed-matter physics, nuclear matter, warm dense matter and cold atomic and molecular gases. This requires not only codes that efficiently solve the Keldysh–Kadanoff–Baym equations but also a quick, easy and reliable use of the entire arsenal of selfenergy approximations. The present article attempted to pave the way for such applications focusing on the latter task.

To this end, we gave a short selfcontained introduction to the theory of NEGF in Sec. 2 and an overview on recent numerical results for the dynamics of finite Hubbard systems in Sec. 3. There a variety of selfenergy approximations was used that included, in addition to the commonly used Hartree–Fock and second-order Born approximation (SOA), also the third-order approximation (TOA), the particle–particle and particle–hole $T$ matrices (TPP, TPH), the $GW$ and the FLEX approximations, and their accuracy was investigated. The first tests concerned the interacting ground-state properties, in particular, the ground-state energy and the spectral function of small Hubbard clusters of varying coupling strength and filling. The best results were obtained for the perturbative approaches, i.e., the second-order and especially the third-order approximations, the latter being applied for the first time to the Hubbard model, in this article. For small and medium fillings, the TOA ground-state energies were, by far, the most accurate, and the TOA outperformed all previously applied approximations. For half filling, though, the TOA exactly agrees with the SOA due to particle–hole symmetry and, consequently, yields no improvement over it.

We then analyzed the spectral properties of small Hubbard clusters by computing the single-particle spectral function via the solution of the full two-time KBE. Here, the best results were achieved with the FLEX approximation, for all filling factors. Additionally, the TOA shows single-particle energy transitions which are not contained in any other approximation. Thus, by taking into account the full set of approximation methods, most of the relevant energy levels of the systems differing by one particle from the analyzed system can be determined.

Finally, the performance of the different selfenergies in various time-dependent setups, was investigated. We showed results for the time evolution following a confinement quench that was motivated by recent experiments with fermionic atoms. The NEGF results using the particle–particle $T$-matrix selfenergy agreed very well with the experiment. The second setup used a one-dimensional charge-density-wave state as initial state. Here, by comparison with DMRG simulations the best results were obtained for third-order selfenergies. The third setup were finite graphene-type honeycomb clusters that were exposed to the impact of energetic ions. Here we compared, among others, second-order Born selfenergies and HF-GKBA simulations. The fourth type of excitation was a rapid weak change of the lattice potential. The subsequent (linear-response) dynamics allowed to compute the spectral function of the system.

Finally, we considered a strong excitation of the system where the lattice potential of one site was changed. Here previous studies have indicated problems in the dynamics of the NEGF, in particular, a strongly damped behavior that is absent in exact-diagonalization calculations [118]. This behavior was found to vary strongly for different selfenergy approximations. We also compared two-time and single-time simulations within the HF-GKBA. The latter almost completely removed the damping.

***Summary of Selfenergy approximations***.   In the second part of this review, a detailed overview of approximation strategies for the single-particle selfenergy within the framework of nonequilibrium Green functions has been presented. Here we followed two strategies. The first is a perturbative approach with respect to the interaction strength that, i.a., gives rise to the Born series. After reproducing the familiar and broadly used second-order Born approximation we derived the third-order selfenergy (TOA). This approximation contains all terms that are of third order in the interaction strength including, in particular, the relevant ladder-type and polarization diagrams. This important new approximation was derived for all relevant system types, starting from a general basis representation. Additionally, for the special cases of a basis, in which the interaction is diagonal, as well as for the fermionic and bosonic Hubbard basis with a scalar interaction, the corresponding selfenergy terms with Keldysh components have been derived.

The second strategy was a non-perturbative approach where the selfenergies are derived using resummations of infinite partial series. Here our starting point was the set of Hedin's equations from which we derived the $GW$ approximation. The closely related set of integral equations for the bare interaction and vertex led us to the particle–particle and the particle–hole $T$ matrix. In each case we presented all details of the formulas that are needed for an efficient numerical implementation. As before, the results were summarized for general basis sets, for the case of a diagonal potential and for the Hubbard basis. The presentation was concluded by a discussion of the fluctuating-exchange (FLEX) approximation that involves the combination of the terms from both $T$ matrices and $GW$.

***Outlook.*** With the set of selfenergies that were introduced in this review a powerful arsenal of approximations is available. In addition to two-time results that are obtained from the full Keldysh–Kadanoff–Baym equations, also the single-time version is available. Here the basis is the Hartree–Fock-GKBA that, in many cases was found to be complementary to the KBE. The HF-GKBA removes the artificial damping for two-time simulations but does not yield spectral information. It has been successfully combined with most of the selfenergies discussed in this paper, which confirms the attractive properties of this approximation. Moreover, comparisons with DMRG benchmark data indicated that the exact result is typically enclosed between KBE and HF-GKBA. This means that, if both simulations are performed independently, accurate predictions are possible even in the absence of benchmark data.

An important further development will consist in combining the GKBA with propagators beyond Hartree–Fock, i.e. correlated propagators [24]. This is expected to improve the spectral content of single-time simulations and bringing the GKBA simulations closer to the exact result.

The presently available selfenergies have been found to work well for arbitrary filling parameters and weakly to moderately correlated Fermi systems, within the Hubbard model this corresponds to $U/J \lesssim 8$. For larger couplings the present implementation failed to converge. Here it will be important to attempt modified implementations to extend the range of accessible coupling strengths. In addition, it is of high interest to derive higher-order selfenergy approximations, along the lines outlined in this article. For example, for half filling, the implementation of the fourth-order terms, following the algorithm presented in Section 4.4, could achieve significant improvements, as was done successfully for homogeneous systems by Gebhard *et. al.* in Ref. [138]. One important obstacle to the application of the fourth-order terms, though, is the numerical downside of a quartic scaling with the propagation time, which will probably limit the applicability to small system sizes and short propagation times. If achievable, though, it opens the way to many new approximation strategies and resummations starting from the fourth-order terms similar to the (third-order-starting) $T$ matrices. Furthermore, the equations for the screened interaction $W$, cf. Eq. (104), and the polarizability $P$, cf. Eq. (105), can be solved exactly in fourth order for a given approximation of the vertex function.

# Appendices

## A. Derivations in full notation

### A.1. Second-order selfenergy contributions

### A.1.1. Direct second-order selfenergy

The first second-order selfenergy term involves $W^{(2)}$, which structurally is given by, cf. Eq. (102),

$$W^{(2)} = W^{\mathrm{ns}}\Big(P^{(0)}, W^{(1)}\Big). \tag{409}$$

The structure of the zeroth-order term of the polarization is

$$P^{(0)} = P\Big(\Gamma^{(0)}\Big). \tag{410}$$

Thus, it is given by

$$P^{(0)}_{ijkl}\Big(z_1, z_2\Big) = \pm \mathrm{i}\hbar \int_{\mathcal{C}} \mathrm{d}z_3 \sum_m G_{im}\Big(z_1, z_3\Big) \tag{411}$$

$$\int_{\mathcal{C}} \mathrm{d}z_4 \sum_n G_{nl}\Big(z_4, z_1\Big) \delta_{\mathcal{C}}\Big(z_3, z_{4+}\Big) \delta_{\mathcal{C}}\Big(z_2, z_4\Big) \delta_{mk} \delta_{jn}$$

$$= \pm \mathrm{i}\hbar G_{ik}\Big(z_1, z_2\Big) G_{jl}\Big(z_2, z_1\Big). $$

Inserting this result into Eq. (409), one arrives at

$$W^{(2)}_{ijkl}\Big(z_1, z_2\Big) = \sum_{mn} w_{imnl}\Big(z_1\Big) \tag{412}$$

$$\int_{\mathcal{C}} \mathrm{d}z_3 \sum_{pq} \Big(\pm \mathrm{i}\hbar G_{np}\Big(z_1, z_3\Big) G_{qm}\Big(z_3, z_1\Big)\Big) \delta_{\mathcal{C}}\Big(z_3, z_2\Big) w_{pjkq}\Big(z_3\Big)$$

and, employing Eq. (104), finally, one has

$$
\begin{aligned}
W_{ijkl}^{(2)}&\big(z_1, z_2\big) \\
&= \pm i\hbar \sum_{mn} w_{imnl}\big(z_1\big) \sum_{pq} G_{np}\big(z_1, z_2\big) G_{qm}\big(z_2, z_1\big) w_{pjkq}\big(z_2\big) .
\end{aligned}
\tag{413}
$$

With this, $\Sigma^{(2),2,0}$ can be calculated as, cf. Eq. (101),

$$
\begin{aligned}
\Sigma_{ij}^{(2),2,0}&\big(z_1, z_2\big) \\
&= i\hbar \int_{\mathcal{C}} \mathrm{d}z_3 \sum_{mpq} \bigg( \pm i\hbar \sum_{rs} w_{irsm}\big(z_1\big) \sum_{tu} G_{st}\big(z_1, z_3\big) G_{ur}\big(z_3, z_1\big) w_{tpqu}\big(z_3\big) \bigg) \\
&\qquad \int_{\mathcal{C}} \mathrm{d}z_4 \sum_n G_{mn}\big(z_1, z_4\big) \delta_{\mathcal{C}}\big(z_4, z_{2^+}\big) \delta_{\mathcal{C}}\big(z_3, z_2\big) \delta_{np} \delta_{qj} .
\end{aligned}
\tag{414}
$$

Evaluating the terms, one arrives at

$$
\begin{aligned}
\Sigma_{ij}^{(2),2,0}&\big(z_1, z_2\big) \\
&= \pm \big(i\hbar\big)^2 \sum_{mn} G_{mn}\big(z_1, z_2\big) \sum_{st} G_{st}\big(z_1, z_2\big) \\
&\qquad \sum_r w_{irsm}\big(z_1\big) \sum_u w_{tnju}\big(z_2\big) G_{ur}\big(z_2, z_1\big) .
\end{aligned}
\tag{415}
$$

### A.1.2. Exchange–correlation second-order selfenergy

The other second-order selfenergy term, $\Sigma_{ij}^{(2),1,1}$, requires the first-order term of the vertex $\Gamma$, the structure of which is

$$
\Gamma^{(1)} = \Gamma\big(\delta\Sigma^{\mathrm{xc},(1)}/\delta G, \Gamma^{(0)}\big) .
\tag{416}
$$

This term involves the functional derivative of $\Sigma^{\mathrm{xc},(1)}$ with respect to $G$. One has

$$
\frac{\delta\Sigma_{ij}^{\mathrm{xc},(1)}\big(z_1, z_2\big)}{\delta G_{rs}\big(z_5, z_6\big)} = \frac{\delta\Sigma_{ij}^{\mathrm{xc},(1),\mathrm{F}}\big(z_1, z_2\big)}{\delta G_{rs}\big(z_5, z_6\big)} .
\tag{417}
$$

Employing Eq. (156), one finds

$$
\begin{aligned}
\frac{\delta\Sigma_{ij}^{\mathrm{xc},(1)}\big(z_1, z_2\big)}{\delta G_{rs}\big(z_5, z_6\big)} &= i\hbar\delta_{\mathcal{C}}\big(z_1, z_2\big) \sum_{mn} w_{injm}\big(z_1\big) \frac{\delta G_{mn}\big(z_1, z_{1^+}\big)}{\delta G_{rs}\big(z_5, z_6\big)} \\
&= i\hbar\delta_{\mathcal{C}}\big(z_1, z_2\big) \delta_{\mathcal{C}}\big(z_1, z_5\big) \delta_{\mathcal{C}}\big(z_1, z_6\big) w_{isjr}\big(z_1\big) ,
\end{aligned}
\tag{418}
$$

where

$$
\frac{\delta G_{ij}\big(z_1, z_2\big)}{\delta G_{mn}\big(z_5, z_6\big)} = \delta_{\mathcal{C}}\big(z_1, z_5\big) \delta_{\mathcal{C}}\big(z_2, z_6\big) \delta_{im} \delta_{jn}
\tag{419}
$$

has been applied. With this,

$$
\Gamma^{(1)}_{ijkl}\big(z_1, z_2, z_3\big) \tag{420}
$$
$$
= \int_{\mathcal{C}} \mathrm{d}z_4 \mathrm{d}z_5 \sum_{mn} \frac{\delta \Sigma^{\mathrm{xc},(1)}_{il}\big(z_1, z_2\big)}{\delta G_{mn}\big(z_4, z_5\big)} \int_{\mathcal{C}} \mathrm{d}z_6 \sum_p G_{mp}\big(z_4, z_6\big)
$$
$$
\int_{\mathcal{C}} \mathrm{d}z_7 \sum_q G_{qn}\big(z_7, z_5\big) \Gamma^{(0)}_{pjkq}\big(z_6, z_7, z_3\big) .
$$

Using Eq. (418), one has

$$
\Gamma^{(1)}_{ijkl}\big(z_1, z_2, z_3\big) \tag{421}
$$
$$
= \mathrm{i}\hbar \delta_{\mathcal{C}}\big(z_1, z_2\big) \sum_{mn} w_{inlm}(z_1) \int_{\mathcal{C}} \mathrm{d}z_6 \sum_p G_{mp}\big(z_1, z_6\big)
$$
$$
\int_{\mathcal{C}} \mathrm{d}z_7 \sum_q G_{qn}\big(z_7, z_1\big) \Gamma^{(0)}_{pjkq}\big(z_6, z_7, z_3\big) .
$$

With Eq. (155), finally,

$$
\Gamma^{(1)}_{ijkl}\big(z_1, z_2, z_3\big) \tag{422}
$$
$$
= \mathrm{i}\hbar \delta_{\mathcal{C}}\big(z_1, z_2\big) \sum_{mn} w_{inlm}(z_1) G_{mk}\big(z_1, z_3\big) G_{jn}\big(z_3, z_1\big)
$$

ensues. Inserting this result yields

$$
\Sigma^{(2),1,1}_{ij}\big(z_1, z_2\big) \tag{423}
$$
$$
= \mathrm{i}\hbar \int_{\mathcal{C}} \mathrm{d}z_3 \sum_{mpq} W^{(1)}_{ipqm}\big(z_1, z_3\big) \int_{\mathcal{C}} \mathrm{d}z_4 \sum_n G_{mn}\big(z_1, z_4\big) \Gamma^{(1)}_{nqpj}\big(z_4, z_2, z_3\big) .
$$

Employing Eqs. (154) and (422), one arrives at

$$
\Sigma^{(2),1,1}_{ij}\big(z_1, z_2\big) \tag{424}
$$
$$
= \mathrm{i}\hbar \sum_{mpq} w_{ipqm}(z_1) \int_{\mathcal{C}} \mathrm{d}z_4 \sum_n G_{mn}\big(z_1, z_4\big) \Gamma^{(1)}_{nqpj}\big(z_4, z_2, z_1\big)
$$
$$
= \big(\mathrm{i}\hbar\big)^2 \sum_{mpq} w_{ipqm}(z_1) \sum_n G_{mn}\big(z_1, z_2\big) \sum_{rs} w_{nsjr}(z_2) G_{rp}\big(z_2, z_1\big) G_{qs}\big(z_1, z_2\big) .
$$

## A.2. Third-order selfenergy contributions

### A.2.1. Third-order term: $\Sigma^{(3),\{3;0,2\},0}_{ij}$

Using Eq. (104), one finds

$$
W^{(3),0,2}_{ijkl}\big(z_1, z_2\big) \tag{425}
$$
$$
= \sum_{mn} w_{imnl}(z_1) \int_{\mathcal{C}} \mathrm{d}z_3 \sum_{pq} \big(\pm \mathrm{i}\hbar G_{np}\big(z_1, z_3\big) G_{qm}\big(z_3, z_1\big)\big) W^{(2)}_{pjkq}\big(z_3, z_2\big) .
$$

Employing Eq. (412) yields

$$W_{ijkl}^{(3),0,2}(z_1, z_2) \tag{426}$$
$$= \sum_{mn} w_{imnl}(z_1) \int_{\mathcal{C}} \mathrm{d}z_3 \sum_{pq} \Big( \pm \mathrm{i}\hbar G_{np}(z_1, z_3) G_{qm}(z_3, z_1) \Big)$$
$$\left( \pm \mathrm{i}\hbar \sum_{rs} w_{prsq}(z_3) \sum_{tu} G_{st}(z_3, z_2) G_{ur}(z_2, z_3) w_{tjku}(z_2) \right).$$

Evalutating and reordering, one has

$$W_{ijkl}^{(3),0,2}(z_1, z_2) \tag{427}$$
$$= (\mathrm{i}\hbar)^2 \sum_{mn} w_{imnl}(z_1) \int_{\mathcal{C}} \mathrm{d}z_3 \sum_{pq} G_{np}(z_1, z_3) G_{qm}(z_3, z_1) \sum_{rs} w_{prsq}(z_3)$$
$$\sum_{tu} G_{st}(z_3, z_2) G_{ur}(z_2, z_3) w_{tjku}(z_2).$$

With this, the first term of the first third-order selfenergy class, $\Sigma^{(3),3,0}$, becomes

$$\Sigma_{ij}^{(3),\{3;0,2\},0}(z_1, z_2) \tag{428}$$
$$= \mathrm{i}\hbar \int_{\mathcal{C}} \mathrm{d}z_3 \sum_{mpq} W_{ipqm}^{(3),0,2}(z_1, z_3) \int_{\mathcal{C}} \mathrm{d}z_4 \sum_n G_{mn}(z_1, z_4) \Gamma_{nqpj}^{(0)}(z_4, z_2, z_3).$$

Using Eq. (155), one has

$$\Sigma_{ij}^{(3),\{3;0,2\},0}(z_1, z_2) \tag{429}$$
$$= (\mathrm{i}\hbar)^3 \sum_{mn} G_{mn}(z_1, z_2) \sum_{rs} w_{irsm}(z_1) \int_{\mathcal{C}} \mathrm{d}z_3 \sum_{tu} G_{st}(z_1, z_3) G_{ur}(z_3, z_1)$$
$$\sum_{vw} w_{tvwu}(z_3) \sum_{xy} G_{wx}(z_3, z_2) G_{yv}(z_2, z_3) w_{xnjy}(z_2).$$

*A.2.2. Third-order term: $\Sigma_{ij}^{(3),\{3;1,1\},0}$*

For the second class of the interaction, $W^{(3),1,1}$, the first-order contribution to the polarizability is needed, which is given by, cf. Eqs. (105) and (422),

$$P_{ijkl}^{(1)}(z_1, z_2) = P_{ijkl}(z_1, z_2)(\Gamma^{(1)}) \tag{430}$$
$$= \pm \mathrm{i}\hbar \int_{\mathcal{C}} \mathrm{d}z_3 \sum_m G_{im}(z_1, z_3) \int_{\mathcal{C}} \mathrm{d}z_4 \sum_n G_{nl}(z_4, z_1) \Gamma_{mjkn}^{(1)}(z_3, z_4, z_2).$$

Employing Eq. (422), one arrives at

$$P_{ijkl}^{(1)}\big(z_1,z_2\big) \tag{431}$$
$$= \pm\big(\mathrm{i}\hbar\big)^2 \int_{\mathcal{C}}\mathrm{d}z_3 \sum_m G_{im}\big(z_1,z_3\big) \sum_n G_{nl}\big(z_3,z_1\big)$$
$$\sum_{pq} w_{mqnp}\big(z_3\big)G_{pk}\big(z_3,z_2\big)G_{jq}\big(z_2,z_3\big)\,.$$

Inserting this result back, one finds, using Eq. (104),

$$W_{ijkl}^{(3),1,1}\big(z_1,z_2\big) = \sum_{mn} w_{imnl}\big(z_1\big) \int_{\mathcal{C}}\mathrm{d}z_3 \sum_{pq} P_{nqpm}^{(1)}\big(z_1,z_3\big)W_{pjkq}^{(1)}\big(z_3,z_2\big)$$
$$= \sum_{mn} w_{imnl}\big(z_1\big) \int_{\mathcal{C}}\mathrm{d}z_3 \sum_{pq}\Big(\pm\big(\mathrm{i}\hbar\big)^2 \int_{\mathcal{C}}\mathrm{d}z_4 \sum_r G_{nr}\big(z_1,z_4\big) \sum_s G_{sm}\big(z_4,z_1\big)$$
$$\sum_{tu} w_{rust}\big(z_4\big)G_{tp}\big(z_4,z_3\big)G_{qu}\big(z_3,z_4\big)\Big)\delta_{\mathcal{C}}\big(z_3,z_2\big)w_{pjkq}\big(z_2\big)\,. \tag{432}$$

After reordering, one has

$$W_{ijkl}^{(3),1,1}\big(z_1,z_2\big) \tag{433}$$
$$= \pm\big(\mathrm{i}\hbar\big)^2 \sum_{mn} w_{imnl}\big(z_1\big) \int_{\mathcal{C}}\mathrm{d}z_3 \sum_r G_{nr}\big(z_1,z_3\big) \sum_s G_{sm}\big(z_3,z_1\big)$$
$$\sum_{tu} w_{rust}\big(z_3\big) \sum_{pq} G_{tp}\big(z_3,z_2\big)G_{qu}\big(z_2,z_3\big)w_{pjkq}\big(z_2\big)\,.$$

With these results, the second term of the class $\Sigma^{(3),3,0}$ is found, using Eq. (101),

$$\Sigma_{ij}^{(3),\{3;1,1\},0}\big(z_1,z_2\big) \tag{434}$$
$$= \mathrm{i}\hbar \int_{\mathcal{C}}\mathrm{d}z_3 \sum_{mpq} W_{ipqm}^{(3),1,1}\big(z_1,z_3\big) \int_{\mathcal{C}}\mathrm{d}z_4 \sum_n G_{mn}\big(z_1,z_4\big)\Gamma_{nqpj}^{(0)}\big(z_4,z_2,z_3\big)\,.$$

Using Eq. (155), one finds

$$\Sigma_{ij}^{(3),\{3;1,1\},0}\big(z_1,z_2\big) \tag{435}$$
$$= \pm\big(\mathrm{i}\hbar\big)^3 \sum_{mn} G_{mn}\big(z_1,z_2\big) \sum_{rs} w_{irsm}\big(z_1\big) \int_{\mathcal{C}}\mathrm{d}z_3 \sum_t G_{st}\big(z_1,z_3\big) \sum_u G_{ur}\big(z_3,z_1\big)$$
$$\sum_{vw} w_{twuv}\big(z_3\big) \sum_{xy} G_{vx}\big(z_3,z_2\big)G_{yw}\big(z_2,z_3\big)w_{xnjy}\big(z_2\big)\,.$$

### A.2.3. Third-order term: $\Sigma_{ij}^{(3),2,1}$

Continuing with the second class $\Sigma^{(3),2,1}$, it is directly worked out by combining Eqs. (412)

and (422),

$$\Sigma_{ij}^{(3),2,1}(z_1, z_2) = i\hbar \int_{\mathcal{C}} dz_3 \sum_{mpq} W_{ipqm}^{(2)}(z_1, z_3) \tag{436}$$

$$\int_{\mathcal{C}} dz_4 \sum_n G_{mn}(z_1, z_4) \Gamma_{nqpj}^{(1)}(z_4, z_2, z_3).$$

Inserting Eq. (422) yields

$$\Sigma_{ij}^{(3),2,1}(z_1, z_2) \tag{437}$$

$$= \pm (i\hbar)^3 \int_{\mathcal{C}} dz_3 \sum_{mrs} w_{irsm}(z_1) \sum_{tu} G_{st}(z_1, z_3) G_{ur}(z_3, z_1) \sum_{pq} w_{tpqu}(z_3)$$

$$\sum_n G_{mn}(z_1, z_2) \sum_{vw} w_{nwjv}(z_2) G_{vp}(z_2, z_3) G_{qw}(z_3, z_2).$$

### A.2.4. Third-order term: $\Sigma_{ij}^{(3),1,\{2;1,1\}}$

For the single contribution to the class $\Gamma^{(2),1,1}$, one finds, employing Eqs. (418) and (422),

$$\Gamma_{ijkl}^{(2),1,1}(z_1, z_2, z_3) = \int_{\mathcal{C}} dz_4 dz_5 \sum_{mn} \frac{\delta \Sigma_{il}^{\text{xc},(1)}(z_1, z_2)}{\delta G_{mn}(z_4, z_5)} \int_{\mathcal{C}} dz_6 \sum_p G_{mp}(z_4, z_6)$$

$$\int_{\mathcal{C}} dz_7 \sum_q G_{qn}(z_7, z_5) \Gamma_{pjkq}^{(1)}(z_6, z_7, z_3). \tag{438}$$

Evaluating the derivative yields

$$\Gamma_{ijkl}^{(2),1,1}(z_1, z_2, z_3) \tag{439}$$

$$= i\hbar \delta_{\mathcal{C}}(z_1, z_2) \sum_{mn} w_{inlm}(z_1) \int_{\mathcal{C}} dz_6 \sum_p G_{mp}(z_1, z_6)$$

$$\int_{\mathcal{C}} dz_7 \sum_q G_{qn}(z_7, z_1) \Gamma_{pjkq}^{(1)}(z_6, z_7, z_3).$$

Employing Eq. (422), one finds

$$\Gamma_{ijkl}^{(2),1,1}(z_1, z_2, z_3) \tag{440}$$

$$= (i\hbar)^2 \delta_{\mathcal{C}}(z_1, z_2) \sum_{mn} w_{inlm}(z_1) \int_{\mathcal{C}} dz_6 \sum_p G_{mp}(z_1, z_6)$$

$$\sum_q G_{qn}(z_6, z_1) \sum_{rs} w_{psqr}(z_6) G_{rk}(z_6, z_3) G_{js}(z_3, z_6).$$

This enables the computation of $\Sigma^{(3),1,\{2;1,1\}}$ with Eqs. (101) and (154),

$$\Sigma^{(3),1,\{2;1,1\}}_{ij}\big(z_1,z_2\big) \tag{441}$$
$$= i\hbar \int_{\mathcal{C}} dz_3 \sum_{mpq} W^{(1)}_{ipqm}\big(z_1,z_3\big) \int_{\mathcal{C}} dz_4 \sum_{n} G_{mn}\big(z_1,z_4\big) \Gamma^{(2),1,1}_{nqpj}\big(z_4,z_2,z_3\big)$$
$$= i\hbar \sum_{mpq} w_{ipqm}\big(z_1\big) \int_{\mathcal{C}} dz_4 \sum_{n} G_{mn}\big(z_1,z_4\big) \Gamma^{(2),1,1}_{nqpj}\big(z_4,z_2,z_1\big).$$

Using Eq. (440), one arrives at

$$\Sigma^{(3),1,\{2;1,1\}}_{ij}\big(z_1,z_2\big) \tag{442}$$
$$= \big(i\hbar\big)^3 \sum_{mpq} w_{ipqm}\big(z_1\big) \sum_{n} G_{mn}\big(z_1,z_2\big) \sum_{rs} w_{nsjr}\big(z_2\big) \int_{\mathcal{C}} dz_3 \sum_{t} G_{rt}\big(z_2,z_3\big)$$
$$\sum_{u} G_{us}\big(z_3,z_2\big) \sum_{vw} w_{twuv}\big(z_3\big) G_{vp}\big(z_3,z_1\big) G_{qw}\big(z_1,z_3\big).$$

*A.2.5. Second-order vertex terms*

For the first terms, one finds

$$\Gamma^{(2),\{2;2,0\},0}_{ijkl}\big(z_1,z_2,z_3\big) = \int_{\mathcal{C}} dz_4 dz_5 \sum_{mn} \frac{\delta \Sigma^{(2),2,0}_{il}\big(z_1,z_2\big)}{\delta G_{mn}\big(z_4,z_5\big)} \int_{\mathcal{C}} dz_6 \sum_{p} G_{mp}\big(z_4,z_6\big)$$
$$\int_{\mathcal{C}} dz_7 \sum_{q} G_{qn}\big(z_7,z_5\big) \Gamma^{(0)}_{pjkq}\big(z_6,z_7,z_3\big) \tag{443}$$
$$= \int_{\mathcal{C}} dz_4 dz_5 \sum_{mn} \frac{\delta \Sigma^{(2),2,0}_{il}\big(z_1,z_2\big)}{\delta G_{mn}\big(z_4,z_5\big)} G_{mk}\big(z_4,z_3\big) G_{jn}\big(z_3,z_5\big).$$

Inserting Eq. (415), the term attains the form

$$\Gamma^{(2),\{2;2,0\},0}_{ijkl}\big(z_1,z_2,z_3\big) = \pm\big(i\hbar\big)^2 \int_{\mathcal{C}} dz_4 dz_5 \tag{444}$$
$$\sum_{mn} \frac{\delta \bigg( \sum_{pq} G_{pq}\big(z_1,z_2\big) \sum_{st} G_{st}\big(z_1,z_2\big) \sum_{r} w_{irsp}\big(z_1\big) \sum_{u} w_{tqlu}\big(z_2\big) G_{ur}\big(z_2,z_1\big) \bigg)}{\delta G_{mn}\big(z_4,z_5\big)}$$
$$G_{mk}\big(z_4,z_3\big) G_{jn}\big(z_3,z_5\big).$$

Evaluating the derivative, one has

$$\Gamma^{(2),\{2;2,0\},0}_{ijkl}\big(z_1,z_2,z_3\big) = \Gamma^{(2),\{2;2,0\},0,A}_{ijkl}\big(z_1,z_2,z_3\big) \tag{445}$$
$$+ \Gamma^{(2),\{2;2,0\},0,B}_{ijkl}\big(z_1,z_2,z_3\big) + \Gamma^{(2),\{2;2,0\},0,C}_{ijkl}\big(z_1,z_2,z_3\big),$$

with

$$\Gamma_{ijkl}^{(2),\{2;2,0\},0,\mathrm{A}}\big(z_1,z_2,z_3\big) \tag{446}$$

$$= \pm\big(\mathrm{i}\hbar\big)^2 \sum_{mn}\sum_{st} G_{st}\big(z_1,z_2\big) \sum_r w_{irsm}\big(z_1\big)$$

$$\sum_u w_{tnlu}\big(z_2\big) G_{ur}\big(z_2,z_1\big) G_{mk}\big(z_1,z_3\big) G_{jn}\big(z_3,z_2\big)$$

and

$$\Gamma_{ijkl}^{(2),\{2;2,0\},0,\mathrm{B}}\big(z_1,z_2,z_3\big) \tag{447}$$

$$\pm\big(\mathrm{i}\hbar\big)^2 \sum_{mn}\sum_{pq} G_{pq}\big(z_1,z_2\big) \sum_r w_{irmp}\big(z_1\big)$$

$$\sum_u w_{nqlu}\big(z_2\big) G_{ur}\big(z_2,z_1\big) G_{mk}\big(z_1,z_3\big) G_{jn}\big(z_3,z_2\big)$$

as well as

$$\Gamma_{ijkl}^{(2),\{2;2,0\},0,\mathrm{C}}\big(z_1,z_2,z_3\big) \tag{448}$$

$$\pm\big(\mathrm{i}\hbar\big)^2 \sum_{mn}\sum_{pq} G_{pq}\big(z_1,z_2\big) \sum_{st} G_{st}\big(z_1,z_2\big)$$

$$w_{insp}\big(z_1\big) w_{tqlm}\big(z_2\big) G_{mk}\big(z_2,z_3\big) G_{jn}\big(z_3,z_1\big)\,.$$

Similarly, one finds

$$\Gamma_{ijkl}^{(2),\{2;1,1\},0}\big(z_1,z_2,z_3\big) = \int_{\mathcal{C}} \mathrm{d}z_4\mathrm{d}z_5 \sum_{mn} \frac{\delta\Sigma_{il}^{(2),1,1}\big(z_1,z_2\big)}{\delta G_{mn}\big(z_4,z_5\big)} \int_{\mathcal{C}} \mathrm{d}z_6 \sum_p G_{mp}\big(z_4,z_6\big)$$

$$\int_{\mathcal{C}} \mathrm{d}z_7 \sum_q G_{qn}\big(z_7,z_5\big) \Gamma_{pjkq}^{(0)}\big(z_6,z_7,z_3\big)\,. \tag{449}$$

Inserting Eq. (424) yields

$$\Gamma_{ijkl}^{(2),\{2;1,1\},0}\big(z_1,z_2,z_3\big) = \big(\mathrm{i}\hbar\big)^2 \int_{\mathcal{C}} \mathrm{d}z_4\mathrm{d}z_5 \tag{450}$$

$$\sum_{mn} \frac{\delta\left(\sum_{prs} w_{irsp}\big(z_1\big) \sum_q G_{pq}\big(z_1,z_2\big) \sum_{tu} w_{qult}\big(z_2\big) G_{tr}\big(z_2,z_1\big) G_{su}\big(z_1,z_2\big)\right)}{\delta G_{mn}\big(z_4,z_5\big)}$$

$$G_{mk}\big(z_4,z_3\big) G_{jn}\big(z_3,z_5\big)\,,$$

which, after evaluation of the derivative, yields

$$\Gamma_{ijkl}^{(2),\{2;1,1\},0}\big(z_1,z_2,z_3\big) = \Gamma_{ijkl}^{(2),\{2;1,1\},0,\mathrm{A}}\big(z_1,z_2,z_3\big) \tag{451}$$

$$+ \Gamma_{ijkl}^{(2),\{2;1,1\},0,\mathrm{B}}\big(z_1,z_2,z_3\big) + \Gamma_{ijkl}^{(2),\{2;1,1\},0,\mathrm{C}}\big(z_1,z_2,z_3\big)\,,$$

with

$$\Gamma_{ijkl}^{(2),\{2;1,1\},0,\mathrm{A}}\big(z_1, z_2, z_3\big) \tag{452}$$

$$= \big(\mathrm{i}\hbar\big)^2 \sum_{mn} \sum_{rs} w_{irsm}\big(z_1\big) \sum_{tu} w_{nult}\big(z_2\big) G_{tr}\big(z_2, z_1\big)$$

$$G_{su}\big(z_1, z_2\big) G_{mk}\big(z_1, z_3\big) G_{jn}\big(z_3, z_2\big)$$

and

$$\Gamma_{ijkl}^{(2),\{2;1,1\},0,\mathrm{B}}\big(z_1, z_2, z_3\big) \tag{453}$$

$$+ \big(\mathrm{i}\hbar\big)^2 \sum_{mn} \sum_{ps} w_{insp}\big(z_1\big) \sum_{q} G_{pq}\big(z_1, z_2\big) \sum_{u} w_{qulm}\big(z_2\big)$$

$$G_{su}\big(z_1, z_2\big) G_{mk}\big(z_2, z_3\big) G_{jn}\big(z_3, z_1\big)$$

as well as

$$\Gamma_{ijkl}^{(2),\{2;1,1\},0,\mathrm{C}}\big(z_1, z_2, z_3\big) \tag{454}$$

$$+ \big(\mathrm{i}\hbar\big)^2 \sum_{mn} \sum_{pr} w_{irmp}\big(z_1\big) \sum_{q} G_{pq}\big(z_1, z_2\big) \sum_{t} w_{qnlt}\big(z_2\big)$$

$$G_{tr}\big(z_2, z_1\big) G_{mk}\big(z_1, z_3\big) G_{jn}\big(z_3, z_2\big).$$

### A.2.6. Third-order terms: $\Sigma_{ij}^{(3),1,2}$

With this result, the corresponding selfenergy terms can be computed,

$$\Sigma_{ij}^{(3),1,\{2;\{2;2,0\},0,\mathrm{A}\}}\big(z_1, z_2\big) \tag{455}$$

$$= \mathrm{i}\hbar \int_{\mathcal{C}} \mathrm{d}z_3 \sum_{mpq} W_{ipqm}^{(1)}\big(z_1, z_3\big) \int_{\mathcal{C}} \mathrm{d}z_4 \sum_{n} G_{mn}\big(z_1, z_4\big) \Gamma_{nqpj}^{(2),\{2;2,0\},0,\mathrm{A}}\big(z_4, z_2, z_3\big)$$

$$= \mathrm{i}\hbar \sum_{mpq} w_{ipqm}\big(z_1\big) \int_{\mathcal{C}} \mathrm{d}z_4 \sum_{n} G_{mn}\big(z_1, z_4\big) \Gamma_{nqpj}^{(2),\{2;2,0\},0,\mathrm{A}}\big(z_4, z_2, z_1\big).$$

Inserting Eq. (446), one arrives at

$$\Sigma_{ij}^{(3),1,\{2;\{2;2,0\},0,\mathrm{A}\}}\big(z_1, z_2\big) = \pm\big(\mathrm{i}\hbar\big)^3 \sum_{mpq} w_{ipqm}\big(z_1\big) \tag{456}$$

$$\int_{\mathcal{C}} \mathrm{d}z_4 \sum_{n} G_{mn}\big(z_1, z_4\big) \sum_{rsuv} G_{uv}\big(z_4, z_2\big) \sum_{t} w_{ntur}\big(z_4\big)$$

$$\sum_{w} w_{vsjw}\big(z_2\big) G_{wt}\big(z_2, z_4\big) G_{rp}\big(z_4, z_1\big) G_{qs}\big(z_1, z_2\big).$$

For the second term, one has

$$\Sigma_{ij}^{(3),1,\{2;\{2;2,0\},0,\mathrm{B}\}}\big(z_1, z_2\big) \tag{457}$$

$$= \mathrm{i}\hbar \sum_{mpq} w_{ipqm}\big(z_1\big) \int_{\mathcal{C}} \mathrm{d}z_4 \sum_{n} G_{mn}\big(z_1, z_4\big) \Gamma_{nqpj}^{(2),\{2;2,0\},0,\mathrm{B}}\big(z_4, z_2, z_1\big).$$

Using Eq. (447) yields

$$\Sigma_{ij}^{(3),1,\{2;\{2;2,0\},0,\mathrm{B}\}}\big(z_1,z_2\big) \tag{458}$$

$$= \pm\big(\mathrm{i}\hbar\big)^3 \sum_{mpq} w_{ipqm}\big(z_1\big) \int_{\mathcal{C}} \mathrm{d}z_4 \sum_n G_{mn}\big(z_1,z_4\big) \sum_{rs}\sum_{tu} G_{tu}\big(z_4,z_2\big)$$

$$\sum_v w_{nvrt}\big(z_4\big) \sum_w w_{sujw}\big(z_2\big) G_{wv}\big(z_2,z_4\big) G_{rp}\big(z_4,z_1\big) G_{qs}\big(z_1,z_2\big)\,.$$

The third term is given by

$$\Sigma_{ij}^{(3),1,\{2;\{2;2,0\},0,\mathrm{C}\}}\big(z_1,z_2\big) \tag{459}$$

$$= \mathrm{i}\hbar \sum_{mpq} w_{ipqm}\big(z_1\big) \int_{\mathcal{C}} \mathrm{d}z_4 \sum_n G_{mn}\big(z_1,z_4\big) \Gamma_{nqpj}^{(2),\{2;2,0\},0,\mathrm{C}}\big(z_4,z_2,z_1\big)\,.$$

With Eq. (448), one finds

$$\Sigma_{ij}^{(3),1,\{2;\{2;2,0\},0,\mathrm{C}\}}\big(z_1,z_2\big) \tag{460}$$

$$= \pm\big(\mathrm{i}\hbar\big)^3 \sum_{mpq} w_{ipqm}\big(z_1\big) \int_{\mathcal{C}} \mathrm{d}z_4 \sum_n G_{mn}\big(z_1,z_4\big) \sum_{rs}\sum_{tu} G_{tu}\big(z_4,z_2\big)$$

$$\sum_{vw} G_{vw}\big(z_4,z_2\big) w_{nsvt}\big(z_4\big) w_{wujr}\big(z_2\big) G_{rp}\big(z_2,z_1\big) G_{qs}\big(z_1,z_4\big)\,.$$

For the other class, one has

$$\Sigma_{ij}^{(3),1,\{2;\{2;1,1\},0,\mathrm{A}\}}\big(z_1,z_2\big) \tag{461}$$

$$= \mathrm{i}\hbar \int_{\mathcal{C}} \mathrm{d}z_3 \sum_{mpq} W_{ipqm}^{(1)}\big(z_1,z_3\big) \int_{\mathcal{C}} \mathrm{d}z_4 \sum_n G_{mn}\big(z_1,z_4\big) \Gamma_{nqpj}^{(2),\{2;1,1\},0,\mathrm{A}}\big(z_4,z_2,z_3\big)$$

$$= \mathrm{i}\hbar \sum_{mpq} w_{ipqm}\big(z_1\big) \int_{\mathcal{C}} \mathrm{d}z_4 \sum_n G_{mn}\big(z_1,z_4\big) \Gamma_{nqpj}^{(2),1,1,\mathrm{A}}\big(z_4,z_2,z_1\big)\,.$$

Inserting Eq. (452) yields

$$\Sigma_{ij}^{(3),1,\{2;\{2;1,1\},0,\mathrm{A}\}}\big(z_1,z_2\big) \tag{462}$$

$$= \big(\mathrm{i}\hbar\big)^3 \sum_{mpq} w_{ipqm}\big(z_1\big) \int_{\mathcal{C}} \mathrm{d}z_4 \sum_n G_{mn}\big(z_1,z_4\big) \sum_{rs}\sum_{tu} w_{ntur}\big(z_4\big)$$

$$\sum_{vw} w_{swjv}\big(z_2\big) G_{vt}\big(z_2,z_4\big) G_{uw}\big(z_4,z_2\big) G_{rp}\big(z_4,z_1\big) G_{qs}\big(z_1,z_2\big)\,.$$

Similarly, the second term reads

$$\Sigma_{ij}^{(3),1,\{2;\{2;1,1\},0,\mathrm{B}\}}\big(z_1,z_2\big) \tag{463}$$

$$= \mathrm{i}\hbar \sum_{mpq} w_{ipqm}\big(z_1\big) \int_{\mathcal{C}} \mathrm{d}z_4 \sum_n G_{mn}\big(z_1,z_4\big) \Gamma_{nqpj}^{(2),\{2;1,1\},0,\mathrm{B}}\big(z_4,z_2,z_1\big)\,.$$

With Eq. (453), one has

$$
\Sigma_{ij}^{(3),1,\{2;\{2;1,1\},0,B\}}\big(z_1,z_2\big)
\tag{464}
$$

$$
= \big(\mathrm{i}\hbar\big)^3 \sum_{mpq} w_{ipqm}(z_1) \int_{\mathcal{C}} \mathrm{d}z_4 \sum_n G_{mn}\big(z_1,z_4\big) \sum_{rs}\sum_{tv} w_{nsvt}\big(z_4\big)
$$

$$
\sum_u G_{tu}\big(z_4,z_2\big) \sum_w w_{uwjr}\big(z_2\big) G_{vw}\big(z_4,z_2\big) G_{rp}\big(z_2,z_1\big) G_{qs}\big(z_1,z_4\big)\,.
$$

For the third term, one finds

$$
\Sigma_{ij}^{(3),1,\{2;\{2;1,1\},0,C\}}\big(z_1,z_2\big)
\tag{465}
$$

$$
= \mathrm{i}\hbar \sum_{mpq} w_{ipqm}(z_1) \int_{\mathcal{C}} \mathrm{d}z_4 \sum_n G_{mn}\big(z_1,z_4\big) \Gamma_{nqpj}^{(2),\{2;1,1\},0,C}\big(z_4,z_2,z_1\big)\,.
$$

Employing Eq. (454), one arrives at

$$
\Sigma_{ij}^{(3),1,\{2;\{2;1,1\},0,C\}}\big(z_1,z_2\big)
\tag{466}
$$

$$
= \big(\mathrm{i}\hbar\big)^3 \sum_{mpq} w_{ipqm}(z_1) \int_{\mathcal{C}} \mathrm{d}z_4 \sum_n G_{mn}\big(z_1,z_4\big) \sum_{rs}\sum_{tv} w_{nvrt}\big(z_4\big)
$$

$$
\sum_u G_{tu}\big(z_4,z_2\big) \sum_w w_{usjw}\big(z_2\big) G_{wv}\big(z_2,z_4\big) G_{rp}\big(z_4,z_1\big) G_{qs}\big(z_1,z_2\big)\,.
$$

## A.3. Resummation approaches: GW approximation

The $GW$ approximation solves Hedin's equation for the screened interaction $W$ according to Eq. (104) with the zeroth-order vertex $\Gamma^{(0)}$. The set of equations is given by the Dyson equation, cf. Eq. (91),

$$
G_{ij}\big(z_1,z_2\big) = G_{ij}^{(0)}\big(z_1,z_2\big)
\tag{467}
$$

$$
+ \int_{\mathcal{C}} \mathrm{d}z_3 \mathrm{d}z_4 \sum_{mn} G_{im}^{(0)}\big(z_1,z_3\big) \Sigma_{mn}\big(z_3,z_4\big) G_{nj}\big(z_4,z_2\big)\,,
$$

the equation for the selfenergy [cf. Eq. (95)]

$$
\Sigma_{ij}\big(z_1,z_2\big) = \Sigma_{ij}^{\mathrm{H}}\big(z_1,z_2\big) + \Sigma_{ij}^{\mathrm{xc}}\big(z_1,z_2\big)\,,
\tag{468}
$$

with

$$
\Sigma_{ij}^{\mathrm{xc}}\big(z_1,z_2\big) = \mathrm{i}\hbar \int_{\mathcal{C}} \mathrm{d}z_3 \sum_{mpq} W_{ipqm}\big(z_1,z_3\big)
\tag{469}
$$

$$
\int_{\mathcal{C}} \mathrm{d}z_4 \sum_n G_{mn}\big(z_1,z_4\big) \Gamma_{nqpj}^{(0)}\big(z_4,z_2,z_3\big)
$$

$$
= \mathrm{i}\hbar \sum_{mp} W_{ipjm}\big(z_1,z_2\big) G_{mp}\big(z_1,z_2\big)\,,
$$

the zeroth-order polarizability, cf. Eq. (411),

$$P_{ijkl}(z_1, z_2) = P_{ijkl}^{(0)}(z_1, z_2) = \pm i\hbar G_{ik}(z_1, z_2) G_{jl}(z_2, z_1),$$ (470)

the zeroth-order vertex, cf. Eq. (155),

$$\Gamma_{ijkl}(z_1, z_2, z_3) = \Gamma_{ijkl}^{(0)}(z_1, z_2, z_3) = \delta_{\mathcal{C}}(z_1, z_{2^+}) \delta_{\mathcal{C}}(z_3, z_2) \delta_{ik} \delta_{jl}$$ (471)

and the screened interaction [cf. Eqs. (102) and (104)]

$$W_{ijkl}(z_1, z_2) = \delta_{\mathcal{C}}(z_1, z_2) w_{ijkl}(z_1) + W_{ijkl}^{\mathrm{ns}}(z_1, z_2),$$ (472)

with

$$W_{ijkl}^{\mathrm{ns}}(z_1, z_2)$$ (473)
$$= \sum_{mn} w_{imnl}(z_1) \int_{\mathcal{C}} \mathrm{d}z_3 \sum_{pq} P_{nqpm}^{(0)}(z_1, z_3) W_{pjkq}(z_3, z_2)$$
$$= \pm i\hbar \sum_{mn} w_{imnl}(z_1) \int_{\mathcal{C}} \mathrm{d}z_3 \sum_{pq} G_{np}(z_1, z_3) G_{qm}(z_3, z_1) W_{pjkq}(z_3, z_2).$$

To solve this set of equations, one has to determine the selfconsistent solution of Eq. (472). Thereto, it is more suitable to eliminate the singular bare interaction by using

$$W_{ijkl}^{\mathrm{ns}}(z_1, z_2) = W_{ijkl}(z_1, z_2) - W_{ijkl}^{\mathrm{bare}}(z_1, z_2)$$ (474)
$$= W_{ijkl}(z_1, z_2) - \delta_{\mathcal{C}}(z_1, z_2) w_{ijkl}(z_1).$$

The selfenergy [cf. Eqs. (468) and (469)] in terms of $W^{\mathrm{ns}}$ is then given by

$$\Sigma_{ij}^{\mathrm{GW}}(z_1, z_2) = \Sigma_{ij}^{\mathrm{H}}(z_1, z_2) + i\hbar \sum_{mp} W_{ipjm}(z_1, z_2) G_{mp}(z_1, z_2)$$
$$= \Sigma_{ij}^{\mathrm{H}}(z_1, z_2) + i\hbar \sum_{mp} w_{ipjm}(z_1) G_{mp}(z_1, z_{1^+}) \delta_{\mathcal{C}}(z_1, z_2)$$
$$+ i\hbar \sum_{mp} W_{ipjm}^{\mathrm{ns}}(z_1, z_2) G_{mp}(z_1, z_2).$$ (475)

Using Eq. (156), the expression simplifies to

$$\Sigma_{ij}^{\mathrm{GW}}(z_1, z_2)$$ (476)
$$= \Sigma_{ij}^{\mathrm{H}}(z_1, z_2) + \Sigma_{ij}^{\mathrm{F}}(z_1, z_2) + i\hbar \sum_{mp} W_{ipjm}^{\mathrm{ns}}(z_1, z_2) G_{mp}(z_1, z_2)$$
$$=: \Sigma_{ij}^{\mathrm{H}}(z_1, z_2) + \Sigma_{ij}^{\mathrm{F}}(z_1, z_2) + \Sigma_{ij}^{\mathrm{GW,corr}}(z_1, z_2).$$

For the screened interaction, one has

$$
W_{ijkl}^{\mathrm{ns}}\big(z_1, z_2\big) \tag{477}
$$
$$
= \pm \mathrm{i}\hbar \sum_{mn} w_{imnl}\big(z_1\big) \sum_{pq} G_{np}\big(z_1, z_2\big) G_{qm}\big(z_2, z_1\big) w_{pjkq}\big(z_2\big)
$$
$$
\pm \mathrm{i}\hbar \sum_{mn} w_{imnl}\big(z_1\big) \int_{\mathcal{C}} \mathrm{d}z_3 \sum_{pq} G_{np}\big(z_1, z_3\big) G_{qm}\big(z_3, z_1\big) W_{pjkq}^{\mathrm{ns}}\big(z_3, z_2\big) .
$$

### A.4. Resummation approaches: T matrix

In contrast to the $GW$ approximation, the $T$ matrix is an approximation, which takes only the bare interaction $w$ into account and aims instead at a good approximation of the bare vertex function $\Lambda$. Thus, its constitutive equations are the Dyson equation,

$$
G_{ij}\big(z_1, z_2\big) = \tag{478}
$$
$$
G_{ij}^{(0)}\big(z_1, z_2\big) + \int_{\mathcal{C}} \mathrm{d}z_3 \mathrm{d}z_4 \sum_{mn} G_{im}^{(0)}\big(z_1, z_3\big) \Sigma_{mn}\big(z_3, z_4\big) G_{nj}\big(z_4, z_2\big) ,
$$

the equation for the selfenergy, cf. Eqs. (95) and (97),

$$
\Sigma_{ij}\big(z_1, z_2\big) = \Sigma_{ij}^{\mathrm{H}}\big(z_1, z_2\big) + \Sigma_{ij}^{\mathrm{xc}}\big(z_1, z_2\big) , \tag{479}
$$

with

$$
\Sigma_{ij}^{\mathrm{xc}}\big(z_1, z_2\big) \tag{480}
$$
$$
= \mathrm{i}\hbar \sum_{mpq} w_{ipqm}\big(z_1\big) \int_{\mathcal{C}} \mathrm{d}z_3 \sum_{n} G_{mn}\big(z_1, z_3\big) \Lambda_{nqpj}\big(z_3, z_2, z_1\big) .
$$

The bare vertex $\Lambda$ is self-consistently given as the solution of

$$
\Lambda_{ijkl}\big(z_1, z_2, z_3\big) = \delta_{\mathcal{C}}\big(z_1, z_{2^+}\big) \delta_{\mathcal{C}}\big(z_3, z_2\big) \delta_{ik} \delta_{jl} \tag{481}
$$
$$
+ \int_{\mathcal{C}} \mathrm{d}z_4 \mathrm{d}z_5 \sum_{mn} \frac{\delta \Sigma_{il}\big(z_1, z_2\big)}{\delta G_{mn}\big(z_4, z_5\big)} \int_{\mathcal{C}} \mathrm{d}z_6 \sum_{p} G_{mp}\big(z_4, z_6\big)
$$
$$
\int_{\mathcal{C}} \mathrm{d}z_7 \sum_{q} G_{qn}\big(z_7, z_5\big) \Lambda_{pjkq}\big(z_6, z_7, z_3\big) .
$$

If these equations are iterated ad infinitum, all selfenergy terms will be generated. To break the circular dependence between Eqs. (480) and (481), the $T$-matrix approximation starts by taking the bare vertex on the right-hand side of Eq. (481) only in zeroth order,

$$
\Lambda_{ijkl}^{(0)}\big(z_1, z_2, z_3\big) = \delta_{\mathcal{C}}\big(z_1, z_{2^+}\big) \delta_{\mathcal{C}}\big(z_3, z_2\big) \delta_{ik} \delta_{jl} , \tag{482}
$$

transforming it into

$$
\Lambda_{ijkl}^{\mathrm{cl}}\bigl(z_1, z_2, z_3\bigr) = \delta_{\mathcal{C}}\bigl(z_1, z_{2^+}\bigr)\delta_{\mathcal{C}}\bigl(z_3, z_2\bigr)\delta_{ik}\delta_{jl} \tag{483}
$$
$$
+ \int_{\mathcal{C}} \mathrm{d}z_4 \mathrm{d}z_5 \sum_{mn} \frac{\delta \Sigma_{il}^{\mathrm{cl}}\bigl(z_1, z_2\bigr)}{\delta G_{mn}\bigl(z_4, z_5\bigr)} G_{mk}\bigl(z_4, z_3\bigr) G_{jn}\bigl(z_3, z_5\bigr).
$$

To arrive at a closed equation, this result is used in Eq. (479), yielding

$$
\Sigma_{ij}^{\mathrm{cl}}\bigl(z_1, z_2\bigr) = \Sigma_{ij}^{\mathrm{H}}\bigl(z_1, z_2\bigr) + \Sigma_{ij}^{\mathrm{xc}}\bigl(z_1, z_2\bigr) \tag{484}
$$
$$
= \Sigma_{ij}^{\mathrm{H}}\bigl(z_1, z_2\bigr) + \mathrm{i}\hbar \sum_{mpq} w_{ipqm}\bigl(z_1\bigr) \int_{\mathcal{C}} \mathrm{d}z_3 \sum_{n} G_{mn}\bigl(z_1, z_3\bigr)\Lambda_{nqpj}^{\mathrm{cl}}\bigl(z_3, z_2, z_1\bigr).
$$

Inserting Eq. (483), one has

$$
\Sigma_{ij}^{\mathrm{cl}}\bigl(z_1, z_2\bigr) = \pm\mathrm{i}\hbar\delta_{\mathcal{C}}\bigl(z_1, z_2\bigr)\sum_{mn} w_{mijn}\bigl(z_1\bigr)G_{nm}\bigl(z_1, z_{1^+}\bigr) \tag{485}
$$
$$
+ \mathrm{i}\hbar\sum_{mn} w_{injm}\bigl(z_1\bigr)G_{mn}\bigl(z_1, z_{1^+}\bigr)\delta_{\mathcal{C}}\bigl(z_1, z_2\bigr)
$$
$$
+ \mathrm{i}\hbar\sum_{mpq} w_{ipqm}\bigl(z_1\bigr)\int_{\mathcal{C}} \mathrm{d}z_3 \sum_{n} G_{mn}\bigl(z_1, z_3\bigr)
$$
$$
\int_{\mathcal{C}} \mathrm{d}z_4 \mathrm{d}z_5 \sum_{rs} \frac{\delta \Sigma_{nj}^{\mathrm{cl}}\bigl(z_3, z_2\bigr)}{\delta G_{rs}\bigl(z_4, z_5\bigr)} G_{rp}\bigl(z_4, z_1\bigr) G_{qs}\bigl(z_1, z_5\bigr).
$$

This term can be restructured to yield

$$
\Sigma_{ij}^{\mathrm{cl}}\bigl(z_1, z_2\bigr) \tag{486}
$$
$$
= \mathrm{i}\hbar\delta_{\mathcal{C}}\bigl(z_1, z_2\bigr)\sum_{mn} G_{mn}\bigl(z_1, z_{1^+}\bigr)w_{injm}^{\pm}\bigl(z_1\bigr)
$$
$$
+ \mathrm{i}\hbar\sum_{mpq} w_{ipqm}\bigl(z_1\bigr)\int_{\mathcal{C}} \mathrm{d}z_3 \sum_{n} G_{mn}\bigl(z_1, z_3\bigr)
$$
$$
\int_{\mathcal{C}} \mathrm{d}z_4 \mathrm{d}z_5 \sum_{rs} \frac{\delta \Sigma_{nj}^{\mathrm{cl}}\bigl(z_3, z_2\bigr)}{\delta G_{rs}\bigl(z_4, z_5\bigr)} G_{rp}\bigl(z_4, z_1\bigr) G_{qs}\bigl(z_1, z_5\bigr),
$$

with the (anti-)symmetrized potential $w_{ijkl}^{\pm}\bigl(z_1\bigr) := w_{ijkl}\bigl(z_1\bigr) \pm w_{jikl}\bigl(z_1\bigr)$. Using Eq. (156) again, one finds

$$
\Sigma_{ij}^{\mathrm{cl}}\bigl(z_1, z_2\bigr) \tag{487}
$$
$$
= \Sigma_{ij}^{\mathrm{H}}\bigl(z_1, z_2\bigr) + \Sigma_{ij}^{\mathrm{F}}\bigl(z_1, z_2\bigr) + \mathrm{i}\hbar \sum_{mpq} w_{ipqm}\bigl(z_1\bigr)\int_{\mathcal{C}} \mathrm{d}z_3 \sum_{n} G_{mn}\bigl(z_1, z_3\bigr)
$$
$$
\int_{\mathcal{C}} \mathrm{d}z_4 \mathrm{d}z_5 \sum_{rs} \frac{\delta \Sigma_{nj}^{\mathrm{cl}}\bigl(z_3, z_2\bigr)}{\delta G_{rs}\bigl(z_4, z_5\bigr)} G_{rp}\bigl(z_4, z_1\bigr) G_{qs}\bigl(z_1, z_5\bigr).
$$

Taking the derivative with respect to $G$, one arrives at

$$\frac{\delta\Sigma_{ij}^{\text{cl}}(z_1, z_2)}{\delta G_{tu}(z_6, z_7)} = i\hbar\delta_{\mathcal{C}}(z_1, z_2)\delta_{\mathcal{C}}(z_1, z_6)\delta_{\mathcal{C}}(z_{1^+}, z_7)w_{iujt}^{\pm}(z_1)$$

$$\left(\frac{\delta\Sigma_{ij}^{\text{cl,A}}(z_1, z_2)}{\delta G_{tu}(z_6, z_7)} + \frac{\delta\Sigma_{ij}^{\text{cl,B}}(z_1, z_2)}{\delta G_{tu}(z_6, z_7)}\right.$$

$$\left. + \frac{\delta\Sigma_{ij}^{\text{cl,C}}(z_1, z_2)}{\delta G_{tu}(z_6, z_7)} + \frac{\delta\Sigma_{ij}^{\text{cl,D}}(z_1, z_2)}{\delta G_{tu}(z_6, z_7)}\right), \tag{488}$$

with

$$\frac{\delta\Sigma_{ij}^{\text{cl,A}}(z_1, z_2)}{\delta G_{tu}(z_6, z_7)} \tag{489}$$

$$= i\hbar\delta_{\mathcal{C}}(z_1, z_6)\sum_{pq}w_{ipqt}(z_1)$$

$$\int_{\mathcal{C}}\mathrm{d}z_4\mathrm{d}z_5\sum_{rs}\frac{\delta\Sigma_{uj}^{\text{cl}}(z_7, z_2)}{\delta G_{rs}(z_4, z_5)}G_{rp}(z_4, z_1)G_{qs}(z_1, z_5),$$

$$\frac{\delta\Sigma_{ij}^{\text{cl,B}}(z_1, z_2)}{\delta G_{tu}(z_6, z_7)} \tag{490}$$

$$= i\hbar\delta_{\mathcal{C}}(z_1, z_7)\sum_{mq}w_{iuqm}(z_1)\int_{\mathcal{C}}\mathrm{d}z_3\sum_n G_{mn}(z_1, z_3)$$

$$\int_{\mathcal{C}}\mathrm{d}z_5\sum_s\frac{\delta\Sigma_{nj}^{\text{cl}}(z_3, z_2)}{\delta G_{ts}(z_6, z_5)}G_{qs}(z_1, z_5)$$

and

$$\frac{\delta\Sigma_{ij}^{\text{cl,C}}(z_1, z_2)}{\delta G_{tu}(z_6, z_7)} \tag{491}$$

$$= i\hbar\delta_{\mathcal{C}}(z_1, z_6)\sum_{mp}w_{iptm}(z_1)\int_{\mathcal{C}}\mathrm{d}z_3\sum_n G_{mn}(z_1, z_3)$$

$$\int_{\mathcal{C}}\mathrm{d}z_4\sum_r\frac{\delta\Sigma_{nj}^{\text{cl}}(z_3, z_2)}{\delta G_{ru}(z_4, z_7)}G_{rp}(z_4, z_1),$$

as well as

$$\frac{\delta \Sigma_{ij}^{\mathrm{cl,D}}\big(z_1, z_2\big)}{\delta G_{tu}\big(z_6, z_7\big)} \tag{492}$$

$$= \mathrm{i}\hbar \sum_{mpq} w_{ipqm}\big(z_1\big) \int_{\mathcal{C}} \mathrm{d}z_3 \sum_n G_{mn}\big(z_1, z_3\big)$$

$$\int_{\mathcal{C}} \mathrm{d}z_4 \mathrm{d}z_5 \sum_{rs} \frac{\delta \Sigma_{nj}^{\mathrm{cl}}\big(z_3, z_2\big)}{\delta G_{rs}\big(z_4, z_5\big)\delta G_{tu}\big(z_6, z_7\big)} G_{rp}\big(z_4, z_1\big) G_{qs}\big(z_1, z_5\big) .$$

Neglecting $\frac{\delta \Sigma_{nj}^{\mathrm{cl}}\big(z_3, z_2\big)}{\delta G_{rs}\big(z_4, z_5\big)\delta G_{tu}\big(z_6, z_7\big)}$ as an approximation, Eq. (488) becomes a closed equation for $\frac{\delta \Sigma^{\mathrm{cl}}}{\delta G}$. The first iteration yields the second-order terms

$$\frac{\delta \Sigma_{ij}^{\mathrm{cl,(2)}}\big(z_1, z_2\big)}{\delta G_{tu}\big(z_6, z_7\big)} \approx \frac{\delta \Sigma_{ij}^{\mathrm{cl,(2),A}}\big(z_1, z_2\big)}{\delta G_{tu}\big(z_6, z_7\big)} \tag{493}$$

$$+ \frac{\delta \Sigma_{ij}^{\mathrm{cl,(2),B}}\big(z_1, z_2\big)}{\delta G_{tu}\big(z_6, z_7\big)} + \frac{\delta \Sigma_{ij}^{\mathrm{cl,(2),C}}\big(z_1, z_2\big)}{\delta G_{tu}\big(z_6, z_7\big)} ,$$

with

$$\frac{\delta \Sigma_{ij}^{\mathrm{cl,(2),A}}\big(z_1, z_2\big)}{\delta G_{tu}\big(z_6, z_7\big)} \tag{494}$$

$$= \big(\mathrm{i}\hbar\big)^2 \delta_{\mathcal{C}}\big(z_1, z_6\big) \delta_{\mathcal{C}}\big(z_2, z_7\big) \sum_{pq} w_{ipqt}\big(z_1\big)$$

$$\sum_{rs} w_{usjr}^{\pm}\big(z_2\big) G_{rp}\big(z_2, z_1\big) G_{qs}\big(z_1, z_2\big)$$

and

$$\frac{\delta \Sigma_{ij}^{\mathrm{cl,(2),B}}\big(z_1, z_2\big)}{\delta G_{tu}\big(z_6, z_7\big)} \tag{495}$$

$$= \big(\mathrm{i}\hbar\big)^2 \delta_{\mathcal{C}}\big(z_1, z_7\big) \delta_{\mathcal{C}}\big(z_2, z_6\big) \sum_{mq} w_{iuqm}\big(z_1\big)$$

$$\sum_n G_{mn}\big(z_1, z_2\big) \sum_s w_{nsjt}^{\pm}\big(z_2\big) G_{qs}\big(z_1, z_2\big)$$

as well as

$$\frac{\delta \Sigma_{ij}^{\mathrm{cl,(2),C}}\big(z_1, z_2\big)}{\delta G_{tu}\big(z_6, z_7\big)} \tag{496}$$

$$= \big(\mathrm{i}\hbar\big)^2 \delta_{\mathcal{C}}\big(z_1, z_6\big) \delta_{\mathcal{C}}\big(z_2, z_7\big) \sum_{mp} w_{iptm}\big(z_1\big)$$

$$\sum_n G_{mn}\big(z_1, z_2\big) \sum_r w_{nujr}^{\pm}\big(z_2\big) G_{rp}\big(z_2, z_1\big) .$$

Note that, for the first iteration, $\frac{\delta \Sigma_{ij}^{\mathrm{cl,D}}(z_1, z_2)}{\delta G_{tu}(z_6, z_7)}$ is exactly equal to zero, thus Eq. (493) is also exact up to second order in $w$. In the following, each of the three terms will be considered separately. To start with, one recognizes that all three terms yield the same first and second-order contributions to the selfenergy, which read

$$\Sigma_{ij}^{\mathrm{cl,(1)}}(z_1, z_2) = \Sigma_{ij}^{\mathrm{H}}(z_1, z_2) + \Sigma_{ij}^{\mathrm{F}}(z_1, z_2) \tag{497}$$

$$\Sigma_{ij}^{\mathrm{cl,(2)}}(z_1, z_2) = (\mathrm{i}\hbar)^2 \sum_{mpq} w_{ipqm}(z_1) \sum_n G_{mn}(z_1, z_2) \tag{498}$$

$$\sum_{rs} w_{nsjr}^{\pm}(z_2) G_{rp}(z_2, z_1) G_{qs}(z_1, z_2)$$

and agree with the exact first and second-order terms, already encountered in Eqs. (158), (415) and (424). The third-order contributions to $\Sigma^{\mathrm{cl}}$ from the second-order terms in Eq. (493) are given by

$$\Sigma_{ij}^{\mathrm{cl,(3),A}}(z_1, z_2) = \mathrm{i}\hbar \sum_{mpq} w_{ipqm}(z_1) \int_{\mathcal{C}} \mathrm{d}z_3 \sum_n G_{mn}(z_1, z_3) \tag{499}$$

$$\int_{\mathcal{C}} \mathrm{d}z_4 \mathrm{d}z_5 \sum_{rs} \frac{\delta \Sigma_{nj}^{\mathrm{cl,(2),A}}(z_3, z_2)}{\delta G_{rs}(z_4, z_5)} G_{rp}(z_4, z_1) G_{qs}(z_1, z_5).$$

Inserting the second-order term yields

$$\Sigma_{ij}^{\mathrm{cl,(3),A}}(z_1, z_2) = (\mathrm{i}\hbar)^3 \sum_{mpq} w_{ipqm}(z_1) \int_{\mathcal{C}} \mathrm{d}z_3 \sum_n G_{mn}(z_1, z_3) \tag{500}$$

$$\sum_{rs} \sum_{tu} w_{ntur}(z_3) \sum_{vw} w_{swjv}^{\pm}(z_2)$$

$$G_{vt}(z_2, z_3) G_{uw}(z_3, z_2) G_{rp}(z_3, z_1) G_{qs}(z_1, z_2).$$

For the second third-order selfenergy term, one has

$$\Sigma_{ij}^{\mathrm{cl,(3),B}}(z_1, z_2) = \mathrm{i}\hbar \sum_{mpq} w_{ipqm}(z_1) \int_{\mathcal{C}} \mathrm{d}z_3 \sum_n G_{mn}(z_1, z_3) \tag{501}$$

$$\int_{\mathcal{C}} \mathrm{d}z_4 \mathrm{d}z_5 \sum_{rs} \frac{\delta \Sigma_{nj}^{\mathrm{cl,(2),B}}(z_3, z_2)}{\delta G_{rs}(z_4, z_5)} G_{rp}(z_4, z_1) G_{qs}(z_1, z_5).$$

which evaluates to

$$\Sigma_{ij}^{\mathrm{cl,(3),B}}(z_1, z_2) = (\mathrm{i}\hbar)^3 \sum_{mpq} w_{ipqm}(z_1) \int_{\mathcal{C}} \mathrm{d}z_3 \sum_n G_{mn}(z_1, z_3) \tag{502}$$

$$\sum_{rs} \sum_{tu} w_{nsut}(z_3) \sum_v G_{tv}(z_3, z_2) \sum_w w_{vwjr}^{\pm}(z_2)$$

$$G_{uw}(z_3, z_2) G_{rp}(z_2, z_1) G_{qs}(z_1, z_3).$$

The third term reads

$$\Sigma_{ij}^{\mathrm{cl},(3),\mathrm{C}}\big(z_1,z_2\big) = \mathrm{i}\hbar \sum_{mpq} w_{ipqm}\big(z_1\big) \int_{\mathcal{C}} \mathrm{d}z_3 \sum_n G_{mn}\big(z_1,z_3\big) \tag{503}$$

$$\int_{\mathcal{C}} \mathrm{d}z_4 \mathrm{d}z_5 \sum_{rs} \frac{\delta \Sigma_{nj}^{\mathrm{cl},(2),\mathrm{C}}\big(z_3,z_2\big)}{\delta G_{rs}\big(z_4,z_5\big)} G_{rp}\big(z_4,z_1\big) G_{qs}\big(z_1,z_5\big) . \tag{504}$$

Using the second-order result, one arrives at

$$\Sigma_{ij}^{\mathrm{cl},(3),\mathrm{C}}\big(z_1,z_2\big) = \mathrm{i}\hbar \sum_{mpq} w_{ipqm}\big(z_1\big) \int_{\mathcal{C}} \mathrm{d}z_3 \sum_n G_{mn}\big(z_1,z_3\big) \tag{505}$$

$$\sum_{rs} \sum_{tu} w_{nurt}\big(z_3\big) \sum_v G_{tv}\big(z_3,z_2\big) \sum_w w_{vsjw}^{\pm}\big(z_2\big)$$

$$G_{wu}\big(z_2,z_3\big) G_{rp}\big(z_3,z_1\big) G_{qs}\big(z_1,z_2\big) .$$

## B. Non-selfconsistent second-order selfenergy contributions

### B.1. General basis

The first two classes are just the same as in the selfconsistent approximation, cf. Eqs. (176) and (183), with the replacement $G \to G^{(0)}$. Likewise, their components follow directly from Eqs. (184) and (185). For the third and fourth class, one needs the contributions to $G^{(1)}$, which are

$$G_{ij}^{(1),\{\mathrm{H},0\},0}\big(z_1,z_2\big) \tag{506}$$

$$= \int_{\mathcal{C}} \mathrm{d}z_3 \mathrm{d}z_4 \sum_{mn} G_{im}^{(0)}\big(z_1,z_3\big) \Sigma_{mn}^{\mathrm{H},0}\big(z_3,z_4\big) G_{nj}^{(0)}\big(z_4,z_2\big)$$

$$= \pm \mathrm{i}\hbar \int_{\mathcal{C}} \mathrm{d}z_3 \sum_{mn} G_{im}^{(0)}\big(z_1,z_3\big) \sum_{pq} w_{pmnq}\big(z_3\big) G_{qp}^{(0)}\big(z_3,z_{3^+}\big) G_{nj}^{(0)}\big(z_3,z_2\big)$$

and

$$G_{ij}^{(1),\{\mathrm{F},0\},0}\big(z_1,z_2\big) \tag{507}$$

$$= \int_{\mathcal{C}} \mathrm{d}z_3 \mathrm{d}z_4 \sum_{mn} G_{im}^{(0)}\big(z_1,z_3\big) \Sigma_{mn}^{\mathrm{F},0}\big(z_3,z_4\big) G_{nj}^{(0)}\big(z_4,z_2\big)$$

$$= \mathrm{i}\hbar \int_{\mathcal{C}} \mathrm{d}z_3 \sum_{mn} G_{im}^{(0)}\big(z_1,z_3\big) \sum_{pq} w_{mqnp}\big(z_3\big) G_{pq}^{(0)}\big(z_3,z_{3^+}\big) G_{nj}^{(0)}\big(z_3,z_2\big) .$$

With these results, the additional non-selfconsistent contributions are

$$\Sigma_{ij}^{(2),\{H,0\},\{1,\{H,0\},0\}}(z_1, z_2) = \pm i\hbar\delta_{\mathcal{C}}(z_1, z_2) \sum_{mn} w_{mijn}(z_1) G_{nm}^{(1),\{H,0\},0}(z_1, z_{1^+})$$

$$= \left(i\hbar\right)^2 \delta_{\mathcal{C}}(z_1, z_2) \sum_{mn} w_{mijn}(z_1) \qquad\qquad (508)$$

$$\int_{\mathcal{C}} dz_3 \sum_{pq} G_{np}^{(0)}(z_1, z_3) \sum_{rs} w_{rpqs}(z_3) G_{sr}^{(0)}(z_3, z_{3^+}) G_{qm}^{(0)}(z_3, z_{1^+}),$$

as well as

$$\Sigma_{ij}^{(2),\{H,0\},\{1,\{F,0\},0\}}(z_1, z_2) = \pm i\hbar\delta_{\mathcal{C}}(z_1, z_2) \sum_{mn} w_{mijn}(z_1) G_{nm}^{(1),\{F,0\},0}(z_1, z_{1^+})$$

$$= \pm\left(i\hbar\right)^2 \delta_{\mathcal{C}}(z_1, z_2) \sum_{mn} w_{mijn}(z_1) \qquad\qquad (509)$$

$$\int_{\mathcal{C}} dz_3 \sum_{pq} G_{np}^{(0)}(z_1, z_3) \sum_{rs} w_{psqr}(z_3) G_{rs}^{(0)}(z_3, z_{3^+}) G_{qm}^{(0)}(z_3, z_{1^+})$$

and

$$\Sigma_{ij}^{(2),\{F,0\},\{1,\{H,0\},0\}}(z_1, z_2) = i\hbar\delta_{\mathcal{C}}(z_1, z_2) \sum_{mn} w_{injm}(z_1) G_{mn}^{(1),\{H,0\},0}(z_1, z_{1^+})$$

$$= \pm\left(i\hbar\right)^2 \delta_{\mathcal{C}}(z_1, z_2) \sum_{mn} w_{nimj}(z_1) \qquad\qquad (510)$$

$$\int_{\mathcal{C}} dz_3 \sum_{pq} G_{mp}^{(0)}(z_1, z_3) \sum_{rs} w_{rpqs}(z_3) G_{sr}^{(0)}(z_3, z_{3^+}) G_{qn}^{(0)}(z_3, z_{1^+}),$$

as well as

$$\Sigma_{ij}^{(2),\{F,0\},\{1,\{F,0\},0\}}(z_1, z_2) = i\hbar\delta_{\mathcal{C}}(z_1, z_2) \sum_{mn} w_{injm}(z_1) G_{mn}^{(1),\{F,0\},0}(z_1, z_{1^+})$$

$$= \left(i\hbar\right)^2 \delta_{\mathcal{C}}(z_1, z_2) \sum_{mn} w_{nimj}(z_1) \qquad\qquad (511)$$

$$\int_{\mathcal{C}} dz_3 \sum_{pq} G_{mp}^{(0)}(z_1, z_3) \sum_{rs} w_{psqr}(z_3) G_{rs}^{(0)}(z_3, z_{3^+}) G_{qn}^{(0)}(z_3, z_{1^+}).$$

The corresponding components are all time-diagonal and read

$$\Sigma_{ij}^{(2),\{H,0\},\{1,\{H,0\},0\},\delta}(t_1) \qquad\qquad (512)$$

$$= \left(i\hbar\right)^2 \sum_{mn} w_{mijn}(t_1) \Bigg($$

$$\int_{t_0}^{t_1} dt_3 \sum_{pq} G_{np}^{(0),>}(t_1, t_3) \sum_{rs} w_{rpqs}(t_3) G_{sr}^{(0),<}(t_3, t_3) G_{qm}^{(0),<}(t_3, t_1)$$

$$+ \int_{t_1}^{t_0} dt_3 \sum_{pq} G_{np}^{(0),<}(t_1, t_3) \sum_{rs} w_{rpqs}(t_3) G_{sr}^{(0),<}(t_3, t_3) G_{qm}^{(0),>}(t_3, t_1) \Bigg)$$

and

$$\Sigma_{ij}^{(2),\{H,0\},\{1,\{F,0\},0\},\delta}(t_1) \tag{513}$$

$$= \pm\left(i\hbar\right)^2 \sum_{mn} w_{mijn}(t_1)\Bigg($$

$$\int_{t_0}^{t_1} dt_3 \sum_{pq} G_{np}^{(0),>}(t_1,t_3) \sum_{rs} w_{psqr}(t_3) G_{rs}^{(0),<}(t_3,t_3) G_{qm}^{(0),<}(t_3,t_1)$$

$$+ \int_{t_1}^{t_0} dt_3 \sum_{pq} G_{np}^{(0),<}(t_1,t_3) \sum_{rs} w_{psqr}(t_3) G_{rs}^{(0),<}(t_3,t_3) G_{qm}^{(0),>}(t_3,t_1)\Bigg),$$

as well as

$$\Sigma_{ij}^{(2),\{F,0\},\{1,\{H,0\},0\},\delta}(t_1) \tag{514}$$

$$= \pm\left(i\hbar\right)^2 \sum_{mn} w_{nimj}(t_1)\Bigg($$

$$\int_{t_0}^{t_1} dt_3 \sum_{pq} G_{mp}^{(0),>}(t_1,t_3) \sum_{rs} w_{rpqs}(t_3) G_{sr}^{(0),<}(t_3,t_3) G_{qn}^{(0),<}(t_3,t_1)$$

$$+ \int_{t_1}^{t_0} dt_3 \sum_{pq} G_{mp}^{(0),<}(t_1,t_3) \sum_{rs} w_{rpqs}(t_3) G_{sr}^{(0),<}(t_3,t_3) G_{qn}^{(0),>}(t_3,t_1)\Bigg)$$

and

$$\Sigma_{ij}^{(2),\{F,0\},\{1,\{F,0\},0\},\delta}(t_1) \tag{515}$$

$$= \left(i\hbar\right)^2 \sum_{mn} w_{nimj}(t_1)\Bigg($$

$$\int_{t_0}^{t_1} dt_3 \sum_{pq} G_{mp}^{(0),>}(t_1,t_3) \sum_{rs} w_{psqr}(t_3) G_{rs}^{(0),<}(t_3,t_3) G_{qn}^{(0),<}(t_3,t_1)$$

$$+ \int_{t_1}^{t_0} dt_3 \sum_{pq} G_{mp}^{(0),<}(t_1,t_3) \sum_{rs} w_{psqr}(t_3) G_{rs}^{(0),<}(t_3,t_3) G_{qn}^{(0),>}(t_3,t_1)\Bigg).$$

### B.2. Diagonal basis

For $w_{ijkl} = \delta_{il}\delta_{jk} w_{ij}$, the non-selfconsistent selfenergy terms attain the form, cf. Eqs. (189) and (191),

$$\Sigma_{ij}^{(2),(2),2,0,0,\text{diagonal}}(z_1,z_2) \tag{516}$$

$$\equiv \Sigma_{ij}^{(2),2,0,\text{diagonal}}(z_1,z_2)\left(G \to G^{(0)}\right)$$

and

$$\Sigma_{ij}^{(2),(2),1,0,1,\text{diagonal}}(z_1,z_2) \tag{517}$$

$$\equiv \Sigma_{ij}^{(2),1,1,\text{diagonal}}(z_1,z_2)\left(G \to G^{(0)}\right),$$

as well as

$$\Sigma_{ij}^{(2),\{H,0\},\{1,\{H,0\},0\},\text{diagonal}}\big(z_1,z_2\big) \tag{518}$$

$$= \big(\mathrm{i}\hbar\big)^2 \delta_{\mathcal{C}}\big(z_1,z_2\big)\delta_{ij}\sum_m w_{mi}\big(z_1\big)$$

$$\int_{\mathcal{C}}\mathrm{d}z_3\sum_p G_{mp}^{(0)}\big(z_1,z_3\big)G_{pm}^{(0)}\big(z_3,z_{1^+}\big)\sum_r w_{rp}\big(z_3\big)G_{rr}^{(0)}\big(z_3,z_{3^+}\big)$$

and

$$\Sigma_{ij}^{(2),\{H,0\},\{1,\{F,0\},0\},\text{diagonal}}\big(z_1,z_2\big) \tag{519}$$

$$= \pm\big(\mathrm{i}\hbar\big)^2 \delta_{\mathcal{C}}\big(z_1,z_2\big)\delta_{ij}\sum_m w_{mi}\big(z_1\big)$$

$$\int_{\mathcal{C}}\mathrm{d}z_3\sum_p G_{mp}^{(0)}\big(z_1,z_3\big)\sum_q w_{pq}\big(z_3\big)G_{pq}^{(0)}\big(z_3,z_{3^+}\big)G_{qm}^{(0)}\big(z_3,z_{1^+}\big).$$

Further,

$$\Sigma_{ij}^{(2),\{F,0\},\{1,\{H,0\},0\},\text{diagonal}}\big(z_1,z_2\big) \tag{520}$$

$$= \pm\big(\mathrm{i}\hbar\big)^2 \delta_{\mathcal{C}}\big(z_1,z_2\big)w_{ij}\big(z_1\big)$$

$$\int_{\mathcal{C}}\mathrm{d}z_3\sum_p G_{ip}^{(0)}\big(z_1,z_3\big)\sum_r w_{rp}\big(z_3\big)G_{rr}^{(0)}\big(z_3,z_{3^+}\big)G_{pj}^{(0)}\big(z_3,z_{1^+}\big)$$

and

$$\Sigma_{ij}^{(2),\{F,0\},\{1,\{F,0\},0\},\text{diagonal}}\big(z_1,z_2\big) \tag{521}$$

$$= \big(\mathrm{i}\hbar\big)^2 \delta_{\mathcal{C}}\big(z_1,z_2\big)w_{ij}\big(z_1\big)$$

$$\int_{\mathcal{C}}\mathrm{d}z_3\sum_p G_{ip}^{(0)}\big(z_1,z_3\big)\sum_q w_{pq}\big(z_3\big)G_{pq}^{(0)}\big(z_3,z_{3^+}\big)G_{qj}^{(0)}\big(z_3,z_{1^+}\big).$$

The components read [cf. Eqs. (191) and (192)]

$$\Sigma_{ij}^{(2),(2),2,0,0,\text{diagonal},\gtrless}\big(t_1,t_2\big) \equiv \Sigma_{ij}^{(2),2,0,\text{diagonal},\gtrless}\big(t_1,t_2\big)\big(G\to G^{(0)}\big), \tag{522}$$

$$\Sigma_{ij}^{(2),(2),1,0,1,\text{diagonal},\gtrless}\big(t_1,t_2\big) \equiv \Sigma_{ij}^{(2),1,1,\text{diagonal},\gtrless}\big(t_1,t_2\big)\big(G\to G^{(0)}\big), \tag{523}$$

as well as [cf. Eq. (512) to Eq. (516)]

$$\Sigma_{ij}^{(2),\{H,0\},\{1,\{H,0\},0\},\text{diagonal},\delta}\big(t_1\big) \tag{524}$$

$$= \big(\mathrm{i}\hbar\big)^2 \delta_{ij}\sum_m w_{mi}\big(t_1\big)\bigg($$

$$\int_{t_0}^{t_1}\mathrm{d}t_3\sum_p G_{mp}^{(0),>}\big(t_1,t_3\big)\sum_r w_{rp}\big(t_3\big)G_{rr}^{(0),<}\big(t_3,t_3\big)G_{pm}^{(0),<}\big(t_3,t_1\big)$$

$$+ \int_{t_1}^{t_0}\mathrm{d}t_3\sum_p G_{mp}^{(0),<}\big(t_1,t_3\big)\sum_r w_{rp}\big(t_3\big)G_{rr}^{(0),<}\big(t_3,t_3\big)G_{pm}^{(0),>}\big(t_3,t_1\big)\bigg)$$

and

$$\Sigma_{ij}^{(2),\{\text{H},0\},\{1,\{\text{F},0\},0\},\delta}\big(t_1\big) \tag{525}$$

$$= \pm\big(\mathrm{i}\hbar\big)^2 \delta_{ij} \sum_m w_{mi}\big(t_1\big)\Big($$

$$\int_{t_0}^{t_1} \mathrm{d}t_3\, G_{mr}^{(0),>}\big(t_1,t_3\big) \sum_{rs} w_{rs}\big(t_3\big) G_{rs}^{(0),<}\big(t_3,t_3\big) G_{sm}^{(0),<}\big(t_3,t_1\big)$$

$$+ \int_{t_1}^{t_0} \mathrm{d}t_3\, G_{mr}^{(0),<}\big(t_1,t_3\big) \sum_{rs} w_{rs}\big(t_3\big) G_{rs}^{(0),<}\big(t_3,t_3\big) G_{sm}^{(0),>}\big(t_3,t_1\big)\Big).$$

Further,

$$\Sigma_{ij}^{(2),\{\text{F},0\},\{1,\{\text{H},0\},0\},\text{diagonal},\delta}\big(t_1,t_2\big) \tag{526}$$

$$= \pm\big(\mathrm{i}\hbar\big)^2 w_{ji}\big(t_1\big)\Big($$

$$\int_{t_0}^{t_1} \mathrm{d}t_3 \sum_p G_{ip}^{(0),>}\big(t_1,t_3\big) \sum_{rs} w_{rp}\big(t_3\big) G_{rr}^{(0),<}\big(t_3,t_3\big) G_{pj}^{(0),<}\big(t_3,t_1\big)$$

$$+ \int_{t_1}^{t_0} \mathrm{d}t_3 \sum_p G_{ip}^{(0),<}\big(t_1,t_3\big) \sum_{rs} w_{rp}\big(t_3\big) G_{rr}^{(0),<}\big(t_3,t_3\big) G_{pj}^{(0),>}\big(t_3,t_1\big)\Big)$$

and

$$\Sigma_{ij}^{(2),\{\text{F},0\},\{1,\{\text{F},0\},0\},\delta}\big(t_1\big) \tag{527}$$

$$= \big(\mathrm{i}\hbar\big)^2 \sum_n w_{ji}\big(t_1\big)\Big($$

$$\int_{t_0}^{t_1} \mathrm{d}t_3\, G_{ir}^{(0),<}\big(t_1,t_3\big) \sum_{rs} w_{rs}\big(t_3\big) G_{rs}^{(0),<}\big(t_3,t_3\big) G_{sj}^{(0),>}\big(t_3,t_1\big)$$

$$+ \int_{t_1}^{t_0} \mathrm{d}t_3\, G_{ir}^{(0),>}\big(t_1,t_3\big) \sum_{rs} w_{rs}\big(t_3\big) G_{rs}^{(0),<}\big(t_3,t_3\big) G_{sj}^{(0),<}\big(t_3,t_1\big)\Big).$$

*B.3. Hubbard basis*

In the Hubbard basis, the non-selfconsistent contributions attain the form

$$\Sigma_{i\alpha j\alpha}^{(2),(2),2,0,0,\text{Hubbard},\mathfrak{b}}\big(z_1,z_2\big) \tag{528}$$

$$\equiv \Sigma_{i\alpha j\alpha}^{(2),2,0,\text{Hubbard},\mathfrak{b}}\big(z_1,z_2\big)\big(G\to G^{(0)}\big)$$

and

$$\Sigma_{i\alpha j\alpha}^{(2),(2),1,0,1,\text{Hubbard},\mathfrak{b}}\big(z_1,z_2\big) \tag{529}$$

$$\equiv \Sigma_{i\alpha j\alpha}^{(2),1,1,\text{Hubbard},\mathfrak{b}}\big(z_1,z_2\big)\big(G\to G^{(0)}\big),$$

as well as

$$\Sigma^{(2),\{H,0\},\{1,\{H,0\},0\},Hubbard,\mathfrak{b}}_{i\alpha j\alpha}\bigl(z_1, z_2\bigr) \tag{530}$$

$$= \bigl(\mathrm{i}\hbar\bigr)^2 \delta_{\mathcal{C}}\bigl(z_1, z_2\bigr)\delta_{ij}\sum_\epsilon U\bigl(z_1\bigr)$$

$$\int_{\mathcal{C}} \mathrm{d}z_3 \sum_p G^{(0)}_{i\epsilon p\epsilon}\bigl(z_1, z_3\bigr) G^{(0)}_{p\epsilon i\epsilon}\bigl(z_3, z_{1^+}\bigr) \sum_\zeta U\bigl(z_3\bigr) G^{(0)}_{p\zeta p\zeta}\bigl(z_3, z_{3^+}\bigr)$$

and

$$\Sigma^{(2),\{H,0\},\{1,\{F,0\},0\},Hubbard,\mathfrak{b}}_{i\alpha j\alpha}\bigl(z_1, z_2\bigr) \tag{531}$$

$$= \bigl(\mathrm{i}\hbar\bigr)^2 \delta_{\mathcal{C}}\bigl(z_1, z_2\bigr)\delta_{ij}\sum_\epsilon U\bigl(z_1\bigr)$$

$$\int_{\mathcal{C}} \mathrm{d}z_3 \sum_p G^{(0)}_{i\epsilon p\epsilon}\bigl(z_1, z_3\bigr) U\bigl(z_3\bigr) G^{(0)}_{p\epsilon p\epsilon}\bigl(z_3, z_{3^+}\bigr) G^{(0)}_{p\epsilon i\epsilon}\bigl(z_3, z_{1^+}\bigr).$$

Further,

$$\Sigma^{(2),\{F,0\},\{1,\{H,0\},0\},Hubbard,\mathfrak{b}}_{i\alpha j\alpha}\bigl(z_1, z_2\bigr) \tag{532}$$

$$= \bigl(\mathrm{i}\hbar\bigr)^2 \delta_{\mathcal{C}}\bigl(z_1, z_2\bigr)\delta_{ij} U\bigl(z_1\bigr)$$

$$\int_{\mathcal{C}} \mathrm{d}z_3 \sum_p G^{(0)}_{i\alpha p\alpha}\bigl(z_1, z_3\bigr) U\bigl(z_3\bigr) \sum_\epsilon G^{(0)}_{p\epsilon p\epsilon}\bigl(z_3, z_{3^+}\bigr) G^{(0)}_{p\alpha i\alpha}\bigl(z_3, z_{1^+}\bigr)$$

and

$$\Sigma^{(2),\{F,0\},\{1,\{F,0\},0\},Hubbard,\mathfrak{b}}_{i\alpha j\alpha}\bigl(z_1, z_2\bigr) \tag{533}$$

$$= \bigl(\mathrm{i}\hbar\bigr)^2 \delta_{\mathcal{C}}\bigl(z_1, z_2\bigr)\delta_{ij} U\bigl(z_1\bigr)$$

$$\int_{\mathcal{C}} \mathrm{d}z_3 \sum_p G^{(0)}_{i\alpha p\alpha}\bigl(z_1, z_3\bigr) U\bigl(z_1\bigr) G^{(0)}_{p\alpha p\alpha}\bigl(z_1, z_{1^+}\bigr) G^{(0)}_{p\alpha i\alpha}\bigl(z_3, z_{1^+}\bigr),$$

for bosons, and

$$\Sigma^{(2),(2),2,0,0,Hubbard,\mathfrak{f}}_{i\alpha j\alpha}\bigl(z_1, z_2\bigr) \tag{534}$$

$$\equiv \Sigma^{(2),2,0,Hubbard,\mathfrak{f}}_{i\alpha j\alpha}\bigl(z_1, z_2\bigr)\bigl(G \to G^{(0)}\bigr)$$

and

$$\Sigma^{(2),(2),1,0,1,Hubbard,\mathfrak{f}}_{i\alpha j\alpha}\bigl(z_1, z_2\bigr) \tag{535}$$

$$\equiv \Sigma^{(2),1,1,Hubbard,\mathfrak{f}}_{i\alpha j\alpha}\bigl(z_1, z_2\bigr)\bigl(G \to G^{(0)}\bigr),$$

as well as

$$\Sigma_{i\alpha j\alpha}^{(2),\{H,0\},\{1,\{H,0\},0\},\text{Hubbard},\mathfrak{f}}\left(z_1,z_2\right) \tag{536}$$

$$= \left(\mathrm{i}\hbar\right)^2 \delta_{\mathcal{C}}\left(z_1,z_2\right)\delta_{ij}\sum_{\epsilon\neq\alpha}U\left(z_1\right)$$

$$\int_{\mathcal{C}}\mathrm{d}z_3\sum_p G_{i\epsilon p\epsilon}^{(0)}\left(z_1,z_3\right)G_{p\epsilon i\epsilon}^{(0)}\left(z_3,z_{1^+}\right)\sum_{\zeta\neq\epsilon}U\left(z_3\right)G_{p\zeta p\zeta}^{(0)}\left(z_3,z_{3^+}\right),$$

$$\Sigma_{i\alpha j\alpha}^{(2),\{H,0\},\{1,\{F,0\},0\},\text{Hubbard},\mathfrak{f}}\left(z_1,z_2\right) \equiv 0 \tag{537}$$

and

$$\Sigma_{i\alpha j\alpha}^{(2),\{F,0\},\{1,\{H,0\},0\},\text{Hubbard},\mathfrak{f}}\left(z_1,z_2\right) \equiv 0\,, \tag{538}$$

$$\Sigma_{i\alpha j\alpha}^{(2),\{F,0\},\{1,\{F,0\},0\},\text{Hubbard},\mathfrak{f}}\left(z_1,z_2\right) \equiv 0\,, \tag{539}$$

for fermions. The components read

$$\Sigma_{i\alpha j\alpha}^{(2),(2),2,0,0,\text{Hubbard},\mathfrak{b},\gtrless}\left(t_1,t_2\right) \tag{540}$$

$$\equiv \Sigma_{i\alpha j\alpha}^{(2),2,0,\text{Hubbard},\mathfrak{b},\gtrless}\left(t_1,t_2\right)\left(G\to G^{(0)}\right),$$

$$\Sigma_{i\alpha j\alpha}^{(2),(2),1,0,1,\text{Hubbard},\mathfrak{b},\gtrless}\left(t_1,t_2\right) \tag{541}$$

$$\equiv \Sigma_{i\alpha j\alpha}^{(2),1,1,\text{Hubbard},\mathfrak{b},\gtrless}\left(t_1,t_2\right)\left(G\to G^{(0)}\right),$$

$$\Sigma_{i\alpha j\alpha}^{(2),(2),2,0,0,\text{Hubbard},\mathfrak{f},\gtrless}\left(t_1,t_2\right) \tag{542}$$

$$\equiv \Sigma_{i\alpha j\alpha}^{(2),2,0,\text{Hubbard},\mathfrak{f},\gtrless}\left(t_1,t_2\right)\left(G\to G^{(0)}\right),$$

$$\Sigma_{i\alpha j\alpha}^{(2),(2),1,0,1,\text{Hubbard},\mathfrak{f},\gtrless}\left(t_1,t_2\right) \tag{543}$$

$$\equiv \Sigma_{i\alpha j\alpha}^{(2),1,1,\text{Hubbard},\mathfrak{f},\gtrless}\left(t_1,t_2\right)\left(G\to G^{(0)}\right),$$

as well as

$$\Sigma_{i\alpha j\alpha}^{(2),\{H,0\},\{1,\{H,0\},0\},\text{Hubbard},\mathfrak{b},\delta}\left(t_1,t_2\right) \tag{544}$$

$$= \left(\mathrm{i}\hbar\right)^2\delta_{ij}\sum_{\epsilon}U\left(t_1\right)\Bigg($$

$$\int_{t_0}^{t_1}\mathrm{d}t_3\sum_p G_{i\epsilon p\epsilon}^{(0),>}\left(t_1,t_3\right)G_{p\epsilon i\epsilon}^{(0),<}\left(t_3,t_1\right)\sum_{\zeta}U\left(t_3\right)G_{p\zeta p\zeta}^{(0),<}\left(t_3,t_3\right)$$

$$+ \int_{t_1}^{t_0}\mathrm{d}t_3\sum_p G_{i\epsilon p\epsilon}^{(0),<}\left(t_1,t_3\right)G_{p\epsilon i\epsilon}^{(0),>}\left(t_3,t_1\right)\sum_{\zeta}U\left(t_3\right)G_{p\zeta p\zeta}^{(0),<}\left(t_3,t_3\right)\Bigg)$$

and

$$\Sigma_{i\alpha j\alpha}^{(2),\{H,0\},\{1,\{F,0\},0\},\text{Hubbard},\mathfrak{b},\delta}\left(t_1,t_2\right) \tag{545}$$
$$= \left(\mathrm{i}\hbar\right)^2 \delta_{ij} \sum_\epsilon U\left(t_1\right)\Big($$
$$\int_{t_0}^{t_1} \mathrm{d}t_3 \sum_p G_{i\epsilon p\epsilon}^{(0),>}\left(t_1,t_3\right) U\left(t_3\right) G_{p\epsilon p\epsilon}^{(0),<}\left(t_3,t_3\right) G_{p\epsilon i\epsilon}^{(0),<}\left(t_3,t_1\right)$$
$$+ \int_{t_1}^{t_0} \mathrm{d}t_3 \sum_p G_{i\epsilon p\epsilon}^{(0),<}\left(t_1,t_3\right) U\left(t_3\right) G_{p\epsilon p\epsilon}^{(0),<}\left(t_3,t_3\right) G_{p\epsilon i\epsilon}^{(0),>}\left(t_3,t_1\right)\Big).$$

Further,

$$\Sigma_{i\alpha j\alpha}^{(2),\{F,0\},\{1,\{H,0\},0\},\text{Hubbard},\mathfrak{b},\delta}\left(t_1,t_2\right) \tag{546}$$
$$= \left(\mathrm{i}\hbar\right)^2 \delta_{ij} U\left(z_1\right)\Big($$
$$\int_{t_0}^{t_1} \mathrm{d}t_3 \sum_p G_{i\alpha p\alpha}^{(0),>}\left(t_1,t_3\right) U\left(t_3\right) \sum_\epsilon G_{p\epsilon p\epsilon}^{(0),<}\left(t_3,t_3\right) G_{p\alpha i\alpha}^{(0),<}\left(t_3,t_1\right)$$
$$+ \int_{t_1}^{t_0} \mathrm{d}t_3 \sum_p G_{i\alpha p\alpha}^{(0),<}\left(t_1,t_3\right) U\left(t_3\right) \sum_\epsilon G_{p\epsilon p\epsilon}^{(0),<}\left(t_3,t_3\right) G_{p\alpha i\alpha}^{(0),>}\left(t_3,t_1\right)\Big)$$

and

$$\Sigma_{i\alpha j\alpha}^{(2),\{F,0\},\{1,\{F,0\},0\},\text{Hubbard},\mathfrak{b},\delta}\left(t_1,t_2\right) \tag{547}$$
$$= \left(\mathrm{i}\hbar\right)^2 \delta_{ij} U\left(t_1\right)\Big($$
$$\int_{t_0}^{t_1} \mathrm{d}t_3 \sum_p G_{i\alpha p\alpha}^{(0),>}\left(t_1,t_3\right) U\left(t_3\right) G_{p\alpha p\alpha}^{(0),<}\left(t_3,t_3\right) G_{p\alpha i\alpha}^{(0),<}\left(t_3,t_1\right)$$
$$+ \int_{t_1}^{t_0} \mathrm{d}t_3 \sum_p G_{i\alpha p\alpha}^{(0),<}\left(t_1,t_3\right) U\left(t_3\right) G_{p\alpha p\alpha}^{(0),<}\left(t_3,t_3\right) G_{p\alpha i\alpha}^{(0),>}\left(t_3,t_1\right)\Big)$$

and

$$\Sigma_{i\alpha j\alpha}^{(2),\{H,0\},\{1,\{H,0\},0\},\text{Hubbard},\mathfrak{f},\delta}\left(t_1,t_2\right) \tag{548}$$
$$= \left(\mathrm{i}\hbar\right)^2 \delta_{ij} \sum_{\epsilon\neq\alpha} U\left(t_1\right)\Big($$
$$\int_{t_0}^{t_1} \mathrm{d}t_3 \sum_p G_{i\epsilon p\epsilon}^{(0),>}\left(t_1,t_3\right) G_{p\epsilon i\epsilon}^{(0),<}\left(t_3,t_1\right) \sum_{\zeta\neq\epsilon} U\left(t_3\right) G_{p\zeta p\zeta}^{(0),<}\left(t_3,t_3\right)$$
$$+ \int_{t_1}^{t_0} \mathrm{d}t_3 \sum_p G_{i\epsilon p\epsilon}^{(0),<}\left(t_1,t_3\right) G_{p\epsilon i\epsilon}^{(0),>}\left(t_3,t_1\right) \sum_{\zeta\neq\epsilon} U\left(t_3\right) G_{p\zeta p\zeta}^{(0),<}\left(t_3,t_3\right)\Big).$$

*B.4. Spin-0 bosons/spin-1/2 fermions*

For the specific bosonic and fermionic cases, one has

$$\Sigma_{ij}^{(2),(2),2,0,0,\text{Hubbard},\mathfrak{b},0}(z_1, z_2) \tag{549}$$
$$\equiv \Sigma_{ij}^{(2),2,0,\text{Hubbard},\mathfrak{b},0}(z_1, z_2)\left(G \to G^{(0)}\right)$$

and

$$\Sigma_{ij}^{(2),(2),1,0,1,\text{Hubbard},\mathfrak{b},0}(z_1, z_2) \tag{550}$$
$$\equiv \Sigma_{ij}^{(2),1,1,\text{Hubbard},\mathfrak{b},0}(z_1, z_2)\left(G \to G^{(0)}\right),$$

as well as

$$\Sigma_{ij}^{(2),\{\text{H},0\},\{1,\{\text{H},0\},0\},\text{Hubbard},\mathfrak{b},0}(z_1, z_2) \tag{551}$$
$$= \Sigma_{ij}^{(2),\{\text{H},0\},\{1,\{\text{F},0\},0\},\text{Hubbard},\mathfrak{b},0}(z_1, z_2)$$
$$= \Sigma_{ij}^{(2),\{\text{F},0\},\{1,\{\text{H},0\},0\},\text{Hubbard},\mathfrak{b},0}(z_1, z_2)$$
$$= \Sigma_{ij}^{(2),\{\text{F},0\},\{1,\{\text{F},0\},0\},\text{Hubbard},\mathfrak{b},0}(z_1, z_2)$$
$$= \left(\mathrm{i}\hbar\right)^2 \delta_{\mathcal{C}}(z_1, z_2)\delta_{ij} U(z_1)$$
$$\int_{\mathcal{C}} \mathrm{d}z_3 \sum_p G_{ip}^{(0)}(z_1, z_3) G_{pi}^{(0)}(z_3, z_{1^+}) U(z_3) G_{pp}^{(0)}(z_3, z_{3^+}),$$

for spin-0 bosons, and

$$\Sigma_{i\alpha j\alpha}^{(2),(2),2,0,0,\text{Hubbard},\mathfrak{f},1/2}(z_1, z_2) \tag{552}$$
$$\equiv \Sigma_{i\alpha j\alpha}^{(2),2,0,\text{Hubbard},\mathfrak{f},1/2}(z_1, z_2)\left(G \to G^{(0)}\right)$$

and

$$\Sigma_{i\alpha j\alpha}^{(2),(2),1,0,1,\text{Hubbard},\mathfrak{f},1/2}(z_1, z_2) \tag{553}$$
$$\equiv \Sigma_{i\alpha j\alpha}^{(2),1,1,\text{Hubbard},\mathfrak{f},1/2}(z_1, z_2)\left(G \to G^{(0)}\right),$$

as well as

$$\Sigma_{i\uparrow j\uparrow}^{(2),\{\text{H},0\},\{1,\{\text{H},0\},0\},\text{Hubbard},\mathfrak{f},1/2}(z_1, z_2) \tag{554}$$
$$= \left(\mathrm{i}\hbar\right)^2 \delta_{\mathcal{C}}(z_1, z_2)\delta_{ij} U(z_1)$$
$$\int_{\mathcal{C}} \mathrm{d}z_3 \sum_p G_{i\downarrow p\downarrow}^{(0)}(z_1, z_3) G_{p\downarrow i\downarrow}^{(0)}(z_3, z_{1^+}) U(z_3) G_{p\uparrow p\uparrow}^{(0)}(z_3, z_{3^+})$$

and

$$\Sigma_{i\downarrow j\downarrow}^{(2),\{H,0\},\{1,\{H,0\},0\},\text{Hubbard},\mathfrak{f},1/2}\big(z_1,z_2\big) \tag{555}$$

$$= \big(\mathrm{i}\hbar\big)^2 \delta_{\mathcal{C}}\big(z_1,z_2\big)\delta_{ij}U\big(z_1\big)$$

$$\int_{\mathcal{C}}\mathrm{d}z_3 \sum_p G_{i\uparrow p\uparrow}^{(0)}\big(z_1,z_3\big)G_{p\uparrow i\uparrow}^{(0)}\big(z_3,z_{1^+}\big)U\big(z_3\big)G_{p\downarrow p\downarrow}^{(0)}\big(z_3,z_{3^+}\big),$$

for spin-1/2 fermions. The corresponding components read

$$\Sigma_{ij}^{(2),(2),2,0,0,\text{Hubbard},\mathfrak{b},0,\gtrless}\big(t_1,t_2\big) \tag{556}$$

$$\equiv \Sigma_{ij}^{(2),2,0,\text{Hubbard},\mathfrak{b},0,\gtrless}\big(t_1,t_2\big)\big(G\to G^{(0)}\big),$$

$$\Sigma_{ij}^{(2),(2),1,0,1,\text{Hubbard},\mathfrak{b},0,\gtrless}\big(t_1,t_2\big) \tag{557}$$

$$\equiv \Sigma_{ij}^{(2),1,1,\text{Hubbard},\mathfrak{b},0,\gtrless}\big(t_1,t_2\big)\big(G\to G^{(0)}\big),$$

$$\Sigma_{i\alpha j\alpha}^{(2),(2),2,0,0,\text{Hubbard},\mathfrak{f},1/2,\gtrless}\big(t_1,t_2\big) \tag{558}$$

$$\equiv \Sigma_{i\alpha j\alpha}^{(2),2,0,\text{Hubbard},\mathfrak{f},1/2,\gtrless}\big(t_1,t_2\big)\big(G\to G^{(0)}\big),$$

$$\Sigma_{i\alpha j\alpha}^{(2),(2),1,0,1,\text{Hubbard},\mathfrak{f},1/2,\gtrless}\big(t_1,t_2\big) \tag{559}$$

$$\equiv \Sigma_{i\alpha j\alpha}^{(2),1,1,\text{Hubbard},\mathfrak{f},1/2,\gtrless}\big(t_1,t_2\big)\big(G\to G^{(0)}\big),$$

as well as

$$\Sigma_{ij}^{(2),\{H,0\},\{1,\{H,0\},0\},\text{Hubbard},\mathfrak{b},0,\delta}\big(t_1,t_2\big) \tag{560}$$

$$= \Sigma_{ij}^{(2),\{H,0\},\{1,\{F,0\},0\},\text{Hubbard},\mathfrak{b},0,\delta}\big(t_1,t_2\big)$$

$$= \Sigma_{ij}^{(2),\{F,0\},\{1,\{H,0\},0\},\text{Hubbard},\mathfrak{b},0,\delta}\big(t_1,t_2\big)$$

$$= \Sigma_{ij}^{(2),\{F,0\},\{1,\{F,0\},0\},\text{Hubbard},\mathfrak{b},0,\delta}\big(t_1,t_2\big)$$

$$= \big(\mathrm{i}\hbar\big)^2 \delta_{ij}U\big(t_1\big)\bigg($$

$$\int_{t_0}^{t_1}\mathrm{d}t_3 \sum_p G_{ip}^{(0),>}\big(t_1,t_3\big)G_{pi}^{(0),<}\big(t_3,t_1\big)U\big(t_3\big)G_{pp}^{(0),<}\big(t_3,t_3\big)$$

$$+ \int_{t_1}^{t_0}\mathrm{d}t_3 \sum_p G_{ip}^{(0),<}\big(t_1,t_3\big)G_{pi}^{(0),>}\big(t_3,t_1\big)U\big(t_3\big)G_{pp}^{(0),<}\big(t_3,t_3\big)\bigg),$$

for spin-0 bosons, and

$$\Sigma_{i\uparrow j\uparrow}^{(2),\{H,0\},\{1,\{H,0\},0\},\text{Hubbard,f},1/2,\delta}\left(t_1,t_2\right) \tag{561}$$

$$= \left(i\hbar\right)^2 \delta_{ij} U\left(t_1\right)\Bigg($$

$$\int_{t_0}^{t_1} dt_3 \sum_p G_{i\downarrow p\downarrow}^{(0),>}\left(t_1,t_3\right) G_{p\downarrow i\downarrow}^{(0),<}\left(t_3,t_1\right) U\left(t_3\right) G_{p\uparrow p\uparrow}^{(0),<}\left(t_3,t_3\right)$$

$$+ \int_{t_1}^{t_0} dt_3 \sum_p G_{i\downarrow p\downarrow}^{(0),<}\left(t_1,t_3\right) G_{p\downarrow i\downarrow}^{(0),>}\left(t_3,t_1\right) U\left(t_3\right) G_{p\uparrow p\uparrow}^{(0),<}\left(t_3,t_3\right)\Bigg),$$

as well as

$$\Sigma_{i\downarrow j\downarrow}^{(2),\{H,0\},\{1,\{H,0\},0\},\text{Hubbard,f},1/2,\delta}\left(t_1,t_2\right) \tag{562}$$

$$= \left(i\hbar\right)^2 \delta_{ij} U\left(t_1\right)\Bigg($$

$$\int_{t_0}^{t_1} dt_3 \sum_p G_{i\uparrow p\uparrow}^{(0),>}\left(t_1,t_3\right) G_{p\uparrow i\uparrow}^{(0),<}\left(t_3,t_1\right) U\left(t_3\right) G_{p\downarrow p\downarrow}^{(0),<}\left(t_3,t_3\right)$$

$$+ \int_{t_1}^{t_0} dt_3 \sum_p G_{i\uparrow p\uparrow}^{(0),<}\left(t_1,t_3\right) G_{p\uparrow i\uparrow}^{(0),>}\left(t_3,t_1\right) U\left(t_3\right) G_{p\downarrow p\downarrow}^{(0),<}\left(t_3,t_3\right)\Bigg),$$

for spin-1/2 fermions.